\pdfoutput=1

\documentclass[
PhD, % Change to PhD to change text at bottom of title page
DIC=off,
12pt,
bibliography=totoc, % Include bibliography in contents
listof = totoc, % Include lists of figures and tables in contents
index= totoc, % Include index in contents
onehalfspacing, % onehalfspacing or doublespacing
a4paper %letterpaper
]
{icldt}

\usepackage[square, numbers, sort&compress]{natbib}
\usepackage[%CVSon, % CVS version in footers
	        todos, % show todos (requires todonotes package)
            hyperref % Use hyperref for links
            ]{ihformat}
            
\makeatletter
\newcommand\appendix@numberline[1]{}%\autodot\ } %#1} %\appendixname\  }
\g@addto@macro\appendix{%
  \addtocontents{toc}{
    \let\protect\numberline\protect\appendix@numberline}%
}
\makeatother
\begin{document}

% Title page is input not included to remove extra page.
% % % % % % % % % % % % % % % % % % % % % % % 
% title.tex - Ian Huston
% $Id: title.tex,v 1.1 2009/12/17 17:33:39 ith Exp $
% % % % % % % % % % % % % % % % % % % % % % % 
% Redefine CVSRevision for this section
\renewcommand{\CVSrevision}{\version$Id: title.tex,v 1.1 2009/12/17 17:33:39 ith Exp $}

\setlength{\tabcolsep}{0in}
\newcommand{\isep}{-2 pt}
\newcommand{\lsep}{-0.5cm}
\newcommand{\psep}{-0.6cm}
\renewcommand{\labelitemii}{$\circ$}
\college{Queen Mary, }
\department{Physics and Astronomy}
\supervisor{Karim Malik and David Mulryne}
\title{Non-linear effects in early Universe cosmology}
\author{Pedro Carrilho}
\date{September 2018}% cth added
%Note Actual title format is done in icldt.cls
% 
%
\hypersetup{pdftitle={Non-linear effects in early Universe cosmology},pdfauthor={Pedro Carrilho}}
% 

% Change as appropriate
\declaration{%
I, Pedro Miguel Greg\'{o}rio Carrilho, confirm that the research included within this thesis is my own work or that where it has been carried out in collaboration with, or supported by others, that this is duly acknowledged below and my contribution indicated. Previously published material is also acknowledged below. 
\newline
\newline
I attest that I have exercised reasonable care to ensure that the work is original, and does not to the best of my knowledge break any UK law, infringe any third party's copyright or other Intellectual Property Right, or contain any confidential material. 
\newline
\newline
I accept that the College has the right to use plagiarism detection software to check the electronic version of the thesis.
\newline
\newline
 I confirm that this thesis has not been previously submitted for the award of a degree by this or any other university. The copyright of this thesis rests with the author and no quotation from it or information derived from it may be published without the prior written consent of the author.
\newline
\newline
Details of collaboration and publications: Part of this work is done in collaboration with Karim Malik, Raquel Ribeiro, John Ronayne, David Mulryne and Tommi Tenkanen. I have made a major contribution to the original research presented in this thesis. It is based on the following papers, all of which have been published: \newline
\begin{itemize} 
	\item \underline{Vector and Tensor contributions to the curvature perturbation at second order} \\
	          P. Carrilho and K. A. Malik, \textit{JCAP 1602 (2016) no.02, 021}, \\
	          Arxiv:1507.06922 [astro-ph.CO]
	\item \underline{Quantum quenches during inflation} \\
	          P. Carrilho and R. H. Ribeiro, \textit{Phys.Rev. D95 (2017) no.4, 043516 }, \\
	          ArXiv:1612.00035  [hep-th]
	\item \underline{Isocurvature initial conditions for second order Boltzmann solvers } \\
	          P. Carrilho and K. A. Malik, \textit{JCAP 1808 (2018) no.08, 020}, \\
	          Arxiv:1803.08939 [astro-ph.CO]
	\item \underline{Attractor Behaviour in Multifield Inflation} \\
	          P. Carrilho, D. Mulryne, J. Ronayne and T. Tenkanen, \textit{JCAP 1806 (2018) no.06, 032 }, \\
	          Arxiv:1804.10489 [astro-ph.CO]
\end{itemize}
\vfill
Signature: Pedro Carrilho
\newline
Date: 21/09/2018}

\maketitle

% % % % % % % % % % % % % % % % % % % % % % % % % 
% Abstract
% % % % % % % % % % % % % % % % % % % % % % % % % 
% abstract.tex - Ian Huston
% $Id: abstract.tex,v 1.2 2009/12/17 17:41:41 ith Exp $
% % % % % % % % % % % % % % % % % % % % % % % % % 
% Redefine CVSRevision for this section
\renewcommand{\CVSrevision}{\version$Id: abstract.tex,v 1.2 2009/12/17 17:41:41 ith Exp $}
\chapter*{Abstract}
\label{ch:abstract}
\addcontentsline{toc}{chapter}{Abstract}
\section*{}
\singlespacing

In this thesis, we discuss several instances in which non-linear behaviour affects cosmological evolution in the early Universe. We begin by reviewing the standard cosmological model and the tools used to understand it theoretically and to compute its observational consequences. This includes a detailed exposition of cosmological perturbation theory and the theory of inflation. We then describe the results in this thesis, starting with the non-linear evolution of the curvature perturbation in the presence of vector and tensor fluctuations, in which we identify the version of that variable that is conserved in the most general situation. Next, we use second order perturbation theory to describe the most general initial conditions for the evolution of scalar perturbations at second order in the standard cosmological model. We compute approximate solutions valid in the initial stages of the evolution, which can be used to initialize second order Boltzmann codes, and to compute many observables taking isocurvature modes into account. We then move on to the study of the inflationary Universe. We start by analysing a new way to compute the consequences of a sudden transition in the evolution of a scalar during inflation. We use the formalism of quantum quenches to compute the effect of those transitions on the spectral index of perturbations. Finally, we detail the results of the exploration of a multi-field model of inflation with a non-minimal coupling to gravity. We study popular attractor models in this regime in both the metric and the Palatini formulations of gravity and find all results for both the power spectrum and bispectrum of fluctuations to closely resemble those of the single-field case. In all systems under study we discuss the effects of non-linear dynamics and their importance for the resolution of problems in cosmology.

% % % % % % % % % % % % % % % % % % % % % % % % %

% % % % % % % % % % % % % % % % % % % % % % % % % 
% Acknowledgements
% % % % % % % % % % % % % % % % % % % % % % 
% acknowledgements.tex - Ian Huston
% $Id: acknowledgements.tex,v 1.2 2009/12/17 17:41:41 ith Exp $
% % % % % % % % % % % % % % % % % % % % % % 
% Redefine CVSRevision for this section
\renewcommand{\CVSrevision}{\version$Id: acknowledgements.tex,v 1.2 2009/12/17 17:41:41 ith Exp $}
\chapter*{Acknowledgements}
\label{ch:acknowledgements}
\addcontentsline{toc}{chapter}{Acknowledgements}

This thesis would not have been possible without the many people who supported and guided me during the last 4 years. I would like to start by thanking Karim Malik for his constant help and guidance at every step of the way, for his patience for our lengthy discussions about cosmology and for giving me the opportunity to pursue some of my own ideas. I also want to thank David Mulryne for his support and for always having an open door to answer questions and discuss cosmology with me. I am also indebted to all the other lecturers and researchers in the Astronomy Unit, particularly those in the Cosmology group, Tim Clifton, Chris Clarkson, Alkistis Pourtsidou and Julian Adamek, from whom I learned a lot.

Most of the work presented here would not exist without the fruitful collaborations developed in the last 4 years. I would particularly like to thank Raquel Ribeiro, for having introduced me to a project that I greatly enjoyed completing and for all her support and generosity. I also wanted to thank Tommi Tenkanen for driving our project in the right direction and also for his advice at many stages of this process. John Ronayne was also instrumental for our collaboration and I thank him not only for his very hard work on it, but also for being a great colleague to discuss physics with.

I want to thank all my fellow PhD students, past and present, for greatly enriching my experience at the Astronomy Unit. You make this place great! I want to start by expressing my gratitude to all the former students who welcomed me to our office and gave me valuable advice on PhD life. I would particularly like to thank my friends Sophia Goldberg and Viraj Sanghai for all our conversations about cosmology and everything else. I want to thank my fellow fourth year colleagues and desk mates Charalambos Pittordis and Shailee Imrith for their friendship and for putting up with me throughout the entirety of our PhD journey together, up to the very last moments. I additionally want to thank Jorge Fuentes and Domenico Trotta for their friendship and for many conversations about the woes of PhD life. I also want to express my gratitude to my colleagues Sanson Poon, Eline De Weerd, Louis Coates, Kit Gallagher, Jessie Durk, Rebeca Carrillo, Francesco Lovascio, Paul Hallam, Clark Baker, Jack Skinner, John Strachan, Sandy Zeng, Usman Gillani and Callum Boocock. I am specially indebted to the Giggs bosons and all others who suffered through playing football with me, for helping me keep my sanity at acceptable levels. I also wanted to thank the group of people who played augmented reality games with me, for the same reason.

I want to thank my long-time friends Miguel Batista, Jorge Mota, Pedro Barros and Jo\~{a}o Esteves for their continued friendship, even living in different countries. 

I would not be here without my parents, whose continued support and love I am grateful for. Their trust in my abilities and constant motivation have kept me going for all my life. I also want to thank my extended family for all their support and for making my trips home always enjoyable.

Lastly, I want to show my deep gratitude to Susana, without whom I would not have been half as happy these past years. I want to thank her for all her patience and support, specially in the last few months, which were more than essential for both my sanity and the completion of this thesis. This thesis is also hers.
\vfill

I acknowledge financial support from a Queen Mary Principal's Research Studentship and a Bolsa de Excel\^{e}ncia Acad\'{e}mica of the Funda\c{c}\~{a}o Eug\'{e}nio de Almeida from 2014 to 2017. I was also supported by the Funda\c{c}\~{a}o para a Ci\^{e}ncia e Tecnologia (FCT) grant SFRH/BD/118740/2016 from 2017 to 2018.
% % % % % % % % % % % % % % % % % % % % % % % % % 

% % % % % % % % % % % % % % % % % % % % % % % % % 
% Table of contents and figures
% % % % % % % % % % % % % % % % % % % % % % % % % 
%\setcounter{tocdepth}{0}
\tableofcontents
%\listoffigures %cth commented
%\listoftables %cth commented
% % % % % % % % % % % % % % % % % % % % % % % % % 

% Set one half spacing now that thesis proper is starting.
\onehalfspacing
% % % % % % % % % % % % % % % % % % % % % % % % % 

% Include files for each chapter here using 
% \include{relative-path-to-file}
% 
% Example
% % % % % % % % % % % % % % % % % % % % % % % % % % % % 
% chapter.tex - Ian Huston
% Sample chapter layout
% % % % % % % % % % % % % % % % % % % % % % % % % % % % 
% Redefine CVSRevision for this section. 
% If you don't want to use CVS tags comment out this line
\renewcommand{\CVSrevision}{\version$Id: chapter.tex,v 1.3 2009/12/17 18:16:48 ith Exp $}

% % % % % % % % % % % % % % % % % % % % % % % % % % % % % % % % 
% =========================================================== %
% % % % % % % % % % % % % % % % % % % % % % % % % % % % % % % % 
\chapter{Introduction}
\label{Ch_intro}
% % % % % % % % % % % % % % % % % % % % % % % % % % % % % % % % 
% =========================================================== %
% % % % % % % % % % % % % % % % % % % % % % % % % % % % % % % % 

Throughout the history of humankind, observations of the Universe have led to many explanations for its origin, size and evolution. However, only in the 20th century, has Cosmology emerged as a physical science, and even more recently have there been observations of sufficient quality to accurately describe it as a precision science. 

A paradigm shift occurred with the development of the theory of General Relativity, by Albert Einstein~\cite{Einstein:1915ca,Einstein:1916vd}. This description of space as a dynamical entity changed our view of the Universe in many ways. It was particularly important for cosmology, since it allowed for the development of models of the Universe in which it was no longer static, such as those studied by Friedmann, Lema\^{i}tre, Robertson and Walker that gave rise to the geometry which now carries their names. The first observations of this changing nature of space were obtained by Slipher, Hubble and many others, who observed that the redshift of galaxies increased with their distance to the Earth~\cite{1917PAPhS..56..403S, 1929PNAS...15..168H,1927ASSB...47...49L}. This first suggestion that the Universe was expanding, was further supported by the discovery of the origin of light elements by Alpher, Bethe and Gamow~\cite{1948PhRv...73..803A} and later by the detection of the cosmic microwave background (CMB) by Penzias and Wilson \cite{1965ApJ...142..419P}. The Big Bang theory was thus fully established as the leading description of the evolution of the Universe.

The shift towards precision cosmology began with the measurements of the temperature anisotropies of the CMB by the Cosmic Background explorer (COBE)~\cite{Smoot:1992td}. This measurement, along with those of the first peak in the angular power spectrum of temperature anisotropies by BOOMERanG and MAXIMA, allowed cosmologists to extract precise information from the CMB and estimate the curvature of the Universe for the first time~\cite{deBernardis:2000sbo,2000ApJ...545L...5H}. This was also the first time that the seeds of structure could be inferred on different scales, suggesting that the origin of all structure we see today is primordial. Further observations of the CMB by the satellite experiments WMAP~\cite{Spergel:2003cb,Hinshaw:2012aka} and Planck~\cite{Ade:2013zuv,Ade:2015xua,Akrami:2018vks}, have increased the precision of the angular power spectrum and have independently found evidence for the existence of cold dark matter (CDM). The presence of this mysterious substance had already been suggested much earlier in astrophysical systems by Zwicky~\cite{1933AcHPh...6..110Z,1937ApJ....86..217Z}, Rubin~\cite{Rubin:1970zza,Rubin:1980zd} and many others, but the significance of its detection at early times is still one of the most crucial observations in its support. 

Besides the CMB, many other sources of data have become important in the last few decades. Surveys of large numbers of galaxies and supernovae and those that measure weak lensing have been essential in developing our current picture of the Universe. In particular, supernova surveys have measured the expansion of the Universe to be accelerating for the last 4 billion years~\cite{Perlmutter:1998np,Riess:1998cb}. Moreover, measurements of the baryon acoustic oscillations (BAOs) by the 2-Degree Field survey (2DF)~\cite{Efstathiou:1990xe,Cole:2005sx}, the Sloan Digital Sky Survey (SDSS) \cite{Anderson:2013zyy} as well as the WiggleZ survey~\cite{2011MNRAS.415.2876B} confirmed this accelerated expansion as well as the existence of dark matter. 

The picture that has emerged is often called the ``Standard Model of Cosmology'', ``Concordance Cosmology'' or the $\Lambda$CDM model. This model is extremely successful in describing all the observations that are currently available, but is somewhat less satisfactory from a theoretical point of view, requiring the addition of dark matter and a cosmological constant, $\Lambda$, whose nature is largely unknown. Furthermore, this model also relies on the fact that the early Universe is very close to homogeneous and isotropic, but with small stochastic inhomogeneities.

These properties of the early Universe can be successfully explained in the framework of Cosmic Inflation. It postulates the existence of a stage of accelerated expansion during the first instants of the Universe, which not only homogenizes the Universe but also generates stochastic perturbations via the enhancement of quantum fluctuations. This idea was proposed by Guth, Starobinsky and Linde and developed by many others to explain the horizon and flatness problems of Big Bang cosmology~\cite{Guth:1980zm,Starobinsky:1980te,Linde:1981mu,Sato:1980yn,Albrecht:1982wi, Linde:1983gd}. The generation of scalar fluctuations that could seed structure was later discovered by Sasaki, Mukhanov and others~\cite{Mukhanov:1981xt,Starobinsky:1982ee,Bardeen:1983qw,Hawking:1982cz,Sasaki:1986hm}, but was not part of the original motivation. This prediction and its subsequent confirmation in CMB observations increased the support for inflation as the model for the early Universe, while other models, such as cosmic strings, were ruled out. Current data have been able to accurately pinpoint the amplitude of primordial fluctuations to be $A_s=(2.141\pm 0.052)\times 10^{-9}$ and have also measured their spectral index to be significantly distinct from scale invariant, with a value of $n_s=0.9681\pm 0.0044$.

Regarding inflation, many questions still remain unanswered, such as how many fields actively took part in inflation, as well as how the accelerated expansion ended in the period called reheating and gave rise to the radiation dominated stage that followed it~\cite{Albrecht:1982mp,Guth:1982cv,Kofman:1994rk}. Furthermore, another outstanding question is whether inflation produced primordial gravitational waves, as they are expected to be generated by a mechanism similar to that for scalar fluctuations, but have so far remained undetected~\cite{Akrami:2018vks}. Another question relates to the statistics of the fluctuations, which are currently measured to be Gaussian, within the experimental uncertainty~\cite{Ade:2015ava}, but non-Gaussianities may exist at small levels and their detection would provide insights into the non-linear dynamics of inflation. Even not accounting for these fundamental issues, many models of inflation exist that fit observations~\cite{Martin:2013tda} and one of the main research questions of inflationary cosmology is to find which microscopic model can best describe the early Universe~\cite{Lyth:1998xn}. This would not only allow us to better understand the history of the cosmos, but would also provide evidence about the fundamental laws that rule it, at scales which are unreachable in lab experiments. 

The aim of this thesis is to contribute to the answer of these questions using techniques ranging from quantum field theory in curved spacetime to cosmological perturbation theory, passing through numerical methods and the analysis of different probes of the early Universe. We will start by describing cosmological perturbation theory in chapter \ref{Ch_CPT}, as it is ubiquitous in all of theoretical cosmology and is essential for the understanding of the concordance model. This technique is indeed crucial to solve the differential equations of General Relativity, as a completely non-perturbative formulation is still far from reach by even the most powerful computers available. As the name implies, it relies on an expansion in small quantities --- the size of the primordial fluctuations --- and allows one to linearize the evolution equations and render them solvable. In chapter \ref{Ch_SMC}, we discuss the theory of inflation in detail and describe its phenomenology and how it is constrained by experiment. We also review the state of the Universe after inflation, particularly from the initial stages of radiation domination, until last scattering. We show some of the techniques used to calculate the evolution of fluctuations during that stage, in the final parts of that chapter. We then move on to the original results in this thesis, which we now briefly motivate.

The gauge invariant curvature perturbation on uniform density hypersurfaces, $\zeta$, is a useful variable when computing predictions from inflation \cite{Bardeen:1983qw,Salopek:1990jq,ABMR,BKMR}. It is well known that for the simplest models of inflation, $\zeta$ is constant in time on super-horizon scales~\cite{Wands:2000dp,Lyth:2004gb,Rigopoulos:2003ak,LV,MW2004}. This means that its value measured from the CMB is very easy to relate to its value during inflation. In Chapter \ref{Ch_zeta2}, we study non-linear corrections to this variable. We look particularly at the effects of corrections related to vector and tensor fluctuations and study different definitions of $\zeta$, not all of which lead to the usual conservation of this variable on super-horizon scales. We investigate which conditions need to be obeyed for conservation at the non-linear level and how to define the curvature perturbation that is conserved in the most general case.

%%%%%%%%%%%%%%%%%%%%%%%%%%%%%%%%%%%%%%%%%%%%%%%%%%%%%%%%%%%%%%%%%%%%%%%%%%%%%%%

Conservation of $\zeta$ is relevant for inflationary scenarios with a single scalar field. However, should there be more than one field active during inflation~\cite{Malik:1998gy,Lyth:2001nq,Lyth:2002my,Gordon:2002gv,Wands:2007bd,Mazumdar:2010sa}, not only will the curvature perturbation not be conserved in general, but the energy of the fluctuations in the inflaton fields will also be unevenly distributed between the different species produced during reheating, generating isocurvature modes~\cite{Linde:1984ti,Langlois:1999dw}. Should that happen, it is important to understand how this affects the CMB as well as the later evolution of cosmological fluctuations~\cite{Suto:1984aa,Kodama:1986fg,Kodama:1986ud}. This is done by analysing the system of differential equations describing the system during radiation domination and finding the most general solution whose amplitude grows in time~\cite{Bucher:1999re,Finelli:2008xh, Paoletti:2008ck,Shaw:2009nf,Liu:2017oey}. This is then applied to Boltzmann solvers to calculate predictions for experiment~\cite{Seljak:1996is,Lewis:1999bs,Lesgourgues:2011re,Zumalacarregui:2016pph,Hu:2013twa}. Constraints on isocurvature modes can then be derived from observations of the CMB and large-structure~\cite{Ade:2015lrj,Akrami:2018odb,Tegmark:2003uf,Cole:2005sx,Beltran:2004uv,Sollom:2009vd}. While these probes have not yet detected isocurvature fluctuations, the so-called compensated isocurvature mode may exist, since it evades most of the constraints at the linear level by not producing an overall matter isocurvature mode~\cite{Grin:2011tf,He:2015msa,Heinrich:2016gqe,Valiviita:2017fbx}. Moreover, the possibility that non-adiabatic modes may have non-Gaussianity could provide an alternative way to measure them~\cite{Linde:1996gt,Langlois:2012tm,Hikage:2012tf}. In summary, non-adiabatic modes are still observationally relevant and their detection could open new windows into the physics of the early Universe.

In Chapter \ref{Ch_iso2}, we update this analysis to the non-linear level and study the most general growing solutions at second order in cosmological fluctuations. We calculate approximate solutions for the initial instants of the evolution for each quadratic combination of linear modes. This can then be used as initial conditions in second-order numerical solvers to investigate non-linear effects of these modes and derive new constraints on the early Universe. This is particularly relevant for observables whose predictions require calculations at non-linear orders in perturbation theory. Examples include the intrinsic bispectrum of the CMB \cite{Pitrou:2010sn,Huang:2012ub,Pettinari:2013he,Pettinari:2014vja}, magnetic field generation during the pre-recombination era \cite{Fenu:2010kh,Maeda:2011uq,Nalson:2013jya,Fidler:2015kkt} and vorticity production \cite{Christopherson:2009bt,Christopherson:2010ek,Christopherson:2010dw,Brown:2011dn}.

%%%%%%%%%%%%%%%%%%%%%%%%%%%%%%%%%%%%%%%%%%%%%%%

We then move backwards in cosmic time and study aspects of inflation. As mentioned above, the fundamental nature of inflation is not completely understood and while the standard picture of single field slow-roll inflation is sufficient, it is important to study the effects of alternatives to the simplest case. A particular set of these alternatives are related to transient phenomena occurring during inflation, which typically break slow-roll~\cite{Stewart:2001cd,Choe:2004zg,Dvorkin:2009ne}, and which we  briefly review in Section \ref{slowrollC3} of Chapter~\ref{Ch_SMC}. In Chapter \ref{Ch_quench}, we study one such case by investigating fast phenomena that arise when there is an almost instantaneous change of the couplings of the system---a \textit{quantum quench}. We aim to model generic scenarios in multi-field models of inflation, in which the field trajectory suddenly changes, effectively modifying the parameters of the potential, such as the masses and couplings of the fields. 

More specifically, we study quenches of scalar fields on a de Sitter spacetime using the non-perturbative large-$N$ approximation~\cite{Cooper:1987sa,Cooper:1994hr,Cao:2001hn,Moshe:2003xn}. These methods are very useful for the study of non-linear effects of the largest scales in de Sitter spacetime~\cite{Riotto:2008mv,Serreau:2011fu,Parentani:2012tx,Starobinsky:1994bd,Cooper:1986wv,Cao:2004mn,Gautier:2015pca} and that is one of the reasons why we use them. In applying these methods to the quench, we will therefore be able to calculate the consequences of this fast transition for these infrared effects. The other reason for using these techniques is that they allow us to estimate the effects of these transitions using analytical methods, which can illuminate their interpretation. Quenches are also extensively studied in flat spacetime for many applications~\cite{Calabrese:2006rx,Sotiriadis:2010si,Hung:2012zr} and we also provide a comparison of their effects in that case with those of a curved spacetime. These results are obtained in the static de Sitter spacetime, but our future goal is to calculate observable consequences of these quenches in a more dynamical model of inflation. This will allow for more realistic predictions that can then be compared to experiment.

%%%%%%%%%%%%%%%%%%%%%%%%%%%%%%%%%%%%%%%

A more standard alternative to single-field inflation is the multi-field case, which we review in Section~\ref{multfieldC3} of Chapter~\ref{Ch_SMC}. In Chapter \ref{Ch_palatini}, we investigate particular multi-field models based on cosmological attractors, i.e. models for which the observables reach universal values in some limit of the parameters~\cite{Kallosh:2013maa,Kallosh:2013tua,Galante:2014ifa,Jarv:2016sow,Kallosh:2013daa,Linde:2016uec,Achucarro:2017ing,Christodoulidis:2018qdw,Racioppi:2018zoy}. We study a model whose attractor behaviour is caused by a non-minimal coupling of the scalars to gravity~\cite{Kaiser:2010ps,Kaiser:2015usz,Schutz:2013fua,Kaiser:2013sna,Kaiser:2012ak,White:2012ya,Kaiser:2010yu,Kallosh:2013maa}. This has the further advantage of being well motivated from a fundamental point of view, since quantum corrections naturally generate such couplings in a curved spacetime~\cite{Birrell:1982ix}. In our extension of these models into the multi-field regime, we also analyse the effects of different formulations of gravity, the standard metric one and the Palatini formulation~\cite{Sotiriou:2008rp,Clifton:2011jh,Nojiri:2017ncd}. These two formulations are known to give different predictions in a non-minimally coupled model and have been extensively studied in that case~\cite{Bauer:2008zj,Tamanini:2010uq,Bauer:2010jg,Rasanen:2017ivk,Tenkanen:2017jih,Fu:2017iqg,Racioppi:2017spw,Markkanen:2017tun,Jarv:2017azx}. Reference~\cite{Jarv:2017azx}, in particular, has shown these formulations to be substantially different in single-field attractor models. This further motivates our study, as we can then test whether multi-field effects have different consequences in these two different formulations of gravity. In principle, this could allow for testing which formulation is correct using early Universe cosmology, a test which may not be possible otherwise.
\\

Finally, we discuss the conclusions reached in this thesis and point towards future research directions, in Chapter \ref{Ch_conclusions}.

%Many more experiments are currently being developed, which will not only substantially increase the precision of past measurements, but will also explore observables that are completely new and probe epochs that were so far not within reach. Examples include the Euclid satellite that will measure millions of galaxies and probe their weak lensing signals and the Square Kilometer Array (SKA), which will explore the Universe using the 21 cm signal from neutral hydrogen, allowing for the investigation of much earlier times than other surveys. These new probes, among many others, will put pressure on theory development to better understand our current picture of the Universe through more precise calculations

% % % % % % % % % % % % % % % % % % % % % % % % % % % % 
% chapter.tex - Ian Huston
% Sample chapter layout
% % % % % % % % % % % % % % % % % % % % % % % % % % % % 
% Redefine CVSRevision for this section. 
% If you don't want to use CVS tags comment out this line
\renewcommand{\CVSrevision}{\version$Id: chapter.tex,v 1.3 2009/12/17 18:16:48 ith Exp $}

% % % % % % % % % % % % % % % % % % % % % % % % % % % % % % % % 
% =========================================================== %
% % % % % % % % % % % % % % % % % % % % % % % % % % % % % % % % 
\chapter{Cosmological Perturbation Theory}
\label{Ch_CPT}
% % % % % % % % % % % % % % % % % % % % % % % % % % % % % % % % 
% =========================================================== %
% % % % % % % % % % % % % % % % % % % % % % % % % % % % % % % %

\section{Introduction}

Perturbation theory is one of the most widely used techniques in physics~\cite{Malik:2008yp}. It allows one to drastically simplify calculations and to study problems which would otherwise be impossible to solve. For it to work, however, it requires the existence of a small quantity, relative to which all others may be compared and which, in a first approximation, may be neglected. These quantities may be parameters of the theory being used, as is often done in quantum mechanics or particle physics, or they may be the dynamical variables themselves, as is the case in cosmological perturbation theory. It is common, however, that the smallness of the dynamical variables is related to a small parameter, as is the case for the relationship between the size of cosmological fluctuations and the energy scale of inflation, as will be shown in Chapter \ref{Ch_SMC}. Regardless of its origin, we shall label the size of the quantity by $\epsilon$ in this chapter.

The general procedure used in perturbation theory starts by expanding all relevant variables in powers of $\epsilon$. For a variable $T$, this expansion is
\begin{equation}
\label{pert_intro}
T=\sum_n{\frac{1}{n!}\epsilon^n T^{(n)}}\equiv T^{(0)}+\epsilon\delta T^{(1)}+\frac12\epsilon^2\delta T^{(2)}+\dots,
\end{equation} 
in which we use the conventional factor $1/n!$ inspired by the Taylor expansion and defined the notation $\delta T^{(n)}$ to distinguish between the perturbations of order $n$ and what we will often call the background part of the variable, $T^{(0)}$. For many applications of perturbation theory, this background part may vanish, but in many cases and in this thesis, it will represent a solution of the system of equations under study in a very simple case in which the symmetries of the problem allow for an exact solution.

As we will see in this chapter, one of the great advantages of perturbation theory is that, in many cases, it allows for a system of equations to be linearized, thus simplifying it considerably. After having solved the background equations and finding $T^{(0)}$, the following step is then to solve this linearized system to find the solution $\delta T^{(1)}$, which should depend on the background solution. To find the next order solution, one then merely needs to solve another linear equation for $\delta T^{(2)}$ that is now sourced by terms quadratic in $\delta T^{(1)}$, which are known at this stage. Even if, at every new order, the solution may be more complicated, this procedure can continue up to arbitrary orders to improve the accuracy of the result to the desired level, as well as to study new effects not present at lower order.

This technique is extremely useful, but care must be taken regarding its validity. Perturbation theory is valid when each new contribution $\delta T^{(n+1)}$ is only a small correction to the previous order non-zero variable, $\delta T^{(n)}$, i.e. when $\epsilon\delta T^{(n+1)}\ll\delta T^{(n)}$, for all values of $n$. If the small quantities under study are the dynamical variables themselves, this may not always be verified, as these variables may grow beyond the size which verifies the previous condition. Cosmological perturbation theory can run into this issue, but only when studying the late Universe on relatively small scales. On large scales or in the early Universe, this problem is not known to occur, at least for the most popular models. We will therefore assume perturbation theory to be valid in all its applications in this thesis.

The study of perturbations in cosmology has a rich history, which we now briefly review. The original studies of perturbations in a cosmological setting were done by Lifshitz~\cite{Lifshitz:1945du,Lifshitz:1963ps}, who first calculated the evolution of density perturbations at linear level. Tomita was the first to perform a similar calculation at the non-linear level~\cite{Tomita1,Tomita2,Tomita3}, computing the second-order density evolution. The gauge invariant formalism for cosmological perturbations was developed by Bardeen in Ref.~\cite{Bardeen:1980kt}, who defined the first gauge invariant perturbations, which now carry his name. This built on more general work by Stewart and Walker, who studied perturbations of general spacetimes in Ref.~\cite{Stewart:1974uz}. Kodama and Sasaki generalized the gauge invariant formulation in Ref.~\cite{KodamaSasaki}, deriving the equations for many different cosmological scenarios, including a multi-fluid system obeying the Boltzmann equation. Gauge invariant perturbation theory was thoroughly studied in Refs.~\cite{Bruni:1996im,Abramo:1997hu}, in which many second-order equations were originally derived in a gauge invariant formulation. Many other works have contributed to the development of cosmological perturbation theory, which we do not mention. The interested reader may find more information in the reviews \cite{Mukhanov:1990me, Durrer:1993db, Ma:1995ey, Malik:2008im}.

This chapter will serve to review cosmological perturbation theory, starting with the more general relativistic perturbation theory in Section~\ref{Sec_RPT} and then applying it to the cosmological setting with the Friedmann-Lema\^itre-Robertson-Walker background in Section~\ref{Sec_CPT}. This exposition will serve also to establish the notation used in the remaining chapters of the thesis and to provide the technical background required to understand those chapters.

Many of the calculations shown in this chapter and in the rest of the thesis were performed using the Mathematica package xPand\footnote{\href{http://www.xact.es/xPand/}{http://www.xact.es/xPand/}} \cite{xPand}, which is built into the tensor calculus package xAct\footnote{\href{http://www.xact.es}{http://www.xact.es}} \cite{xAct}.

\section{Relativistic Perturbation Theory}\label{Sec_RPT}

Relativistic perturbation theory is the perturbative technique used in the context of relativistic theories of gravity, such as Einstein's general relativity. It is adequate for these theories because it takes into account the fundamental symmetry of the theory in its formulation, i.e. diffeomorphism invariance, and is therefore the correct perturbative treatment to study tensor fields on Lorentzian manifolds. In what follows, we specialize to the study of general relativity, but much of what is described is also valid in more general settings.

\subsection{General Relativity}

The theory of general relativity (GR) was developed by Albert Einstein in 1915 and describes spacetime as a manifold with its curvature determined by the matter content present in the spacetime. It is based on (pseudo)-Riemaniann geometry and we will briefly review its main points here.

The fundamental dynamical variable in GR is the metric tensor, $g_{\mu\nu}$, which defines infinitesimal distances between points in spacetime
\begin{equation}
ds^2=g_{\mu\nu}dx^\mu dx^\nu\,.
\end{equation}
It is a symmetric, invertible tensor and we use the $(-,+,+,+)$ convention for its signature. The curvature can be calculated from the metric tensor by defining the Levi-Civita connection, $\n_\mu$, compatible with the metric and with components determined by the Christoffel symbols 
\begin{equation}
\Gamma^\alpha_{\mu\nu}=\frac12g^{\alpha\beta}\left(g_{\beta\mu,\nu}+g_{\beta\nu,\mu}-g_{\mu\nu,\beta}\right)\,,\label{connectionC2}
\end{equation}
and obtaining the Riemann tensor associated with it
\begin{equation}
R^\alpha_{\ \beta\mu\nu}=\Gamma^\alpha_{\beta\nu,\mu}-\Gamma^\alpha_{\beta\mu,\nu}+\Gamma^\alpha_{\mu\sigma}\Gamma^\sigma_{\beta\nu}-\Gamma^\alpha_{\nu\sigma}\Gamma^\sigma_{\beta\mu}\,.
\end{equation}
The Ricci tensor, $R_{\mu\nu}$, and the Ricci scalar, $R$, are contractions of the Riemann curvature tensor
\begin{equation}
R_{\mu\nu}=R^\alpha_{\ \mu\alpha\nu}\,,\ \ R=g^{\mu\nu}R_{\mu\nu}\,,
\end{equation}
and their combination defines the Einstein tensor by
\begin{equation}
G_{\mu\nu}=R_{\mu\nu}-\frac12 R g_{\mu\nu}\,.
\end{equation}
Besides being symmetric, this tensor has the important property of being divergence-free, $\nabla_\mu G^{\mu\nu}=0$ due to the Bianchi identities. This is the reason why the Einstein tensor is used in the Einstein field equations
\begin{equation}
G_{\mu\nu}=8\pi G T_{\mu\nu}\,,\label{EFES}
\end{equation}
since the stress-energy tensor of matter, $T_{\mu\nu}$, must also be divergence-free to preserve local conservation of energy and momentum. The constant $G$ is the Newtonian constant of gravity. We choose units for which the speed of light, $c$, is set to unity. 

The field equations can also be derived from the action
\begin{equation}
\label{SEH}
S=\int{\text{d}^4x\sqrt{-g} \left[\frac{R}{16\pi G}+\mathcal{L}_m\right]} \,,
\end{equation}
in which $g$ is the determinant of the metric, used here to define the invariant volume measure and $\mathcal{L}_m$ is the matter Lagrangian. The first term of the action in Eq.~\eqref{SEH} is called the Einstein-Hilbert action and describes the gravitational dynamics. In the standard metric formulation of gravity, the Einstein equations are derived from this action by varying it with respect to the metric tensor\footnote{Other formulations exist, which give rise to the same equations of motion, such as the Palatini formulation, which is studied in Chapter~\ref{Ch_palatini}.}. Both the Einstein-Hilbert action and the field equations are invariant under diffeomorphisms, which is equivalent to saying that the theory is described by tensors, which are independent of the choice of coordinate system or basis, by definition. This symmetry is extremely important and has consequences for the development of a consistent perturbation theory, as we show in the next subsection.

From the action, Eq.~\eqref{SEH}, one can see that the matter Lagrangian, $\mathcal{L}_m$, is related to the stress-energy tensor via
\begin{equation}
\label{TmnfromL}
T_{\mu\nu}=-2 \frac{\delta\mathcal{L}_m}{\delta g^{\mu\nu}}+g_{\mu\nu}\mathcal{L}_m \,.
\end{equation}
In full generality, one can decompose the stress energy tensor into more familiar variables by choosing a set of observers represented by a time-like unit vector field, $u^\mu$. The resulting decomposition is given by (\cite{EMM})
\be
\label{Tmn}
T_{\mu\nu}=\rho u_\mu u_\nu+P h_{\mu\nu}+q_{\mu}u_{\nu}+q_{\nu}u_{\mu}+ \pi_{\mu\nu}\,,
\ee
in which $h_{\mu\nu}=g_{\mu\nu}+u_\mu u_\nu$ is the projection tensor orthogonal to $u^{\mu}$. The remaining variables are the following observer-dependent physical quantities: 
\begin{align}
\text{Energy density}\ &-\ \rho=T_{\mu\nu}u^\mu u^\nu,\\
\text{Pressure}\ &-\ P=T_{\mu\nu}h^{\mu\nu}/3,\\
\text{Energy flux}\ &-\ q^\alpha=-T_{\mu\nu}h^{\mu\alpha}u^{\nu},\\
\text{Anisotropic stress}\ &-\ \pi^{\alpha\beta}=h^{\mu\alpha}h^{\nu\beta}T_{\mu\nu}-P h^{\alpha\beta}.
\end{align}
These quantities obey the constraints $q_\alpha u^\alpha=0$, $\pi_{\alpha\beta} u^\alpha=0$ and $\pi^\mu_\mu=0$, which follow from their definitions. In many cases, the frame is chosen such that the energy flux vanishes, $q^\mu=0$. This is the so-called energy frame, which will be used throughout this thesis. In this frame the matter degrees of freedom described by $q^\mu$ are thus transferred to the observer's 4-velocity, since it is now constrained to follow the flow of the matter to conserve the vanishing energy flux. Any other frame choice is possible and this procedure is always covariant, in the sense that the quantities generated are the same in all systems of coordinates. They are, however, different for different observers, which implies that all quantities are defined with respect to a particular observer, which must also be known in order to make predictions about observables related to those quantities.

The equations of motion for these fluid quantities can be derived from the conservation of the stress-energy tensor, 
\be
\n_\mu T^{\mu\nu}=0\,.\label{cdTmn}
\ee
However, this only gives rise to four equations, one for each value of the free index $\nu$, and there are, in total, ten functional degrees of freedom in the stress-energy tensor. Therefore, to completely describe the evolution of the system, one typically requires more information. 

In many cases it is possible to use a perfect fluid in its energy frame to describe the matter in the system. In that case, the anisotropic stress vanishes and the number of independent variables of the system reduces to five. To further reduce it to four, the same number as the conservation equations, one must still use an equation of state to relate the pressure to the other variables. A barotropic equation of state, $P=P(\rho)$, is often used and can successfully describe many fluids relevant for cosmology. Alternatively, even when the fluid is not perfect, the anisotropic stress can take a form that depends only on the other fluid parameters, such as when it is well represented by shear viscosity.

Alternatively, if a system has a Lagrangian formulation, such as the one in Eq.~\eqref{SEH}, the equations of motion can be derived directly from the Lagrangian and then converted to fluid variables, if necessary. This is the case for field theories, such as electromagnetism and scalar field theories, used to describe inflation. 

Another option, for systems of many particles, is to use the kinetic theory description for each component in terms of their distribution functions, $f$, defined as the number of particles per unit of phase space. The distribution function obeys the Boltzmann equation
\be
\label{BoltzEq}
\frac{\text{d}f}{\text{d}\lambda}=p^\mu\frac{\p f}{\p x^\mu}+\frac{\text{d}p^\mu}{\text{d}\lambda}\frac{\p f}{\p p^\mu}=C[f]\,,
\ee	
in which $\lambda$ is an affine parameter along the trajectories of the particles, $p^\mu$ is the particle's 4-momentum and $C[f]$ is the collision term, representing the interactions between different particles. If no interactions exist, the collision term vanishes and this equation is called the Liouville equation. One can show that the Liouville equation includes the conservation of the stress-energy tensor when integrated with respect to the 4-momentum (\cite{EMM})
\be
\int{\frac{\text{d}^3p}{E}p^\nu\frac{\text{d}f}{\text{d}\lambda}}=0\Leftrightarrow\n_\mu T^{\mu\nu}=0\,,
\ee	
since the relation between the distribution function and the stress-energy tensor is given by
\be
\label{Eq_TMNBoltz}
\int{\frac{\text{d}^3p}{E}p^\mu p^\nu f}=T^{\mu\nu}\,,
\ee
in which $E=-p^\mu u_\mu$ is the energy of the particles as seen by an observer with 4-velocity $u_\mu$ and $\text{d}^3p/E$ is the invariant measure in momentum space. However, the Liouville (or Boltzmann) equation includes more information and can be used to completely describe the system. We shall discuss it further in Section~\ref{Sec_CPT}.

If matter is composed of particles that can be approximated as test particles, its evolution can also be described using the geodesic equation, 
\be
p^\mu\n_\mu p^\nu=0\,,\label{geodesic}
\ee
written here in terms of the 4-momentum vector. This is often useful for computing the evolution of matter if the system under study is composed only of a few particles or if calculations are to be performed numerically. The usefulness of the geodesic equation is far more general, however, than to describe the matter degrees of freedom. It is mostly applied to study the geometry of a spacetime by analysing the trajectories of test particles, such as massless particles that obey
\be
p^\mu p_\mu=0\,,
\ee
or massive ones, with mass $m$,
\be
p^\mu p_\mu=-m^2\,.
\ee

A useful tool in solving problems in General Relativity is the definition of a tetrad basis. This basis consists of a set of four vector fields, $e_{\ul{a}}^\mu$, which span a non-coordinate basis for the vectors. The inverse tetrad, $e_\mu^{\ul{a}}$, is also defined and is a basis for one-forms. Its components are given by
\be
e_{\ul{a}}^\mu e^{\ul{b}}_\mu=\delta_{\ul{a}}^{\ul{b}}\,,\ \ e_{\ul{a}}^\mu e^{\ul{a}}_\nu=\delta_\nu^\mu\,.
\ee
Any tensor can be represented in this basis, with its components given by suitable contractions with the tetrad components. For example, the components of the stress-energy tensor in the tetrad basis are
\be
T^{\ul{a}}_{\ul{b}}=e_\nu^{\ul{a}}e^\mu_{\ul{b}}T^{\nu}_\mu\,.
\ee
The covariant derivative of a tensor in this basis is given by
\be
\n_{\ul{c}}T^{\ul{a}}_{\ul{b}}=\p_{\ul{c}}T^{\ul{a}}_{\ul{b}}+\Omega^{\ \ul{a}}_{\ul{c}\ \ul{d}}T^{\ul{d}}_{\ul{b}}-\Omega^{\ \ul{d}}_{\ul{c}\ \ul{b}}T^{\ul{a}}_{\ul{d}}\,,
\ee
in which we have defined the directed derivative in this basis as $\p_{\ul{c}}\equiv e^\mu_{\ul{c}}\p_\mu$ and we introduced the affine connection components $\Omega^{\ \ul{a}}_{\ul{b}\ \ul{c}}$. The directed derivatives do not commute and their commutator is given by\footnote{The commutator of two operators $A$ and $B$ is defined as $[A,B]=AB-BA$.}
\be
[\p_{\ul{a}},\p_{\ul{b}}]=\lb d^{\ul{c}}_{\ \ul{b} \ul{a}}-d^{\ul{c}}_{\ \ul{a} \ul{b}}\rb\,,
\ee
with the tetrad derivative, $d^{\ul{c}}_{\ \ul{a} \ul{b}}$, given by
\be
d^{\ul{c}}_{\ \ul{a} \ul{b}}\equiv e_\nu^{\ul{c}}e^\mu_{\ul{b}}\frac{\p e^\nu_{\ul{a}}}{\p x^\mu}\,.
\ee
With vanishing torsion, as we are assuming, the affine connection coefficients are given by
\be
\Omega^{\ \ul{a}}_{\ul{b}\ \ul{c}}=\frac12\lb d^{\ul{a}}_{\ \ul{c} \ul{b}}-d^{\ul{a}}_{\ \ul{b} \ul{c}}+d^{\ \ \ul{a}}_{\ul{c} \ul{b}}-d^{\ \ul{a}}_{\ul{c}\ \ul{b}}+d^{\ \ \ul{a}}_{\ul{b} \ul{c}}-d^{\ \ul{a}}_{\ul{b}\ \ul{c}}\rb\,,
\ee
in which some of the indices of the tetrad derivative terms have been raised and lowered with the metric. The relationship between this affine connection and the Levi-Civita connection defined above is given by
\be
\Omega^{\ \ul{a}}_{\ul{b}\ \ul{c}}=d^{\ul{a}}_{\ \ul{c} \ul{b}}+e_\lambda^{\ul{a}}e^\mu_{\ul{c}}e^\nu_{\ul{b}}\Gamma^{\lambda}_{\mu\nu}\,.
\ee

While one can always choose any basis, a particularly useful one is that for which the metric evaluates to the Minkowski metric
\begin{equation}
\label{Eq_tet_mink}
g_{\mu\nu}e_{\ul{a}}^\mu e_{\ul{b}}^\nu=\eta_{\ul{a}\ul{b}}\,.
\end{equation}
This way, one transfers all the information in the metric to the tetrad fields. However, the four vectors $e_{\ul{a}}^\mu$ include more degrees of freedom (16) than the metric (10), which represent 3 Lorentz boosts and 3 rotations. This extra freedom is often fixed by aligning one of the tetrad fields with the 4-velocity of a chosen observer or with some direction that is relevant to the physical system in question. Additionally, one commonly fixes the remaining rotational freedom by stating that the tetrad does not rotate with respect to some set of directions intrinsic to the system. We discuss this further below, when we deal with the cosmological case. 

Tetrads are particularly useful for systems obeying the Boltzmann equation, as the collision term can be directly written in its Minkowski form. We use them below in Section \ref{Sec_CPT} when we write the Boltzmann equation in a cosmological setting.\\

The Einstein equations are non-linear partial differential equations for the metric tensor and for that reason, they are very difficult to solve in general scenarios. The exact solutions that do exist are for fairly simple systems with particular symmetries, such as  Minkowski spacetime, for which there is no curvature anywhere, or for black holes, which have spherical or axial symmetry. The exact solutions that we focus on in this thesis belong to the Friedmann-Lema\^itre-Robertson-Walker (FLRW) family of spacetimes. They are spacetimes with homogeneous and isotropic spatial slices, which makes them suitable for situations in which the cosmological principle is valid. Their line element is given by
\begin{equation}
\label{FLRW}
ds^2=a^2(\tau)\left(-d\tau^2+\frac{dr^2}{1-K r^2} + r^2 \lb d\theta^2+\sin^2\theta d\varphi^2 \rb\right)\,,
\end{equation}
which we have written here in terms of conformal time $\tau$ and in spherical coordinates. The function $a(\tau)$ is the scale factor and must obey evolution equations derived from the Einstein equations; $K$ represents the constant curvature of the homogeneous spatial slices. We will study this solution in detail in Section~\ref{Sec_CPT}. The next subsection will detail the perturbative techniques used to solve the Einstein equations when no exact solution can be found.

\subsection{Perturbing spacetime}

Perturbation theory in a relativistic setting gives rise to interesting issues related to the fact that spacetime itself is perturbed. One must therefore make sure that the formalism is adapted to the geometric nature of the problem and is covariant. We therefore follow Refs.~\cite{Stewart:1974uz,1964rgt..book.....D,Malik:2008yp,Malik:2008im,Malik:2012dr}.

The first step in this procedure is to identify the exact solution of the Einstein field equations that approximates the system under study. For cosmology, this is the FLRW solution, but here we will attempt to be fully general and call that solution the background solution with the background metric $g^{(0)}_{\mu\nu}$. The solution describes the background manifold, $\mathcal{M}_0$. The physical spacetime, represented by the manifold $\M$, is then approximated by the perturbed manifold, which is part of a one-parameter family of manifolds $\M_\epsilon$, with $\epsilon$ being the small parameter defining the perturbative scheme. All of these 4-dimensional manifolds are embedded in a 5-dimensional manifold $\mathcal{N}$. We can then define a diffeomorphism $\phi_\epsilon:\M_0\rightarrow\M_\epsilon$, which identifies points in $\M_0$ to those in $\M_\epsilon$. It is also useful to define a vector field $X$ in the tangent bundle of $\mathcal{N}$, whose integral curves, $\gamma$, intersect each of the manifolds of the family $\M_\epsilon$, thus generating the diffeomorphism $\phi_\epsilon$, by identifying each interception in $\M_\epsilon$ to a point in the background manifold $\M_0$.

Given a tensor field $T$, its Taylor expansion around any point in $\M_0\subset\mathcal{N}$, along the integral curve, $\gamma$, is given by
\be
\label{gen_exp}
T_\phi\equiv\phi^*_\epsilon T_\epsilon=e^{\epsilon \Ld_X}T|_0=T_0+\epsilon(\Ld_X T)|_0+O(\epsilon^2)~,
\ee
in which $T_\epsilon$ is the tensor field $T$ evaluated at the manifold $\M_\eps$, $\Ld_X$ is the Lie derivative along the vector $X$ and $\phi^*$ denotes the pull-back of the diffeomorphism $\phi$, which is used to evaluate the result on $\M_0$. The use of the exponential of the Lie derivative is simply a shorthand for the Taylor expansion, but will be useful below to simplify certain calculations. We also introduce the notation $T_\phi$ to distinguish the pullback of $T$ from the tensor itself. Labeling perturbations as $\delta T$, one can separate the full result order by order as (omitting pull-backs)
\be
T_\phi=T^{(0)}+\delta T^{(1)}+\frac12\delta T^{(2)}+\dots,\label{pert}
\ee
so that $\delta T^{(n)}=\epsilon^n(\Ld_X^n T)|_0$, in which we use a similar notation to Eq.~\eqref{pert_intro}, but have absorbed the $\epsilon$ parameters into the perturbations and have omitted the subscript $\phi$ from the perturbations, for simplicity. Note that all quantities are evaluated in $\M_0$ and will therefore be written in terms of the coordinates of the background manifold.

\subsubsection{Gauge Transformations}

The choice of vector field $X$ in Eq.~\eqref{gen_exp} and corresponding diffeomorphism $\phi$ is not unique, since there is no unique way to identify points in two manifolds. This choice is called the gauge choice and $X$ is called the generator of that gauge. As we will see below, perturbed quantities defined in different gauges will not be equal. This is not surprising, as quantities in one gauge are evaluated at different points from quantities in another gauge. It is useful, therefore, to relate quantities in different gauges and to establish ways to fix the chosen gauge. Two approaches exist for doing just that, called the active and passive approaches. They differ by the choice of manifold on which to focus. The active approach focuses on each point in the perturbed manifold $\M_\eps$ and compares tensors in the corresponding points in $\M_0$ using different gauge generators. The passive approach does the opposite, it begins with points on $\M_0$ and evaluates tensors at different points in the perturbed manifold. They are equivalent and lead to the same formulas for relating perturbations in different gauges and, for that reason, we expose only the active approach here. 

We begin by defining a new gauge generator, $Y$, and its corresponding diffeomorphism, $\psi$. The idea is then to compare the pullbacks of the tensor $T_\epsilon$ via the two diffeomorphism $\phi$ and $\psi$. The composition of the two diffeomorphisms $\phi$ and $\psi$ results in another diffeomorphism $\Phi:\M_0\rightarrow\M_0$ given by $\Phi_\eps=\phi_{-\eps}\circ\psi_\eps$, which now relates the two points in the background manifold, $\M_0$, that correspond to the same point in the perturbed manifold, $\M_\epsilon$. The gauge transformation for the tensor $T_\phi$ is simply the pullback with the composite diffeomorphism, which is given by\footnote{
This relation is easy to demonstrate using the identity $T=(\phi^*_\epsilon)^{-1}\phi^*_\epsilon T$ and applying $\psi^*_\epsilon$, leading to $T_\psi=\psi^*_\epsilon(\phi^*_\epsilon)^{-1}T_\phi$.
}
\be
\label{Def_gauge}
T_\psi=\Phi_\eps^*T_\phi=e^{\eps \Ld_Y}e^{-\eps \Ld_X}T_\phi=\exp\lb\sum_{n=1}^\infty\frac{\eps^n}{n!}\Ld_{\xi^{(n)}}\rb T_\phi~,
\ee
in which the last step is a consequence of the Baker-Campbell-Hausdorff (BCH) formula and the $\xi^{(n)}$ are given by\footnote{The vectors are interpreted here as operators $X=X^\mu\p_\mu$. This implies that a commutator of two vectors $X$ and $Y$ is given, in terms of the Lie derivative, by $[X,Y]=\Ld_X Y$.} 
\be
\xi^{(1)}=Y-X,\ \xi^{(2)}=\lbs X,Y\rbs,\ \xi^{(3)}=\frac12\lbs X+Y,\lbs X,Y\rbs\rbs,\ \text{etc}.
\ee
This is the general gauge transformation rule for any tensorial quantity $T$ and is, therefore, the expression which allows one to relate two different choices of the generating vector, labeled by the gauge transformation vector 
\be
\xi=\sum_{n=1}^\infty\frac{\eps^n}{n!}\xi^{(n)}\,.
\ee
A common way to express these quantities in different gauges is to drop the subscripts $\phi$ and $\psi$ and use instead $\wt{T}=T_\psi$ and $T=T_\phi$. We shall now follow this convention when writing most equations for gauge transformations. Absorbing the perturbation parameter $\eps$ into each $\xi^{(n)}$, one finds the following transformation rules at each order of perturbations, up to second order:
\begin{align}
&\wt{T}^{(0)}=T^{(0)}~,\label{gauge0}\\
&\wt{\delta T}^{(1)}=\delta T^{(1)}+\Ld_{\xi^{(1)}} T^{(0)}~,\label{gauge1}\\
&\wt{\delta T}^{(2)}=\delta T^{(2)}+\lb\Ld_{\xi^{(2)}} +\Ld^2_{\xi^{(1)}} \rb T^{(0)}+2\Ld_{\xi^{(1)}} \delta T^{(1)}~.\label{gauge2}
\end{align}
It is interesting to note that, should a tensorial quantity vanish up to some order $n$, the order $n+1$ quantity is automatically invariant under any gauge transformation. At linear level, this result is called the Stewart-Walker lemma~\cite{Stewart:1974uz} and for a general order, we will label it the generalised Stewart-Walker lemma. As is shown later, most cosmological quantities are not gauge invariant, i.e. they depend on which gauge was chosen to start with. However, once a gauge is fixed, all quantities in that gauge are well defined. 

The process of fixing a gauge is often based on choosing an appropriate number of tensor fields and giving some constraints on their perturbations. As an example, suppose one had a scalar field $\ph$, whose background value $\ph^{(0)}$ is not constant. One can (partially) fix a gauge by deciding that its perturbations vanish, i.e. by forcing $\ph$ to obey the symmetries of the background manifold. In the language described above, one is simply choosing to map the points in $\M_\epsilon$ to points in $\M_0$ for which the value of $\ph$ is the same, which is certainly possible. Making this choice along with similar ones for three other complementary variables (in 4 dimensions), eliminates the freedom in choosing gauge generators. The mapping between the background and perturbed manifolds is completely determined, and thus all perturbations are uniquely defined. By this we also mean that these perturbations defined in the $\delta\ph=0$ gauge are gauge invariant, in the sense demonstrated by the following 1-dimensional example: the gauge transformation required to reach the $\delta\ph=0$ gauge from any other gauge is fixed by the gauge conditions:
\be
\wt{\delta\ph}^{(1)}=0\Rightarrow \xi^{(1)} =-\frac{\delta \ph^{(1)}}{\dot{\ph}^{(0)}}\,,
\ee 
in which we used only the first order transformation and a dot represents the derivative in the only direction available. The perturbations of another scalar quantity, $\Theta$, in the $\delta\ph=0$ gauge are given by
\be
\wt{\delta\Theta}^{(1)}=\delta\Theta^{(1)}-\frac{\dot{\Theta}^{(0)}}{\dot{\ph}^{(0)}}\delta \ph^{(1)}\,.
\ee
One can easily show that the right-hand side of the expression above is gauge invariant. This can be generalised to arbitrary dimensions and so we conclude that variables in a fixed gauge give expressions for gauge-invariant quantities. This is the most common method for generating gauge invariants, but it is also possible to do it simply by finding combinations of variables whose transformations do not include any terms with $\xi$. 

An alternative way to find gauge invariants is to make use of the symmetries of the background. Suppose the background is invariant under translations in a direction $w^\mu$. Then, any derivatives of background quantities in that direction must vanish. Using again the scalar variable $\ph$, this means that $\Upsilon\equiv w^\mu\p_\mu\ph$ vanishes in the background. Then, by the Stewart-Walker lemma, the first-order perturbation $\delta\Upsilon^{(1)}$ is gauge invariant. This so-called covariant formalism~\cite{Ellis:1989jt} is more difficult to implement at higher orders~\cite{Clarkson:2011qk,Clarkson:2011td}, but can be useful for finding gauge-invariant quantities which are unconnected to any specific gauge.

The Einstein field equations are invariant under any gauge transformation, as is any equation relating tensors, since it can always be rewritten as 
\be
G_{\mu\nu}-8\pi G T_{\mu\nu}=0\,,
\ee
and the right-hand-side ($0$) is obviously invariant. This has the consequence that the Einstein field equations can always be written equivalently in any gauge or with any choice of gauge-invariant variables. Another consequence of this symmetry is that all quantities that can possibly be observed must be gauge invariant, because there is no way for the equations to have information about the gauge in which they were used.

This gauge symmetry is different from the diffeomorphism invariance of the fundamental theory. Indeed, as seen through the examples above, the perturbations of a diffeomorphism-invariant quantity are not gauge-invariant. The reason for that is the requirement that the background is split from the perturbation, as we will see in the following example. Let $U\in \M_\epsilon$ be a point in the perturbed manifold. The scalar field $\ph$ at point $U$ can only take one value, $\ph(U)$, which is independent of which coordinate system one chooses to represent the point $U$ in. This is an example of diffeomorphism or coordinate invariance.  Now consider two points, $P$ and $Q$ on the background manifold $\M_0$ that are mapped to $U$ via two different gauges, given respectively by the diffeomorphisms $\phi$ and $\psi$. Assume also that the field $\ph=\ph^{(0)}$ is different in both these points $P$ and $Q$ in the background. Then, it becomes obvious that two perturbations $\delta\ph$ can be defined,
\begin{align}
\delta\ph(P)=(\phi_\epsilon^*\ph)(P)-\ph^{(0)}(P)\,,\\
\wt{\delta\ph}(Q)=(\psi_\epsilon^*\ph)(Q)-\ph^{(0)}(Q)\,,
\end{align}
which represent the perturbation at the same physical point, $U$. The pullbacks of $\ph$ have the same value and are equal to $\ph(U)$, but their Taylor expansions are different, because they are expanded around different points. This is the reason why, even though $\ph$ is diffeomorphism independent, the perturbations at each order are different. This is true, in spite of the sum of all perturbations being exactly the same. %Indeed, the only physical variable is $\ph(U)$, both the background and the perturbation are only mathematical constructs, which are, however, extremely helpful in solving the equations. Gauge invariant quantities are simply those combinations that are independent of the choice of mapping up to the specified order and for that reason are the only quantities that can be calculated unambiguously.

%In many cases, gauge transformations are understood as coordinate transformations. This relation is true in the sense that the choice of labeling a point in $\M_\epsilon$ by different points in $\M_0$ is equivalent to choosing different coordinates for the point in $\M_\epsilon$. %continue

Some points must be made about the gauge transformation of quantities in a tetrad basis. To find how the tetrad itself transforms under a gauge transformation, we use the gauge transformation rule for the metric
\be
g_\psi=\Phi_\eps^* g_\phi\,,
\ee
and substitute in the relation of the metric to the tetrad vectors
\be
\eta_{\ul{a}\ul{b}}\mathbf{e}^{\ul{a}}_\psi\otimes \mathbf{e}^{\ul{b}}_\psi=\eta_{\ul{c}\ul{d}}(\Phi_\eps^*\mathbf{e}^{\ul{c}}_\phi)\otimes (\Phi_\eps^*\mathbf{e}^{\ul{d}}_\phi)\,.
\ee
The general solution is 
\be
\label{gaugetetrad}
\mathbf{e}^{\ul{a}}_\psi=\Lambda_{\ \ul{b}}^{\ul{a}}~\Phi_\eps^*\mathbf{e}^{\ul{b}}_\phi\,,
\ee
in which the matrix $\Lambda$ represents a Poincar\'{e} transformation and therefore obeys
\be
\label{LorMink}
\eta_{\ul{a}\ul{c}}\Lambda_{\ \ul{b}}^{\ul{a}}\Lambda_{\ \ul{d}}^{\ul{c}}=\eta_{\ul{b}\ul{d}}\,.
\ee
It is necessary to apply this transformation to the pull-back of the tetrad because we would like the tetrads in both gauges to be similarly aligned, i.e. the choices made to constrain their extra freedom to rotations and boosts must be the same. Those choices, along with Eq.~\eqref{LorMink} completely restrict the components of the matrix $\Lambda$ and allow one to calculate it from the gauge transformation rule, Eq.~\eqref{gaugetetrad}. We shall do this below for the cosmological case. A direct way to find $\Lambda$ consists of simply inverting Eq.~\eqref{gaugetetrad} to find
\be
\label{invgaugetetrad}
\Lambda_{\ \ul{b}}^{\ul{a}}=\mathbf{e}^{\ul{a}}_\psi\cdot\Phi_\eps^*\mathbf{e}_{\phi\ul{b}}\,.
\ee
This method requires advance knowledge of the transformation properties of the tetrad, but can be useful if one just needs the Lorentz transformation for a different purpose, such as to calculate tensor components more easily, as we now describe.

Components of tensors in a tetrad basis have a slightly different transformation rule, due to the fact that they are not written in the coordinate basis of the background, but in terms of the tetrad pulled-back from the physical manifold $M_\epsilon$. This implies that the gauge transformations for the components of a vector $V$ are
\be
V_\psi^{\ul{a}}=V_\psi\cdot \mathbf{e}^{\ul{a}}_\psi=(\Phi_\eps^*V_\phi)\cdot (\Lambda_{\ \ul{b}}^{\ul{a}}~\Phi_\eps^*\mathbf{e}^{\ul{b}}_\phi)\,,
\ee
and using the rules of the pull-back one finds
\be
V_\psi^{\ul{a}}%=\Lambda_{\ \ul{b}}^{\ul{a}}~\Phi_\eps^*(V_\phi\cdot \mathbf{e}^{\ul{b}}_\phi)
=\Lambda_{\ \ul{b}}^{\ul{a}}~\Phi_\eps^*V_\phi^{\ul{b}}\,,
\ee
and since the components $V^{\ul{b}}$ are scalars, their pull-back is simply given by
\be
\Phi_\eps^*V_\phi^{\ul{b}}=\exp{\lb\xi^\mu\p_\mu\rb}V_\phi^{\ul{b}}\,.
\ee
This implies that to find the gauge transformation of contravariant tensor components, we only have to multiply them by the appropriate number of Lorentz transformation matrices and use the usual gauge transformation rules for scalars. Covariant components transform with the inverse matrix, which we denote by $\Lambda_{\ul{b}}^{\ \ul{a}}$. For example, a tensor field with components $T^{\ul{a}}_{\ul{b}}$ transforms as
\be
\label{gaugeTtetrad}
\wt{T}^{\ul{a}}_{\ul{b}}=\Lambda_{\ \ul{c}}^{\ul{a}}\Lambda_{\ul{b}}^{\ \ul{d}}\exp{\lb\xi^\mu\p_\mu\rb}T^{\ul{c}}_{\ul{d}}\,.
\ee

For tensors which are not fields, such as the momentum of particles $p^\mu$ (or $p^{\ul{a}}$), the perturbations and gauge transformations are not defined in this way. Not being fields, the Taylor expansion is not defined. However, these tensors may still be mapped from the physical manifold, $\M_\epsilon$, to the background $\M_0$ using the usual pull-back operation. This is important for calculating quantities derived from the distribution function of a species, such as the stress-energy tensor described in Eq.~\eqref{Eq_TMNBoltz}, since those quantities involve integrations over the momentum. Therefore, the gauge transformation of the 4-momentum $p^{\ul{a}}$ is its pull-back, which for the components in the tetrad basis is given by
\be
\wt{p}^{\ul{a}}=\Lambda_{\ \ul{c}}^{\ul{a}}\Phi_\eps^*p^{\ul{c}}\,,
\ee
in which the last part just means the pull-back of its components. This transformation is exactly equivalent to that for a general vector, but the last term is not expanded, as that is not possible for vectors defined only at a point. What one may do is write the pull-back of the components as the components evaluated at a different point, i.e. $(\Phi_\eps^*p^{\ul{c}})(Q)=p^{\ul{c}}(P)$, in which, once more, the points $Q$ and $P$ are related by the map $\Phi_\epsilon$.\footnote{In fact, all such momenta are pull-backs of the 4-momentum in the tangent space of $\M_\epsilon$. Thus, all integrals of the distribution function are always integrals over the momentum evaluated in the physical manifold.}
Using the fact that the distribution function, $f$, is a scalar, we can now show that Eq.~\eqref{gaugeTtetrad} is obeyed by the stress-energy tensor given in Eq.~\eqref{Eq_TMNBoltz} when written in the tetrad basis:
\be
\wt{T}^{\ul{a}}_{\ul{b}}=\int{\frac{\text{d}^3\wt{p}}{\wt{E}}\wt{p}^{\ul{a}} \wt{p}_{\ul{b}} \wt{f}}=\Lambda_{\ \ul{c}}^{\ul{a}}\Lambda_{\ul{b}}^{\ \ul{d}}\int{\frac{\text{d}^3p}{E}p^{\ul{c}} p_{\ul{d}} \exp{\lb\xi^\mu\p_\mu\rb}f}\,,
\ee
in which all momenta are evaluated at the same point and thus we omitted their pull-backs and the measure $\text{d}^3p/E$ is invariant under Lorentz transformations and therefore no extra $\Lambda$ terms arise from it.\\

An interesting, but expected, property of gauge transformations is that they form a group, under the composition operation. To show this, we have to check that these transformations satisfy the group criteria: closure, associativity, invertibility and the existence of an identity element. The identity criterion is obviously satisfied, as, in the language of the right-hand-side of Eq.~\eqref{Def_gauge}, we can use $\xi=0$. For invertibility, we must show that there exists a vector $\sigma$ such that
\be
e^{\Ld_\sigma}e^{\Ld_\xi}=1\,.
\ee
One can easily show that this is satisfied for $\sigma=-\xi$, using the fact that, in that case, the operators commute. Associativity is inherited from the associativity of Lie derivatives. The last issue is closure, which simply states that a combination of two gauge transformations is another gauge transformation. In other words, we must show that a vector $\upsilon$ exists, such that
\be
e^{\Ld_\sigma}e^{\Ld_\xi}=e^{\Ld_\upsilon}\,.
\ee
Using the BCH formula on the left-hand-side we can see that this is satisfied if
\be
\upsilon=\xi+\sigma+\frac12 [\sigma,\xi]+\frac1{12}[\sigma-\xi,[\sigma,\xi]]+\dots\,,
\ee
so that, up to second order its components are
\begin{align}
\label{ups1}
\upsilon^{(1)}=\xi^{(1)}+\sigma^{(1)}\,,\\
\label{ups2}
\upsilon^{(2)}=\xi^{(2)}+\sigma^{(2)}+[\sigma^{(1)},\xi^{(1)}]\,.
\end{align}
One can take this further. Should the generators of the gauges related by $\xi$ be $X$ and $Y$ as before, and the generator of the third gauge be $Z$, then the second gauge transformation changes from $Y$ to $Z$ and the composition of the two is a transformation from the gauge labeled by $X$ to that labeled by $Z$. One can show, from Eqs.~\eqref{ups1} and \eqref{ups2} that $\upsilon$ is related to $X$ and $Z$ in the correct way:
\be
\upsilon^{(1)}=Z-X\,,\ \upsilon^{(2)}=[X,Z]\,.
\ee
This shows that gauge transformations form a group. This is, in fact, essential for these transformations to be well defined, since, if they were not a group, no gauge-invariants could exist, and consequently no well defined results could be calculated.

%Smallness of gauge transformations?

Now that we have developed all the necessary formalism for dealing with perturbations of spacetime, we will now apply it to the background solution most commonly used in cosmology --- the Friedmann-Lema\^itre-Robertson-Walker (FLRW) solution --- in the next section. 

\section{Perturbations in FLRW}\label{Sec_CPT}

The FLRW line element, given in Eq.~\eqref{FLRW}, corresponds to a family of solutions with homogeneous and isotropic spatial slices. Furthermore, in the coordinates chosen here, the components of the metric depend only on time, which in much of this thesis is represented by the conformal time coordinate, $\tau$. The conversion to cosmic time, $t$, is given by
\be
t=\int{a \text{d}\,\tau}\,.
\ee
We begin by describing the equations of motion for the scale factor $a(\tau)$ and the matter variables at the background level. We then introduce perturbations to this solution, working only in the flat case, i.e. $K=0$ in Eq.~\eqref{FLRW}. We show the perturbed evolution equations for both metric and matter perturbations as well as their gauge transformations. We conclude with the perturbed Liouville term of the Boltzmann equation and a derivation of the equation for the anisotropic stress.

\subsection{Background}

The Einstein equations give rise to only two independent equations for the scale factor $a(\tau)$, of which only one is dynamical. Before showing them, it is useful to define the Hubble rate, $H$, given by
\be
H=\frac{\dot{a}}{a}\,,
\ee
in which a dot over a quantity represents a derivative with respect to cosmic time $t$. The conformal Hubble rate is similarly given by
\be
\Hh=\frac{a'}{a}=a H\,,
\ee
where a prime denotes the derivative with respect to conformal time. The Friedmann equation is a constraint for the conformal Hubble rate and is given by
\be
\label{Eq_Fried}
\Hh^2=\frac{8\pi G}{3}a^2\rho-K\,.
\ee
The only other independent part of the Einstein field equations can be found from their trace and is given by
\be
\Hh'=-\frac{4\pi G}{3}a^2(\rho+3P)\,.
\ee
To simplify notation, we have used the symbol of the variable to denote its background value, i.e. $\rho=\rho^{(0)}$. We have assumed that the matter is well described by a perfect fluid at the background level and that the frame used to project the stress-energy tensor is the energy frame, as mentioned above. The conservation of the stress-energy tensor gives another dynamical equation which is not independent of the two Einstein equations:
\be
\rho'=-3\Hh(\rho+P)\,.
\ee
Many solutions to these equations have been found in particularly simple cases, such as when a single fluid dominates the energy density and has the simple equation of state
\be
P=w\rho\,,
\ee
with $w$ constant. Solving for $\rho$ one finds
\be
\rho(a)=\rho_0 a^{-3(1+w)}\,,
\ee
with $\rho_0$ an integration constant, often set to the value of $\rho$ today. The particular cases of interest are those with zero curvature, $K=0$ and with specific equations of state for radiation ($w=1/3$), matter ($w=0$) and vacuum energy ($w=-1$). The corresponding solutions for $a(\tau)$ are
\begin{align}
a(\tau)=\sqrt{\frac{8\pi G \rho_0}{3}}\tau\,,&\ w=\frac13\,,\label{Eq_rad_dom}\\
a(\tau)=\frac{2\pi G \rho_0}{3}\tau^2\,,&\ w=0\,,\label{Eq_matt_dom}\\
a(\tau)=-\sqrt{\frac{3}{8\pi G \rho_0}}\frac1\tau\,,&\ w=-1\,,\label{Eq_dS1}
\end{align}
in which we have assumed expanding initial conditions ($a'>0$). As will be made clear below, many more solutions exist, with fluid mixtures or with scalar fields, which cannot always be found analytically. We leave that discussion to the next chapter.

The components of the tetrad basis vectors and one-forms are easy to find for the background solution, since the metric is diagonal. While non-unique, the simplest tetrad that describes the FLRW metric is that for which all basis vectors are aligned with the coordinate directions. It is given by
\be
e^{\ul{a}}_\mu=a \delta^a_\mu\,,\ \ \ e_{\ul{a}}^\mu=\frac1a \delta_a^\mu\,.
\ee

As will become apparent in Chapters~\ref{Ch_SMC} and \ref{Ch_iso2}, we need to solve the Boltzmann equation to describe both photons and neutrinos at different stages of the evolution of the Universe. We will treat both species as being composed of massless particles, even though this is only an approximation for neutrinos. Furthermore, we will describe these species at a stage in which the energy transfer between them and other species is nearly negligible, and certainly so at the background level. For that reason, it is enough to use only the Liouville equation at that level. For massless species, the 4-momentum can be written in the tetrad basis as $p^{\underline{a}}=(p,p n^i)$, in which $p$ is the magnitude of the 3-momentum and the direction vector obeys $n_in^i=1$. The Liouville equation reduces to 
\be
\frac{\p f}{\p \tau}=\Hh p \frac{\p f}{\p p}\,.
\ee
To simplify it, we have used the geodesic equation, Eq.~\eqref{geodesic}, at the background level,
\be
\frac{\dd p}{\dd \tau}=-p\Hh\,,\ \ \frac{\dd n^i}{\dd \tau}=0\,.\label{geo0}
\ee
The Liouville equation has a very general solution --- $f=f(p a)$ --- but for particles in equilibrium, the distribution function is given by the well-known Bose-Einstein distribution
\be
f_{\text{BE}}(p,\tau)\propto\frac{1}{\exp\lb \frac{p}{T(\tau)}\rb-1}\,,\label{BEf}
\ee
for photons, and the Fermi-Dirac distribution
\be
f_{\text{FD}}(p,\tau)\propto\frac{1}{\exp\lb \frac{p}{T(\tau)}\rb+1}\,,
\ee
for neutrinos. In both cases $T(\tau)$ is the temperature, which decays with expansion as $T\propto a^{-1}$, to satisfy the Liouville equation.

It is also useful to define the redshift, $z$, of photons traveling through the expanding Universe. It is clear from the background geodesic equation, Eq.~\eqref{geo0}, that the energy of photons, $E=p$, obeys $E\propto a^{-1}$. Given the proportionality relation between energy and frequency, $E\propto\nu$, one concludes that the frequency of a photon shifts towards the red as the Universe expands. This defines the redshift, $z$, as the relative change in frequency from emission of a photon in the past ($\nu_0$) to its reception on Earth at the current time ($\nu$). This is given by
\be
z\equiv\frac{\nu_0-\nu}{\nu}=a^{-1}-1\,,
\ee
in which $a$ is the scale factor at the time of emission and we are assuming that $a=1$ at the present time. We see therefore that the redshift, $z$, of a distant source of light is a good proxy for the relative size of the Universe. Since in most standard cosmological models $a$ is a monotonic function of time, both the scale factor and $z$ can be used to describe the time of events in the past, as we shall do in this thesis.

\subsection{Scalar-Vector-Tensor decomposition}

Before writing down the equations of motion for the metric perturbations, we first discuss a way to decompose them according to their transformation properties --- the scalar-vector-tensor (SVT) decomposition.

We perform a (3+1) decomposition of spacetime parametrising each spatial hypersurface with conformal time $\tau$ in a similar way as in the Arnowitt-Deser-Misner formalism~\cite{Arnowitt:1962hi}. This implies that vectors and tensors have temporal and spatial components. A generic vector is given by
\be
V^\mu=(V^0,V^i)\,,
\ee
and the component $V^0$ is a 3-scalar on the spatial slices, while $V^i$ is a 3-vector. This procedure generalizes to higher rank tensors. To respect the isotropy of the background spacetime, $V^i$ must be zero at that level, while $V^0$ is always non-zero for non-vanishing vectors.

Beyond this decomposition, it is also useful to split the remaining degrees of freedom further into scalars, vectors and tensors. This allows for the decoupling of the equations for the different components, at first order. The spatial part of $V^\mu$ is then decomposed as 
\be
V^i=V^{,i}+V^i_\text{v}\,,
\ee
with $V^i_\text{v}$ being divergence-free. The scalar $V$ is related to the divergence of $V^i$, while the divergence-free vector is related to its curl, i.e.,
\be
V^i_{\ ,i}=\n^2 V\,,\ \ \epsilon^{mli}\epsilon_{ijk}V^{j,k}_{\ \ ,l}=\n^2V^m_\text{v}\,.
\ee

The metric is decomposed as
\begin{align}
g_{00}=&-a^2\lb 1+2\phi\rb\,,\label{g00CPT}\\
g_{i0}=&a^2 B_i=a^2\lb B_{,i}-S_i\rb\,,\\
g_{ij}=&a^2\lb\delta_{ij}+2C_{ij}\rb\,,\label{gijCPT}
\end{align}
in which $\phi$ is the perturbation to the lapse, $B$ and $S_i$ are,
respectively, the scalar and vector parts of the shift and $C_{ij}$ is
the perturbation to the spatial part of the metric. $C_{ij}$ is further decomposed as
\begin{equation}
\label{gij}
C_{ij}=-\psi\delta_{ij}+E_{,ij}+F_{(i,j)}+h_{ij}\,,
\end{equation}
in which $\psi$ is the curvature perturbation in this metric convention \cite{Mukhanov:1990me,Malik:2008im}\footnote{Other conventions can also be used, as will be discussed in detail in Chapter \ref{Ch_zeta2}.}, $E$ and $F_i$ are, respectively, a scalar and a vector part of the spatial metric and $h_{ij}$ is the tensor potential, representing gravitational waves. Both $F_i$ and $S_i$ are divergence-free,
\be
F^i_{,i}=0\,,\ S^i_{,i}=0\,,
\ee
and $h_{ij}$ is both divergence-free and traceless,
\be
h^i_{j,i}=0\,,\ h^i_{i}=0\,.
\ee

The 4-velocity is decomposed in a similar way to the generic vector shown above, but it is useful to introduce factors of the scale factor, $a$, in the definition of the perturbations. Furthermore, an observer's 4-velocity must obey 
\be
u_\mu u^\mu=-1\,,
\ee
which implies one can find an expression for the $u^0$ component in terms of $u^i$ and the metric. The final result, valid up to second order, is
\begin{align}
&u^{0}=a^{-1}\left(1-\phi+\frac32\phi^2+\frac12v_i v^i+v^i \lb B_{,i}-S_i\rb\right)\,,\\
&u^{i}=a^{-1}v^i=a^{-1}\lb v^{,i}+v^i_\text{v}\rb\,.
\end{align}
We have slightly abused the notation and used $v_i$ to mean $\delta_{ij} v^{j}$. This simplification of notation is used throughout the thesis for most spatial quantities, as will be mentioned again when appropriate. %TODO: Consider removing this or putting it again somwhere else

Regarding the stress-energy tensor, one could decompose it in a similar way to the metric, but, as we have already introduced a decomposition based on the 4-velocity in Eq.~\eqref{Tmn}, we choose the standard option of decomposing the fluid variables $\rho$, $P$ and $\pi_{\mu\nu}$ instead. For the 4-scalars, we simply write the perturbations by explicitly separating them from the background:
\begin{align}
\rho=\rho^{(0)}+\delta\rho\,,\\
P=P^{(0)}+\delta P\,,
\end{align}
in which we have written the superscript on the background quantities for clarity, but will omit them in the rest of the text. The decomposition of the anisotropic stress tensor is complicated by its constraints, $\pi_{\alpha\beta} u^\alpha=0$ and $\pi^\mu_\mu=0$. For this reason, its components also depend on the velocity fluctuations, as well as the metric. Up to second order in fluctuations, they are given by
\begin{align}
\pi_{00}=0,\ \ \ \pi_{i0}=-2\pi_{ij} v^j\,,\nonumber\\
\pi_{ij}=a^2\lbs\Pi_{ij}+\Pi_{(i,j)}+\Pi_{,ij}-\frac{1}{3}\delta_{ij}\n^2\Pi\rbs+\frac43\delta_{ij}\pi_{kl}C^{kl}\,,
\end{align}
in which we have defined the scalar, $\Pi$, vector, $\Pi_i$, and tensor, $\Pi_{ij}$, parts of the anisotropic stress.

The tetrad basis vectors, $e_{\ul{a}}^\mu$, defined in the previous section can now be calculated for the perturbed FLRW spacetime. Before that, we must fix the superfluous degrees of freedom that the basis vectors contain. We align $e_{\ul{0}}^\mu$ with the vector parallel to the time direction, implying that $e_{\ul{0}}^i=0$.\footnote{This is the choice of Refs.~\cite{Pettinari:2013he, Beneke:2010eg, Senatore:2008vi}, which we follow. Alternatively, some authors \cite{Pitrou:2008hy,Naruko:2013aaa} choose the inverse tetrad to obey $e^{\ul{0}}_i=0$ from the requirement that it is orthogonal to spatial hypersurfaces. %(TODO: CHECK what Challinor and Tim do and cite them.)
} To fix the remaining degrees of freedom, we first note, that at the background level, it is possible to define a coordinate induced tetrad, in which each basis vector is aligned with a coordinate direction, i.e. $e_{\ul{a}}\propto\delta^\mu_{\ul{a}}\p_\mu$, with the kronecker delta enforcing a correspondence between the spacetime indices and the tetrad indices. This complete alignment is no longer possible for the perturbed tetrad, but one can still choose its basis vectors to have the same orientation as in the background and the same index correspondence with the coordinate indices. One can then impose the weaker alignment condition $e_{\ul{i}}^j=e_{\ul{j}}^i$. This fixes the rotation of each tetrad basis vector with respect to the background tetrad in a ``democratic'' way, as opposed to aligning a specific direction. We now use Eq.~\eqref{Eq_tet_mink} to compute the remaining components of the tetrad basis. Up to second order, they are given by
\begin{align}
e_{\ul{0}}^0=\frac1a\lb1-\phi+\frac32\phi^2\rb\,,&\ \ e^{\ul{0}}_0=a\lb1+\phi-\frac12\phi^2\rb\,,\nonumber\\
e_{\ul{0}}^i=0\,,&\ \ e^{\ul{0}}_i=-a B_i (1-\phi)\,,\\
e_{\ul{i}}^0=\frac1a B_j\lb(1-2\phi)\delta_i^j-C^j_i\rb\,,&\ \ e^{\ul{i}}_0=0\,,\nonumber\\
e_{\ul{i}}^j=\frac1a\lb\delta_i^j-C_i^j+\frac32 C_{ik} C^{jk}-\frac12 B_i B^j\rb\,,&\ \ e^{\ul{i}}_j=a\lb\delta^i_j+C^i_j-\frac12 C^{ik} C_{jk}+\frac12 B^i B_j\rb\,.\nonumber
\end{align}

The components of the 4-momentum for massless particles in the tetrad basis are split into a 3-momentum magnitude and a direction via
\be
p^{\ul{a}}=(p,p n^i)\,,
\ee
in which the direction vector $n^i$ is normalized, i.e., $n_in^i=1$. When calculating integrals of the distribution function, we use this split of the momentum to separate the angular integrations from those in the momentum magnitude. One such integral defines the brightness fluctuation $\Delta$~\cite{Pettinari:2013he,Lewis:2002nc},
\begin{equation}
\label{Deltadef}
\Delta(\tau,\vec x,\vec n)=\frac{\int{\text{ d}p\, p^3 (f(\tau,\vec x,p,\vec n)- f^{(0)}(\tau,p))}}{\int{\text{d}p\, p^3 f^{(0)}(\tau,p)}}\,,
\end{equation}
in which we have subtracted the background value of the distribution function, $f^{(0)}$. We will see in Chapter~\ref{Ch_SMC}, that the brightness fluctuations are related to the temperature fluctuations. 

Integrations in the angular directions are often taken into account by decomposing the distribution function or the brightness fluctuation into spherical harmonics. In this thesis and following Ref.~\cite{Carrilho:2018mqy}, we introduce a different projection in terms of tensors, which is similar to that of Kodama and Sasaki~\cite{KodamaSasaki}. To be concrete, we integrate the brightness fluctuation, $\Delta$, with different numbers of direction vectors, $n^i$, using the projectors given by
\begin{equation}
\label{projn}
\mathcal{P}_N^{i_1\cdots i_N}=\int{\frac{\text{d}\Omega}{4\pi} n^{i_1}\cdots n^{i_N}}\,.
\end{equation}
The resulting integrations generate a set of 3-tensors which we call \emph{brightness tensors}, shown here up to rank 3,
\begin{align}
\Delta_0=\mathcal{P}_0 [\Delta]=&\int{\frac{\text{d}\Omega}{4\pi} \Delta(\tau,\vec x,\vec n) }\,,\label{Delta0CPT}\\
\Delta^{i}=\mathcal{P}_1^i [\Delta]=&\int{\frac{\text{d}\Omega}{4\pi} n^i \Delta(\tau,\vec x,\vec n) }\,,\\
\Delta^{ij}=\mathcal{P}_2^{ij} [\Delta]=&\int{\frac{\text{d}\Omega}{4\pi} n^i n^j \Delta(\tau,\vec x,\vec n) }\,,\\
\Delta^{ijk}=\mathcal{P}_3^{ijk} [\Delta]=&\int{\frac{\text{d}\Omega}{4\pi} n^i n^j n^k \Delta(\tau,\vec x,\vec n) }\,.
\end{align}
Note that these tensors appear to describe more degrees of freedom than the usual multipoles. For example, $\Delta^{ij}$ is a symmetric 3-tensor, thus having in total 6 components, while the usual $\ell=2$ multipoles only represent $2\ell+1=5$ degrees of freedom. This discrepancy can be understood by noticing that the brightness tensors are related amongst each other. The extra d.o.f. in this example is actually in the trace of $\Delta^{ij}$, which is obviously equal to $\Delta_0$, since $n_in^i=1$. Therefore, it is the traceless part of each of these tensors that includes the same information as the usual multipoles. For that reason, it is useful to also define traceless brightness tensors:
\begin{align}
\Delta_T^{ij}=&\Delta^{ij}-\frac13\delta^{ij}\Delta_0\,,\\
\Delta_T^{ijk}=&\Delta^{ijk}-\frac{3}{5}\delta^{(ij}\Delta^{k)}\,,\\
\Delta_T^{ijkl}=&\Delta^{ijkl}-\frac{6}{7}\delta^{(ij}\Delta_T^{kl)}-\frac{1}{5}\delta^{(ij}\delta^{kl)}\Delta_0\,,\\
\Delta_T^{ijklm}=&\Delta^{ijklm}-\frac{10}{9}\delta^{(ij}\Delta_T^{klm)}-\frac{3}{7}\delta^{(ij}\delta^{kl}\Delta^{m)}\label{Delta3CPT}\,.
\end{align}

These quantities can be related to the components of the stress-energy tensor using Eq.~\eqref{Eq_TMNBoltz} and the conversion from the coordinate to the tetrad basis. The brightness tensors up to rank 2 can be written as
\begin{align}
\label{d0CPT}
&\Delta_0=-\frac{\delta T^{0}_{\ 0}-B_i T^{i}_{\ 0}}{\rho}\,,\\
\label{d1CPT}
&\Delta^{i}=-\frac{T^{j}_{\ 0}}{\rho}(\delta^i_j(1-\phi)+C^i_j)\,,\\
\label{d2CPT}
&\Delta^{\ \  i}_{T\, j}=\frac{1}{\rho}\left(T^{k}_{\ l}\left(\delta_{j}^l\delta_{k}^i-\frac13 \delta_{j}^i\delta_{k}^l+\delta_{j}^l C^{i}_{k}-\delta_{k}^i C^{l}_{j}\right)+T^k_{\ 0}\lb\delta^i_k B_j-\frac13 \delta^i_j B_k\rb\right)\,,
\end{align}
in which $\rho$ is the background energy density of the appropriate massless species. The variables can also be converted into the usual fluid variables in the desired frame. The quantity $\Delta_0$ is related to the density perturbation, while $\Delta^i$ is related to the fluid velocity and $\Delta^{\ \  i}_{T\, j}$ can be used to represent the anisotropic stress. At first order in fluctuations, they are proportional, but, at higher orders, frame effects can introduce further complications into their explicit relations.

The brightness tensors are also decomposed into their scalar, vector and tensor parts. For the rank 1 and 2 tensors, we use the same decomposition as for the velocity and anisotropic stress, respectively:
\begin{equation}
\Delta^{i}=\Delta_1^{,i}+\Delta_{1v}^i\,,\label{D1isvt}
\end{equation}
\begin{equation}
\Delta_T^{ij}=\Delta_{2}^{,ij}-\frac{1}{3}\delta^{ij}\n^2\Delta_2+\Delta_{2v}^{(i,j)}+\Delta_{2t}^{ij}\label{D2isvt}\,.
\end{equation}
The labels $v$ and $t$ denote the transverse vector and transverse and traceless tensor parts. As for the rank 3 tensor, there are, in total, 7 degrees of freedom split into one scalar, one vector, one rank 2 tensor and one rank 3 tensor. They are defined via
\begin{equation}
\Delta_T^{ijk}=\Delta_3^{,ijk}-\frac35\delta^{(ij}\n^2\Delta_3^{,k)}+\Delta_{3v}^{(i,jk)}-\frac15\delta^{(ij}\n^2\Delta_{3v}^{k)}+\Delta_{3t}^{(ij,k)}+\Delta_{3T}^{ijk}\,.\label{D3isvt}
\end{equation}
Higher rank tensors could be similarly decomposed, but, for brevity, we do not do so here. It should be noted that the rank 3 transverse traceless tensor, $\Delta_{3T}^{ijk}$, is often ignored, because its evolution equations are not sourced at the linear level, since no fundamental field exists with spin 3 and no linear mechanism exists for exciting this mode. The same applies for higher rank tensors. At the non-linear level, however, all those tensors would be sourced by combinations of lower order tensors and would thus be generated.

\subsection{Gauge Transformations}

We use the SVT decomposition also for the gauge transformation vector, $\xi^\mu$, resulting in
\be
\xi^\mu=(\xi^0,\xi^i)=(\alpha,\beta^{,i}+\gamma^i)\,.
\ee

The gauge transformations are derived from Eqs.~\eqref{gauge0}-\eqref{gauge2}, up to second order in perturbations. For 4-scalars, such as the energy density, one finds
\be
\wt{\delta\rho}=\delta\rho+\alpha\rho'+ \frac12\alpha\lb\rho''\alpha+\rho'\alpha'+2\delta\rho'\rb+\frac12\lb2\delta\rho+\rho'\alpha\rb_{,k}\lb\beta^{,k}+\gamma^k\rb\,,
\ee
and for any other scalar, such as the pressure $P$, one only has to substitute all the $\rho$ and $\delta\rho$ for the desired background and perturbations of the 4-scalar in question.

For the velocity fluctuations $v$ and $v^i_\text{v}$, we use the rules to transform the 4-velocity $u^\mu$ and split the result in the same way. The resulting transformations are
\begin{align}
\wt{v}=&v-\beta'+\frac12 \n^{-2}\Xv_{\ ,k}^{\,k}\,,\\
\wt{v_\text{v}}^i=&v_\text{v}^i-\gamma^{i~\prime}+\frac12 \Xv^{\,i}+\frac12 \n^{-2}\Xv_{\ ,ki}^{\,k}\,,
\end{align}
with the second-order parts written in terms of $\Xv^{\,i}$, which is given by
\begin{align}
\label{defXvi}
\Xv^i \equiv&\ \xi^{i\prime}\left(2\phi+\alpha'+2\Hh\alpha\right)-\alpha\xi^{i\prime\prime}
\nonumber\\
&
-\xi^k\xi^{i\prime}_{,k}+\xi^{k\prime}\xi^i_{,k}+2\alpha\left(v^{i\prime}-\Hh v^{i}\right)+2v^i_{,k}\xi^k-2v^k\xi^i_{,k}\,.
\end{align}

The transformations for the metric quantities are obtained from the gauge transformation rules applied to the metric tensor. From the time-time component one finds the following gauge transformation for the perturbation to the lapse
\begin{align}
\label{transphi2}
\wt{\phi} &= \phi+\Hh\alpha+\alpha'+\frac12\alpha\left[\alpha''+5\Hh{\alpha}' +\left(\Hh'+2\Hh^2\right)\alpha +4\Hh\phi+2\phi'\right]\nonumber \\
&+{\alpha}'\left({\alpha}'+2\phi\right)+\frac12\xi^{k}\left({\alpha}'+\Hh{\alpha}+2\phi\right)_{,k}\nonumber \\
&+\frac12\xi^{k\prime}\left[\alpha_{,k}-2B_{k}-\delta_{kl}\xi^{l\prime}\right]\,.
\end{align}
We have used here the slightly longer notation $\delta_{kl}\xi^{l}$, whereas this is often written as $\xi_{k}$, in the literature~\cite{Malik:2008im}. We chose the form used above to avoid confusion with the spatial component of the covariant vector, which can also be defined as $\xi_\nu=g_{\mu\nu}\xi^\mu$ and would give a different result. This choice is made throughout the thesis and will appear in most calculations involving second-order gauge transformations.

Using the space-time component, one finds the transformations for $B$ and $S^i$, which are given by
\begin{align}
\label{B2itrans}
\widetilde {B}&=B+\beta'-\alpha+\XB^i_{\ ,i}\,,\\
\widetilde {S^i}&=S^i-\gamma^{i~\prime}-\XB^i+\n^{-2}\XB^{k,i}_{\ ,k}\,,
\end{align}
with the non-linear terms given by
\begin{align}
\label{defXBi}
\XB^i
\equiv&
\Big[
\left(2\Hh B^{i}+B^{i\prime}\right)\alpha
+B^i_{,k}\xi^k-2\phi\alpha^{,i}+B_{k}\xi^{k,i}
+B^{i}\alpha'+2 C^i_{k}\xi^{k\prime}
 \Big]\nonumber\\
&+2\Hh\alpha\left(\xi^{i\prime}-\alpha^{,i}\right)
+\frac12\left[\alpha_1'\left(\xi^{i\prime}-3\alpha^{,i}\right)\right.
+\alpha\left(\xi^{i\prime\prime}-\alpha^{,i\prime}\right)\nonumber\\
&+\xi^{k\prime}\left(\xi^i_{,k}+2\delta_{kl}\xi^{l,i}\right)
+\xi^k\left(\xi^{i\prime}_{,k}-\alpha^{,i}_{,k}\right)
\left.-\alpha_{,k}\xi^{k,i}\right]\,.
\end{align}

Finally, the transformations of the components of the spatial metric are given by
\begin{align}
\label{psi2g}
\wt\psi &= \psi-\Hh\alpha-\frac14\X^i_{\ i}+\frac14\n^{-2}\X^{ij}_{\ \ ,ij}\,,\\
\wt E &= E+\beta+\frac34\n^{-2}\n^{-2}\X^{ij}_{\ \ ,ij}-\frac14\n^{-2}\X^{i}_{\ i}\,,\\
\wt F_i &= F_i+\gamma_i+\n^{-2}\X_{ij}^{\ \ ,j}-\n^{-2}\n^{-2}\X^{jk}_{\ ,jki}\,,\\
\wt h_{ij} &= h_{ij}+\frac12\X_{ij}+\frac14\lb\n^{-2}\X_{kl}^{\ \ ,kl}-\X_{k}^{\ k}\rb\delta_{ij}\\
&+\frac14\n^{-2}\n^{-2}\X^{kl}_{\ ,klij}+\frac14\n^{-2}\X^{k}_{\ k,ij}-\frac12\n^{-2}\lb\X_{ik\ \ ,j}^{\ ,k}+\X_{jk\ \ ,i}^{\ ,k}\rb\,,\nonumber
\end{align}
with $\X_{ij}$ given by
\begin{align}
\label{Xijdef}
\X^{ij}\equiv &\ \Big[\left(\Hh^2+\frac{a''}{a}\right)\alpha^2+\Hh\left(\alpha\alpha'+\alpha_{,k}\xi^{~k}\right)\Big] \delta^{ij}\\
&
+2\Big[\alpha\left(C^{ij\prime}+2\Hh C^{ij}\right)+C^{ij}_{,k}\xi^{~k}+C^{i}_k\xi^{k,j}+C_{k}^{j}\xi^{k,i}\Big]+\left(B^{i}\alpha^{,j}+B^{j}\alpha^{,i}\right)
\nonumber\\
&
+2\Hh\alpha\left( \xi^{i,j}+\xi^{j,i}\right)-\alpha^{,i}\alpha^{,j}+\delta_{kl}\xi^{k,i}\xi^{l,j}
\nonumber\\
&+\frac12\left[\alpha\left( \xi^{i,j\prime}+\xi^{j,i\prime} \right)+\left(\xi^{i,j}_{,k}+\xi^{j,i}_{,k}\right)\xi^{k}+\xi^{i}_{,k}\xi^{k,j}+\xi^{j}_{,k}\xi^{k,i}+\xi^{i\prime}\alpha^{,j}+\xi^{j\prime}\alpha^{,i}\right]\nonumber
\,.
\end{align}

We can see that all metric perturbations are gauge dependent already at first order, except for the tensor perturbation, $h_{ij}$. This can be explained in terms of the Stewart-Walker lemma by computing the expansion of the Weyl tensor, $C^\alpha_{\ \beta\mu\nu}$, the traceless part of the Riemann curvature tensor. This tensor vanishes at the background level, and must therefore be invariant at first order. This can be used to find many other gauge-invariants involving scalar and vector potentials by splitting the Weyl tensor into those parts. Performing the tensor projection, for example, of $C^0_{\ ij0}$, one can independently conclude that $h_{ij}$ 
is invariant.

The anisotropic stress tensor is also gauge-invariant at first order, but this is no longer true at second order. Its gauge transformations are given by
\begin{align}
\wt\Pi &=\Pi+\frac32\n^{-2}\n^{-2}\Xpi^{kl}_{,kl}\,,\\
\wt\Pi_i &=\Pi_i+2\n^{-2}\Xpi^{k}_{i,k}-2\n^{-2}\n^{-2}\Xpi^{kl}_{,ikl}\,,\\
\wt\Pi_{ij} &=\Pi_{ij}+\Xpi_{ij}+\frac12\delta_{ij}\n^{-2}\Xpi^{kl}_{,kl}-2\n^{-2}\Xpi^{k}_{(i,j)k}+\frac12\n^{-2}\n^{-2}\Xpi^{kl}_{,ijkl}\,,
\end{align}
with $\Xpi$ given by
\begin{align}
\label{XPi}% I have to define \xi_i and other vectors and tensors which are defined in a different way. This is so that I don't confuse them with the spatial part of the corresponding 4-vector.
\Xpi_{ij}\equiv &\ \frac{1}{a^2}\lb\alpha \pi_{ij}'-\frac23\pi_{kl}\xi^{k,l}\delta_{ij}+2\pi_{k(i}\xi^{k}_{,j)}+\xi^k\pi_{ij,k}\rb
\,.
\end{align}

The transformation properties of the metric potentials could also have been found by studying the gauge transformations of the tetrad variables, given by Eq.~\eqref{gaugetetrad}. In order to do that, we must find the components of the Lorentz transformation matrix $\Lambda$. Using the constraints defining our tetrad ($e_{\ul{0}}^i=0$ and $e_{\ul{i}}^j=e_{\ul{j}}^i$) and Eq.~\eqref{LorMink}, we find the components of $\Lambda$ and its inverse to be, up to first order
\begin{align}
\Lambda_{\ul{0}}^{\ \ul{0}}=1\,,&\ \ \Lambda_{\ \ul{0}}^{\ul{0}}=1\,,\\
\Lambda_{\ul{0}}^{\ \ul{i}}=\xi^{i\prime}\,,&\ \ \Lambda_{\ \ul{0}}^{\ul{i}}=-\xi^{i\prime}\,,\\
\Lambda_{\ul{i}}^{\ \ul{0}}=\delta_{ij}\xi^{j\prime}\,,&\ \ \Lambda_{\ \ul{i}}^{\ul{0}}=-\delta_{ij}\xi^{j\prime}\,,\\
\Lambda_{\ul{i}}^{\ \ul{j}}=\delta_i^j+\frac12(\xi^j_{,i}-\delta_{ik}\xi^{k,j})\,,&\ \ \Lambda_{\ \ul{i}}^{\ul{j}}=\delta_i^j+\frac12(\delta_{ik}\xi^{k,j}-\xi^j_{,i})\,.
\end{align}
These can be used to calculate the gauge transformations for the brightness tensors defined above. This is only necessary for brightness tensors of rank 3 and above, since for the lower rank tensors, these transformations can be calculated using those for the stress-energy tensor. The transformation for the rank 3 brightness tensor is given by
\be
\wt{\Delta}^{ijk}=\Delta^{ijk}+\xi^\mu \Delta^{ijk}_{\ \ \ ,\mu}+3\delta^{(i}_l\delta^j_r\delta\Lambda^{\ul{k})}_{\ \ul{s}}\Delta^{lrs}-3\delta^{(i}_l\delta^j_r\xi^{k)\prime}\Delta^{lr}+\xi^{l\prime}\Delta^{ijk}_{\ \ \ l}\,,
\ee
in which $\delta\Lambda^{\ul{k}}_{\ \ul{s}}$ is the perturbed part of $\Lambda^{\ul{k}}_{\ \ul{s}}$.

\subsubsection{Notable gauges and invariants}

Many gauges have become popular in the literature and this thesis makes use of several different ones. We now describe their definitions and compute some of the gauge-invariant quantities that arise from them.

A gauge that is prolific in inflationary theory is the uniform density gauge. It is often defined only with one condition --- $\delta\rho=0$ --- and, as we will see in Chapter~\ref{Ch_zeta2}, many different sets of gauge conditions can be used to fix the remaining gauge freedom. Most of Chapter~\ref{Ch_zeta2} is dedicated to the gauge-invariant curvature perturbation on uniform density hypersurfaces, $\zeta$. As will be made clear in that chapter, many versions of this variable can be defined, especially at second order. We show here only the first-order version, which agrees with our definition of the spatial metric:
\be
\label{zeta1}
\zeta\equiv-\psi-\Hh\frac{\delta\rho}{\rho'}\,.
\ee
This variable is used because it has interesting conservation properties on large scales, as shall be made clear below. For that reason, it is in terms of this variable that many inflationary observables are calculated and we also use it in our discussions of the theory of inflation throughout this thesis.

A similarly useful gauge is the comoving gauge. It is defined by 
\begin{align}
v=B=v_\text{v}^i=0\,.
\end{align}
In single-field inflation, this gauge is equivalent to setting the scalar field perturbations, $\delta\varphi$, to zero, since $\delta\varphi\propto v+B$. This allows for the description of the system in terms of metric variables only, in a similar way to the uniform density gauge. This similarity is further confirmed when comparing the gauge-invariants constructed in both gauges. In this comoving gauge, one defines the comoving curvature perturbation,
\be
\mathcal{R}\equiv\psi-\Hh(v+B)\,,
\ee
and in slow-roll models of single field inflation it can be shown to be approximately equal to $-\zeta$, on large scales. Because of this, these variables are both used in the literature to describe the scalar modes produced during inflation. The symbol $\zeta$ is also sometimes used to mean $\mathcal{R}$, and both quantities are often just called ``curvature perturbation'', without reference to the particular gauge in which they were defined.

We now describe flat gauge. Its name derives from the fact that, in this gauge, the spatial slices have as flat a metric as possible. Its definition is thus given by
\begin{align}
\psi=E=F^i=0\,.
\end{align} 
The only perturbation that remains non-zero in the spatial metric is the tensor part, which is gauge-invariant at first order and can never be eliminated by a gauge choice. This gauge is very common within multi-field inflation, as in that case, it allows one to use only the perturbations of the scalar fields to describe the full system to the desired accuracy on large scales. An interesting gauge-invariant quantity which is defined by this gauge is the energy density perturbation on flat hypersurfaces, whose first-order expression is
\be
\label{rhoflat1}
\delta\rho_\text{f}\equiv\delta\rho+\frac{\rho'}{\Hh}\psi=-\frac{\rho'}{\Hh}\zeta\,.
\ee
The relation with $\zeta$ is what makes this variable interesting, as it makes it easy to calculate $\zeta$ from the knowledge of the energy density, which is a function only of the scalar fields active during inflation.

A very popular gauge for studying the post-inflationary Universe is longitudinal gauge. This gauge is also often called conformal Newtonian gauge, and is defined by the following conditions on two scalar variables,
\begin{align}
B=E=0\,.
\end{align} 
If the problem under study only involves scalars, this choice is sufficient and turns out to diagonalise the metric, making many calculations simpler. When extended to include vector degrees of freedom, this gauge is often called Poisson gauge. Two possible definitions exist in the literature, with the choice
\be
F^i=0\,,
\ee
being the most common~\cite{Bombelli:1994zh,Pettinari:2014vja,Beneke:2010eg,Adamek:2015eda}. It is motivated by the similarity with the Coulomb gauge of electromagnetism ($\nabla\cdot A=0$), since its gauge conditions are equivalent to $B^i_{\ ,i}=0$ and $C^{ij}_{T\ ,i}=0$, with $C_T^{ij}$ being the traceless part of the perturbations of the spatial metric, $C^{ij}$. The alternative choice,
\be
S^i=0\,,
\ee
is also used and is based on the requirement that the contravariant vector orthogonal to spatial hypersurfaces has a vanishing spatial part \cite{Malik:2008im}. Among the various gauge-invariant quantities arising in this gauge are the Bardeen potentials~\cite{Bardeen:1980kt}, given by
\begin{align}
\Phi\equiv&\ \phi+\Hh(B-E')+(B-E')'\,,\\
\Psi\equiv&\ \psi-\Hh(B-E')\,.\label{Bardeenpsi}
\end{align}
These quantities were the first gauge-invariants to be explicitly calculated and have been used in the literature for a very long time. They have the property of simplifying one of the equations of motion considerably as one can verify by substituting them into Eq.~\eqref{Eqpsiphi} below. This gauge also has the advantage of nearly mimicking the evolution equations of Newtonian cosmology on short scales, at least at first order.\footnote{Other gauges exists in which this is also true. The $N$-body gauge \cite{Fidler:2016tir} is particularly suitable for making the connection between Newtonian and relativistic cosmology.}

The last gauge we discuss here is synchronous gauge. In it, there exists a set of observers following geodesics for whom proper time coincides with cosmic time, $dt=a d\tau$, which is the reason for its name. It is defined by the choices
\begin{align}
\label{synchdef1}
\phi=B=S^i=0\,,
\end{align} 
and is also very popular in the literature, having been used in many well known numerical solvers \cite{Ma:1995ey,Seljak:1996is,Lewis:1999bs,Lesgourgues:2011re}. A well known issue occurs with this gauge, as the conditions that define it in Eq.~\eqref{synchdef1} are not sufficient to fully determine the gauge and therefore, some residual gauge freedom remains. The first-order gauge generators necessary to convert from another gauge into synchronous gauge are given by
\begin{align}
&\alpha^{(1)}=-\frac1a \left(\int{a \phi^{(1)} \text{d}\tau}-C_\alpha^{(1)}(x^i)\right)\,,\\
&\beta^{(1)}=\int{\left(\alpha^{(1)}-B^{(1)}\right)\text{d}\tau}+C_\beta^{(1)}(x^i)\,,\label{beta1}\\
&\gamma_{i}^{(1)}=\int{S_i^{(1)}\text{d}\tau}+C_{\gamma\,i}^{(1)}(x^i)\,.
\end{align}
The functions $C_\beta$ and $C_\gamma^i$ are constant in time and can be fixed by a choice of coordinates at the initial hypersurface~\cite{Malik:2008im} and would only affect initial conditions of the variables $E$ and $F$, which are not relevant for the dynamics. The function $C_\alpha$, however, can affect the definition of many other variables and can generate so-called gauge modes, when solving the differential equations of the system. To avoid this, $C_\alpha$ can be unambiguously chosen by setting the initial velocity perturbation of some species to zero. It can easily be checked, by using Eqs. \eqref{gauge2} and \eqref{beta1}, that, at first order, this constant is given by
\begin{equation}
C_\alpha^{(1)}=a(\tau_0)(v_s^{(1)}(\tau_0)+B^{(1)}(\tau_0))\,,
\end{equation}
in which $\tau_0$ is the initial time, and $v_s$ is the velocity of a certain species. This species is often chosen to be cold dark matter, since in synchronous gauge, the Euler equation for CDM is given by
\begin{equation}
v_c^{(1)\prime}+\Hh v_c^{(1)}=0\,,
\end{equation}
which implies that if $v_c^{(1)}=0$ at any time, it must be zero at all other times. Therefore, this choice not only fixes the gauge, but is also more economical in that there is one fewer equation to be solved. At second order, the situation is very similar. The second-order gauge generators are formally given by
\begin{align}
&\alpha^{(2)}=-\frac1a \left(\int{a \left(\phi^{(2)} +\X_\phi\right)\text{d}\tau}-C_\alpha^{(2)}(x^i)\right)\,,\\
&\n^2\beta^{(2)\prime}=\n^2\left(\alpha^{(2)}-B^{(2)}\right)-\X_{B\ ,k}^{\ k}\,,
\end{align}
while the gauge transformation of the dark matter velocity is
\begin{equation}
\n^2\wt{v_c^{(2)}}=\n^2\left(v_c^{(2)}-\beta^{(2)\prime}\right)+\X_{v_c\ ,k}^{\ k}\,.
\end{equation}
The terms denoted by $\X_X$ are the quadratic parts of the gauge transformations, given in Eqs.~\eqref{transphi2}, \eqref{defXBi} and \eqref{defXvi}. From this, we can see that we can also determine the constant function $C_\alpha^{(2)}$, by setting the second-order dark matter velocity to zero, giving
\begin{equation}
\n^2C_\alpha^{(2)}=a(\tau_0)\left(\n^2\left(v_c^{(2)}(\tau_0)+B^{(2)}(\tau_0)\right)+\X_{B\ ,k}^{\ k}(\tau_0)+\X_{v_c\ ,k}^{\ k}(\tau_0)\right)\,.
\end{equation}
As was already true at first order, the Euler equation for $v_c^{(2)}$ also constrains it to be zero at all times should it be zero at any instant, and given that $v_c$ was also chosen to be zero at first order. From this we can conclude that synchronous gauge is effectively equivalent to a ``dark matter-comoving'' gauge, specified by the conditions $v_c=B=0$. This equivalence is complete, as one can then derive the remaining synchronous gauge condition, $\phi=0$, by noting that the dark matter Euler equation (i.e. $\n_\beta T_c^{i\beta}=0$), in that gauge, is simply a constraint, which is satisfied at both orders by $\phi^{(1)}=0$ and $\phi^{(2)}=0$. This equivalence also demonstrates that synchronous gauge (with zero dark matter velocity) is well defined, in spite of the non-locality in time of its gauge generator, $\alpha$.

\subsection{Evolution equations}\label{EvoEqsFLRW}

\subsubsection{Einstein Equations}

We will now show the perturbed Einstein field equations, Eq.~\eqref{EFES}, up to second order. We also split the equations into their scalar, vector and tensor parts and write them without specifying any gauge. We begin with the time-time Einstein equation:
\begin{align}
&\n^2\psi-\Hh\n^2B+\Hh\n^2E'-3\Hh\psi'-3\Hh^2\phi-\frac32\Hh^2\delta=\XNL^0_0\,,\label{EEq00CPT}
\end{align}
in which we introduced the density contrast $\delta\equiv\delta\rho/\rho$ and we collected all the non-linear terms in $\XNL^0_0$, which is given by
\begin{align}
\label{TTNL}
\XNL^0_0=&\ (\Hh^2-\Hh')v_i (v^i+B^i)+\frac32 \Hh^2 B^i B_i-6\Hh^2\phi^2+2\Hh C'\phi\\\nonumber
&+B^{i,j}\lb\frac12(\delta_{ij}C-C_{ij})'-2\Hh (C_{ij}+\phi\delta_{ij})+\frac14(B_{(i,j)}-\delta_{ij}B^k_{,k})\rb\\\nonumber
&+B^{i}\lb C_{[j,i]}^{j\prime}+\Hh C_{,i}-2\Hh C_{i,j}^{j}+\Hh \phi_{,i}+\frac14 \Hh B_{[i,j]}^{,j}\rb\\\nonumber
&+\frac14\lb C_{ij}'C^{ij\prime}-\lb C^{\prime}\rb^2 \rb+2\Hh C_{ij}C^{ij\prime}-C^{ij}C_{,ij}+2C^{ij}C^{k}_{i,jk}-C^{ij}C^{\ \ ,k}_{ij,k}\\\nonumber
&+\frac14C_{,j}C^{,j}+C^{j}_{i,j}C^{i\ ,k}_{k}-C_{,i}C^{i\ ,k}_{k}+\frac12 C_{ij,k}C^{ik,j}-\frac34 C_{ij,k}C^{ij,k}\,,
\end{align}
in which $C\equiv C^k_k$. This short-hand will also be used in the other equations to label their respective non-linear contributions, $\XNL^0_i$ and $\XNL^i_j$, much like the variables introduced in the gauge transformations above. It should be said, however, that they are not components of any tensor and that the notation used is only meant to convey the fact that they are derived from the Einstein equations with one covariant index and one contravariant index. We shall not provide the reader with explicit expressions for the remaining non-linear parts here, as they become too cumbersome for this presentation. However, we do write down a simplified version of these equations in Chapters~\ref{Ch_zeta2} and \ref{Ch_iso2}, when they are required for the calculations in question.

The space-time equation results in a scalar equation,
\begin{align} 
&\psi'+2\Hh\phi-2(\Hh^2-\Hh') (v+B)=\n^{-2}\XNL^{0,i}_{i}\,,\label{EEqi0CPT}
\end{align}
and a vector equation,
\begin{align} 
&\n^2F_i'+\n^2S_i + 4(\Hh^2-\Hh')(v_i-S_i)=4 \lb\XNL^0_i-\n^{-2}\XNL^{0,j}_{j,i}\rb\,.
\end{align}

The spatial part of the Einstein equation, like the spatial metric, is composed of two scalar parts
\begin{align} 
&\psi''+2\Hh\psi'+\Hh\phi'+(\Hh^2+2\Hh')\phi-8\pi G a^2 \lb\frac12\delta P+\frac13 \n^2 \Pi\rb=\n^{-2}\XNL^{i,j}_{j,i}\,,\\
&E''-B'+2\Hh(E'-B)+\psi-\phi-8\pi G a^2 \Pi=3\n^{-2}\n^{-2}\XNL^{i,j}_{j,i}-\n^{-2}\XNL^{i}_{i}\,,\label{Eqpsiphi}
\end{align}
one vector part
\begin{align} 
&F_i''+S_i'+2\Hh(F_i'+S_i)-8\pi G a^2 \Pi_i=4\n^{-2}\XNL^{k}_{i,k}-4\n^{-2}\n^{-2}\XNL^{k,l}_{l,ki}\,,
\end{align}
and a tensor part
\begin{align}
\label{hijEq}
h^{i\prime\prime}_j+2\Hh h^{i\prime}_{j}-\n^2 h^{i}_{j}-8\pi G a^2 \Pi^{i}_{j}=&\ 2\XNL^{i}_{j}+\delta^i_j\n^{-2}\XNL^{k,l}_{l,k}\\
&-4\n^{-2}\XNL^{k}_{(i,j)k}+\n^{-2}\n^{-2}\XNL^{k,l}_{l,ijk}\,.\nonumber
\end{align}
We can clearly see in all equations above, one of the advantages of the SVT decomposition --- the scalars, vectors and tensors do not couple to each other at first order, implying that one can solve their respective equations independently of the others. At second order, this is no longer exactly true, as the second-order equations are sourced by combinations of first-order scalars, vectors and tensors. This can be seen clearly in the non-linear part of the time-time equation, shown in Eq.~\eqref{TTNL}, in which all types of couplings exist. However, the second-order parts of variables continue not to mix, so one can still evolve them independently. 

\subsubsection{Conservation of the stress-energy tensor}

We now show the equations derived from the covariant conservation of the stress-energy tensor, Eq.~\eqref{cdTmn}. The time component is given by
\begin{align} 
&\delta'-\frac{\rho+P}{\rho} \lb3\psi'-\n^2(v+E')\rb-3\Hh(P\delta-\delta P)=\X_{T}^{\ 0}\,,\label{enecons}
\end{align}
with $\X_{T}^{\ 0}$ the non-linear source of the equation. The spatial component gives the generalization of the Euler equation, which can be further split into a scalar and a vector component. The scalar equation is
\begin{align} 
\label{vsTeq}
&(v+B)'+\lb1-3\frac{P'}{\rho'}\rb\Hh(v+B)+\phi+\frac1{\rho+P}\lb \delta P+\frac23\n^2\Pi\rb=\n^{-2}\X_{T\ ,i}^{\ i}\,,
\end{align}
while the vector one is given by
\begin{align}
\label{vVTeq} 
&(v_\text{v}^i-S^i)'+\lb1-3\frac{P'}{\rho'}\rb\Hh(v_\text{v}^i-S^i)+\frac1{2(\rho+P)}\n^2\Pi^i=\X_{T}^{\ i}-\n^{-2}\X_{T\ ,j}^{\ j,i}\,,
\end{align}
where, once again, $\X_{T}^{\ i}$ encodes the non-linear terms.

\subsubsection{Boltzmann Equation}

The Boltzmann equation, given by Eq.~\eqref{BoltzEq}, is a partial differential equation in both the spacetime position and the momentum. While, this can be solved directly, in principle, it is easier to solve equations for the brightness tensors defined above. This implies projecting the Boltzmann equation by integrating over the momentum. One first performs the same integration in momentum as that used to define $\Delta$ in Eq.~\eqref{Deltadef}. The resulting equation is then projected using the projectors given in Eq.~\eqref{projn} to generate equations for each of the brightness tensors. While this procedure gives rise to an infinite number of equations for an infinite hierarchy of brightness tensors, often only a finite number of them are required to compute observables to the specified degree of accuracy. Furthermore, as mentioned above, for many systems, such as perfect fluids, only a finite number of brightness tensors are non-zero. 

Here we shall present the derivation of the equation for the rank-2 brightness tensor for massless species, representing the evolution of the anisotropic stress. It will serve as an example for the method described above, as well as being one of the equations used in Chapter \ref{Ch_iso2}. 

The derivation begins with the rewriting of the Boltzmann equation in terms of conformal time, instead of the affine parameter used in Eq.~\eqref{BoltzEq}. This is done by dividing by $p^0$. Furthermore, we represent the momentum dependence of the distribution function in the tetrad basis, as this will simplify the derivation of the collision term for photons, to be done below. The resulting equation is
\begin{equation}
\label{boltzeq1}
\frac{\p f}{\p \tau}+\frac{\p f}{\p x^i}\frac{\dd x^i}{\dd \tau}+\frac{\p f}{\p p}\frac{\dd p}{\dd \tau}+\frac{\p f}{\p n^i}\frac{\dd n^i}{\dd \tau}=\frac{1}{p^0} C[f]\,.
\end{equation}
The left-hand-side is commonly called the Liouville term and we shall focus on it now. The right-hand side is the collision term and its description is left for Chapter~\ref{Ch_SMC} as it depends crucially on the properties of the interactions between the species of interest, which we are not specifying here.

The Liouville term can be further simplified using the geodesic equation, Eq.~\eqref{geodesic}. For simplicity, we will now proceed by calculating quantities in synchronous gauge. In this gauge, the velocity of particles is given by
\begin{equation}
\frac{\dd x^i}{\dd \tau}=\frac{p^i}{p^0}=\left(\delta_j^i-C^i_{\ j}\right)n^j\,.
\end{equation}
From the geodesic equation, we get
\begin{equation}
\frac{1}{p}\frac{\dd p}{\dd \tau}=-\left[\Hh \delta_{kl}+C_{kl}'-C^{i}_{\ k}C_{il}'-C^{i}_{\ l}C_{ik}'\right]n^k n^l\,,
\end{equation}
and
\begin{equation}
\frac{d n^i}{d \tau}=-\left(\delta^{ik}-n^in^k\right)\left[C_{kl}'n^l+n^jn^l\lb C_{jk,l}-C_{jl,k}\rb\right]\,.
\end{equation}
Substituting these into the Liouville term and integrating it over the momentum magnitude one finds the following equation for $\Delta$,
\begin{align}
\label{Boltzint}
\Delta'+\p_i\Delta\left(\delta^i_j-C^i_{\ j}\right)n^j+4(1+\Delta) n^k n^l \left(C_{kl}'-C^{i}_{\ k}C_{il}'-C^{i}_{\ l}C_{ik}'\right)&\\
-\frac{\p \Delta}{\p n^i}\left(\delta^{ik}-n^in^k\right)\left[C_{kl}'n^l+n^jn^l\lb C_{jk,l}-C_{jl,k}\rb\right]&=\frac{\int{\text{ d}p\, p^3\frac{1}{p^0} C[f]}}{\int{\text{d}p\, p^3 f^{(0)}(\tau,p)}}\nonumber\, ,
\end{align}
in which we used the following identities 
\begin{align}
\frac{1}{\int{\text{d}p\, p^3 f^{(0)}(\tau,p)}}\int{\text{ d}p\, p^3 p \frac{\p f}{\p p}}=-4(1+\Delta)\,,\\
\frac{1}{\int{\text{d}p\, p^3 f^{(0)}(\tau,p)}}\int{\text{ d}p\, p^3 \frac{\p f}{\p \tau}}=\frac{\p \Delta}{\p \tau}-4\Hh(1+\Delta)\,.
\end{align}
Equation~\eqref{Boltzint} is a sufficient representation of the Liouville term for the purposes of cosmological perturbation theory. As mentioned above, this equation exhibits a dependance on the direction of particles, but one can now apply the projectors in Eq.~\eqref{projn} to find equations for the brightness tensors. For this example, we apply $\mathcal{P}_2^{ij}$ to find the left-hand-side of the equation for the rank-2 brightness tensor: 
\begin{align}
\label{Bright2wtr}
&\Delta^{ij\prime}+\Delta^{ijk}_{\ \ \ ,l}\left(\delta^l_k-C^l_{\ k}\right) -C_{kl}'\lb \Delta^{ijkl}-\Delta^{il}\delta^{kj}-\Delta^{ij}\delta^{kl}-\Delta^{jl}\delta^{ki}\rb\\
&+2C_{r[k,l]}\lb \Delta^{ilr}\delta^{kj}+\Delta^{jlr}\delta^{ki}+\Delta^{ijl}\delta^{kr}\rb\nonumber\\
&+\frac{8}{15}\left[C^{ij}-C_k^iC^{kj}+\frac12\delta^{ij}(C-C_{kl}C^{kl})\right]'=\frac{\int{\text{d}p\frac{\text{d}\Omega}{4\pi}n^i n^j}\, p^3\frac{1}{p^0} C[f]}{\int{\text{d}p\, p^3 f^{(0)}(\tau,p)}}\,.\nonumber
\end{align}
We can see that both in this final equation and in the previous one we have kept the collision term to illustrate which operations were performed on the original equation. To derive this equation for $\Delta^{ij}$ we used some identities for the integrals of the direction vectors,
\begin{align}
\int{\frac{\text{d}\Omega}{4\pi}n^i n^j}=\frac13 \delta^{ij}\,,
\int{\frac{\text{d}\Omega}{4\pi}n^i n^j n^k n^l}=\frac{1}{15}\lb\delta^{ij}\delta^{kl}+\delta^{ik}\delta^{jl}+\delta^{il}\delta^{jk}\rb\,.
\end{align}
The integrals of terms including $\p \Delta/\p n^i$ were obtained via integration by parts and can be shown to obey the general formula
\be
\int{\frac{\text{d}\Omega}{4\pi}n^{i_1} ... n^{i_M}\left(\delta^{jk}-n^jn^k\right)\frac{\p \Delta}{\p n^j}}=(2+M)\Delta^{ki_1...i_M}-M
\delta^{k(i_1}\Delta^{i_2...i_M)}\,.
\ee
To conclude the derivation, one would now subtract the trace from Eq.~\eqref{Bright2wtr} to find the equation for the traceless brightness tensor of rank 2, which more accurately represents the anisotropic stress. A version of this equation will be shown in Chapter~\ref{Ch_iso2}.

We can draw some conclusions from Eq.~\eqref{Bright2wtr}. We notice that, already at the linear level, this equation for $\Delta^{ij}$ depends on the rank-3 tensor $\Delta^{ijk}$, so to completely solve it, one would also need the equation for the latter variable. At second order, we also see that this problem is aggravated as there is a dependence also on the rank-4 tensor $\Delta^{ijkl}$. This confirms the well known fact that the system of equations that arises from the projections of the Boltzmann equation form a hierarchy that is not closed, i.e. the equation for the tensor of rank $N$ will depend at least on that of rank $N+1$ for any $N$. Fortunately, their contributions are less and less important, the further in rank they are from the variable of interest. For instance, to calculate the solution for the rank-2 tensor, setting the rank-3 tensor to zero would greatly impact the result, but setting the rank-10 tensor to zero, would have a much smaller effect. Therefore, what is usually done is to choose a certain value of $N=N^*$ for which the brightness tensors of rank $N>N^*$ are approximated analytically, and one then solves the remaining system of equations numerically. The alternative option of setting certain tensors to zero would, in fact, introduce so-called reflection effects into the final result, which should be avoided. Very high-rank tensors are not usually needed for most applications, as there exist approximate formulas which permit one to calculate high-rank tensors from the knowledge of a few of the lower-rank ones. This is done through a formal solution to the Boltzmann hierarchy, called the line-of-sight formula, which has been very important for the development of calculations of CMB anisotropies, which we discuss in the next chapter.

% % % % % % % % % % % % % % % % % % % % % % % % % % % % 
% chapter.tex - Ian Huston
% Sample chapter layout
% % % % % % % % % % % % % % % % % % % % % % % % % % % % 
% Redefine CVSRevision for this section. 
% If you don't want to use CVS tags comment out this line
\renewcommand{\CVSrevision}{\version$Id: chapter.tex,v 1.3 2009/12/17 18:16:48 ith Exp $}

% % % % % % % % % % % % % % % % % % % % % % % % % % % % % % % % 
% =========================================================== %
% % % % % % % % % % % % % % % % % % % % % % % % % % % % % % % % 
\chapter{Early Universe Cosmology}
\label{Ch_SMC}
% % % % % % % % % % % % % % % % % % % % % % % % % % % % % % % % 
% =========================================================== %
% % % % % % % % % % % % % % % % % % % % % % % % % % % % % % % % 

In this chapter, we describe the physics of the early Universe and the methods used to make predictions from that epoch. We will begin with a review of inflation, describing its motivations and key features in Section \ref{Sec_Inf}. We then describe the evolution of the Universe after inflation in \ref{Sec_CMB}, briefly reviewing the cosmological history, as given by the $\Lambda$CDM model, focusing on the stages leading up to recombination. We also describe the evolution of linear perturbations, from the radiation dominated universe until the present time and how they depend on the primordial fluctuations.

\section{Inflation}\label{Sec_Inf}

\subsection{Introduction}

The theory of inflation was originally developed to explain the so-called problems of Big Bang cosmology. There were originally three of these problems, which can be understood as issues of fine tuning of the initial conditions for the usual radiation dominated stage of cosmology. We only describe the flatness and horizon problems, for brevity and due to the reduced relevance of the third one, the monopole problem.

The so-called flatness problem is an issue with the initial curvature of the Universe. To see this, one can rewrite the Friedmann equation, Eq.~\eqref{Eq_Fried}, in terms of the density parameter $\Omega=8\pi G\rho/3H^2$ as 
\be
\Omega(\tau)-1=\frac{K}{\Hh^2}\,.
\ee
If the conformal Hubble rate $\Hh$ is decaying, as happens for matter obeying the strong energy condition, $\rho+3P>0$, then $\Omega$ will move away from $1$ as the Universe evolves. However, observations reveal that $\Omega$ is very close to unity at the present time, which implies that $\Omega-1$ must have been extremely fine-tuned to be close to $0$ in the early Universe if matter always obeys the condition for deceleration.

Another issue is the horizon problem. The particle horizon is defined as the largest distance a particle could have travelled from an initial time to a later one. The comoving particle horizon is given by
\be
r_h=\int_0^t\frac{dt'}{a(t')}=\int_{-\infty}^{\log a} \frac{d\log \tilde{a}}{\Hh(\tilde{a})}\,,
\ee
in which we have assumed that the scale factor vanishes at the initial time $t=0$. We note that this is an integral over the conformal Hubble radius, $\Hh^{-1}$. We see once more that, if $\Hh$ is decaying, the largest contributions to the comoving horizon are those from the later time being considered. This implies that in the early Universe, the horizon was far smaller than it is today, thus limiting causal contact between larger regions. However, observations of the CMB today show that its temperature is nearly isotropic over scales which should not have been in contact in the early Universe. Again, this appears to require a large fine-tuning of the initial conditions of the Universe, which is usually undesirable.

In both cases we see that it is the requirement that the conformal Hubble radius $\Hh^{-1}$ grows at all times that gives rise to these fine tuning issues. Should it decay with time for a sufficient amount of time, $\Omega$ would approach $1$ and the comoving horizon would receive a very large contribution from the early Universe. This would allow the regions that were in causal contact in the past to be larger, as well as reducing the spatial curvature to the vanishing values observed. Introducing an epoch of accelerated expansion --- inflation --- before the radiation domination era gives precisely this decrease of the conformal Hubble radius and solves all these problems.

The requirement that inflation last for a large enough time to solve the problems above can be translated into a condition on the number of times the size of the Universe increases by a factor of $e$. This is what is commonly called the number of e-folds and is given by
\be
N=\log \frac{a_{\text{end}}}{a}=\int_t^{t_\text{end}}{H dt'}\,,
\ee
with the subscript 'end' referring to the end of inflation. The solution of the problems described above requires $N\gtrsim 60$. During this time one must have $\Hh'>0$ or equivalently, the slow-roll parameter $\epsilon$, defined as
\be
\epsilon\equiv-\frac{\dot{H}}{H^2}=-\frac{\text{d}\log H}{\text{d}N}\,,
\ee
must obey $\epsilon<1$. As already mentioned above, this implies a violation of the strong energy condition for the matter dominating the Universe during inflation. We show in the next section that a scalar field can violate that condition and thus inflate the Universe, hence making scalar field theory a candidate for a viable model of the early Universe. 

\subsection{Scalar field dynamics}

The dynamics of a scalar field are encoded in its action. For a canonical scalar field, $\ph$, it is given by
\be
\label{ph_act}
S_\ph=\int\text{d}^4x\sqrt{-g}\lbs-\frac12 \p_\mu\ph\p^\mu\ph-V(\ph)\rbs\,,
\ee
in which $V(\ph)$ is the scalar potential, which will determine most of the dynamics. A more general action will be written down in Chapter \ref{Ch_palatini}, including multiple non-canonical fields. However, the canonical single-field case will suffice for this review of inflation. 
The stress-energy tensor for this action can be found by applying Eq.~\eqref{TmnfromL} to this case, giving
\be
T_{\mu\nu}=\p_\mu\ph\p_\nu\ph-g_{\mu\nu} \lb\frac12\p_\mu\ph\p^\mu\ph+V(\ph)\rb\,.
\ee
At the background level, the energy density and pressure are given by
\begin{align}
\rho=\frac12\dot\ph^2+V(\ph)\,,\\
P=\frac12\dot\ph^2-V(\ph)\,.
\end{align}
It is now clear that the strong energy condition can be violated by this field if $V>(\dot\ph)^2$, i.e. if the field is moving sufficiently slowly along its potential. This is what gives the name to the slow-roll parameter, $\epsilon$, as the condition on the time derivative of $\ph$ is equivalent to $\epsilon<1$, the condition for successful inflation. In order for inflation to last for the required number of e-folds, it is also necessary that $\epsilon$ is small for a sufficient amount of time. The $\eta$ parameter is then defined as
\be
\eta=\epsilon-\frac{1}{2\epsilon}\frac{\text{d}\epsilon}{\text{d}N}\,,
\ee
to measure the rate of change of $\epsilon$. Many more slow-roll parameters are defined, for the higher derivatives of 
$H$, but we will not require them for this introduction. These parameters measure the deviation of the background spacetime from a pure de Sitter spacetime, which is given by the solution in Eq.~\eqref{Eq_dS1} and for which the Hubble rate $H$ is constant. Most models of inflation have a long phase of slow-roll evolution in which all slow-roll parameters are much smaller than unity and a solution can be found perturbatively around the de Sitter solution. To see this in more detail, let us look at the evolution equation for the scalar field. It is obtained through the variation of the action, Eq.~\eqref{ph_act}, with respect to the scalar field and is called the Klein-Gordon equation,
\be
\Box\ph-\p_{\ph} V(\ph)=0\,,
\ee
in which the 4-dimensional D'Alembert operator is given by $\Box=\n_\mu\n^\mu$ and $\p_\ph$ is the derivative with respect to the scalar field. At the background level, this equation is given by
\be
\ddot\ph+3H\dot\ph+\p_\ph V=0\,.
\ee
Together with the Friedmann equation, which in this case is given by 
\be
\label{FriedSca}
H^2=\frac{8\pi G}{3} \lb\frac12\dot\ph^2+V(\ph)\rb\,,
\ee
these equations constitute the full system to solve at the background level for a single, canonically normalized, scalar field. Solving them exactly for a generic potential is often impossible and numerical techniques are usually employed. However, the perturbative approach mentioned above can be used to simplify the solution of the system considerably. This is the so-called slow-roll approximation, in which an expansion in slow-roll parameters is made. This assumes that the slow-roll conditions
\be
\label{SRconds}
\epsilon \ll 1\,,\ \ |\eta|\ll 1\,,
\ee
hold true throughout the inflationary stage. As described above, the first condition guarantees inflation and allows one to approximate
\be
H^2\approx \frac{8\pi G}{3} V(\ph)\,.
\ee
The condition on $\eta$ allows for sufficient inflation and justifies neglecting the second time derivative of $\ph$ when comparing it to $H\dot\ph$, giving a simplified version of the Klein-Gordon equation,
\be
\dot\ph\approx-\frac{\p_\ph V}{3H}\,.
\ee

When these conditions are approximately valid, it is also useful to define the potential slow-roll parameters, given by
\begin{align}
\epsilon_V=\frac{1}{16\pi G}\lb\frac{\p_\ph V}{V}\rb^2\,,\\
\eta_V=\frac{1}{8\pi G}\frac{\p_\ph^2 V}{V}\,.
\end{align}
These parameters are small as a consequence of the original slow-roll conditions, Eq.~\eqref{SRconds}, and can be used as expansion parameters for a perturbative solution of the equations of motion. One can calculate the expressions for the original slow-roll parameters perturbatively, in terms of the potential ones. At first order in slow-roll, this is given by
\begin{align}
\epsilon\approx\epsilon_V\,,\ \eta\approx \eta_V-\epsilon_V\,.
\end{align}
Using this, one can calculate an approximate number of e-folds by using
\be
\label{Nev}
N\approx\int_{\ph_{\text{end}}}^\ph \frac{\text{d}\ph}{\sqrt{2\epsilon_V}}\,,
\ee
whose calculation just depends on the potential. The requirement that the number of e-folds is greater than $60$ then constrains the parameters of the potential through the formula above.

\subsection{Generation of fluctuations}\label{Secgenfluc}

We now turn to the study of quantum fluctuations around the homogeneous background of inflation. We review the calculation of their spectrum and show their dependence on the slow-roll parameters defined above. We begin by writing the perturbed Klein-Gordon equation at first order, in the gauge given by the conditions
\be
\delta\ph=0\,,\ E=0\,,\ F^i=0\,.
\ee
It is then an equation for the curvature perturbation $\psi$, which in this gauge equals $\mathcal{R}$, the comoving curvature perturbation. After eliminating the other metric potentials, $\phi$ and $B$, the resulting equation is
\be
\label{psieqinfl}
\psi''-\n^2\psi+\lb\Hh+\frac{2\ph''}{\ph'}+\frac{8\pi G\ph'^2-2\Hh'}{2\Hh}\rb\psi'=0\,.
\ee
This can be rewritten in terms of the Sasaki-Mukhanov variable~\cite{Sasaki:1986hm,Mukhanov:1981xt}, $v=z\psi$, with $z=a\ph'/\Hh$, giving
\be
v''-\lbs\n^2+\frac{z''}{z}\rbs v=0\,.
\ee
One now performs a Fourier transform
\be
v(\tau,\vec{x})=\int{\frac{\text{d}^3k}{(2\pi)^3}v_k(\tau) e^{i\vec{k}\cdot\vec{x}}}\,,
\ee
finding
\be
\label{MSeqk}
v_k''+\lbs k^2-\frac{z''}{z}\rbs v_k=0\,.
\ee
We now see that this is the equation for a harmonic oscillator with a time-dependent frequency. We note that, if $z''/z>k^2$ an instability arises, increasing the amplitude of the fluctuations enormously. Since $z''/z\sim \Hh^2$, this instability happens approximately when the size of the Hubble radius equals the length scale in question, $k^{-1}$. And, given that the Hubble radius is decreasing throughout inflation, this enhancement of fluctuations will happen to smaller and smaller scales until the end of inflation. Or, from a different point of view, at sufficiently early times, each scale $k$ is deep inside the horizon ($k^2\gg z''/z$) and will eventually exit the horizon, while being amplified.

So far we have treated these fluctuations classically, which is a good approximation as they exit the horizon. However, their evolution while still inside the horizon is quantum mechanical, and thus, one must quantize the Mukhanov-Sasaki variable, $v$, to account for the sub-horizon evolution. We simply perform the usual canonical quantization, by elevating $v$ and its conjugate momentum $v'$ to operators, via
\be
v(\tau,\vec{x})=\int{\frac{\text{d}^{3}k}{(2 \pi)^{3}}\left(a_k v_k(\tau) e^{i\vec{k}\cdot\vec{x}}+a_k^\dagger (v_k(\tau))^* e^{-i\vec{k}\cdot\vec{x}}\right)}\,,
\ee
with $a_k$ and $a_k^\dagger$ being the anihilation and creation operators obeying
\be
[a_k,a_q^\dagger]=(2\pi)^3\delta^{(3)}(\vec k-\vec q)\,,\ [a_k,a_q]=0\,,\ [a_k^\dagger,a_q^\dagger]=0\,.
\ee
Canonical normalization of the operators $v$ and $v'$ implies
\be
[v(\tau,\vec{x}),v'(\tau,\vec{y})]=i\hbar\delta^{(3)}(\vec x-\vec y)\,,\ [v(\tau,\vec{x}),v(\tau,\vec{y})]=0\,,\ [v'(\tau,\vec{x}),v'(\tau,\vec{y})]=0\,.
\ee
This in turn constrains the normalization of the mode functions to obey
\be
\label{v_k_conds}
v_k v_k^{*\,\prime}-v_{-k}^* v_{-k}^{\prime}=i\hbar\,,\ |v_k|^2=|v_{-k}|^2\,,\ |v_k'|^2=|v_{-k}'|^2 \,.
\ee

The vacuum state $|0\rangle$ is defined by
\be
a_k|0\rangle=0\,,
\ee
and is not unique. It can, however, be specified by requiring that the vaccum is the state with minimum energy. This definition is not so clear when the system is time-dependent, but can be accommodated, if one considers the sub-horizon limit, in which $z''/z$ is negligible when compared to $k^2$. This implies that the initial fluctuation $v_k$ is given by
\be
\label{ini_BD}
v_k(\tau\rightarrow-\infty)=\frac{e^{-ik\tau}}{\sqrt{2k}}\,.
\ee
This choice of vacuum state is called the \emph{Bunch--Davies Vacuum}~\cite{Chernikov:1968zm,Schomblond:1976xc,1978RSPSA.360..117B}. Excited states have also been considered and we will show a mechanism for effectively creating them, in Chapter \ref{Ch_quench}. 

Together with the normalization conditions given in Eq.~\eqref{v_k_conds}, the initial conditions in Eq.~\eqref{ini_BD} completely fix the freedom of the functions $v_k(\tau)$ and allow one to find a unique solution. Finding such a solution analytically is not straightforward and is impossible in most cases. However, if the slow-roll conditions are valid, an approximation can be found for $z''/z$, which is given by
\be
\frac{z''}{z}\approx\frac{2+6\epsilon-3\eta}{\tau^2}\,.
\ee
Substituting this into the Mukhavov-Sasaki equation, Eq.~\eqref{MSeqk}, one finds the following approximate solution
\be
v_k(\tau)=\sqrt{-\tau}\lb c_1(k)H^{(1)}_\nu(-k\tau)+c_2(k)H^{(2)}_\nu(-k\tau)\rb\,,
\ee
in which $H^{(i)}_\nu$ are Hankel functions of the $i$th kind of order $\nu$, given by
\be
\nu\approx\frac32+2\epsilon-\eta\,.
\ee
Both $\epsilon$ and $\eta$ were approximated to constants to calculate this result. The $k$-dependent coefficients $c_1$ and $c_2$ are to be determined by the initial conditions and the normalization constraints. That procedure results in the following solution
\be
v_k(\tau)=\sqrt{\frac{\pi}{4}}e^{i(2\nu+1)\frac{\pi}{4}}\sqrt{-\tau} H^{(1)}_\nu(-k\tau)\,.\label{SolSMC3}
\ee

Since the origin of these fluctuations is quantum mechanical, they form a stochastic field, whose realization cannot be predicted. What can be calculated are the correlation functions of such a field. In this case, these are defined as vacuum expectation values of collections of operators,
\be
\<ABC...Z\>\equiv\<0|ABC...Z|0\>\,.
\ee
The most important one is the power spectrum, which is given by the two-point correlation function. For the comoving curvature perturbation $\mathcal{R}$, we define the power spectrum $P_{\R}$ by
\be
\langle \R_k \R_q \rangle=P_{\R}(k)(2\pi)^3\delta^{(3)}(\vec{k}+\vec{q})\,.
\ee
Using the fact that $\psi=\R$ in the gauge used here and the relation between $\psi$ and $v$, it is straightforward to show that
\be
P_{\R}(k)=\frac{|v_k|^2}{2a^2\epsilon}\,.
\ee
Continuing with the assumption of constant slow-roll parameters, we find the power spectrum to be\footnote{In the approximation of constant slow-roll parameters, we have $\eta=\epsilon$, so the time-dependence shown here disappears. This is expected, as we will show below that $\R$ is conserved after horizon crossing.}
\be
P_{\R}(k)=\frac{\bar{H}^2}{4\epsilon M_{\text{Pl}}^2 k^3} k^{-4\epsilon+2\eta}\tau^{2(\eta-\epsilon)}\,,
\ee
in which $M_{\text{Pl}}^2=(8\pi G)^{-1}$ is the reduced Planck mass and $\bar{H}$ is defined by the solution for $a(\tau)$ in this slow-roll regime:
\be
a(\tau)=\frac{1}{\bar{H}(-\tau)^{\frac{1}{1-\epsilon}}}\,.
\ee
From this result, we can immediately read off the spectral index
\be
n_s-1\equiv\frac{\text{d}\log (k^3P_\R(k))}{\text{d}\log k}=-4\epsilon+2\eta\,.\label{nsC3}
\ee
All values of the slow-roll parameters are evaluated at horizon crossing, as the approximation that they are constant is not expected to last for the entire duration of inflation (except for exponential inflation). Evaluating the power spectrum at horizon crossing allows us to recover the standard de Sitter result, by rewriting $\bar{H}$ in terms of the Hubble rate at horizon crossing $H_*$,
\be
\bar{H}^2=k^{2\epsilon}H_*^2(\epsilon-1)^2\,.
\ee
This results in
\be
P_\R(k)=\frac{H_*^2}{4\epsilon_* M_{\text{Pl}}^2 k^3}(1-\epsilon_*)^2\,,
\ee
which differs slightly from the standard result because it takes into account higher order contributions in slow-roll. We have now concluded the calculation of the spectrum of scalar fluctuations generated during inflation. This is one of the most important results in inflationary theory, since these fluctuations will later act as the seeds of structure in the late Universe, as we discuss below. 

This result also shows that the size of cosmological perturbations is initially determined by the energy scale of inflation, $H$. Since this energy scale must be much smaller than the Planck mass, we can conclude that cosmological fluctuations are initially small. As mentioned in the beginning of Chapter~\ref{Ch_CPT}, this is very important for the validity of perturbation theory, as a perturbative expansion would otherwise be impossible. Furthermore, should the result above have a very large amplitude, its validity would be questionable, since perturbation theory was employed to derive it. Given its smallness, we can conclude that our approach is consistent.
\\

Another key prediction of inflation is the generation of primordial gravitational waves, which we now review. The mechanism for their amplification is very similar to that of scalar fluctuations, and they also originate from vacuum fluctuations. To see that, let us start by rewriting the linear version of Eq.~\eqref{hijEq}, the evolution equation for tensor modes, in Fourier space. We expand the tensor fluctuations by factoring out the polarization tensor, $\epsilon_{ij}^s$, resulting in
\be
h_{ij}(\tau,\vec{x})=\int{\frac{\text{d}^{3}k}{(2 \pi)^{3}}\sum_{s=+,\times}\epsilon_{ij}^s(k)h_k^s(\tau)e^{i\vec{k}\cdot\vec{x}}}\,,
\ee
with $\epsilon_{ij}^s(k)$ obeying $\epsilon^{i\ s}_{i}(k)=k^i\epsilon_{ij}^s(k)=0$ and $+,\times$ represent the two possible polarizations of the tensor modes. In these variables, Eq.~\eqref{hijEq} becomes
\be
h_k^{s\,\prime\prime}+2\Hh h_k^{s\,\prime}+k^2h_k^{s}=0\,,
\ee
which is very similar to the equation for the curvature perturbation, Eq.~\eqref{psieqinfl}. The procedure to calculate the power spectrum is therefore also very similar. The canonical variable that one quantizes is $v^s_k=a h^s_k/2$ and the equivalent Mukhanov-Sasaki equation is 
\be
\label{MSeqkT}
v_k^{s\prime\prime}+\lbs k^2-\frac{a''}{a}\rbs v_k^s=0\,.
\ee
Following the same procedures as before, one finds the power spectrum to be
\be
P_t(k)=2P_h(k)=\frac{4}{k^3}\frac{H_*^2}{M_{\text{Pl}}^2}\,,
\ee
in which we include the contributions from the two polarizations. The spectral index is given by
\be
n_t\equiv\frac{\text{d}\log (k^3P_t(k))}{\text{d}\log k}=-2\epsilon\,.
\ee
The relative size of tensor fluctuations is measured by the tensor to scalar ratio $r$. This is defined as the ratio of power spectra and is given by
\be
r\equiv\frac{P_t}{P_\R}=16\epsilon\,.\label{rC3}
\ee
We can now see that $n_t$ and $r$ must be proportional to each other in a slow-roll scenario. Should both be measured in the future, one could test whether inflation happened in a slow-roll regime.
\\

So far, we have not described the statistics of the stochastic field of perturbations beyond the two-point function. This is sufficient in cases in which the fluctuations have Gaussian statistics, since all other correlation functions are either zero or completely determined by the two-point function. However, in more general cases, all correlation functions may be independent and understanding them can illuminate the statistics of the fluctuations, which in turn are dependent on the fundamental physics of inflation. The first correlation function of interest is the three-point function of scalar perturbations, which defines the bispectrum, $B_\R$,
\be
\langle\R_{k_1}\R_{k_2}\R_{k_3}\rangle=(2\pi)^3\delta^{(3)}(\vec{k_1}+\vec{k_2}+\vec{k_3}) B_\R(k_1,k_2,k_3)\,.\label{bispC3}
\ee
A detection of a non-zero bispectrum would be a signal of non-Gaussianity, since this correlation function vanishes for a Gaussian distribution. One of the aims of many future experiments is to measure the effects of a finite primordial bispectrum, as it would reveal much about the physics of the early Universe. A common way to parametrize non-Gaussianity is by defining $f_{\text{NL}}$ via
\be
f_{\text{NL}}(k_1,k_2,k_3)=\frac{5}{6}\frac{B_\R(k_1,k_2,k_3)}{P_\R(k_1)P_\R(k_2)+P_\R(k_1)P_\R(k_3)+P_\R(k_2)P_\R(k_3)}\,.
\ee

Since $f_{\text{NL}}$ is a function of three wave-vectors, which are constrained by the Dirac delta function, it is useful to describe this dependence in terms of different triangle configurations. The most common ones are the squeezed ($k_1\approx k_2\gg k_3$), the equilateral ($k_1\approx k_2 \approx k_3$) and the folded/flattened ($k_1\approx k_2 \approx k_3/2$) configurations~\cite{Lewis:2011au}. Different models of the early Universe predict different shapes of non-Gaussianity, which peak at the different configurations. 

The prediction from single-field slow-roll inflation with a Bunch-Davies vacuum is generically that $f_{\text{NL}}$ is small in all configurations. This is illustrated by a result named the Maldacena consistency relation~\cite{Maldacena:2002vr, Creminelli:2004yq} and given by
\be
\lim_{k_3\rightarrow 0}f_{\text{NL}}(k_1,k_2,k_3)=\frac{5}{12}(1-n_s)\,.\label{maldacenaC3}
\ee
Since $1-n_s$ is $O(\epsilon,\eta)$, this would imply that the detection of a substantial $f_{\text{NL}}$ in the squeezed configuration would rule out single-field slow-roll inflation. Furthermore it has recently been shown that even this result is too optimistic, as the \emph{observed} $f_{\text{NL}}$ actually vanishes in this limit due to observer effects~\cite{Tanaka:2011aj,Pajer:2013ana,Bravo:2017gct}. This can be explained by the fact that a curvature fluctuation on a scale larger than the horizon would not be observed, as it amounts to a constant re-scaling of the background scale factor. For this reason, such a large-scale fluctuation must not be correlated to those on smaller scales, implying that the squeezed limit must be zero. Therefore any measurement of non-Gaussianity in the squeezed limit would invalidate single-field inflation.
\\

The results shown in this section have all been evaluated at horizon crossing, but are valid until the end of inflation and beyond. The reason for that is that both the curvature perturbation $\R$ and the tensor amplitude $h$ are conserved quantities at super-horizon scales~\cite{Wands:2000dp}, as we now show. We begin by noting that $\R$ is related to $-\zeta$, defined in Eq.~\eqref{zeta1}, via
\be
-\zeta=\R+\frac{k^2}{\Hh^2}\frac{2\rho}{3(\rho+P)}\Psi\,,
\ee 
in which $\Psi$ is one the Bardeen potentials shown in Eq.~\eqref{Bardeenpsi}. We conclude here, that on large scales, $k\ll\Hh$, and assuming $\R\sim\Psi$, the second term is negligible and $\R=-\zeta$. Thus one has only to prove conservation of $\zeta$. 

The equation of motion for $\zeta$ can be derived from the energy conservation equation, Eq.~\eqref{enecons}, and is given by
\be
\label{z1evo}
\zeta'=\frac13k^2(v+E')-\Hh\frac{\delta P_{\text{nad}}}{\rho+P}~,
\ee
with $\delta P_{\text{nad}}$ being the non-adiabatic pressure perturbation. On super-horizon scales, the first term is negligible, while the second one vanishes for adiabatic fluctuations, such as those of a single scalar field undergoing slow-roll evolution. We conclude therefore that $\zeta$ is conserved on super-horizon scales, which automatically implies $\R$ is also conserved, given their similarity. We thus justify the evaluation of the power spectrum at horizon crossing, since it will stop evolving after that point.
\\

We have concluded that to compute predictions from inflationary models, such as the spectral index, one must find the values of the slow-roll parameters at horizon crossing. To do this, it is simpler to compute them using the potential slow-roll parameters $\epsilon_V$ and $\eta_V$ as they can be found by simply taking derivatives of the potential. A further step must be taken, however, to find the field value at which to evaluate the derivatives of the potential. This can be done by writing the number of e-folds as a function of the field value, using Eq.~\eqref{Nev}. For example, for the Starobinsky model, whose potential in the Einstein frame is given by
\be
V(\ph)=\Lambda^4\lb1-e^{-\sqrt{\frac23}\frac{\ph}{M_{\text{Pl}}}}\rb^2\,,
\ee
the potential slow-roll parameters are
\be
\epsilon_V=\frac{4}{3}\lb1-e^{\sqrt{\frac23}\frac{\ph}{M_{\text{Pl}}}}\rb^{-2}\approx\frac{4}{3}e^{-2\sqrt{\frac23}\frac{\ph}{M_{\text{Pl}}}}\,,
\ee
and 
\be
\eta_V=-\frac{4}{3}e^{-\sqrt{\frac23}\frac{\ph}{M_{\text{Pl}}}}\lb1-2e^{-\sqrt{\frac23}\frac{\ph}{M_{\text{Pl}}}}\rb\lb1-e^{-\sqrt{\frac23}\frac{\ph}{M_{\text{Pl}}}}\rb^{-2}\approx-\frac{4}{3}e^{-\sqrt{\frac23}\frac{\ph}{M_{\text{Pl}}}}\,,
\ee
while the number of e-folds before the end of inflation is
\be
N(\ph)\approx\frac{3}{4}e^{\sqrt{\frac23}\frac{\ph}{M_{\text{Pl}}}}\,.
\ee
This can be inverted very easily to give the following results for the spectral index, $n_s-1$ and tensor-to-scalar ratio $r$,
\be
n_s-1=-\frac{2}{N}\,,\ \ r=\frac{12}{N^2}\,.\label{StaroC3}
\ee
Substituting in $N=60$, required by the solution of the horizon and flatness problems, results in $n_s=0.967$ and $r=0.003$, which are among the values that better fit the data collected so far~\cite{Akrami:2018odb}. The same procedure followed here can be used to compare predictions of many single-field slow-roll inflationary models with experiment as has been done, for example, in the reviews \cite{Lyth:1998xn,Martin:2013tda}.

\subsection{Multi-field inflation}\label{multfieldC3}

A very common extension of the inflationary scenario discussed here is the introduction of additional scalar fields~\cite{Malik:1998gy,Lyth:2001nq,Lyth:2002my,Gordon:2002gv,Wands:2007bd,Mazumdar:2010sa}. These scenarios are often richer in phenomenology than the single-field case and also correspondingly more difficult to compute accurately, which is why numerical methods are usually unavoidable~\cite{Elliston:2013afa,Dias:2016rjq,Mulryne:2016mzv}. The addition of extra fields is also somewhat motivated from top-down physical theories, such as string theory, in which many scalar fields appear naturally. Furthermore, even in the Standard Model of particle physics, the Higgs field is present, and should it not be the inflaton, it would be a second scalar present during inflation.\footnote{See, however, Refs. \cite{Bezrukov:2007ep,Enckell:2018kkc} for the case in which the Higgs is the inflaton.}

Let us introduce a single extra scalar to exemplify some of the effects of multi-field inflation. The action of the scalars is then
\be
\label{phch_act}
S_{\ph}=\int\text{d}^4x\sqrt{-g}\lbs-\frac12 \p_\mu\ph_I\p^\mu\ph^I-V(\ph_I)\rbs\,,
\ee
in which a sum is implied in the repeated field indices, which are labeled with capital roman letters. The potential $V(\ph_I)$ may now include interaction terms between the two fields. 

At the background level, it is often useful to define the total field velocity as
\be
\dot\ph=\sqrt{\dot\ph_I\dot\ph^I}\,,
\ee
so that the Friedmann equation can still be written as in Eq.~\eqref{FriedSca}. The background trajectories are now two dimensional and will generically be substantially different, given different initial conditions. This typically does not occur in single-field inflation, since an attractor is reached in most cases~\cite{Salopek:1990jq,Liddle:1994dx}. However, if the two-field potential has a heavy direction, i.e. $\p_2^2V$ is very large, for example, then all trajectories will eventually be directed to the minimum in that direction. Often the evolution after that is very similar to the single-field case.

We also split the field fluctuations along the directions parallel and perpendicular to the background direction given by $\dot\ph$. We therefore define
\be
\delta\ph\equiv\delta\ph_I\frac{\dot\ph^I}{\dot\ph}\,,\ \ \delta S\equiv\delta\ph_I e^I_s\,, \label{dSdefC3}
\ee
with $e^I_s$ the unit vector perpendicular to $\dot\ph^I$, which we will call the entropic direction, since fluctuations in that direction are non-adiabatic entropy perturbations. In multi-field scenarios, it is more common to use flat gauge, $\psi=0$, than the gauge used above, in order to treat all fields equally. However, it is still useful to relate the field fluctuations with the curvature perturbation $\R=-\zeta$~\cite{Dias:2014msa,CNM}. At linear order, this relation is given by
\be
\R=\frac{H}{\dot\ph}\delta\ph=\frac{\Hh}{\ph'}\delta\ph\,,
\ee
which we have written also in terms of quantities in conformal time. We can now rewrite the evolution equation for $\zeta$, Eq.~\eqref{z1evo}, in terms of $\R$, on large scales as (\cite{Malik:2008im,Turzynski:2014tza})
\be
\R'=\frac{2\Hh}{\ph'}\theta'\delta S\,,
\ee
with $\theta'$ the angular velocity in field space,
\be
\theta'=-a^2\frac{\p_sV}{\ph'}\,,\label{thetapC3}
\ee
which is given in terms of the derivative of the potential with respect to the entropic direction, $\p_s V$. As its name indicates, $\theta'$ parametrizes how fast the field trajectory turns. Should the field follow a linear trajectory in field space, then we may conclude that $\R$ is conserved.

Let us now analyse the evolution of entropy perturbations, $\delta S$. We do that by projecting the perturbed Klein-Gordon equations in the entropic direction $e_s^I$. On large scales, that equation is given by (\cite{Turzynski:2014tza,Garcia-Saenz:2018ifx})
\be
\delta S''+2\Hh\delta S'+a^2m_{s}^2\delta S=0\,,
\ee
with the effective mass, $m_s^2$, given by
\be
m_s^2=\p_s^2V+3\lb\frac{\theta'}{a}\rb^2\,.\label{EffmassC3}
\ee
We can conclude from here that the entropy fluctuations may be substantially damped, if the effective mass $m_s$ is very large. This is the case when there is a heavy direction in field space, since $\p_s^2V$ is very large, as mentioned above. If the turning rate $\theta'$ is small, then even the fluctuations generated in this case are very similar to those arising in single-field inflation. However, if the opposite is true, and the turning rate is larger than $\p_s^2V$, then even with small entropy fluctuations, $\delta S\propto 1/\theta'$, the sourcing of curvature fluctuations is still efficient, since $\R'\propto\theta'\delta S\sim1$. Therefore, the results change with respect to the expectation of the single-field case, even if $m_s$ is large. In the absence of a heavy direction, there is no general result and substantial non-adiabatic fluctuations may be generated. Consequently the curvature perturbation will not be conserved on super-horizon scales.

Many more interesting effects occur when two or more scalar fields are active during inflation. One of them is the possibility of generating substantial non-Gaussianity, since the existence of multiple active fields during inflation avoids the Maldacena consistency relation. This is a very distinctive feature, and would be effective at discerning this scenario from the single-field, slow-roll case. Furthermore, the presence of non-adiabatic fluctuations and the related evolution of $\zeta$ on large scales gives rise to a different tensor-to-scalar ratio than predicted in the single-field case. In particular, the relation between $r$ and $n_t$ is modified and it can be used to test these models. Furthermore, entropy fluctuations can excite isocurvature modes after inflation, which can leave an imprint on the later Universe. We will study these isocurvature modes in Chapter~\ref{Ch_iso2} at second order in perturbations. 

Further modifications of the multi-field scenario can also enrich their phenomenology. In Chapter~\ref{Ch_palatini} we study such an inflation model, in which we add a non-minimal coupling to gravity. As we will see below, this effectively gives rise to a modified kinetic term, which can generate interesting effects such as a curved field space. Many other effects can arise, which break the slow-roll assumption. In the next section we discuss some of the scenarios in which that happens.

\subsection{Breaking slow-roll}\label{slowrollC3}

Another simple modification to the single-field slow-roll scenario described above is to allow for a temporary break of slow-roll before the end of inflation. This can occur in many different situations, primarily if the potential has sharp features, such as a step or a bump, in small regions in field space~\cite{1992JETPL..55..489S,Adams:2001vc,Ashoorioon:2006wc,Ashoorioon:2008qr,Hazra:2010ve,Adshead:2011jq}. Alternatively, in models with non-canonical kinetic terms, the feature may also be in the effective sound speed of fluctuations, instead of the potential~\cite{Chung:1999ve,Khoury:2008wj,Achucarro:2012fd,Achucarro:2013cva,Konieczka:2014zja,Mooij:2015cxa}. These situations can also be created in a multi-field setting, by introducing fast changes in the inflationary trajectory~\cite{Konieczka:2014zja}, or generating coherent oscillations in the entropic direction~\cite{Chen:2014joa,Chen:2014cwa,Chen:2015lza,Chen:2016qce} and also by changing the effective mass of the entropic direction~\cite{Joy:2007na}, the latter of which is related to our study of a quantum quench in Chapter~\ref{Ch_quench}.

All these situations have in common the fact that some quantity changes on a time scale faster than the Hubble rate. This has the consequence of generating an amplification or dampening of the fluctuations in the scales that crossed the horizon when the feature was being traversed. Using the single-field case as an example, let us note what happens if the quantity $z''/z$ changes rapidly but then returns to its previous value. This is what happens in the case of a small step in the potential around $\ph\approx \ph^*$, which we can parametrize via
\be
V(\ph)=V_0(\ph)\lbs 1+A\tanh\lb\frac{\ph-\ph^*}{B}\rb\rbs\,.
\ee
For most of the field evolution, $z''/z$ is very similar to $a''/a$. However, when the field traverses the step, it accelerates, so that $z''/z$ increases for a short time and then decreases below $a''/a$ when the field decelerates again. Scales which have already exited the horizon when this occurs are not affected by this rapid change in $z''/z$, as are scales for which $k\gg z''/z$. However, the evolution of intermediate scales is modified since the ratio between $k$ and $z''/z$ changes rapidly. For example, a scale that would exit the horizon during the transition, is now amplified earlier, but then re-enters the sub-horizon regime while the field decelerates, only to leave it again shortly after. This temporary oscillatory phase lasts different amounts of time for perturbations of different scales, generating a modulation in the amplitude of their power spectrum. This modulation decays with $k$ as smaller scales are progressively less affected by the feature. The detailed analysis and explanation of this scenario is given in Ref.~\cite{Adams:2001vc}, in which numerical calculations reveal the oscillations described and how they depend on the parameters of the feature.

There have been some hints of these features in observations of the CMB, albeit with low statistical significance. However, their detection would provide us with new insights into the physics of the early Universe and is even conjectured to allow for a distinction between the inflationary scenario and other alternatives~\cite{Chen:2016qce}.
\\

We have now concluded our exposition of inflation and will now briefly describe the evolution of the Universe after inflation, beginning at the stage of reheating and proceeding with the evolution of the Universe until the generation of anisotropies in the CMB.

\section{Post-inflation evolution}\label{Sec_CMB}

\subsection{From reheating to nucleosynthesis}

The inflationary stage described in the previous section must have ended at some point, at least in a patch that included our observable Universe. Therefore, the inflationary potential must be such that, after some time of approximately slow-roll evolution, a more rapid stage ensues, in which the slow-roll parameter $\epsilon$ grows and reaches unity. At this time, the accelerated expansion stops and the comoving horizon begins growing. Consequently, fluctuations are no longer amplified. Beyond that point in time, the energy density of the inflaton(s) must be transferred to other fields, which are, or eventually decay to, the known particles of the Standard Model of particle physics. This stage is called reheating \cite{Albrecht:1982mp,Guth:1982cv,Kofman:1994rk} and is one of the least understood stages of the evolution of the Universe. It is often modeled by assuming that the inflaton potential has a minimum around which the field oscillates. A coupling with other fields is then introduced as an effective decay rate, $\Gamma_{\ph R}$, which converts the energy in the inflaton into radiation via the equation,
\be
\rho_\ph'+3(\Hh+\Gamma_{\ph R})\rho_\ph=0\,.
\ee
Many different models exist that describe reheating and attempt to estimate when it happens, how many e-folds it lasts and how efficient it is. The question regarding its length in time is important because it influences when the fluctuations measured today crossed the horizon, described by the number of e-folds, $N$, given above. This is a theoretical uncertainty in most models of inflation and predictions are usually calculated for $50<N<60$. Fortunately, as we have seen in the example of Starobinsky inflation above, predictions are often not very sensitive to small variations of $N$ and this uncertainty is not so relevant.

Furthermore, if only a single field is responsible for driving inflation, the large-scale curvature perturbations generated are unaffected by the reheating stage, since $\zeta$ is conserved on super-horizon scales. The fluctuations in $\zeta$ are distributed equally between the different species and only an adiabatic mode survives. On the other hand, if there are multiple fields active until the end of inflation, $\zeta$ may no longer be conserved and the detailed physics of reheating can play a role in its evolution. Moreover, all  fields have to decay to the Standard Model species, which implies that their perturbations will be distributed in non-trivial ways among the perturbations of the energy density of different species, generating isocurvature perturbations. The way in which this happens is not straightforward to estimate and is often very model dependent.

After reheating, the Universe enters a radiation dominated stage, described, at the background level, by the solution given in Eq.~\eqref{Eq_rad_dom}. Most species are expected to quickly reach a state of thermal equilibrium with a very high temperature, $T_{\text{reh}}>$~TeV, since their densities are expected to be high enough to ensure their frequent interaction. If all species reach this state of equilibrium, then isocurvature modes decay during this stage~\cite{Weinberg:2004kf}, unless they are sourced by some other means.

The expansion of the Universe causes its temperature to decay as $T\propto 1/a$, allowing phase transitions to occur. Given our lack of knowledge of particle physics above the TeV scale, it is not impossible that many phase transitions happened at very early times, when the temperature was larger than that scale or even during inflation. This may include a Grand Unification phase transition, in which the symmetry unifying the strong and the electroweak interactions was broken; as well as a mechanism for baryogenesis and leptogenesis. Again, many models exist to explain these phenomena, but we will not describe them here.

The first phase transition that is known to have occurred is the electroweak phase transition around the temperature of $100$ GeV. At this stage, Standard model particles acquired masses and the electromagnetic interaction splits from the weak force. The weak interaction probability, $\sigma_w$, then began decaying as $\sigma_w\propto T^2$ and the corresponding interaction rate of electrons with neutrinos, $\Gamma_\nu=n_e\sigma_w$, now behaved as $\Gamma_\nu\propto T^5$, assuming the electrons are still relativistic with $n_e\propto T^3$. The expansion rate, given by the Hubble parameter, changes with temperature as $H\propto T^2$ during the radiation dominated stage. At some point, after the temperature has fallen sufficiently, the interaction rate falls below the expansion rate, making interactions increasingly rare. Therefore, from that point on, at temperatures lower than about $T\sim 1$ MeV, neutrinos can no longer maintain equilibrium with the electrons and consequently with all other interacting species. Soon after, the temperature drops below the electron mass, $m_e\approx 511$ keV and electrons efficiently annihilate with positrons. Their number density drops further and this also contributes to the complete decoupling of neutrinos from the remaining plasma of electrons and protons. Furthermore, positron-electron annihilation produces many photons, and leads to an increase in their temperature. Since neutrinos are decoupled from the remaining plasma, their temperature does not change at this stage and is thus kept lower that the photon temperature throughout their evolution. Neutrinos then propagate freely, thus forming the Cosmic neutrino background (C$\nu$B), which has never been detected. Their free streaming also has important consequences for the evolution of perturbations, since their distribution can now become anisotropic, as we will see in the next section.

While the above was happening in the lepton sector, another phase transition happened in the quark sector --- the QCD phase transition. This occurred at a temperature of around $T\sim100$~MeV, and, after this transition, the quark-gluon plasma dissipated, and the quarks became confined in hadrons and mesons. Eventually, most baryons decay to form protons and neutrons, with a ratio of abundances determined by their mass difference $\Delta m=1.293$ MeV, via $n/p=e^{-\Delta m/T}$.
This ratio is maintained by their frequent interactions via the weak force, but is nevertheless decreasing due to the decay of temperature with the expansion of the Universe. However, the interaction rate of weak interactions between protons and neutrons falls below the Hubble rate at around $T\approx 0.7$~MeV and then the neutron-to-proton ratio is frozen at the value $n/p\approx 1/6$. Due to their higher mass, neutrons then decay into protons via beta decay with a lifetime of around $880$ seconds, until the temperature falls below that required for forming nuclei, $T\sim 0.1$ MeV, at which point the neutron-to-proton ratio has decreased to $n/p\approx 1/7$. This is the starting point of Big Bang Nucleosynthesis (BBN), whose relatively low temperature is due to the small size of the baryon-to-photon ratio, $\eta_{b\gamma}\sim 10^{-10}$, which delays efficient nuclei formation until the temperature drops well below their binding energy. When nuclei do begin forming, protons and neutrons go through a chain of of reactions, creating Deuterium and Helium-3, until most neutrons become bound in Helium-4 nuclei, since it is the most stable light element. The mass fraction of Helium-4 relative to that of all baryons is then easy to estimate with a counting argument to be approximately $1/4$. Other abundances of the light elements can also be predicted using more refined calculations \cite{PitrouEtal2018} and the agreement of these predictions with measurements is a key piece of evidence of the Big Bang model of cosmology.

The origin of dark matter in the early Universe may have followed a similar pathway as the other species. The hypothesis that dark matter is made of weakly interacting massive particles (WIMPs) postulates that the weak interaction keeps dark matter in equilibrium in the early Universe at a very high temperature. WIMPs then decouple when their interaction rate falls below the freeze-out temperature, being non-relativistic at that stage. This then fixes their abundance and comparing that to observations allows one to derive relations between the interaction rate and the mass of the WIMP. However, the unknown nature of dark matter and the failure of its direct detection in current experiments, implies that very little can be confirmed about its origin and formation mechanism. In particular, it may have never been in equilibrium at early times, such as happens in models describing dark matter as composed of feebly interacting massive particles (FIMPs) \cite{Bernal:2017kxu}. Many other models of dark matter exist, based, for example on axions~\cite{Marsh:2015xka} or even primordial black holes~\cite{Carr:1974nx}. Due to their very different formation mechanisms, and the fact that many of them may contribute to the total dark matter energy density, it is difficult to say with certainty how dark matter was formed and how it affected the early Universe, besides through its action on the Universe as a cold species with negligible interactions.

%As mentioned above, if complete thermal equilibrium is reached, isocurvature fluctuations decay. However, this decay can be avoided if dark matter has never been in equilibrium or even if other species did not interact for a sufficiently long time, a situation which would be made more likely if reheating ended at a later time. Then isocurvatures may not decay completely and their effects might be measurable at later times.

\subsection{Recombination and the CMB}\label{RecCMB}

\subsubsection{Background evolution and thermodynamics}

After neutrino decoupling and shortly after nucleosynthesis has ran its course, at $z\sim 10^8$, the majority of the matter in the Universe is composed of: nearly massless neutrinos that are free-streaming, cold dark matter behaving as dust and the baryon-photon plasma, composed of electrons, ions and photons interacting via Compton and Coulomb interactions. Radiation is still the dominant component of the Universe and hence the background expansion rate still obeys the solution given by Eq.~\eqref{Eq_rad_dom}. However, since the energy density of radiation dilutes faster than that of non-relativistic matter, composed of both dark matter and baryons, it is inevitable that the Universe becomes matter dominated after some time. This occurs at a temperature $T\sim1$ eV, corresponding to a redshift $z\approx3300$.

The photons are kept in equilibrium with the remaining plasma due to Compton interactions. However, since their temperature is now smaller than the masses of both ions and electrons, these interactions do not cause the energy of the photons to change appreciably. They are thus well described by the non-relativistic limit of Compton scattering, called Thomson scattering. The cross-section for the interactions between photons and electrons is $\sigma_T=4.328\times10^{-29}\ \text{m}^2$, while that for the corresponding interaction with protons is smaller by a factor of the square of their mass ratio, $(m_e/m_p)^2\sim 10^{-7}$. The dominant interactions of photons are therefore those with electrons, whose interaction rate is $n_e\sigma_T$, representing the inverse of the mean time a photon travels between scatterings.

The interaction rate of Thomson scattering is sufficiently large to keep the plasma in equilibrium for a very long time. So long, that if nothing else were to occur, the interaction rate would not fall below the expansion rate, $H$, until a redshift $z\sim 40$. However, before that, electrons begin combining with protons efficiently to form atoms in the process called recombination. After this stage, almost all electrons become bound in atoms and no longer interact with the photons. This sudden drop in the free electron number density, $n_e$, causes the Thomson interaction rate to sharply decrease and fall below the expansion rate. Photons are then decoupled from the baryon fluid and begin streaming freely. The point in which photons last scattered happens at this stage and thus the Cosmic Microwave Background is formed, which is observed today to have a temperature $T=2.35\times 10^{-4}$ eV$=2.725$ K. 

The temperature and redshift at which decoupling happened are important quantities and are difficult to estimate analytically, due to the complicated non-equilibrium physics of the process and the need to describe the different energy levels of the hydrogen atom. This is usually done using numerical codes such as RECFAST~\cite{Seager:1999bc} and HyRec~\cite{AliHaimoud:2010dx}, which accurately compute the ionisation history using only a few energy levels. However, an order of magnitude estimate can be obtained analytically and we shall briefly describe it now, beginning with a simplified description of recombination and then estimating the decoupling temperature and redshift.

The quantity that controls recombination is the free electron fraction, given by
\be
x_e=\frac{n_e}{n_b}\,,
\ee
with $n_b$ the baryon number density, which is approximately equal to the total number density of electrons, due to the neutrality of the Universe and if we neglect the contribution from helium atoms. We also assume that the only reaction that occurs is 
\be
e^-\ +\ p^+\ \leftrightarrow\ H\ +\ \gamma\,,
\ee 
since it is the dominant reaction for production of hydrogen. Under these conditions, the Saha equation can describe the evolution of $x_e$ during equilibrium. It is given by (\cite{dodelson2003modern,Pettinari:2014vja})
\be
\frac{x_e^2}{1-x_e}=\frac{\sqrt{\pi}}{4\sqrt{2}\zeta(3)}\frac{1}{\eta_{b\gamma}}\lb\frac{m_e}{T}\rb^{3/2} e^{-E_H/T}\,,
\ee
with $E_H=13.6$ eV being the binding energy of hydrogen. We see that besides $E_H$, the other parameter that controls the evolution is the baryon-to-photon ratio $\eta_{b\gamma}$. Similarly to what was already described above for nucleosynthesis, the fact that $\eta_{b\gamma}$ is very small, implies that the temperature must fall far below the binding energy, $E_H$, for $x_e$ to deviate significantly from its initial value of $x_e=1$. Close to recombination, $x_e$ falls rapidly due to the exponential factor in the Saha equation. Estimating the start of recombination as the moment when $x_e=0.5$, results in a recombination temperature of $T=0.32$ eV and a corresponding redshift of $z=1360$. Under the approximations used here, $x_e$ then decays to zero exponentially. This solution is, however, not very accurate after the first instants of recombination. This is because it does not include the non-equilibrium effects of an expanding Universe. The main effect is the freeze out of recombination, when the rate of the reaction above falls below the expansion rate. Estimates from numerical solvers of the ionisation history show that the free electron fraction asymptotes to a constant value of $x_e\sim 10^{-3}$, which is not reached until much later at a redshift of order $100$.

To compute when the CMB was formed, one would need to go further and calculate when decoupling happens. This can be estimated by comparing the interaction rate of Thomson scattering and the Hubble rate, but even this simple estimate would require a numerical solution for $x_e$, so we will not describe the details here. That estimate is also not very accurate for last scattering, as it would return a value of $z\approx 900$ when $\sigma_T n_e=H$ \cite{Pettinari:2014vja}. A more accurate estimate is obtained by asking instead at what redshift a photon is most likely to have last scattered. This probability is described by the visibility function 
\be
g(\tau)=-\kappa' e^{-\kappa}\,,
\ee
in which $\kappa$ is the optical depth, defined via the integral of the interaction rate. Its derivative in conformal time is therefore
\be
\kappa'=-a n_e \sigma_T\,,
\ee
which is different from the interaction rate quoted above by a factor of $a$, due to the change to conformal time. The function $g(\tau)$ can be shown numerically to peak at the redshift $z_{\text{LSS}}=1100$, which defines the \emph{last scattering surface}. This is the redshift at which the CMB was formed.

After last scattering, photons do not interact very often and essentially only redshift on their way to Earth, due to cosmic expansion. Because of this, and the fact that, during recombination, the energy exchanges between photons and electrons are too small, photons maintain their equilibrium spectrum, as given by Eq.~\eqref{BEf} with $T\propto a^{-1}$. This was measured to high precision by the COBE satellite, thus confirming this prediction~\cite{Mather:1993ij,Fixsen:1996nj}. Their energy density simply drops off as radiation with $\rho_\gamma\propto a^{-4}$.

The free electrons that remain still interact frequently with photons until much later, keeping their temperature matched to the photon temperature. This is because of the much larger number of photons with respect to that of electrons, which maintains equilibrium only for the least abundant of the two species. However, most baryons are now in the form of hydrogen atoms, which are fully decoupled from the remaining species. The background evolution of their energy density is that of dust, $\rho_b\propto a^{-3}$.

This concludes our short review of the background evolution of the baryon-photon plasma. Regarding the other species, in the standard model of cosmology, cold dark matter simply evolves as dust, with $\rho_{c}\propto a^{-3}$, as do the baryons. The neutrinos evolve as radiation, with $\rho_\nu\propto a^{-4}$, at least until their effective temperature is smaller than their masses of $O(\text{eV})$, when they become non-relativistic.

\subsubsection{Evolution of perturbations and CMB anisotropies}

We have just described above the background and thermal evolution of the baryon-photon plasma, as well as that of cold dark matter and neutrinos. Primordial fluctuations from the very early Universe are transferred to all of these species, generating density and velocity fluctuations or, more generally, fluctuations of their distribution functions. The study of the evolution of these perturbations can provide great insights into the evolution of the Universe, and allows us to predict the spectrum of anisotropies of the CMB. This is one of the key observations of modern cosmology, as it contains a very large amount of information about the primordial Universe, as well as the cosmic expansion and the contents of the Universe. For this reason, we briefly review here the methods used to compute the anisotropy spectrum.

The first issue we address is that of the state of perturbations at the start of this stage, after electron-positron annihilation, which provides the initial conditions for their evolution. As briefly mentioned above, the allocation of the primordial fluctuations among each species depends on the character of the mode under consideration, i.e., whether it is an adiabatic or an isocurvature mode. Initially, the adiabatic mode has a non-zero curvature perturbation, $\zeta\propto-\R$, and all entropy fluctuations $S_{sr}$ vanish\footnote{Note that these quantities are different from the the entropy fluctuations, $\delta S$, defined in multi-field inflation in Eq.~\eqref{dSdefC3}. In spite of their probable connection due to the generating mechanism of isocurvature, it is unlikely that these two quantities are equal in most cases, so we chose to distinguish them explicitly by using the different notation, $S_{sr}$.}. These are given by
\be
S_{sr}=\frac{\delta_s}{1+w_s}-\frac{\delta_r}{1+w_r}\,,
\ee
with $\delta_s$ and $w_s$, respectively, the density contrast and equation of state parameter of each species. For an isocurvature perturbation, the opposite is true, with the curvature perturbation vanishing initially and one or more of the entropy fluctuations being finite.\footnote{A more complete definition and description of all the possible isocurvature modes is given in Chapter~\ref{Ch_iso2}, in which these modes are studied in great detail and up to second order.} 

A prediction of the relative sizes of each of these modes is non-trivial, not only because many different models of inflation exist, but also because the evolution between reheating and the stage under study here is not straightforward. On the one hand, we are fairly confident that the adiabatic mode exists and is conserved throughout its super-horizon evolution, and its amplitude has been measured. On the other hand, the isocurvature modes may be generated by multi-field inflation, but may later decay substantially, if the species they relate to reaches equilibrium. Their size when species decouple is therefore the result of a competition of their size after reheating and how much they decayed over their evolution. This is generally unpredictable, unless one has a very detailed model of the entire evolution. Here and in the rest of this thesis, we take the agnostic view and study all isocurvature modes and their possible contributions to the evolution of fluctuations.

The evolution equations for the fluctuations of the relevant species are essentially given in Section \ref{EvoEqsFLRW} of Chapter \ref{Ch_CPT}, except for the absence of collision terms. Neutrinos and dark matter are not interacting during the stage of interest and therefore will obey the Liouville equation and the conservation of their individual stress-energy tensors. Furthermore, due to the high rate of Coulomb interactions between electrons and ions, we will assume that they form a single fluid of baryons, even before recombination, and thus we will not need to know the collision term for those interactions. Photons and charged particles, however, do interact very strongly via Thomson scattering and their collision term must be calculated in order to describe the evolution of their perturbations accurately. The calculation of the collision term for photons was performed in detail up to second order in Ref.~\cite{Dodelson:1993xz} and we now reproduce the main steps.

The collision rate $C[f](\vec{p})$ is defined as the rate of change of the number of particles with momentum $\vec{p}$. The reaction in question is Thomson scattering,
\be
\gamma(\vec{p})\ +\ e^-(\vec{q})\ \leftrightarrow\ \gamma(\vec{p}')\ +\ e^-(\vec{q}')\,,
\ee
where we have explicitly labeled all momenta. The collision rate is an integral over all possible momenta that contribute to create or destroy a photon of momentum $\vec{p}$. It is given by
\begin{align}
C[f](\vec{p})=&\ \int\frac{\text{d}^3q}{(2\pi)^3 E_q}\int\frac{\text{d}^3p'}{(2\pi)^3 E_{p'}}\int\frac{\text{d}^3q'}{(2\pi)^3 E_{q'}}\left|\mathcal{M}\right|^2\delta^{(3)}(\vec{p}+\vec{q}-\vec{p}'-\vec{q}')\\
&\times \delta(E_p+E_q-E_{p'}-E_{q'})\lbs f_{p'}g_{q'}(1+f_p)(1-g_q)-f_{p}g_{q}(1+f_{p'})(1-g_{q'})\rbs\,,\nonumber
\end{align}
in which $E_x$ is the energy of the particle labeled by momentum $x$ in the tetrad frame, $f$ is the distribution function of photons, $g$ is the distribution function of electrons and $\left|\mathcal{M}\right|^2$ is the Thomson interaction amplitude. The delta functions enforce energy and momentum conservation. This expression is rather general and no approximation has been used. However, to transform it into a more useful form, we eliminate the momentum $\vec{q}'$ with the delta function, and we perturb the result in two different ways. First we expand the photon distribution function $f$ in cosmological fluctuations as $f_p=f_p^{(0)}+\delta f_p$. Second, we expand all other quantities in powers of the energy transfer,
\be
\frac{E_q-E_{q'}}{T}\approx \frac{q}{m_e}\,,
\ee
which as mentioned above, is very small at these temperatures, in which the electron momentum $q$ is non-relativistic. The Thomson amplitude is, in this approximation,
\be
\left|\mathcal{M}\right|^2=6\pi\sigma_T m_e^2\lb1+\cos^2\theta\rb\,,
\ee
where $\cos\theta=\vec{n}\cdot\vec{n}'$ is the cosine of the angle between the photon momenta. The electron distribution function for momentum $\vec{q}'$ is expanded around the one for the ingoing momentum $\vec{q}$ in powers of the energy transfer. After this expansion, the expression for $C[f]$ includes integrals of $g_q$ multiplied by several powers of the electron momentum $q$. Given that we know the electron distribution function to be of the Maxwell-Boltzmann form, these integrals are simple to calculate in terms of the moments of the distribution, such as the free electron number density $n_e$ and the average velocity of electrons, $v_e$. At first order, the result is
\begin{align}
C[f]=-\frac{3}{8}n_e\sigma_T p\int{\frac{\text{d}\Omega'}{4\pi}(3+\cos 2\theta)\lbs \delta f(p,\vec{n})-\delta f(p,\vec{n}')\right.}\\
\left.+p(\vec{n}-\vec{n}^{\prime})\cdot\vec{v}_{e}f^{(0)\prime}(p)\rbs\,.\nonumber
\end{align}
To obtain the source terms for the equations of the brightness tensors, one then has to integrate this result over the momentum $p$ and over the angular directions with different powers of the direction vector $n^i$. This results in the following expressions for the collision terms up to rank 2 brightness tensors:
\begin{align}
&\frac{\int{\frac{\text{d}\Omega}{4\pi}}\int{\text{d}p\, p^3\frac{1}{p^0} C[f]}}{\int{\text{d}p\, p^3 f^{(0)}(\tau,p)}}=0\,,\label{Col01st}\\
&\frac{\int{\frac{\text{d}\Omega}{4\pi}n^i}\int{\text{d}p\, p^3\frac{1}{p^0} C[f]}}{\int{\text{d}p\, p^3 f^{(0)}(\tau,p)}}=\frac43\kappa'(v_\gamma^i-v_e^i)\,,\label{Col11st}\\
&\frac{\int{\frac{\text{d}\Omega}{4\pi}n^i n^j}\int{\text{d}p\, p^3\frac{1}{p^0} C[f]}}{\int{\text{d}p\, p^3 f^{(0)}(\tau,p)}}=\frac{9}{10}\kappa'\Delta_{\gamma T}^{ij}\,.
\end{align}
We have included the factor $1/p^0$ multiplying each collision rate, since that is what appears when the Boltzmann equation is written in terms of conformal time, as seen in Eq.~\eqref{boltzeq1}. We conclude from here that, at first order in perturbations, the energy conservation equation for photons is not sourced by collisions, while the the momentum conservation equation has a source that depends on the velocity difference between the photons and the baryons. The evolution of anisotropic stress is only sourced by itself.

At very early times, when the interaction rate is very high, the collision terms drive the evolution to make them vanish, as any deviation from this generates a very strong source in the equations. Therefore, at sufficiently early times, one may use the \emph{tight coupling approximation}:
\be
v_\gamma^i=v_e^i\equiv v_{b\gamma}^i\,,\ \Delta_{\gamma T}^{ij}=0\,.\label{TCAdef}
\ee
This implies that we will require one fewer equation to describe the evolution, as there will be a single Euler equation for the baryon-photon plasma. Note, however, that since the collision term does not affect the energy conservation equation, one still has to evolve two equations for the evolution of $\delta_\gamma$ and $\delta_b$.

The same arguments used to show the vanishing of the anisotropic stress can also be used to show that all higher rank brightness tensors are zero in this approximation. This means that the photon fluid acts as a perfect fluid with interactions at early times. Furthermore, this also implies that no anisotropies are generated in this fluid. It is only around last scattering that anisotropies are created, since there are no more interactions to stop photons from free streaming. The small inhomogeneities in the gravitational potentials and the photon energy density are then transformed into anisotropies at recombination and this is, in essence, what we later observe in the CMB. We shall briefly review the calculation of the spectrum of anisotropies below.
 
Before going into the details of the anisotropy generation, we briefly mention how dark matter and neutrinos evolve. Dark matter behaves like dust, having negligible pressure and anisotropic stress. Its evolution therefore follows Eqs.~\eqref{enecons} and \eqref{vsTeq} with $P=\delta P=\Pi=0$. This has the effect that dark matter clusters according to the gravitational field being sourced by all species. The neutrinos follow the same equations but with $P_\nu=\rho/3$ and $\delta P=\delta\rho/3$. Because they have no interactions, their anisotropic stress is not suppressed, as it is initially in the photon fluid. However, because they did interact strongly in the past, anisotropies only start being generated after their decoupling, which happens only slightly earlier than the epoch under analysis here. Furthermore, anisotropies only grow when sourced by inhomogeneities, hence, only after the fluctuations re-enter the Hubble horizon and ``see'' an inhomogeneous Universe, can the anisotropic stress be generated. All this is encoded in Eq.~\eqref{Bright2wtr} without collision term and more generally in the Liouville equation for neutrinos. In spite of it being initially negligible for the scales of interest, neutrinos eventually contribute with a source of anisotropic stress in the pre-recombination Universe and this has an effect on the gravitational potentials, through the space-space Einstein equation, Eq.~\eqref{Eqpsiphi}.

We now move on to the estimation of the CMB anisotropies, which are later measured in the temperature field. We provide a simplified description, following Ref.~\cite{Challinor:2004bd}, but change much of the notation and do the calculations in a different gauge. 

The temperature fluctuations $\Theta$ are defined by a modification of the background photon distribution function, given by
\be
f(\tau,\vec{x},p,\vec{n})=\lbs\exp\lb \frac{p}{T^{(0)}(\tau)\lbs1+\Theta(\tau,\vec{x},p,\vec{n})\rbs}\rb-1\rbs^{-1}\,.
\ee
These temperature perturbations can easily be shown to be related to the brightness fluctuations, $\Delta$, defined in Chapter~\ref{Ch_CPT}, via
\be
\Delta_\gamma=4\Theta+6\Theta^2\,,\label{tempDel}
\ee
and are thus an equivalent way of describing the perturbed Boltzmann equation.\footnote{The relation shown above is obtained given a certain definition of temperature, which is in this case associated to the first moment of the distribution function, the brightness. At second order the different definitions of temperature do not coincide and one must choose a definition carefully. For more details, the reader is directed to Refs.~\cite{Pettinari:2014vja, Pitrou:2014ota}.} 

We aim to compute the fluctuations of $\Theta$ today and to calculate their angular power spectrum, since this is what is measured in the CMB. This is defined by
\be
C_\ell^{TT}\equiv\frac{1}{4\pi}\int{\text{d}\Omega\text{d}\Omega' P_\ell(n^in_i')\langle\Theta(\vec{n})\Theta(\vec{n}')\rangle}\,,
\ee
in which $P_\ell$ are Legendre polynomials of order $\ell$ and all variables are evaluated at the present time and at the position of the Earth. The Legendre polynomials, $P_\ell(x)$ always include their argument raised up to the power $\ell$, i.e. $x^\ell$. This implies that to calculate the spectrum at a value $\ell$ requires knowledge of the brightness tensors up to rank $\ell$. Therefore, if one is interested in predicting the CMB power spectrum up to $\ell$ of order $1000$, one needs to evolve at least the same number of differential equations for photons plus those for the other species and the Einstein equations. A numerical solution is thus very computationally intensive and the original codes written for that task, such as COSMICS \cite{Ma:1995ey}, could take several days to compute the spectrum. Fortunately, a different method exists, using the so-called line-of-sight formalism~\cite{Seljak:1996is}. It uses a different way to solve the Boltzmann equation, which we now describe. 

We begin by rewriting the momentum integrated Boltzmann equation, Eq.~\eqref{Boltzint}, at first order and in terms of the temperature fluctuation, $\Theta$,
\be
\Theta'+n^i\p_i\Theta-\kappa'\Theta=-n^in^jC'_{ij}-\kappa'\lbs n^iv_{b\,i}+\frac{3}{4}\int{\frac{\text{d}\Omega'}{4\pi}(1+(n^in_i')^2)\Theta(\vec{n}')}\rbs\,,\label{Boltzeq1LOS}
\ee
which we wrote in synchronous gauge, as before, and included the collision term derived above. A line of sight is then defined as a null curve linking a point in which a photon was emitted ($E$) to the point where it was received ($R$). We parametrise this curve by the affine parameter $\lambda$ and we see that its tangent vector can be written as $(1,n^i)$ in the tetrad basis. Using these facts, we can conclude that the first two terms in Eq.~\eqref{Boltzeq1LOS} can be re-written as
\be
\Theta'+n^i\p_i\Theta=\frac{\text{d}\Theta}{\text{d}\lambda}\,.
\ee
Given this, and the fact that $\kappa'=\text{d}\kappa/\text{d}\lambda$, the full equation can be formally solved by an integral over the variable $\lambda$, given by
\be
\lbs\Theta e^{-\kappa}\rbs_R=\lbs\Theta e^{-\kappa}\rbs_E+\int_E^R{\text{d}\lambda e^{-\kappa}S}\,,
\ee
with $S$ being the right-hand side of Eq.~\eqref{Boltzeq1LOS} and $E$, $R$ being, once more, the two ends of the line of sight. Given that we are interested in calculating $\Theta$ at our current position on Earth, we assign that position to point $R$, for which the optical depth, $\kappa$, vanishes. Emission occurs at very early times, when $\kappa\gg 1$, so we can neglect the first term on the right-hand side. Performing the angular integral in the source term, we find
\be
\lbs\Theta\rbs_R= \int_{E}^R{\text{d}\lambda\lbs -\kappa'e^{-\kappa}\lb\Theta_0+\frac34\Theta_T^{ij}n_in_j+ n^iv_{b\,i}\rb- e^{-\kappa}n^i n^j C'_{ij}\rbs}\,,
\ee
in which we have used the temperature tensors of rank 0 ($\Theta_0$) and rank 2 ($\Theta_T^{ij}$) that are related to the brightness tensors via the same relation as in Eq.~\eqref{tempDel}. It now becomes clear how this result leads to a huge simplification of the calculation of the spectrum of anisotropies, since one only needs to compute two of the brightness tensors to calculate the full temperature fluctuation today. The authors of this method developed the Boltzmann solver CMBFAST~\cite{Seljak:1996is,Zaldarriaga:1997va,Zaldarriaga:1999ep}, which improved the computation time of the anisotropy spectrum by several orders of magnitude. This method is now used in all modern Boltzmann solvers, including the linear codes CAMB~\cite{Lewis:1999bs}, CLASS~\cite{Lesgourgues:2011re}, CMBEasy~\cite{Doran:2003sy}, PyCosmo~\cite{Refregier:2017seh} and the second order codes SONG~\cite{Pettinari:2014vja} and CMBQuick.

Having now described the line-of-sight formalism, we now have all the ingredients to accurately calculate the power spectrum of the CMB. To conclude this chapter, let us now summarize the steps of a complete calculation. They are
\begin{itemize}
\item{Compute the background evolution;}
\item{Compute the ionization history;}
\item{Initialize perturbations with adiabatic or isocurvature modes;}
\item{Evolve the equations for the perturbations in Fourier space;}
\item{Calculate the sources for the line-of-sight integral;}
\item{Compute $C_\ell$.}
\end{itemize}
This is the typical procedure followed by a Boltzmann solver, although many details have been omitted.

Boltzmann codes often do far more than just computing the $C_\ell$s, being able to calculate also the polarization of the CMB, as well as its lensing at late time. They are also able to calculate the matter power spectrum at a range of scales, along with the galaxy number count spectrum and that of the weak lensing potential. Second order codes are able to go even further and can calculate the intrinsic bispectrum of the CMB, as well as the generation of vorticity and cosmic magnetic fields. In Chapter~\ref{Ch_iso2}, we discuss how to extend these calculations at second order to the case of isocurvature initial conditions. 

\renewcommand{\CVSrevision}{\version$Id: chapter.tex,v 1.3 2009/12/17 18:16:48 ith Exp $}

\def\be{\begin{equation}}
\def\ee{\end{equation}}
\def\bea{\begin{eqnarray}}
\def\eea{\end{eqnarray}}
\def\lb{\left(}
\def\rb{\right)}
\def\lbs{\left[}
\def\rbs{\right]}
\def\lbc{\left\{}
\def\rbc{\right\}}
\def\p{\partial}
\def\n{\nabla}
\def\dkk{\p_k\p^k}
\def\sh{{\sigma}}
\def\L{{\pounds}}
\def\R{{{\cal{R}}}}
\def\Va{{V_\alpha}}
\def\Q{{\cal{Q}}}
\def\ka{\kappa}
\def\eps{{\epsilon}}
\def\dr{{\delta\rho}}
\def\vp{{\varphi}}
\newcommand\dvp[1]{{\delta\varphi_{#1}}}
\newcommand\dvpI[1]{{\delta\varphi_{{#1}I}}}
\newcommand\dvpK[1]{{\delta\varphi_{{#1}K}}}
\newcommand\dvpM[1]{{\delta\varphi_{{#1}M}}}
\newcommand\dvpN[1]{{\delta\varphi_{{#1}N}}}
\newcommand\dvpL[1]{{\delta\varphi_{{#1}L}}}
\newcommand\dvpJ[1]{{\delta\varphi_{{#1}J}}}
\newcommand\dU[1]{{\delta U_{#1}}}
\def\S{{\cal{S}}}
\def\H{{\cal H}}
\def\cs2{c_{\text{s}}^2}
\def\U0{{\bar U_0}}
\def\wt{\widetilde}
\def\dT{{\delta{\bf T}_1}}
\def\dTT{{\delta{\bf T}_2}}
\def\drho{{\delta\rho_1}}
\def\drhorho{{\delta\rho_2}}
\def\dP{{\delta P_1}}
\def\dPP{{\delta P_2}}
\def\dij{\delta_{ij}}
\def\12{\frac{1}{2}}
\def\BkBk{{B_{1,k}B_{1,}^{~k}}}
\def\ppij{{\p^{-1}_i\p^{-1}_j}}
\def\dvpdvpKll{\delta\vp_{1K,l}\delta\vp_{1K,}^{~~~~l}}
\def\Xkdvk{{\sum_K X_K \delta\vp_{1K}}}
\def\M{{\cal{M}}}
\def\k{{\bf{k}}}
\def\q{{\bf{q}}}
\def\kvi{{{k^i}}}
\def\qvi{{{q^i}}}
\def\pvi{{{p^i}}}
%

%Gauge subscripts KAM

\def\tom{{\text{tom}}}
\def\syn{{\text{syn}}}
\def\com{{\text{com}}}
\def\fg{{\text{flat}}}
\def\lg{{\ell}}
\def\udg{{\delta\rho}}

%gauge invariants
\def\J{{J}}
\def\A{{A}}
\def\V{{V}}
\def\W{{W}}
\def\U{{\Upsilon}}

\def\X{{\cal{X}}}
\def\Xv{{{\cal{X}}_{\text{v}}}}

\newcommand\eq[1]{Eq.~(\ref{#1})}
\newcommand\eqs[1]{Eqs.~(\ref{#1})}
%
%

% % % % % % % % % % % % % % % % % % % % % % % % % % % % % % % % 
% =========================================================== %
% % % % % % % % % % % % % % % % % % % % % % % % % % % % % % % % 
\chapter{The evolution of the curvature perturbation in the presence of vectors and tensors}
\label{Ch_zeta2}
% % % % % % % % % % % % % % % % % % % % % % % % % % % % % % % % 
% =========================================================== %
% % % % % % % % % % % % % % % % % % % % % % % % % % % % % % % % 

\section{Introduction}

In this chapter, we study the evolution of the curvature perturbation at second order. We have already seen in Chapter~\ref{Ch_SMC} that, for simple inflation models, this quantity is conserved on super-horizon scales at the linear level. Here we study the effect of non-linearities on that result, which include the mode coupling between scales and between scalars, vectors and tensors. We start by
reviewing the different versions of the gauge-invariant curvature
perturbation on uniform density hypersurfaces and show how they are
related. We then derive the evolution equation for each convention and
compare the results. Besides the scalar contributions we also keep all
vector and tensor contributions, as well as the anisotropic
stress. Finally, we take the large-scale limit and check for the
conditions of existence of conserved quantities.\\

The chapter is organized as follows. In the next section, we present
the different conventions for the metric perturbations and give the
necessary gauge transformations. The different definitions of $\zeta$
are given in Section \ref{GI}, along with a number of auxiliary gauge
invariant quantities. A derivation of the evolution of $\zeta^{(2)}$ is
presented in Section \ref{evolution}. We then present our conclusions
in Section \ref{conclusion}.

\section{Definitions of the spatial metric}
\label{Conv}

In this initial section, we build on the treatment of cosmological perturbation theory developed in Chapter~\ref{Ch_CPT} and introduce four different ways to split the the metric tensor into perturbations. These vary in the way the spatial part of the
metric is arranged. The version that we will use in most of the
calculations below takes the form given by Eqs.~\eqref{g00CPT}--\eqref{gijCPT}, which we now reproduce:
\begin{align}
g_{00}=&-a^2\lb 1+2\phi\rb\,,\\
g_{i0}=&a^2\lb B_{,i}-S_i\rb\,,\\
g_{ij}=&a^2\lbs\delta_{ij}+2C_{ij}\rbs\,.
\end{align}
The first convention we will treat is defined by arranging $C_{ij}$
as in Eq.~\eqref{gij}, i.e.
\begin{equation}
\label{Cijnew}
C_{ij}=-\psi\delta_{ij}+E_{,ij}+F_{(i,j)}+h_{ij}\,.
\end{equation}
This is the metric convention used by Mukhanov, Feldman and Brandenberger in Ref. \cite{Mukhanov:1990me} and Malik and Wands in Ref. \cite{Malik:2008im}, for example.

This first convention for $\psi$ can be understood, at first order, as the perturbation to the intrinsic curvature, as explained in Appendix \ref{app3}. As we will see, the other conventions do not have this property, but can be generally understood as perturbations to the scale factor $a(t)$. Appendix \ref{app3} also contains a definition of the scale factor from the extrinsic curvature, which is more easily relatable to the versions of $\psi$ given below.

A variation from the form given in Eq.~\eqref{Cijnew} consists of collecting the trace
of $C_{ij}$ in a single variable, here denoted by $\psi_T$. This split was used, e.g., by Bardeen in Ref.~\cite{Bardeen:1980kt} and also by Kodama and Sasaki in Ref.~\cite{KodamaSasaki}, where $\psi_T$ was denoted by $H_L$. The
perturbation to the spatial part of the metric becomes
\be
\label{gijT}
C_{ij}=-\psi_T\delta_{ij}+E_{,ij}-\frac13\delta_{ij}\n^2E+F_{(i,j)}+h_{ij}\,,
\ee
which, upon comparison with the previous convention, Eq.~\eqref{Cijnew},
shows that the new curvature perturbation $\psi_T$ is related to $\psi$ at all
orders via
\be
\label{psT}
\psi_T=\psi-\frac13\n^2E\,.
\ee

The third kind of decomposition of $g_{ij}$ we will treat is similar
to the second one, Eq. \eqref{gijT}, but factors out the determinant
of the spatial part of the metric, instead of the trace. This is the decomposition used by Salopek and Bond in Ref. \cite{Salopek:1990jq} and also by Maldacena in Ref. \cite{Maldacena}. It can be
written as
\be
\label{gijD}
g_{ij}=a^2e^{2\psi_D}[e^{\omega}]_{ij}\,,
\ee
in which $\omega$ is a traceless tensor and $\psi_D$ is the curvature
perturbation of interest in this convention, defined by
\be
e^{6\psi_D}\equiv \det (g_{ij}/a^2)\,.
\ee 
% NEW STUFF
This quantity is usually interpreted as being a perturbation to the number of e-folds \cite{Salopek:1990jq}, $N$, given by $N=\ln a-\psi_D-\psi_D^2$. A related interpretation would be to think of it as a perturbation to the volume of spatial hypersurfaces, as it is proportional to the determinant of the spatial metric.
%%%%
It can be shown \cite{Malik:2008im}, that,
up to second order, $\psi_D$ is related to the other conventions by the
following expressions,
\begin{align}
\psi_{D}=&-\psi_{T}-\frac13C_{ij}C^{ij}=\\
=&-\psi+\frac13\n^2E-\psi^2-\frac13h_{ij}h^{ij}
+\frac23\psi\n^2E\nonumber\\
&-\frac23 h^{ij} \lb E_{,ij}+F_{i,j}\rb-\frac13 F_{(i,j)}F^{j,i}-\frac13E_{,ij}\lb2 F^{i,j}+E^{,ij}\rb\,.
\end{align}

The fourth convention is not a variation of $g_{ij}$ per se, but only a different way of defining the curvature perturbation. As with the third definition, Eq. \eqref{gijD}, we factor out the determinant of the spatial part of the metric, but in this case, we use the inverse metric to do so. Therefore, it is now defined as
\be
\label{gijI}
g^{ij}=a^{-2}e^{-2\psi_{I}}[e^{\omega_I}]^{ij},
\ee
in which, again, $\omega_I$ is a traceless tensor and $\psi_I$ is the new version of the curvature perturbation, determined by $e^{-6\psi_I}\equiv \det (g^{ij} a^2)$.
% NEW STUFF
To our knowledge, this is the first time this definition has been used in the literature. Concerning its interpretation, it can still be seen as a perturbation to the scale factor and we find it to be equal to the integrated expansion, when the latter is evaluated in a comoving threading (see Appendix \ref{app3} for more details).
%%%%%%%%%%
Comparing this new version of $\psi$ to the original one, we find the following relation
\begin{align}
\psi_{I}=&-\psi+\frac13\n^2E-\psi^2-\frac13h_{ij}h^{ij}+\frac23\psi\n^2E+\frac16\lb B_{,i}^{\vphantom{,i}}-S_{i}\rb\lb B^{,i}-S^{i}\rb\nonumber\\
&-\frac23 h^{ij} \lb E_{,ij}+F_{i,j}\rb-\frac13 F_{(i,j)}F^{j,i}-\frac13E_{1,ij}\lb2 F^{i,j}+E^{,ij}\rb.
\end{align}

We will use these four conventions to define different versions of
the gauge-invariant curvature perturbation in the next section.

\section{Gauge-invariant quantities}
\label{GI}

The method we use to generate gauge-invariant variables is described in Chapter~\ref{Ch_CPT} and Refs.~\cite{Malik:2008im,Malik:2008yp}, and starts with
performing a gauge transformation on a variable of interest,
e.g.~$\psi$. One then substitutes the gauge generator components
$\xi^\mu$ with those obtained by solving a gauge fixing constraint,
e.g.~$\widetilde{\dr}=0$. The end result is a gauge-invariant
quantity, e.g.~the curvature perturbation in uniform density
hypersurfaces, $\zeta$. We apply this method for the quantities of
interest in the subsections below.

\subsection{Curvature perturbation on uniform density hypersurfaces}

The focus of this chapter is the curvature perturbation on uniform
density hypersurfaces $\zeta$. As was already mentioned in Chapter~\ref{Ch_CPT} above, it is
defined to be equal to $-\psi$ in the gauge in which the density field
is uniform ($\widetilde{\dr}=0$). Starting with our first convention
for the metric, Eq.~\eqref{Cijnew}, this condition is sufficient to fully
construct $\zeta$ at first order as (\cite{Bardeen:1983qw,Wands:2000dp})
\be
\label{z1}
\zeta^{(1)}\equiv-\psi^{(1)}-\Hh\frac{\dr^{(1)}}{\rho'}\,.
\ee
However, at second order, one is also forced to specify the first-order gauge to define this curvature perturbation unambiguously. For
this convention of the metric tensor, Eq.~\eqref{Cijnew}, we will use the
following gauge conditions to define $\zeta^{(2)}$ (\cite{MW2004})
\be
\zeta^{(2)}\equiv-\widetilde{\psi^{(2)}}~, \text{ if }\widetilde{\dr^{(2)}}=\widetilde{\dr^{(1)}}=\widetilde{E^{(1)}}=0~,~~\widetilde{F^{(1)}_i}=0\,.
\ee
These add a flat threading ($\widetilde{E^{(1)}}=0\,,\ \widetilde{F^{(1)}_i}=0$) to the uniform density gauge (often defined only with $\widetilde{\dr}=0$). The general
formal expression for $\zeta^{(2)}$ is given in Ref.~\cite{Malik:2008im}. In full detail,
the formula is rather complicated and we write it here with the
r.h.s.~evaluated in flat gauge,
\begin{align}
\label{z2}
\zeta^{(2)}=&-\frac{\H}{\rho'}\dr^{(2)}+\frac{1}{\rho^{\prime 2}}\lb 2\H^2+\H'-\H\frac{\rho''}{\rho'}\rb\dr^{(1)\,2}+\frac{2\H}{\rho^{\prime2}}\dr^{(1)}\dr^{(1)\prime}\nonumber\\
&-\frac 1 {2 \rho^{\prime 2}} \dr_{,k}^{(1)}\dr^{(1),k}-\frac 1 {\rho'}\lb B_{,k}^{(1)}-S_{k}^{(1)}\rb\dr^{(1),k}\nonumber\\
&+\nabla^{-2}\lbc\frac12\lbs\frac 1{\rho'^2}\dr^{(1),i}\dr^{(1),j}+\frac 2{\rho'}\dr^{(1),(i}\lb B^{(1),j)}-S^{(1)\,j)}\rb\rbs_{,ij}\right.\nonumber\\
&\left.+\frac{1}{\rho'}\lb h^{(1)\,ij\prime}+2\H h^{(1)\,ij}\rb \dr_{,ij}^{(1)}\rbc\,.
\end{align}
We can see that, in contrast with the first-order result, the second
order $\zeta$ is much harder to relate to density perturbations in
flat gauge, given the presence of vectors and tensors. In spite of
this, this expression is still useful in writing the gauge-invariant
curvature perturbation in terms of multiple scalar fields, as is done
in Refs.~\cite{CNM,Dias:2014msa}.
\\

Let us now move to the second convention of the metric,
Eq.~\eqref{gijT}. In this case, $\widetilde{\dr}=0$ is no longer a
sufficient gauge condition to define an invariant, even at first
order; one must also specify the scalar part of the threading, due to
the inclusion of $E$ in the definition of $\psi_T$ (see
Eq.~\eqref{psT}). The extra condition we choose here is
$\widetilde{v^{(1)}}=0$, which results in the following
expression\footnote{An alternative choice would be
$\widetilde{E^{(1)}}=0$, but that would simply result in the expression
for the original metric convention, as
$\widetilde{\psi_{T}^{(1)}}=\widetilde{\psi^{(1)}}$ in that case.}
\be
\label{zT1w}
\zeta_{T}^{(1)}=-\psi_{T}^{(1)}-\frac{\Hh}{\rho'}\dr^{(1)}+\frac{1}{3}\n^2\int v^{(1)} d\tau\,,
\ee
in which the integral in conformal time is indefinite. The
introduction of these integrals is the disadvantage of using the gauge
condition, $\widetilde{v^{(1)}}=0$. This might be problematic, as this
condition only sets the gauge up to an arbitrary function of the
spatial coordinates, which, in turn, might spoil the gauge invariance of
the new variable. In spite of this, it is possible to construct a gauge
invariant quantity, by defining it to be
\be
\label{zT1}
\zeta_{T}^{(1)}\equiv\zeta^{(1)}+\frac{1}{3}\n^2\int \J^{(1)} d\tau\,.
\ee
with $\J^{(1)}$ being the gauge-invariant velocity on flat hypersurfaces, defined by
\be
\label{j1}
\J^{(1)}=E^{(1)\prime}+v^{(1)}\,.
\ee
While the integral in Eq.~\eqref{zT1} is still indefinite, the integrand is gauge invariant and, therefore, this is the
definition we use.

At second order, one sets the second order gauge in the same way,
i.e.~$\wt{\dr^{(2)}}=\widetilde{v^{(2)}}=0$ and, to avoid additional issues
with indefinite integrals, one can choose
$\widetilde{\dr^{(1)}}=\widetilde{E^{(1)}}=\widetilde{F^{(1)}_i}=0$ for the first
order gauge fixing. With this choice, we find
\be
\label{zT2}
\zeta_{T}^{(2)}=\zeta^{(2)}+\frac{1}{3}\n^2\int \J^{(2)} d\tau,
\ee
in which $\J^{(2)}$ is the second order equivalent of $\J^{(2)}$ in this
gauge, i.e.~it equals $E^{(2)\prime}+v^{(2)}$ in the gauge obeying
$\widetilde{\dr^{(1)}}=\widetilde{E^{(1)}}=\widetilde{F^{(1)}_i}=0$. As is visible
in the expression above, Eq.~\eqref{zT1}, the only variable of
interest is $\n^2\J^{(2)}$ and hence, for shortness of presentation, that
is all we show below, with the r.h.s. evaluated in flat gauge
\begin{align}
\label{j2}
\n^2\J^{(2)}=&\n^2v^{(2)}+\frac{2}{\rho'}\lbs\dr^{(1)}\lb\H(v_{V}^{(1)\,i}+v^{(1),i})-v_{V}^{(1)\,i\prime}-v^{(1)\prime,i}\rb\rbs_{,i}\\
&+\lbs\frac{\dr_{,i}^{(1)}\dr^{(1),i}}{2\rho'^2}+\frac{\lb B^{(1),i}-S^{(1)\,i}\rb \dr_{,i}^{(1)}}{\rho'}\right.\nonumber\\
&\left.+\nabla^{-2}\lbc-\frac32\lbs\frac 1{\rho'^2}\dr^{(1),i}\dr^{(1),j}+\frac 2{\rho'}\dr^{(1),(i}\lb B^{(1),j)}-S^{(1)\,j)}\rb\rbs_{,ij}\right.\right.\nonumber\\
&\left.\left.-\frac{3}{\rho'}\lb h^{(1)\,ij\prime}+2\H h^{(1)\,ij}\rb \dr_{,ij}^{(1)}\rbc\rbs'\,.\nonumber
\end{align}
As we will see in Section \ref{evolution}, this quantity is
relevant regardless of the choice of convention for the metric, as it
will appear in the evolution equation for the curvature perturbation.
\\

Let us now turn to the third convention of the metric,
Eq.~\eqref{gijD}. For this case, $\zeta_D$ will be defined as being
equal to $\psi_D$ instead of $-\psi_D$, in order to keep the same sign
as $\zeta$. Starting at first order, we see that we get either
$\zeta_{D}^{(1)}=\zeta^{(1)}$ or $\zeta_{D(v)}^{(1)}=\zeta_{T}^{(1)}$, depending on
whether we choose $\widetilde{E^{(1)}}=0$ or $\widetilde{v^{(1)}}=0$,
respectively, for fixing the threading. The second order result is
more interesting, as there is no gauge fixing for which it is equal to
either of the other definitions above. In the most conservative case,
the choice of gauge fixing is $\widetilde{\dr^{(2)}}=\widetilde{E^{(2)}}=0$ at
second order and
$\widetilde{\dr^{(1)}}=\widetilde{E^{(1)}}=\widetilde{F^{(1)}_i}=0$ at first
order. This results in\footnote{This result is well known in the case without tensors. See, for example, Refs. \cite{BKMR,Dias:2014msa,LV}.}
\be
\label{zD2}
\zeta_{D}^{(2)}=\zeta^{(2)}-\frac23h_{ij}^{(1)}h^{(1)\,ij}-2(\zeta^{(1)})^2\,.
\ee
A different gauge fixing is $\widetilde{\dr^{(2)}}=\widetilde{v^{(2)}}=0$ and
$\widetilde{\dr^{(1)}}=\widetilde{v^{(1)}}=\widetilde{v_{V\,i}^{(1)}}=0$, for which
the result is
\begin{align}
\label{zD2v}
\zeta_{D(v)}^{(2)}=&\zeta^{(2)}+\frac13 \int \n^2\J^{(2)}d\tau-\frac23h^{(1)}_{ij}h^{(1)\,ij}-2(\zeta^{(1)})^2+2\zeta^{(1)}_{,i}\int\lb \J^{(1),i}+\V^{(1)\,i}\rb d\tau\nonumber\\
&+\frac23\int\lbc\lbs\lb \J^{(1),i}+\V^{(1)\,i}\rb\U^{(1)}\rbs_{,i} +\n^2\J_{,i}^{(1)}\int \lb \J^{(1),i}+\V^{(1)\,i}\rb d\tau'\rbc d\tau\,,
\end{align}
in which $\V^{(1)\,i}$ is the gauge-invariant velocity vector perturbation
in flat hypersurfaces and $\U$ is the gauge-invariant lapse
perturbation in uniform density hypersurfaces. In a general gauge,
these quantities are given by
\begin{align}
&\V^{(1)\,i}=v_{V}^{(1)\,i}+F^{(1)\,i\prime},\\
&\U^{(1)}=\phi^{(1)}-\H\frac{\dr^{(1)}}{\rho'}-\lb\frac{\dr^{(1)}}{\rho'}\rb'\,.
\end{align}

For the fourth version of the curvature perturbation, Eq. \eqref{gijI}, the procedure is very similar to the one for the third convention. As in the previous case, the first-order quantities obey $\zeta_{I}^{(1)}=\zeta^{(1)}$ or $\zeta_{I(v)}^{(1)}=\zeta_{T}^{(1)}$, depending on whether $\widetilde{E^{(1)}}=0$ or $\widetilde{v^{(1)}}=0$ is chosen for setting the threading. At second order, the results are
\begin{align}
\label{zI2}
\zeta_{I}^{(2)}=&\ \zeta^{(1)}-\frac23h_{ij}^{(1)}h^{(1)ij}-2(\zeta^{(1)})^2\\
&+\frac13\lb \W_{i}^{(1)}-\V_{i}^{(1)}+\A^{(1)}_{,i}-\J^{(1)}_{,i}\vphantom{\A^{(1),i}}\rb\lb \W^{(1)\,i}-\V^{(1)\,i}+\A^{(1),i}-\J^{(1),i}\rb\,,\nonumber
\end{align}
if the gauge is fixed with $\widetilde{\dr^{(2)}}=\widetilde{E^{(2)}}=0$ and $\widetilde{\dr^{(1)}}=\widetilde{E^{(1)}}=\widetilde{F^{(1)}_i}=0$, and 
\begin{align}
\label{zI2v}
\zeta_{I(v)}^{(2)}=&\ \zeta^{(2)}+\frac13 \int \n^2\J^{(2)}d\tau-\frac23h_{ij}^{(1)}h^{(1)\,ij}-2(\zeta^{(1)})^2\\
&+\frac13\lb \W^{(1)}_{i}+\A^{(1)}_{,i}\vphantom{\A_{1}^{,i}}\rb\lb \W^{(1)i}+\A^{(1),i}\rb+2\zeta^{(1)}_{,i}\int\lb \J^{(1),i}+\V^{(1)\,i}\rb d\tau\nonumber\\
&+\frac23\int\lbc\lbs\lb \J^{(1),i}+\V^{(1)\,i}\rb\U^{(1)}\rbs_{,i} +\n^2\J^{(1)}_{,i}\int \lb \J^{(1),i}+\V^{(1)i}\rb d\tau'\rbc d\tau\,,\nonumber
\end{align}
when the gauge choice is  $\widetilde{\dr^{(2)}}=\widetilde{v^{(2)}}=0$ and $\widetilde{\dr^{(1)}}=\widetilde{v^{(1)}}=\widetilde{v_{Vi}^{(1)}}=0$. The new first-order gauge-invariant quantities that appear are the vector velocity in zero shift gauge, $\W^{(1)}_i$, and the
momentum perturbation in uniform density gauge, $\A^{(1)}$. They are given
by
\begin{align}
&\W^{(1)}_i=v^{(1)}_{Vi}-S^{(1)}_i,\\
&\A^{(1)}=v^{(1)}+B^{(1)}+\frac{\dr^{(1)}}{\rho'}.
\end{align}

\subsection{Non-adiabatic pressure}

One of the quantities determining the evolution of the curvature
perturbation is the non-adiabatic
pressure \cite{Mollerach:1989hu,Wands:2000dp,nonad,KodamaSasaki,Bardeen:1980kt} as we have already mentioned in Chapter \ref{Ch_SMC}. It is defined as the
deviation from the adiabatic relation as
\be
\label{nad}
\delta P=c_\text{s}^2\delta\rho+\delta P_\text{nad}\,,
\ee
with $c_{\text s}$ the adiabatic sound speed defined as $c_{\text
s}^2=P'/\rho'$. At first order, this definition automatically
generates a gauge-invariant quantity, but, at second order, this is
not sufficient and one can define many quantities that reproduce the definition, Eq.~\eqref{nad}, when particular gauge choices are made. Our first choice is to
define a gauge-invariant quantity in the gauge in which
$\widetilde{\dr^{(1)}}=\widetilde{E^{(1)}}=\widetilde{F^{(1)}_i}=0$. In a
general gauge, this quantity is given by
\begin{align}
\label{nad2}
\delta P_{\text{nad}}^{(2)}=&\delta P^{(2)}-c_{\text s}^2\dr^{(2)}-\frac{2}{\rho'}\dr^{(1)}\delta P^{(1)\prime}+\lb\frac{P''}{\rho'^2}-\frac{P'\rho''}{\rho'^3}\rb(\dr^{(1)})^2\nonumber\\
&+\frac{2c_s^2}{\rho'}\dr^{(1)}\dr^{(1)\prime}-2\lb F^{(1)\,i}+E^{(1),i}\rb\delta P_{\text{nad},i}^{(1)}\,,
\end{align}
which we still name $\delta P_{\text{nad}}$, for simplicity. With the different choice of threading,
$\widetilde{v^{(1)}}=\widetilde{v_{Vi}^{(1)}}=0$, one finds instead the quantity
\begin{align}
\label{nad2v}
\delta P_{\text{nad}\, (v)}^{(2)}=\delta P_{\text{nad}}^{(2)}+2\delta P_{\text{nad},i}^{(1)}\int\lb \V^{(1)\,i}+\J^{(1),i}\rb d\tau\,.
\end{align}
For a barotropic fluid, with $P=P(\rho)$, both expressions vanish, as
can be easily checked by evaluating them in their defining gauge,
i.e.~with $\dr^{(1)}=E^{(1)}=F^{(1)}_i=0$.

The quantities presented so far include the full set of gauge-invariant quantities required for the full derivation of the evolution equations below.

\section{Evolution equations}\label{evolution}

In this section, we present the derivation of the evolution equations
for all versions of $\zeta$. Our strategy consists of calculating the
derivative of expression \eqref{z2} and using only the perturbed
energy-momentum conservation equations up to second order to simplify
the result. Lastly, we substitute the gauge dependent variables for
gauge-invariant ones, using the expressions found in the previous
section, to arrive at our final result. Having found the result for
$\zeta^{(2)}$ in the original convention of the metric, Eq.~\eqref{Cijnew},
we then rewrite the evolution equation in terms of the different
definitions of $\zeta$.

\subsection{Fluid equations}

As shown in Chapter~\ref{Ch_CPT}, energy-momentum conservation, $\n_\nu T^{\mu\nu}=0$, governs the
evolution of the fluid density and velocity, through Eqs.~ \eqref{enecons}, \eqref{vsTeq} and \eqref{vVTeq}. We
reproduce these evolution equations here, order by order, evaluating them in flat gauge, for brevity of presentation. 

The first-order energy conservation equation is given by
\be
\label{evorho1}
\dr^{(1)\prime}+3\H\lb\dr^{(1)}+\delta P^{(1)}\rb+\lb\rho+P\rb\n^2v^{(1)}=0~\,,
\ee
while momentum conservation is
\be
\label{evov1}
\delta P^{(1)}_{,k}+(\rho+P)\lbs Z^{(1)\prime}_{k}+\phi^{(1)}_{,k}+\lb1
-3 c_s^2\rb\H Z^{(1)}_{k}\rbs+\frac23 \n^2\Pi^{(1)}_{,k}+\frac12 \n^2\Pi^{(1)}_{k}=0\,,
\ee
where the momentum perturbation $Z^{(1)}_{k}$ is given by
\be
Z^{(1)}_{k}=v_{V\,k}^{(1)}-S^{(1)}_{k}+B^{(1)}_{,k}+v^{(1)}_{,k}\,.
\ee
At second order, we only require the energy conservation equation, which is
\begin{align}
\label{evorho2}
\delta\rho^{(2)\prime}=&-3\H\lb\dr^{(2)}+\delta P^{(2)}\rb-\lb\rho+P\rb\n^2v^{(2)}-2\lb\delta P^{(1)}+\dr^{(1)}\rb \n^2 v^{(1)}\nonumber\\
&-2\dr^{(1)}_{,k}\lb v^{(1)\,k}_{V}+v^{(1),k}\rb-2 \delta P^{(1)}_{,k} Z^{(1)\,k}-Z^{(1)\,k}\lb\frac43 \n^2\Pi^{(1)}_{,k}+\n^2\Pi^{(1)}_{k}\rb\nonumber\\
&-\lb \rho+P\rb \lbs\vphantom{\lb v_{V1}^{\ \ k}+v_1^{\ ,k}\rb} 4 Z^{(1)\prime}_{k} Z^{(1)\,k}+2\lb 1-3 c_s^2\rb \H Z^{(1)}_{k}Z^{(1)\,k} + 2\phi^{(1)}_{,k}Z^{(1)\,k}\right.\nonumber\\
&\left.+2\phi^{(1)}_{,k}\lb v_{V}^{(1)\,k}+v^{(1),k}\rb+2\phi^{(1)} \n^2 v^{(1)}-4h^{(1)\prime}_{ij}h^{(1)\,ij}\rbs\\
&-2\lb h^{(1)\prime}_{ij}+v^{(1)}_{V\,i,j}+v^{(1)}_{,ij}\rb\lb\Pi^{(1)\,ij}+\Pi^{(1)\,(i,j)}+\Pi^{(1),ij}-\frac13\delta^{ij}\n^2\Pi^{(1)}\rb\,.\nonumber
\end{align}
The above equations are sufficient to derive evolution equations
for the curvature perturbation at first and at second order \cite{Wands:2000dp}.

\subsection{Evolution of the curvature perturbation}\label{EvoCurv}

We can now derive the evolution equation for the curvature
perturbation on uniform density hypersurfaces. We follow the strategy
stated at the beginning of this section. At first order, the result is
well known to be
\be
\label{z1evoc4}
\zeta^{(1)\prime}=-\frac13\n^2\J^{(1)}-\H\frac{\delta P^{(1)}_{\text{nad}}}{\rho+P}~,
\ee
where only the first-order energy conservation equation was used. On
large scales (``$\n\rightarrow0$") and in the absence of non-adiabatic
pressure, one finds the familiar conservation equation $\zeta^{(1)\prime}=0$, which was used in Chapter \ref{Ch_SMC} to justify the evaluation of the spectrum of inflationary perturbations at horizon crossing.

For the other conventions for the curvature perturbation, $\zeta^{(1)}_{T}$, $\zeta^{(1)}_{D(v)}$ and $\zeta_{I(v)}^{(1)}$, the evolution equation at first order is the
same and is given by
\be
\label{zetaT1p}
\zeta_{T}^{(1)\prime}=-\H\frac{\delta P^{(1)}_{\text{nad}}}{\rho+P}~,
\ee
which shows these versions of $\zeta^{(1)}$ are conserved at all scales,
when non-adiabatic pressure is negligible \cite{Lyth:2004gb,LM}.

At second order, the complexity increases. The detailed procedure to obtain the
final result is as follows: use the energy conservation equation at
first (Eq.~\eqref{evorho1}) and second order (Eq.~\eqref{evorho2}) to
substitute for $\dr^{(1)\prime}$ and $\dr^{(2)\prime}$ and substitute
$\lb\frac43 \n^2\Pi^{(1)}_{,k}+\n^2\Pi^{(1)}_{k}\rb$ with the momentum
conservation equation, Eq.~\eqref{evov1}. The last step is to use the
defining expressions of the gauge invariants to eliminate all gauge
dependent variables. The final result is given by\footnote{Note the
absence of inverse Laplacians. That is explained by an exact
cancellation between the terms in $\zeta^{(2)\prime}$ and those in $\n^2J^{(2)}$,
as can be shown by comparing equations \eqref{j2} and \eqref{z2}.}
\begin{align}
&\lb-\zeta^{(2)}+2(\zeta^{(1)})^2-\frac13\lb \W^{(1)}_{i}+\A^{(1)}_{,i}\vphantom{\A_{1}^{,i}}\rb\lb \W^{(1)\,i}+\A^{(1),i}\rb+\frac23h^{(1)}_{ij}h^{(1)\,ij}\rb'=\\
&\frac13\n^2\J^{(2)}+\H\frac{\delta P^{(2)}_{\text{nad}}}{\rho+P}-2\H\lb\frac{\delta P^{(1)}_{\text{nad}}}{\rho+P}\rb^2+\frac{2}{3}\lbs\U^{(1)}\lb \V^{(1)\,i}+\J^{(1),i}\rb\rbs_{,i}+2\zeta^{(1)}_{,i}\lb \V^{(1)\,i}+\J^{(1),i}\rb\nonumber\\
&-\frac{2\H}{\rho'}\lb\Pi^{(1)}_{ij}+\Pi^{(1)}_{(i,j)}+\Pi^{(1)}_{,ij}-\frac{1}{3}\delta_{ij}\n^2\Pi^{(1)}\rb\lb h^{(1)\,ij\prime}+\V^{(1)\,i,j}+\J^{(1),ij}\rb\nonumber.
\end{align}
%%%%%%%%%%%%%%%%%% NEW STUFF %%%%%%%%%%%%%%%%%%%%%%%%%%%%55
We are now able to identify the different terms that source the evolution of $\zeta^{(2)}$. We note, in particular, the appearance of vector and tensor source terms as well as the anisotropic stress which did not appear at first order in this equation\footnote{Note however, that the scalar part of the anisotropic stress tensor would source the evolution of $\zeta$ at first order by acting on the evolution of $\n^2J$. This can be seen more clearly by deriving Eq.~\eqref{z1evoc4} and using the momentum conservation equation, Eq.~\eqref{evov1}, to substitute for $\n^2J$:
\begin{align}
\zeta^{(1)\prime\prime}+\H\zeta^{(1)\prime}-\frac{P'}{3(\rho+P)}\n^2A^{(1)}-\frac13\n^2\Phi^{(1)}+\left(\H\frac{\delta P_{\text{nad}}^{(1)}}{\rho+P}\right)'\\
+\H^2\frac{\delta P^{(1)}_{\text{nad}}} {\rho+P}-\frac{\n^2\delta P^{(1)}_{\text{nad}}}{3 (\rho+P)}-\frac{2}{9 (\rho+P)}\n^2\n^2\Pi^{(1)}=0\,,\nonumber
\end{align}
in which $\Phi$ is one of the Bardeen potentials, given in terms of the variables in this chapter as $\Phi^{(1)}=\Upsilon^{(1)}+\H(A^{(1)}-J^{(1)})+(A^{(1)}-J^{(1)})'$.
}.
%%%%%%%%%%%%%%%%%%%%%%%%%%%%%%%%%%%%%%%%%%%%%%%%%%%%%%%%%%5

We are now in the position to substitute for the other versions of
$\zeta$ and find their evolution equations. For $\zeta^{(2)}_{T}$, we find
\begin{align}
&\lb-\zeta^{(2)}_{T}+2(\zeta^{(1)})^2-\frac13\lb \W^{(1)}_{i}+\A^{(1)}_{,i}\vphantom{\A_{1}^{,i}}\rb\lb \W^{(1)\,i}+\A^{(1),i}\rb+\frac23h^{(1)}_{ij}h^{(1)\,ij}\rb'=\\
&\H\frac{\delta P^{(2)}_{\text{nad}}}{\rho+P}-2\H\lb\frac{\delta P^{(1)}_{\text{nad}}}{\rho+P}\rb^2+\frac{2}{3}\lbs\U^{(1)}\lb \V^{(1)\,i}+\J^{(1),i}\rb\rbs_{,i}+2\zeta^{(1)}_{,i}\lb \V^{(1)\,i}+\J^{(1),i}\rb\nonumber\\
&-\frac{2\H}{\rho'}\lb\Pi^{(1)}_{ij}+\Pi^{(1)}_{(i,j)}+\Pi^{(1)}_{,ij}-\frac{1}{3}\delta_{ij}\n^2\Pi^{(1)}\rb\lb h^{(1)ij\prime}+\V^{(1)\,i,j}+\J^{(1),ij}\rb\nonumber,
\end{align}
while $\zeta_{D}^{(2)}$ evolves as
\begin{align}
&\lb-\zeta_{D}^{(2)}-\frac13\lb \W^{(1)}_{i}+\A^{(1)}_{,i}\vphantom{\A_{1}^{,i}}\rb\lb \W^{(1)\,i}+\A^{(1),i}\rb\rb'=\\
&\frac13\n^2\J^{(2)}+\H\frac{\delta P_{\text{nad}}^{(2)}}{\rho+P}-2\H\lb\frac{\delta P_{\text{nad}}^{(1)}}{\rho+P}\rb^2+\frac{2}{3}\lbs\U^{(1)}\lb \V^{(1)\,i}+\J^{(1),i}\rb\rbs_{,i}+2\zeta^{(1)}_{D,i}\lb \V^{(1)\,i}+\J^{(1),i}\rb\nonumber\\
&-\frac{2\H}{\rho'}\lb\Pi^{(1)}_{ij}+\Pi^{(1)}_{(i,j)}+\Pi^{(1)}_{,ij}-\frac{1}{3}\delta_{ij}\n^2\Pi^{(1)}\rb\lb h^{(1)\,ij\prime}+\V^{(1)\,i,j}+\J^{(1),ij}\rb\nonumber,
\end{align}
and the result for $\zeta_{D(v)}^{(2)}$ is
\begin{align}
&\lb-\zeta^{(2)}_{D(v)}-\frac13\lb \W^{(1)}_{i}+\A^{(1)}_{,i}\vphantom{\A_{1}^{,i}}\rb\lb \W^{(1)\,i}+\A^{(1),i}\rb\rb'=\H\frac{\delta P^{(2)}_{\text{nad}(v)}}{\rho+P}-2\H\lb\frac{\delta P_{\text{nad}}^{(1)}}{\rho+P}\rb^2\nonumber\\
&-\frac{2\H}{\rho'}\lb\Pi^{(1)}_{ij}+\Pi^{(1)}_{(i,j)}+\Pi^{(1)}_{,ij}-\frac{1}{3}\delta_{ij}\n^2\Pi^{(1)}\rb\lb h^{(1)\,ij\prime}+\V^{(1)\,i,j}+\J^{(1),ij}\rb.
\end{align}
The simplest evolutions equations are found for the $\zeta_{I}^{(2)}$ and $\zeta_{I(v)}^{(2)}$ versions of the gauge-invariant curvature perturbation. They are given by
\begin{align}
&\lb-\zeta_{I}^{(2)}+\frac13\lb \V^{(1)}_{i}+\J^{(1)}_{,i}\vphantom{\A^{(1),i}}\rb\lb \V^{(1)\,i}+\J^{(1),i}-2\W^{(1)\,i}-2\A^{(1),i}\rb \rb'=\\
&\frac13\n^2\J^{(2)}+\H\frac{\delta P^{(2)}_{\text{nad}}}{\rho+P}-2\H\lb\frac{\delta P_{\text{nad}}^{(1)}}{\rho+P}\rb^2+\frac{2}{3}\lbs\U^{(1)}\lb \V^{(1)\,i}+\J^{(1),i}\rb\rbs_{,i}+2\zeta^{(1)}_{I,i}\lb \V^{(1)\,i}+\J^{(1),i}\rb\nonumber\\
&-\frac{2\H}{\rho'}\lb\Pi^{(1)}_{ij}+\Pi^{(1)}_{(i,j)}+\Pi^{(1)}_{,ij}-\frac{1}{3}\delta_{ij}\n^2\Pi^{(1)}\rb\lb h^{(1)\,ij\prime}+\V^{(1)\,i,j}+\J^{(1),ij}\rb,\nonumber
\end{align}
and 
\begin{align}
\label{zI2p}
&-\zeta_{I(v)}^{(2)\prime}=\H\frac{\delta P_{\text{nad}(v)}^{(2)}}{\rho+P}-2\H\lb\frac{\delta P_{\text{nad}}^{(1)}}{\rho+P}\rb^2\\
&-\frac{2\H}{\rho'}\lb\Pi^{(1)}_{ij}+\Pi^{(1)}_{(i,j)}+\Pi^{(1)}_{,ij}-\frac{1}{3}\delta_{ij}\n^2\Pi^{(1)}\rb\lb h^{(1)\,ij\prime}+\V^{(1)\,i,j}+\J^{(1),ij}\rb\nonumber.
\end{align}
This final expression, like its first-order version, Eq. \eqref{zetaT1p}, shows that, in the absence of non-adiabatic pressure and anisotropic stress, this version of the curvature perturbation is conserved on all scales. While this is interesting, in order for this result to be useful, one would likely be forced to estimate the integrals in the defining expression for $\zeta_{I}^{(2)}$, Eq. \eqref{zI2v}. This is not likely to be straightforward, given the indeterminate nature of the integrals. This evolution equation matches the results of Ref. \cite{Enqvist:2006fs} for the integrated expansion in the absence of anisotropic stress, obtained in the covariant approach.

\subsection{Large scale approximation}

Here we perform the large scale approximation, by neglecting all terms with spatial derivatives in the equations above \footnote{This is generally well motivated in the case of some metric potentials, as one expects the perturbed metric to approach the background metric on large scales \cite{Lyth:2004gb}, and we will assume the same is true for the matter variables, including the anisotropic stress. Should this assumption not hold for the particular model under study, then the results in this section are not valid and one should use the full results from section \ref{EvoCurv}.}. We begin by showing the expressions for the different versions of the curvature perturbation in this approximation, evaluated in flat gauge. 

Both $\zeta^{(2)}$ and $\zeta_{T}^{(2)}$ are approximated by
\begin{align}
\label{z2largescale}
\zeta^{(2)}=\zeta_{T}^{(2)}=-\frac{\H}{\rho'}\dr^{(2)}+\frac{1}{\rho'^2}\lb 2\H^2+\H'-\H\frac{\rho''}{\rho'}\rb(\dr^{(1)})^2+\frac{2\H}{\rho'^2}\dr^{(1)}\dr^{(1)\prime}\,,
\end{align}
while the large scale limit for $\zeta_{D}^{(2)}=\zeta_{D(v)}^{(2)}$ is
\begin{align}
\label{z2Dlargescale}
\zeta_{D}^{(2)}=-\frac{\H}{\rho'}\dr^{(2)}+\frac{1}{\rho'^2}\lb \H'-\H\frac{\rho''}{\rho'}\rb(\dr^{(1)})^2+\frac{2\H}{\rho'^2}\dr^{(1)}\dr^{(1)\prime}-\frac23h^{(1)}_{ij}h^{(1)\,ij}\,,
\end{align}
and the limits of $\zeta_{I}^{(2)}$ and $\zeta_{I(v)}^{(2)}$ are
\begin{align}
\label{z2Ilargescale}
\zeta_{I}^{(2)}=&\ -\frac{\H}{\rho'}\dr^{(2)}+\frac{1}{\rho'^2}\lb \H'-\H\frac{\rho''}{\rho'}\rb(\dr^{(1)})^2+\frac{2\H}{\rho'^2}\dr^{(1)}\dr^{(1)\prime}\nonumber\\
&-\frac23h^{(1)}_{ij}h^{(1)\,ij}+\frac13 S^{(1)}_{i} S^{(1)\,i}\,,\\
\zeta_{I(v)}^{(2)}=&\ -\frac{\H}{\rho'}\dr^{(2)}+\frac{1}{\rho'^2}\lb \H'-\H\frac{\rho''}{\rho'}\rb(\dr^{(1)})^2+\frac{2\H}{\rho'^2}\dr^{(1)}\dr^{(1)\prime}\nonumber\\
&-\frac23h^{(1)}_{ij}h^{(1)\,ij}+\frac13 \W^{(1)}_{i} \W^{(1)\,i}\,.
\end{align}
These expressions agree with similar ones obtained through the $\delta N$ formalism, where comparison is possible (see Ref.~\cite{Lyth:2004gb}).

The large scale limit simplifies the evolution equations to
\begin{align}
\lb-\zeta^{(2)}+2(\zeta^{(1)})^2-\frac13\W^{(1)}_{i}\W^{(1)\,i}\right.&\left.+\frac23h^{(1)}_{ij}h^{(1)\,ij}\rb'=\nonumber\\
&\H\frac{\delta P_{\text{nad}}^{(2)}}{\rho+P}-2\H\lb\frac{\delta P_{\text{nad}}^{(1)}}{\rho+P}\rb^2-\frac{2\H}{\rho'}\Pi^{(1)}_{ij}h^{(1)\,ij\prime}\,,
\end{align}
for $\zeta^{(2)}$, here representing both the original $\zeta^{(2)}$ and $\zeta_{T}^{(2)}$;
\be
\lb-\zeta_{D}^{(2)}-\frac13\W^{(1)}_{i}\W^{(1)i}\rb'=\H\frac{\delta P_{\text{nad}}^{(2)}}{\rho+P}-2\H\lb\frac{\delta P_{\text{nad}}^{(1)}}{\rho+P}\rb^2-\frac{2\H}{\rho'}\Pi^{(1)}_{ij}h^{(1)\,ij\prime}\,,
\ee
for the evolution of both $\zeta_{D}^{(2)}$ and $\zeta_{D(v)}^{(2)}$;
\begin{align}
\lb-\zeta_{I}^{(2)}+\frac13\V^{(1)}_{i}\lb\V^{(1)\,i}\right.\right.&\left.\left.-2\W^{(1)i}\rb\vphantom{\frac13}\rb'=\nonumber\\
&\H\frac{\delta P_{\text{nad}}^{(2)}}{\rho+P}-2\H\lb\frac{\delta P_{\text{nad}}^{(1)}}{\rho+P}\rb^2-\frac{2\H}{\rho'}\Pi^{(1)}_{ij}h^{(1)\,ij\prime}\,,
\end{align}
for $\zeta_{I}^{(2)}$ and
\begin{align}
-\zeta_{I(v)}^{(2)\prime}=\H\frac{\delta P_{\text{nad}}^{(2)}}{\rho+P}-2\H\lb\frac{\delta P_{\text{nad}}^{(1)}}{\rho+P}\rb^2-\frac{2\H}{\rho'}\Pi^{(1)}_{ij}h^{(1)\,ij\prime}\,.
\end{align}
for $\zeta_{I(v)}^{(2)}$. 
Note that, in all cases above, the
pairs are equal in the large scale approximation, except for $\zeta_{I}^{(2)}$ and $\zeta_{I(v)}^{(2)}$, which have a different contribution from vector perturbations. From this result, one can see that, even in the absence of the scalar non-adiabatic pressure, $\delta P_{\text{nad}}$,
neither curvature perturbation is conserved,
\be
\lb-\zeta^{(2)}+\frac23h^{(1)}_{ij}h^{(1)\,ij}-\frac13 W^{(1)\,i} W^{(1)}_{i}\rb'=-\zeta_{I(v)}^{(2)\prime}
=-\frac{2\H}{\rho'}\Pi^{(1)}_{ij}h^{(1)\,ij\prime}\,.
\ee
However, if the traceless, transverse part of the anisotropic stress, $\Pi^{(1)}_{ij}$, is negligible, $\zeta_{I(v)}^{(2)}$ is in fact conserved
\be
\zeta_{I(v)}^{(2)\prime}=\lb\zeta_{D}^{(2)}+\frac13 W^{(1)\,i}W^{(1)}_{i}\rb'=\lb\zeta^{(2)}-\frac23h^{(1)}_{ij}h^{(1)\,ij}+\frac13 W^{(1)i} W^{(1)}_{i}\rb'=0\,.
\ee
%%% NEW STUFF %%%%%%%%
Although $\zeta_{I(v)}^{(2)}$ is exactly conserved, the difference between $\zeta_{I(v)}^{(2)\prime}$ and $\zeta_{D}^{(2)\prime}$ only depends on vector perturbations, which are usually negligible. Moreover, using the vector part of the momentum conservation equation, Eq. \eqref{evov1}, in the absence of anisotropic stress, we find the evolution of $W^{(1)}_i$ is given by
\be
W^{(1)\prime}_{i}+\H (1-3c_s^2) W^{(1)}_{i}=0.
\ee
Thus, this vector perturbation is conserved during radiation domination ($c_s^2=1/3$) and, as a consequence, $\zeta_{D}^{(2)}$ is exactly conserved during that epoch. In the general case, we may therefore write the evolution of $\zeta_{D}^{(2)}$ on large scales as 
\be
\zeta_{D}^{(2)\prime}=-\frac23\H (1-3c_s^2)W^{(1)\,i}W^{(1)}_{i}\,,
\ee
showing again that it may only have an appreciable evolution if the vector modes are large.
%%%%%%%%%%%%%%%%%%%%5

The evolution equations simplify further in Einstein gravity, as, in the absence of
anisotropic stress, tensor modes stop evolving and hence
this new conservation law converges fairly quickly to the conservation
of $\zeta^{(2)}$ itself.
Therefore, for Einstein gravity, all versions of the curvature perturbation are conserved up to second order on
large scales, if both the non-adiabatic pressure and the
anisotropic stress are negligible. However, should the evolution of vectors and tensors be appreciable, the version of $\zeta$ which is conserved is $\zeta_{I(v)}$, i.e., the version defined by the determinant of $g^{ij}$ and by using a comoving threading to fix the gauge.

\section{Conclusion}\label{conclusion}

We obtained the evolution equation for the curvature perturbation at
second order in cosmological perturbation theory, valid on all scales. With the inclusion of vectors, tensors and anisotropic stress,
this result allows for high precision calculations of correlation functions on all scales. We derive this for six different definitions of $\zeta$, based on several different splits of the spatial metric and on various choices of the defining gauge. The results for the evolution equations show a substantial difference in apparent complexity, being simpler when the threading defining $\zeta$ was chosen to be the comoving one, i.e. $\widetilde{v}^i=0$. Eq. \eqref{zI2p} for the evolution of $\zeta_{I(v)}$ is particularly short, but its usefulness is unclear due to the existence of indefinite time integrals in the definitions of $\zeta_{I(v)}^{(2)}$ and $\delta P_{\text{nad}(v)}^{(2)}$. On the other hand, for the versions of $\zeta$ for which the threading was chosen with $\widetilde E=0$, or the original $\zeta$, the definitions include inverse Laplacians (see Eq. \eqref{z2}). In both cases, non-locality is present in some form, either in time or in space, and there is no version of the curvature perturbation which evades both of these issues. However, in both cases, the difficulties of the calculation are resolved by solving additional differential equations, both of which require boundary conditions. In the case of the inverse Laplacian, the equation to solve is a Poisson equation, which only depends on first-order quantities at a single time, while for the case of the integrals in time, knowledge of the full time evolution of second order quantities is required ($\n^2J^{(2)}$ in Eq. \eqref{zI2v}, for example). This seems to render the quantities without integrals in time more amenable for situations that require the calculation of $\zeta$ from its definition, such as when its value is evaluated from the value of scalar field or density perturbations. In any case, all these issues disappear in the large scale approximation, for which the inverse Laplacian term in question has a well defined limit and the integrals vanish.

Moreover, we found that, on large scales, the
evolution of $\zeta$ is sourced by the transverse traceless part of the anisotropic stress tensor, as well as non-adiabatic pressure. Both quantities must therefore be negligible for any version of $\zeta$ to be
conserved. Furthermore, the version of the curvature perturbation which is exactly conserved is the one based on the determinant of $g^{ij}$ and comoving threading, $\zeta_{I}^{(v)}$, Eq. \eqref{zI2v}. Other definitions may evolve with the evolution of tensor and vector modes, should such an evolution be allowed by the theory of gravitation under study. For General Relativity, however, vector perturbations are usually very small and the evolution of tensor modes is negligible in the absence of anisotropic stress; therefore all versions of the curvature perturbation are approximately conserved on large scales.
%

%%%%%% NEW STUFF %%%%%%%%%%
The results presented here are valid as long as the energy and momentum conservation equations, Eqs. \eqref{evorho1}, \eqref{evov1} and \eqref{evorho2}, are satisfied. This will be true if the stress-energy tensor is covariantly conserved, i.e. $\n_\mu T^{\mu\nu}=0$, and the connection is the Levi-Civita connection (i.e. no torsion is present). This is the case in GR, but also in other theories, such as Massive Gravity and Bigravity \cite{D'Amico:2011jj,DeFelice:2014nja}. The latter theories are interesting in this context, as the tensor modes evolve differently due to the non-zero mass of the graviton \cite{Fasiello:2015csa} and therefore, $\zeta_D$ and $\zeta_I$ would be the only versions of the curvature perturbation that are conserved. 

Furthermore, the usefulness of these results may be extended to theories of gravity for which $\n_\mu \tilde{T}^{\mu\nu}\neq0$, in which $\tilde{T}^{\mu\nu}$ represents here the r.h.s. of the field equations of that theory. This is possible if one can perform a conformal transformation to the Einstein frame and apply the same ideas to the stress-energy tensor that arises as the r.h.s. of the new field equations. The difference between our standard scenario and a modified one is that the effective matter quantities defined in one of the frames, would not have the same physical significance as the ones we use in this work. Therefore, in those modified situations it may be less trivial to clearly say when the curvature perturbation is conserved, as, e.g. the effective $\delta P_{\text{nad}}$ may not be negligible in both frames when the true matter perturbations are adiabatic. The same could apply to the anisotropic stress.
%%%%%%%%%%%%%%%%%%%%%%%%%%%%

Previous results on the subject of conserved quantities have not included
anisotropic stress \cite{Enqvist:2006fs} and have either done the calculations fully in the
large scale approximation \cite{Lyth:2004gb} or used a different quantity \cite{LV,LV2,LV3}.

% % % % % % % % % % % % % % % % % % % % % % % % % % % % 
% chapter.tex - Ian Huston
% Sample chapter layout
% % % % % % % % % % % % % % % % % % % % % % % % % % % % 
% Redefine CVSRevision for this section. 
% If you don't want to use CVS tags comment out this line
\renewcommand{\CVSrevision}{\version$Id: chapter.tex,v 1.3 2009/12/17 18:16:48 ith Exp $}

% % % % % % % % % % % % % % % % % % % % % % % % % % % % % % % % 
% =========================================================== %
% % % % % % % % % % % % % % % % % % % % % % % % % % % % % % % % 
\chapter{Isocurvature initial conditions at second order}
\label{Ch_iso2}
% % % % % % % % % % % % % % % % % % % % % % % % % % % % % % % % 
% =========================================================== %
% % % % % % % % % % % % % % % % % % % % % % % % % % % % % % % % 

\section{Introduction}

In this chapter, we calculate the initial evolution of cosmological fluctuations at second order in the presence of isocurvature modes. These calculations are essential for initializing Boltzmann codes at second order~\cite{Pettinari:2014vja} and thus to calculate observables with the required accuracy for comparing with experiment. We begin in Section \ref{defs} by introducing the multi-fluid system we use in the remainder of the chapter. In Section \ref{diffsys}, we describe the general differential system under study and how to split its perturbative solutions into different parts. After that, we introduce a clear definition of the isocurvature basis in Section \ref{isocurv} as used in previous literature and present our results for the initial time evolution in synchronous gauge in Section \ref{inievo}. We then discuss our results and conclude in Section \ref{conclusionCiso}. We also consider gauge transformations of our results into Poisson gauge, but leave that for Appendix \ref{gaugetr}.

\section{Cosmological perturbation theory for a multi-fluid system}\label{defs}

In this first section, we introduce the multi-fluid system that will be used in the rest of the chapter. We follow most of the notation and conventions introduced above in Chapters~\ref{Ch_CPT} and \ref{Ch_SMC}. In particular, the metric is expanded as in Eqs~\eqref{g00CPT}--\eqref{gijCPT}, with the same definition of the spatial metric, Eq.~\eqref{gij}. As for the total stress-energy tensor, we choose, once again, the energy
frame to represent it so that it is given by Eq.~\eqref{Tmn} with $q_\mu=0$, which we reproduce here,
\be
T_{\mu\nu}=\lb P+\rho\rb u_\mu u_\nu+P g_{\mu\nu}+\pi_{\mu\nu}\,.
\ee
Its perturbative expansion is the same as in Chapter \ref{Ch_CPT}. We define here the variable $\sigma$ to represent the scalar anisotropic stress. It is given by, at all orders,
\be
\sigma^{(i)}=-\frac1{2\rho}\n^2\Pi^{(i)}\,.
\ee
This variable is more appropriate in this context as it is more directly linked to the conventions used in the literature and, as we shall see below, has growing mode solutions.

The stage of the evolution of the Universe we study in this chapter is the radiation dominated epoch at the time following neutrino decoupling and electron-positron annihilation, the same epoch that was described in Section~\ref{RecCMB} of Chapter~\ref{Ch_SMC}. At this stage, (in the $\Lambda$CDM model) there are four matter species that are present in the Universe, namely, neutrinos ($\nu$), photons ($\gamma$), baryons ($b$) and cold dark matter ($c$). We construct the total stress-energy tensor by adding those of each species, labelled by the index $s$,
\be
T^{\alpha\beta}=\sum_s T_s^{\alpha\beta}\,,
\ee
which are given by
\begin{align}
&T_c^{\alpha\beta}=\rho_c u_c^\alpha u_c^\beta\,,\\
&T_b^{\alpha\beta}=\rho_b u_b^\alpha u_b^\beta\,,\\
&T_\gamma^{\alpha\beta}=\frac43\rho_\gamma u_\gamma^\alpha u_\gamma^\beta+\frac13 \rho_\gamma g^{\alpha\beta}\,,\\
&T_\nu^{\alpha\beta}=\frac43\rho_\nu u_\nu^\alpha u_\nu^\beta+\frac13 \rho_\gamma g^{\alpha\beta}+\pi_\nu^{\alpha\beta}\,.
\end{align}
It is clear from these expressions that only neutrinos have anisotropic stress, as it is assumed that photons are tightly coupled with baryons at this time, and dark matter is too cold to have appreciable anisotropic stress. As we have shown in Chapter~\ref{Ch_SMC}, these conditions are sufficient to set the anisotropic stress of those species to zero. Note as well that all species have been written in their specific energy frames given by each 4-velocity vector $u_s^\alpha$. This implies that the calculation of the total fluid quantities, such as the total energy density, is not a simple sum of those variables defined in each frame. We perform this calculation by projecting the stress-energy tensors of each species into a global energy frame, labelled by the 4-velocity vector $u^\mu$. After this change of frame, we find the total energy density, pressure and anisotropic stress are given by
\begin{align}
\label{rhotot}
\rho=&\gamma_c^2\rho_c+\gamma_b^2\rho_b+\frac{4\gamma_\gamma^2-1}{3}\rho_\gamma+\frac{4\gamma_\nu^2-1}{3}\rho_\nu+\pi_\nu^{\alpha\beta}u_\alpha u_\beta \,,\\
P=&\frac{\gamma_c^2-1}{3}\rho_c+\frac{\gamma_b^2-1}{3}\rho_b+\frac{4\gamma_\gamma^2-1}{9}\rho_\gamma+\frac{4\gamma_\nu^2-1}{9}\rho_\nu +\frac13\pi_\nu^{\alpha\beta}u_\alpha u_\beta\,,\\
\pi^{\alpha\beta}=&\pi_\nu^{\alpha\beta}-\frac13(g^{\alpha\beta}+4u^{\alpha}u^{\beta})\pi_\nu^{\mu\lambda}u_\mu u_\lambda\nonumber\\
&+\sum_s(1+w_s)\left(\frac{1-\gamma_s^2}{3}g^{\alpha\beta}+\frac{1-4\gamma_s^2}{3}u^{\alpha}u^{\beta}+u_s^{\alpha}u_s^{\beta}\right)\rho_s\,,\label{eqpitot}
\end{align}
while the 4-velocity of the energy frame can be related to that of each fluid by solving the following equation for $u^\alpha$:
\be
\sum_s(1+w_s)\rho_s \gamma_s (u_s^\alpha-\gamma_s u^\alpha)-\pi_\nu^{\alpha\beta}u_\beta-\pi_\nu^{\mu\beta}u_\mu u_\beta u^\alpha=0\,,
\ee
which is obtained from the energy frame condition, i.e. by setting the momentum density vector $q^\alpha$ to zero. In the absence of neutrino anisotropic stress, one would find the following solution for $u^\alpha$:
\be
u^\alpha=\frac{\sum_s(1+w_s)\rho_s \gamma_s u_s^\alpha}{\sum_s(1+w_s)\rho_s \gamma_s^2}\,.
\ee
This result is still correct at first order, but is not sufficient at second order. In all expressions above, $w_s=P_s/\rho_s$ is the equation of state parameter and $\gamma_s$ is the Lorentz factor for changing between the energy frame and each species' rest frame, which is given by 
\be
\gamma_s=-u_s^\lambda u_\lambda.
\ee
All these equations are fully covariant and are therefore valid at all orders in perturbation theory. In the following we will use them at second order.

\subsection{Evolution equations}

To describe the evolution of this system we assume Einstein gravity,
\be
G^{\alpha\beta}=8\pi G T^{\alpha\beta}\,,
\ee
and describe the evolution of each fluid by:
\begin{align}
&\n_\beta T_\gamma^{\alpha\beta}=C^\alpha_{\gamma b}\,,\\
&\n_\beta T_\nu^{\alpha\beta}=0\,,\\
&\n_\beta T_b^{\alpha\beta}=-C^\alpha_{\gamma b}\,,\\
&\n_\beta T_c^{\alpha\beta}=0\,.
\end{align}
where we have included the interaction of photons with baryons, represented by $C^\alpha_{\gamma b}$ and given in Eqs~\eqref{Col01st} and \eqref{Col11st}, at first order. However, we will assume the tight coupling approximation (TCA) is valid, which, as described in Chapter~\ref{Ch_SMC}, means that the velocity of the photons and baryons is equal. For the case of the neutrinos, we also introduce an equation for the anisotropic stress, which is derived from the Liouville equation. We shall write these equations below in their perturbed versions. We write only the second-order equations as the first-order ones can be obtained straightforwardly by setting all the non-linear terms to zero. We also simplify our presentation by including only scalar equations as we are only studying second-order scalar modes sourced by first-order scalars. We leave the study of vector and tensor modes for future work.

Regarding the gauge choice, we write all equations in the synchronous gauge, as defined in Eq.~\eqref{synchdef1}. The reason for this choice is historical, as most literature in this field was developed in synchronous gauge, making it easier to compare our results with past ones. Furthermore, this historical fact has led most experimentalists to use the synchronous gauge definitions when constraining primordial initial conditions, which adds to our motivation to use this gauge. This will be further clarified below, in Section \ref{isocurv}. We follow the arguments of Chapter~\ref{Ch_CPT} and fix the extra gauge freedom mentioned by choosing the initial velocity field of cold dark matter to be zero, which also has the further advantage of simplifying the differential system, as the dark matter velocity is constrained to be zero at all times by the equations of motion. 

We begin by writing the field equations for the two scalar potentials available in synchronous gauge. The only ones we require are the constraint equations, given in Eqs.~\eqref{EEq00CPT} and \eqref{EEqi0CPT}, which we reproduce here in synchronous gauge and in terms of the four species under study,
\begin{align}
&\n^2\psi+\Hh\n^2E'-3\Hh\psi'-\frac32\Hh^2\sum_s{\Omega_s\delta_{s}}=6\Hh\psi\psi'-\frac32(\psi')^2-4\psi\n^2\psi-\frac32\psi_{,i}\psi^{,i}\nonumber\\
&-2(\psi\n^2E)'+\psi'\n^2E'+\n^2E_{,i}\psi^{,i}+\n^2E\n^2\psi+\psi_{,ij}E^{,ij}-\frac14\n^2E'\n^2E'\label{EEq00}\\
&+\frac14\n^2E_{,i}\n^2E^{,i}+2\Hh E'_{,ij}E^{,ij}+\frac14E'_{,ij}E^{\prime,ij}-\frac14E_{,ijk}E^{,ijk}+\frac32\Hh^2\sum_s{(1+w_s)\Omega_s v_{s,i}v_{s}^{,i}}\,,\nonumber
\end{align}
and
\begin{align} 
&\psi'-\frac32\Hh^2\sum_s{(1+w_s)\Omega_s v_{s}}=-2(\psi\n^2\psi)'-4\psi'_{,i}\psi^{,i}+\n^2E'_{,i}\psi^{,i}+\frac12\n^2E'\n^2\psi\nonumber\\
&+\n^2E\n^2\psi'+\psi'_{,ij}E^{,ij}+\frac12\psi_{,ij}E^{\prime\,,ij}+\frac12\n^2E'_{,i}\n^2E^{,i}-\frac12E'_{,ijk}E^{,ijk}\label{EEq0i}\\
&-\frac34\Hh^2\sum_s{\Omega_s(1+w_s)\left[2\left((\delta_{s}-2\psi)v_{s}^{,i}\right)_{,i}+(v_{s,i}E^{,ij})_{,j}\right]}\nonumber\\
&-\Omega_\nu\Hh^2\left[(\sigma_{\nu}v_{\nu}^{,i})_{,i}-3(\n^{-2}\sigma_{\nu}^{,ij} v_{\nu,i})_{,j}\right]\,,\nonumber
\end{align}
in which $\Omega_s=8\pi G \rho_s/3H^2$ is the standard density parameter for each species, $\delta_s$ in the density contrast for each species, defined by $\delta_s=\delta\rho_s/\rho_s$, $v_s$ is the corresponding velocity fluctuation and $\sigma_\nu$ represents the scalar part of the neutrino anisotropic stress. The energy conservation equations for the fluids can be derived from Eq.~\eqref{enecons} and are given by
\begin{align}
&\delta_{s}'-(1+w_s)\left(3\psi'-\n^2(E'+v_{s})\right)=2(1+w_s)\left(3\psi\psi'-(\psi\n^2E)'+E'_{,ij}E^{,ij}\right)\nonumber\\
&+\delta_{s}\delta_{s}'-(1+w_s)v_{s}^{,i}\left(2v'_{s,i}+\delta_{s,i}-3\psi_{,i}+\n^2E_{,i}+(1-3w_s)\Hh v_{s,i}\right)\label{Eqdelta}\\
&+\frac23 \delta_s^\nu \left[2\sigma_{\nu,i}v_{\nu}^{,i}-\sigma_{\nu}\n^2(E'+v_{\nu})+3\n^{-2}\sigma_{\nu,ij}(E'+v_{\nu})^{,ij}\right]\,,\nonumber
\end{align}
where we have assumed that each fluid has a constant equation of state and have aggregated all possible cases for the four species under study. The quantity $\delta_s^\nu$, appearing the last line of Eq~\eqref{Eqdelta}, is the Kronecker delta symbol and is unrelated to the density contrast. 

Concerning the momentum conservation equations, we only have to write them for the neutrinos and the photon-baryon plasma. This is due to having chosen the synchronous gauge, which allows one to set the cold dark matter velocity to zero to fix the residual gauge conditions. Furthermore, since we assume the TCA is valid, there is only one equation for the common velocity of photons and baryons, $v_{b\gamma}$. This equation is obtained by summing the two momentum conservation equations for baryons and photons and is given by
\begin{align}
&\n^2\left[(3\Omega_b+4\Omega_\gamma)v_{b\gamma}'+\Omega_\gamma\delta_{\gamma}+3\Omega_b\Hh v_{b\gamma}\right]=-4\Omega_\gamma\left(\delta_{\gamma}v_{b\gamma}^{\prime,i}\right)_{,i}-3\Omega_b\left(\delta_{b}v_{b\gamma}^{\prime,i}\right)_{,i}\nonumber\\
&+v_{b\gamma}^{,i}\left[\Omega_\gamma\left(4\psi_{,i}-\frac{20}{3}\n^2E'_{,i}-\frac{8}{3}\n^2v_{b\gamma,i}\right)+\Omega_b\left(6\psi_{,i}-6\n^2E'_{,i}-3\n^2v_{b\gamma,i}-3\Hh\delta_{b,i}\right)\right]\nonumber\\
&-2\Omega_\gamma\left(\psi\delta_{\gamma}^{,i}-E^{,ij}\delta_{\gamma,j}\right)_{,i}+\n^2v_{b\gamma}\left[\Omega_\gamma\left(4\psi'+\frac43\n^2E'+\frac43\n^2v_{b\gamma}\right)+\Omega_b\left(6\psi'-\frac12\delta_{b}\right)\right]\nonumber\\
&-v_{b\gamma}^{,ij}\left(4\Omega_\gamma+3\Omega_b\right)\left(2E_{,ij}'+v_{b\gamma,ij}\right)\,,
\end{align}
while the one for neutrinos is given by
\begin{align}
&\n^2\left[v_{\nu}'+\frac14\delta_{\nu}+\sigma_{\nu}\right]=\frac12\left(\delta_{\gamma}^{,i}(E_{,ij}-\psi\delta_{ij})-v_{b\gamma}^{,i}(4E_{,ij}+2v_{b\gamma,ij})\right)^{,j}\nonumber\\
&-\left((\delta_{\nu}'-5\psi'+\n^2E'+\n^2 v_{\nu})v_{\nu}^{,i}-\delta_{\nu}v_{\nu}^{\prime\,,i}\right)_{,i}\label{Eqvnu2}\\
&+\left(\psi\sigma_{\nu,i}+\frac12\psi_{,i}\sigma_{\nu}-\frac32\psi^{,j}\n^{-2}\sigma_{\nu,ij}-\frac12(\sigma_{\nu}v_{\nu,i}-3v_{\nu}^{,j}\n^{-2}\sigma_{\nu,ij})'\right)^{,i}\nonumber\\
&-\frac12\left(\frac23\n^2E\sigma_{\nu,i}+E_{,ij}\sigma_{\nu}^{,j}-\frac43\n^2E_{,i}\sigma_{\nu}+5E^{,jk}\n^{-2}\sigma_{\nu,ijk}+4\n^2E^{,j}\n^{-2}\sigma_{\nu,ij}\right)^{,i}\nonumber\,.
\end{align}
The equation for $\sigma_\nu$ is derived from the Liouville equation, as explained in Section \ref{EvoEqsFLRW}. The final equation was already partially given in Eq.~\eqref{Bright2wtr}. Here, we take the traceless part of that equation and set the collision term to zero. This gives
\begin{align}
\label{Boltz}
&\Delta_{T\,ij}'+\left(\Delta_{T\,ijk,l}-\frac15\left(\frac23\delta_{ij}\delta_k^{r}-\delta_{kj}\delta_i^{r}-\delta_{ik}\delta_j^r\right)\Delta_{r,l}\right)(\delta^{kl}-C^{kl})-\Delta_{T\,ij}^{\ \ \ kl}E'_{,kl}\nonumber\\
&-4\Delta_{T\,ij} \psi'-\frac{10}{21}\delta_{ij}\Delta_T^{kl}E'_{,kl}+\frac17\left(6\Delta_{T\,ij}\n^2E'+5\Delta_{T\,i}^{\ k}E^{\prime}_{\,,jk}+5\Delta_{T\,j}^{\ k}E^{\prime}_{\,,ik}\right)+\frac{8}{15}\Delta_0E^{\prime}_{\,,ij}\nonumber\\
&-\frac{8}{45}\Delta_0\delta_{ij}\n^2E'-\left(4\Delta_{T\,ij}^{\ \ k}+\frac15\left(\frac23\delta_{ij}\delta^{ks}-\delta^k_{j}\delta_i^{s}-\delta_{i}^{k}\delta_j^s\right)\Delta_s\right)\psi_{,k}\nonumber\\
&+\frac{8}{15}\left[C_{ij}-C_{ki}C^{k}_{j}-\frac13\delta_{ij}(C^k_k-C_{kl}C^{kl})\right]'=0\,,
\end{align}
in which the $\Delta$ variables are perturbations to the momentum integrated distribution function of neutrinos, defined in Eqs.~\eqref{Deltadef} and \eqref{Delta0CPT}--\eqref{Delta3CPT}. The first three brightness tensors are related to the stress-energy tensor via Eqs.~\eqref{d0CPT}--\eqref{d2CPT}, and we rewrite those relations here, in synchronous gauge:
\begin{align}
\label{d0}
&\Delta_0=-\frac{\delta T^{\ 0}_{\nu\ 0}}{\rho_\nu}\,,\\
\label{d1}
&\Delta^{i}=-\frac{T^{\ j}_{\nu\ 0}}{\rho_\nu}(\delta^i_j+C^i_j)\,,\\
\label{d2}
&\Delta^{\ \  i}_{T\, j}=\frac{1}{\rho_\nu}\left(T^{\ k}_{\nu\ l}-\frac13\delta^k_{\ l}T^{\ r}_{\nu\ r}\right)\left(\delta_{\ j}^l\delta_{\ k}^i+\delta_{\ j}^l E^{,i}_{\ ,k}-\delta_{\ k}^i E^{,l}_{\ ,j}\right)\,,
\end{align}
in which $\rho_\nu$ is the background neutrino energy density. Because we are only dealing with scalar modes, we compute the scalar part of Eq.~\eqref{Boltz} by applying the differential operator $\partial^i\partial^j$. Due to its complexity, we refrain from showing the final evolution equation for $\sigma_\nu$ here. It can be calculated straightforwardly from the scalar equation by using the conversion from the scalar part of $\Delta^{\ \  i}_{T\, j}$ to $\sigma_\nu$, which we give below. We will also display the relations between the other scalar fluid variables and the scalar parts of the brightness tensors, defined in Eqs.~\eqref{D1isvt}--\eqref{D3isvt}. They are given by
\begin{equation}
\Delta_0=\delta_\nu+\frac43v_{\nu,i}v_{\nu}^{,i}\,,
\end{equation}
\begin{equation}
\n^2\Delta_1=\frac43\n^2v_{\nu}+\p_i\left[\left(\frac43(\delta_\nu-\psi)\delta^i_j+\frac43 E^{,i}_{,j}+\frac{1}{\rho_\nu}(\Pi_{\nu,j}^{,i}-\frac13\delta^i_j\n^2\Pi_\nu) \right)v_\nu^{,j}\right]\,,
\end{equation}
\begin{align}
\n^2&\n^2\Delta_2=-2\n^2\sigma_{\nu}+\p_i\p^j\left[2v_{\nu,j}v_{\nu}^{,i}-\frac23v_{\nu,k}v_{\nu}^{,k}\delta^i_j+\frac6{\rho_\nu}\psi\left(\Pi_{\nu,j}^{,i}-\frac13\delta^i_j\n^2\Pi_\nu\right)\right.\\
&\left.-\frac{1}{\rho_\nu}\left(\frac32\Pi_{\nu,jk}E^{,ki}+\frac32\Pi_{\nu,k}^{,i}E_{,j}^{,k}-\n^2\Pi_{\nu}E_{,j}^{,i}+\left(\frac13\n^2\Pi_{\nu}\n^2E-\Pi_{\nu,kl}E^{,kl}\right)\delta^i_j\right)\right]\,.\nonumber
\end{align}

This concludes the description of the evolution equations. In the next sections we will describe this differential system in general and provide details about its formal solution.

\section{Differential System}\label{diffsys}

It is straightforward to show, after applying a Fourier transform, that the differential system presented in the previous section can be described by the following generic equation at any specific non-background order:
\be
\label{diffsystem}
\mathcal{D}_\tau X=Q(\tau)\,,
\ee
in which $\mathcal{D}_\tau$ is a \textbf{\emph{linear}} differential operator, $X$ is a vector including all the variables to evolve and $Q(\tau)$ includes all the non-linear terms, which act as a source at orders higher than the first, while at the linear level we have $Q^{(1)}=0$, by definition. For example, at second order, the source term is a convolution of squares of the first-order (or linear) solutions, 
\be
Q^{(2)}(\tau,\vec k)\supset\int_q X^{(1)}(\vec q-\vec k)X^{(1)}(\vec q)\,,
\ee
in which we introduce the notation
\be
\int_q =\int\frac{\text{d}^3q}{(2\pi)^3}\,.
\ee

In order to solve such a system, one begins by solving the first-order equations. Being linear, the solutions to those equations can be written as a sum of particular solutions, the number of which is the same as the dimension of the solution space, $D$. The solution can therefore be written as
\be
X^{(1)}(\tau,\vec k)=\sum_{i=1}^D{\mathcal{T}_i(\tau,\vec k) I_i^{(1)}(\vec k)}\,,
\ee
in which $\T_i(\tau,\vec k)$ are transfer functions and $I_i(\vec k)$ represent the initial conditions of certain variables of interest. These variables will be called the defining variables of a mode, since they are non-zero only when a specific mode is present. Each of the $\T_i$ is a vector (just like $X$) while each of the $I_i$ is a scalar. The $I_i$ are usually random variables which encode all the statistical information of the initial conditions, and, given that the evolution of the transfer functions is classical, they will allow us to calculate the statistics of $X^{(1)}$ at any time. The fact that each of the $\T_i(\tau)$ is an independent solution of the differential system also means that we can separate the numerical solution of the equations mode by mode, solving each one separately and later calculating the required statistics by summing all the modes. This is especially useful, since it allows for a solution of the equations without the need to specify the amplitude of each initial condition, leaving those parameters to be constrained by experiment.

At second order, the general solution is
\be
X^{(2)}(\tau,\vec k)=\sum_{i}{\mathcal{T}_i(\tau,\vec k) I_i^{(2)}(\vec k)}+\sum_{i,j}{\int_{\vec k_1,\vec k_2}\mathcal{T}^{(2)}_{ij}(\tau,\vec k,\vec k_1,\vec k_2) I_i^{(1)}(\vec k_1)I_j^{(1)}(\vec k_2)}\,,
\ee
in which the first term is the homogeneous solution to Eq.~\eqref{diffsystem}, i.e. it is the same solution as the first-order one, only with different coefficients $I_i^{(2)}$. Given that fact, the total solution, up to this order, can be written as
\begin{align}
X&(\tau,\vec k)=X^{(1)}(\tau,\vec k)+\frac12X^{(2)}(\tau,\vec k)\\
&=\sum_{i}{\mathcal{T}_i(\tau,\vec k) \lb I_i^{(1)}(\vec k)+\frac12I_i^{(2)}(\vec k)\rb}+\frac12\sum_{i,j}{\int_{k_1,k_2}\mathcal{T}^{(2)}_{ij}(\tau,\vec k,\vec k_1,\vec k_2) I_i^{(1)}(\vec k_1)I_j^{(1)}(\vec k_2)}\,,\nonumber
\end{align}
which shows that one can absorb the term $I_i^{(2)}$ into the first-order part, $I_i^{(1)}$, or, equivalently, setting $I_i^{(2)}=0$. In this case the defining variables, $I_i=I_i^{(1)}+\frac12I_i^{(2)}$, are set by the initial conditions of the full $X$ and not just its first-order part. This is also more natural, as, many times, the initial conditions will not be split into different orders, unless they have different properties, such as non-Gaussianity. An alternative scenario is to write $I_i^{(2)}$ as a sum of $I_i^{(1)} I_j^{(1)}$, effectively including it into the second term above. This is also equivalent to the previous case, because nothing constrains $\T^{(2)}_{ij}$ from including terms proportional to $\mathcal{T}_i$.

To numerically solve the differential system in question one may also separate the solution of the different transfer functions $\T^{(2)}_{ij}$, in order to find solutions which are valid for any values of the amplitude of the initial conditions. To see why this split can be performed, we begin by analysing the source $Q(\tau,k)$. It can also be written in terms of the defining variables as:
\be
Q^{(2)}(\tau,\vec k)=\sum_{i,j}{\int_{k_1,k_2}\mathcal{S}_{ij}(\tau,\vec k,\vec k_1,\vec k_2) I_i(\vec k_1)I_j(\vec k_2)}\,,
\ee
in which $\mathcal{S}_{ij}$ are the equivalent of transfer functions for the source terms $Q^{(2)}$. It can be shown, due to the linearity of the differential system, that there is a particular solution to the second-order system which is a sum of the solutions of similar systems with the source $Q^{(2)}$ substituted for each of the terms in the sum above. Hence, to find the evolution of each $\T^{(2)}_{ij}$ one needs only to solve those similar systems in which only the $\{i,j\}$ defining variables are non-zero.

The question that we are concerned with in this chapter is that of the initial evolution of $\T^{(2)}_{ij}$, to be used in setting up its numerical evolution. The aim is to find an approximation to the transfer functions that is valid when all Fourier modes of interest are still super-horizon during the radiation dominated Universe. In the following section, we precisely define the isocurvature basis. 

\section{Definition of isocurvature basis}\label{isocurv}

In the radiation dominated Universe and after electron-positron annihilation at $z\sim10^8$, the species that are relevant are (nearly) massless neutrinos, the dark matter fluid and the tightly coupled baryon-photon plasma. In the case that those species can be represented by barotropic perfect fluids, one can show that the total number of evolving scalar degrees of freedom is 8. This is due to the fact that, for each fluid, the perturbed energy conservation equation and the momentum conservation equation allow us to derive a second-order ODE (in $k$-space). In an appropriate gauge, such as flat gauge~\cite{Christopherson:2010ek}, one may use the Einstein constraint equations to
eliminate the metric potentials, and arrive at a system only in terms of fluid quantities, such as energy densities, pressures, etc. To close the system, one uses the barotropic and perfect nature of the fluids to set the entropy and anisotropic stress fluctuations to zero. Finally, one specifies an equation of state, relating pressure and energy density, which results in a second-order ODE for the density perturbation of each fluid. Thus, for each barotropic perfect fluid there are 2 independent modes, hence 8 in total. The situation is slightly different in synchronous gauge, which we use here. In that case, one of the metric potentials cannot be completely eliminated from the final equations in terms of the density perturbations. Therefore an extra equation for that potential is required, which appears to increase the number of degrees of freedom to 9. This is a peculiarity of this gauge, for which the coordinate freedom has not been exhausted. The 9\textsuperscript{th} mode is in fact a gauge mode, which can be eliminated by setting the initial velocity field of the dark matter fluid to zero, which we do, as mentioned already when this gauge was defined. Beyond the usual 8, more modes can arise if, like the neutrinos, the fluids are not perfect. However, it is unlikely that those modes are present if the fluid has been tightly coupled in the past, as such a stage brings any anisotropic stress to negligible values. After decoupling, an anisotropic stress perturbation will be generated, but only after horizon re-entry.

However, as is well known in the literature \cite{Bucher:1999re}, only 5 of the 8 modes are growing modes in the standard case. This reduction from the total 8 degrees of freedom is due, firstly, to tight coupling, which forces the velocities of baryons and photons to be equal, or, in other words, constrains the mode generated by their difference to be a rapidly decaying mode. Two more modes are also decaying modes, which are usually due to the presence of non-zero initial total velocity and dark matter velocity. In synchronous gauge, however, the dark matter velocity has already been set to zero using the extra gauge freedom, so some other variable must be responsible for generating a decaying solution. This can be found by analysing the first-order versions of Eqs.~\eqref{EEq00} and \eqref{EEq0i} and noting there that, since $\Hh\approx \tau^{-1}$, the terms proportional to $\Hh^2$ will generate decaying modes if initially non-zero. Those terms are proportional to the total density contrast and the total velocity, and thus we conclude that those are the quantities which need to be set to zero to eliminate the corresponding decaying modes at first order. Hence the dark matter decaying mode has been substituted by a total density decaying mode. While they appear unconnected, these two results can be related by the Einstein equations. Using the gauge transformations in Chapter~\ref{Ch_CPT}, one can show that the dark matter velocity in Poisson gauge is equal to the synchronous gauge potential $E'$. Using again the Einstein equations in synchronous gauge, Eqs.~\eqref{EEq00} and \eqref{EEq0i}, we can relate the potential $E'$ to the total density contrast and the total velocity, as follows
\begin{equation}
\n^2E'=-\frac{1}{\Hh}\n^2\psi+\frac92\Hh^2\sum_s{(1+w_s)\Omega_s v_{s}}+\frac32\Hh\sum_s{\Omega_s\delta_{s}}\,,
\end{equation}
and again, since $\Hh\approx \tau^{-1}$, the term with $\psi$ is initially zero, showing that there is a direct relationship between the initial value of $E'$ and those of the total density contrast and velocity. We conclude then, that the conditions for the absence of decaying modes can be written in terms of any two of the three quantities shown above: the total velocity, density contrast or the metric potential $E'$.

The five remaining independent modes are usually represented in the so-called isocurvature basis, in which one defines an adiabatic mode and 4 isocurvature modes: dark matter, baryon and neutrino density isocurvatures as well as the neutrino velocity isocurvature, which are labelled in accordance to the defining variable, $I_i$ that is non-zero in each mode. All observational evidence points towards the adiabatic mode being the dominant one and that is why it is used to define this basis. The other modes could possibly be split in different ways, but we stick here to the conventions of the literature, as this parametrisation is commonly used in observational studies.

At second order, an interesting issue arises. Looking again at Eqs.~\eqref{EEq00} and \eqref{EEq0i}, we see that we actually require $\delta=v=0$ \emph{and} $v'=0$ at the initial time, otherwise the time-space equation still generates a decaying solution, since it depends on $\tau^{-2}$. At first order, however, the condition on the derivative is a consequence of the original conditions, $\delta=v=0$, as can be shown by checking the total momentum conservation equation:
\begin{align}
v'+(1-3c_s^2)\Hh v+\frac{1}{3(1+w)}\left(\delta-\Omega_M\delta_M-4\sigma\right)=0\,,\label{totvpeq}
\end{align}
in which $w=P/\rho$ is the equation of state parameter for the total fluid and $c_s^2=P'/\rho'$ is the adiabatic sound speed. To show that this implies $v'=0$ when $\delta=v=0$, we first note that, initially, the matter density parameter obeys $\Omega_M \propto \tau$ and as a consequence the term with $\delta_M$ is negligible initially (at $\tau\approx0$). The second and crucial step is noticing that the total anisotropic stress, represented by $\sigma$, is initially zero at first order, because it is proportional to the neutrino anisotropic stress. Only the terms with $v$ and $\delta$ are left, thus showing that the conditions $\delta=v=0$ imply $v'=0$ at first order. At second order, this second point is no longer true, since the total anisotropic stress depends on the velocity fluctuations of each species, as can be shown from Eq.~\eqref{eqpitot}, and these are not zero initially in all cases\footnote{Contributions from non-linear terms appearing in the second-order version of Eq.~\eqref{totvpeq} are not important for this argument as they can be shown to be initially zero for all possible growing modes at first order.}. Therefore, the requirements for non-decaying solutions are not satisfied at second order with only two conditions, another one is needed. The extra condition one requires to avoid decaying modes is, in practice, that the neutrino velocity is initially zero. This is because, with a vanishing total velocity as well as no dark matter velocity, the common velocity of baryons and photons is constrained to be proportional to the neutrino velocity. Setting it to zero, implies all initial velocities are zero and hence the initial anisotropic stress at second order also vanishes, avoiding the decaying contribution. Since the neutrino velocity mode is the only linear growing mode that (by definition) has a non-zero neutrino velocity, that is the mode which would generate decaying contributions at second order. For this reason, we choose not to perform any calculations at second order with the neutrino velocity mode. We now describe the standard way of performing the general decomposition, including the description of the neutrino velocity mode, for completeness.

We begin with the adiabatic mode. It is defined to be the mode whose initial conditions have vanishing entropy perturbations and vanishing velocity for all species. At first order, the gauge invariant relative entropy perturbation is given by (\cite{Malik:2002jb})
\be
S_{sr}=3 (\zeta_s-\zeta_r)\,,
\ee
in which $r$ and $s$ label the species in question and $\zeta_s$ is the partial curvature perturbation of species $s$, which is given by
\be
\label{zetai}
\zeta_{s}=-\psi+\frac{\delta_s}{3(1+w_s)}\,,
\ee
where we have assumed that energy transfer is negligible. In order to define any general mode one must give five initial conditions, as that is the number of growing modes present in the system. However, for each mode, we wish to leave one of those initial conditions free so that it may later be fixed by measurement of its correlation functions. Thus, we only present four conditions for each mode. For the adiabatic one, the conditions are, in terms of the relative entropies:
\begin{gather}
\label{Adcond1}
S_{c\gamma}|_{\tau=0}=S_{\nu\gamma}|_{\tau=0}=S_{b\gamma}|_{\tau=0}=S_{c\nu}'|_{\tau=0}=0\,.
\end{gather}
In synchronous gauge, in which these conditions were originally defined, the adiabatic mode is given in terms of density contrasts and the neutrino velocity:
\begin{gather}
\label{Adcond2}
\delta_{c}|_{\tau=0}=\delta_{\nu}|_{\tau=0}=\delta_{b}|_{\tau=0}=v_\nu|_{\tau=0}=0\,.
\end{gather}
We can show that these conditions are equivalent to the ones for the entropies as $\delta_\gamma|_{\tau=0}=0$ due to the total density contrast being set to zero to avoid decaying modes. The defining variable in this case is $\psi|_{\tau=0}=-\zeta|_{\tau=0}$.

For the isocurvature modes, instead of the initial entropy being zero, these modes require the initial curvature perturbation, $\zeta$, to vanish. The different density isocurvature modes are then distinguished from each other by the fact that at least one of the density contrasts (or neutrino velocity) is initially non-zero. 

We summarize here all the conditions for the isocurvature modes at first order in perturbation theory, written in synchronous gauge:

\textbf{Baryon isocurvature:}
\begin{align}
\label{bicond}
&\delta_{c}|_{\tau=0}=\delta_{\nu}|_{\tau=0}=\psi|_{\tau=0}=v_{\nu}|_{\tau=0}=0\,,\\
&\text{Defining variable: }\delta_b.\nonumber
\end{align}

\textbf{Cold dark matter isocurvature:}
\begin{align}
\label{cdicond}
&\delta_{b}|_{\tau=0}=\delta_{\nu}|_{\tau=0}=\psi|_{\tau=0}=v_{\nu}|_{\tau=0}=0\,,\\
&\text{Defining variable: }\delta_c.\nonumber
\end{align}

\textbf{Neutrino Density Isocurvature:}
\label{nidcond}
\begin{align}
&\delta_{c}|_{\tau=0}=\delta_{b}|_{\tau=0}=\psi|_{\tau=0}=v_{\nu}|_{\tau=0}=0\,,\\
&\text{Defining variable: }\delta_\nu.\nonumber
\end{align}

\textbf{Neutrino Velocity Isocurvature:}
\begin{align}
\label{vcond}
&\delta_{c}|_{\tau=0}=\delta_{b}|_{\tau=0}=\delta_{\nu}|_{\tau=0}=\psi|_{\tau=0}=0\,,\\
&\text{Defining variable: }v_\nu.\nonumber
\end{align}

As with the adiabatic mode, similar conditions can be defined with other gauge invariant variables, such as the partial curvature perturbations $\zeta_s$. For example, a new set of conditions would be obtained simply by substituting every $\delta_s$ for the corresponding $\zeta_s$ and $\psi$ for the total $\zeta$. However, the new modes would not form a orthogonal basis in initial condition space, since choosing the $\zeta_s$ as defining variables would imply that the adiabatic mode contains a contribution from each of the density isocurvatures. The choice we present above is only one choice of variables which generate an orthogonal basis for the solution space. Many other choices are certainly possible, but this is the one used in the original literature \cite{Bucher:1999re}. For example, one could also use the same variables, but defined in a different gauge, such as Poisson gauge. While this is an equivalent choice, the results for the initial solutions below would be different, as would the primordial spectra to be constrained by experiment. Using the variables in synchronous gauge avoids having to perform such a conversion.

The conditions at second order are now already automatically set by stating that the Eqs.~\eqref{Adcond2}--\eqref{vcond} apply to the ``non-perturbative" variables and not only to their first-order parts. This is because, by definition, when we choose the component of the vector $X$ to be one of the defining variables, we have:
\be
\label{zero2nd}
I_i(\tau,\vec k)=\sum_{j}{\mathcal{T}^i_j(\tau,\vec k) I_j(\vec k)}+\sum_{m,j}{\int_{k_1,k_2}\mathcal{T}^{i}_{mj}(\tau,\vec k,\vec k_1,\vec k_2) I_m(\vec k_1)I_j(\vec k_2)}\,,
\ee
and thus, the obvious condition of equality, $I_i=I_i$, forces $\T^i_j=\delta^i_j$, as well as $\mathcal{T}^{i}_{mj}=0$, when the index $i$ corresponds to a defining variable. So, the condition is simply that the initial second-order part of the defining variables is exactly zero, for all cases. The choice of defining variables plays a crucial role in the form of the results, as it determines which variables one chooses to be initially zero at second order. A different choice would result in equivalent results, but with a different functional form.

An additional condition must be set regarding the metric potential $E$. At linear order, the initial value of $E$ is not relevant for the evolution of the other quantities, but at second order, this is not the case, i.e. the first-order $E|_{\tau=0}$ does appear in the quadratic source terms and would seem to influence the evolution. However, it can be shown that the initial condition of $E$ (or the value of $E$ at any one time) can be fixed by the labelling of the spatial coordinates at that time~\cite{Malik:2008im}. Therefore, it is fully consistent to set $E|_{\tau=0}=0$ and that is what we do throughout.

With these conditions, one is now able to calculate the initial time evolution for the transfer functions for each part of the solution. This will be done in the next section.

Before showing those results, a few important points must be made regarding the adiabatic nature of the second-order solutions. Firstly, it should be noted that, at second order, the different linear modes mix together. Thus, what we will later call the second-order adiabatic solution is the one which is sourced by quadratic combinations of adiabatic linear modes only. Other solutions exist which are sourced by one adiabatic component and another isocurvature one. We will label all those solutions, \emph{mixed modes}. The second point is that, when this ``adiabatic mode" is defined in this way, it is not obvious that the entropy perturbation, which we define by\footnote{This equation in derived by finding a second-order gauge-invariant quantity which reproduces the linear result and depends only on the two density contrasts in question. This is not a unique definition for the entropy fluctuation, but is sufficient for the purposes of the discussion here.}
\begin{align}
S_{sr}^{(2)}=&\frac{\delta_s^{(2)}}{1+w_s}-\frac{\delta_r^{(2)}}{1+w_r}-\frac{2+w_s+w_r}{(1+w_s)^2}\left(\delta_s^{(1)}\right)^2\\\nonumber
&+\frac{2}{1+w_s}\delta_s^{(1)}\delta_r^{(1)}+\frac{2}{3(1+w_s)\mathcal{H}}\delta_s^{(1)}\left(\frac{\delta_s^{(1)\,\prime}}{1+w_s}-\frac{\delta_r^{(1)\,\prime}}{1+w_r}\right)\,,
\end{align}
should vanish at second order, since this condition was not enforced in any way. In spite of this, all the non-linear terms vanish since all the first-order $\delta_i$ are initially zero when the mode is adiabatic. By the arguments following Eq.~\eqref{zero2nd}, we know that all second-order densities are zero initially, except for the photon density, which is unconstrained by those arguments. However, the presence of a total density contrast can also be shown to generate decaying contributions at second order. Therefore, since we are not considering decaying solutions, by Eq.~\eqref{rhotot}, the photon density contrast is zero at second order as long as all first-order velocities are zero. The solution considered here obeys this condition and is thus truly adiabatic.

In different gauges, the vanishing of the entropies may require different conditions for the density contrasts, particularly if they do not vanish initially at the linear level. For example, Ref.~\cite{Pettinari:2014vja} uses the following conditions, which should be valid in a general gauge, at second order:
\begin{gather}
\label{Adconds2}
\delta_{c}^{(2)}|_{\tau=0}=\delta_{b}^{(2)}|_{\tau=0}=\frac34\delta_{\gamma}^{(2)}|_{\tau=0}-\frac{3}{16}\left(\delta_{\gamma}^{(1)}\right)^2|_{\tau=0}\,,\delta_{\nu}^{(2)}|_{\tau=0}=\delta_{\gamma}^{(2)}|_{\tau=0}\,.
\end{gather}
This is however somewhat more complicated than the second-order initial conditions shown after Eq.~\eqref{zero2nd} and even harder to generalize for the other modes. This further stresses the advantages of working in the same gauge as that in which the defining variables are constructed, as doing otherwise would result in unnecessarily complicated conditions.

Similar arguments apply to the solutions sourced by isocurvature modes. Again, it is not obvious that the gauge invariant curvature perturbation, $\zeta$, will always vanish for all isocurvature solutions, for the same reasons as above. For reference, in the large scale limit, $\zeta$ is given by
\begin{align}
\zeta^{(2)}=&-\psi^{(2)}+\frac{\delta^{(2)}}{3(1+w)}-\frac{1+3w}{9(1+w)^2}\left(\delta^{(1)}\right)^2\\\nonumber
&-\frac{4}{3(1+w)} \delta^{(1)}\psi^{(1)}+\frac{2}{3(1+w)\mathcal{H}}\delta^{(1)}\left(-\psi^{(1)\prime}+\frac{\delta^{(1)\prime}}{3 (1+w)}\right)\,,
\end{align}
where, for brevity, we are presenting only the variable which is invariant under changes of slicing (i.e. gauge transformations involving the time variable only). This is the variable that includes the terms relevant on large scales, as all others would vanish in that limit. We can see that it depends only on the total density contrast, $\delta$, and not on the individual ones for each species. As explained above, $\delta$ is zero for growing solutions, which added to the choice that $\psi|_{\tau=0}=0$ for isocurvatures, results in $\zeta^{(2)}=0$, confirming that all solutions sourced only by isocurvatures are also true isocurvature modes.

\section{Approximate initial time evolution}\label{inievo}

In order to calculate the initial evolution for each partial solution, we expand every variable in powers of $\tau$:\footnote{To make this expansion well defined, one should use a dimensionless expansion parameter, instead of $\tau$, which has dimensions of time (or length, with $c=1$). In practice, as will be clear in the results, the expansion parameter will either be $k\tau$, $k_i\tau$ or $\omega\tau$, with $\omega\equiv\Omega_M \mathcal{H}/\sqrt{\Omega_R}$. The first two are very small for modes deep outside the horizon, while the last one is small for sufficiently early times, given that the constant $\omega$ is $O(10^{-3}) \text{Mpc}^{-1}$. Thus, the expansion in $\tau$ is correct as long as $\tau$ is sufficiently small.}
\be
X=X_0+X_1 \tau +X_2 \tau^2+X_3 \tau^3+\cdots
\ee
This assumes we are neglecting decaying modes, as before. To find the solutions for each mode we apply one of the initial conditions given in Eqs.~\eqref{Adcond2}-\eqref{vcond} to the expansion of the variables $\{\psi,\delta_b,\delta_c,\delta_\nu,v_\nu\}$, generating a series of constraints on specific $X_I$. This constrained expansion is then substituted into the evolution equations, Eqs.~\eqref{EEq00}-\eqref{Boltz}, resulting in a set of algebraic equations for the coefficients, $X_I$, order by order in $\tau$. This will describe the initial solution to the equations of motion for each growing mode. We begin by applying this procedure at first order and recover the results found in Refs.~\cite{Bucher:1999re,Shaw:2009nf}. We substitute those results into the second-order equations of motion and apply the same procedure to find the initial evolution for the second-order transfer function. This is the final step to obtain our main results, which we show below.

We begin, however, by giving an example at linear order. We show here the results for the sum of the two matter isocurvature modes in synchronous gauge:
\begin{align}
\psi=&R_c\left(-\frac{1}{6}\omega\tau+\frac{1}{16}(\omega\tau)^2\right)\delta_c^0+R_b\left(-\frac{1}{6}\omega\tau+\frac{1}{16}(\omega\tau)^2\right)\delta_b^0\,,\nonumber\\
E=&\left(R_c\frac{15-4 R_\nu}{72(15+2 R_\nu)}\omega\tau^3\right)\delta_c^0+\left(R_b\frac{15-4 R_\nu}{72(15+2 R_\nu)}\omega\tau^3\right)\delta_b^0\,,\nonumber\\
\delta_{c}=&\left(1-\frac{R_c}{2}\omega\tau+\frac{3R_c}{16}(\omega\tau)^2\right)\delta_c^0+R_b\left(-\frac{1}{2}\omega\tau+\frac{3}{16}(\omega\tau)^2\right)\delta_b^0\,,\nonumber\\
\delta_{b}=&\left(-\frac{R_c}{2}\omega\tau+\frac{3R_c}{16}(\omega\tau)^2\right)\delta_c^0+\left(1-\frac{R_b}{2}\omega\tau+\frac{3R_b}{16}(\omega\tau)^2\right)\delta_b^0\,,\nonumber\\
\delta_{\gamma}=&\left(-\frac{2R_c}{3}\omega\tau+\frac{R_c}{4}(\omega\tau)^2\right)\delta_c^0+\left(-\frac{2R_b}{3}\omega\tau+\frac{R_b}{4}(\omega\tau)^2\right)\delta_b^0\,,\nonumber\\
\delta_{\nu}=&\left(-\frac{2R_c}{3}\omega\tau+\frac{R_c}{4}(\omega\tau)^2\right)\delta_c^0+\left(-\frac{2R_b}{3}\omega\tau+\frac{R_b}{4}(\omega\tau)^2\right)\delta_b^0\,,\nonumber\\
%v_c=0\,,\\
v_{\gamma b}=&\left(\frac{R_c }{12}\omega\tau^2\right)\delta_c^0+\left(\frac{R_b }{12}\omega\tau^2\right)\delta_b^0\,,\nonumber\\
v_{\nu}=&\left(\frac{R_c }{12}\omega\tau^2\right)\delta_c^0+\left(\frac{R_b }{12}\omega\tau^2\right)\delta_b^0\,,\label{matteriso1}\\
\sigma_\nu=&\left(-\frac{R_c}{6(15+2 R_\nu)}k^2\omega\tau^3\right)\delta_c^0+\left(-\frac{R_b}{6(15+2 R_\nu)}k^2\omega\tau^3\right)\delta_b^0\,,\nonumber
\end{align}
in which $\omega\equiv\Omega_M \mathcal{H}/\sqrt{\Omega_R}$, $R_c=\Omega_c/\Omega_M$, $R_\nu=\Omega_\nu/\Omega_R$, $R_\gamma=\Omega_\gamma/\Omega_R$ and the $\Omega_s$ are the usual density parameters. We have also used the total matter and total radiation density parameters, respectively given by $\Omega_M=\Omega_c+\Omega_b$ and $\Omega_R=\Omega_\gamma+\Omega_\nu$. This implies that $R_c+R_b=1$ as well as $R_\nu+R_\gamma=1$. Moreover, we have abbreviated the initial values of the cold dark matter and baryon density contrasts, $\delta_c|_{\tau=0}$ and $\delta_b|_{\tau=0}$, to $\delta_c^0$ and $\delta_b^0$ for simplicity of notation. We do this for all other defining variables in all modes presented below.

This example is particularly useful because it also allows us to analyse a combination of modes called the compensated isocurvature mode~\cite{Grin:2011tf}. This mode is defined by the choice of initial conditions for which all variables cancel in the equations above, except the matter density contrasts. It is given by the following condition
\begin{equation}
\label{CIP}
\delta_b^0=-\frac{R_c}{R_b}\delta_c^0\,.
\end{equation}
When the initial conditions are exactly related in this way, no other variables are generated at linear order. As we will later verify, this is no longer true at second order, due to mode mixing.

Another property that we can see in this example is that, at first order in perturbation theory, there is a hierarchy between the brightness tensors in terms of their order in $\tau$: it is clear here, that $\delta_\nu\gg v_\nu\gg \sigma_\nu$. This can be shown using the evolution equations for those variables --- the first-order versions of Eqs.~\eqref{Eqvnu2}--\eqref{Boltz} --- from which one deduces that $v_\nu\propto \int \delta_\nu \text{d}\tau$ and $\sigma_\nu\propto \int v_\nu\text{d}\tau$. This implies that one can safely neglect the higher rank brightness tensors, as they will certainly be smaller than the ones shown. At second order, this is not so straightforward, as all variables are sourced by non-linear terms, which do not have to obey such a hierarchy. In order to test this, all the results below include one extra variable, the scalar part of the rank-3 brightness tensor, $\Delta_3$. Should this variable be of the same order in $\tau$ as $\sigma_\nu$, one may assume that all other brightness tensors are of a similar size. Should that be the case, they may not be negligible, since they may affect the evolution of all other variables. In practice, as we show below, none of the modes under study suffer from this problem and this hierarchy is preserved.

We now present the second-order results for all growing modes, excluding the neutrino velocity mode, as it includes decaying contributions at second order, as discussed above. In all of the results shown, we abuse the notation and use the names of the variables to denote the transfer functions multiplied by the defining variables (for example $\psi^{(2)}=\T_{ij}I_iI_j$) i.e. we show only the integrand of the second-order part of the variable. We begin by showing the pure adiabatic solutions and show the results for the isocurvature modes after that by ``activating" each of the four linear growing modes separately.

\subsection{Pure adiabatic mode}

We find the following results for the initial evolution at second order and at leading order in $\tau$, when including only the quadratic source composed by the adiabatic first-order solutions, in synchronous gauge:
\begin{align}
\psi^{(2)}=&-\frac{4R_\nu k^2(3k^2+k_1^2+k_2^2)+5\left(3(k_1^2-k_2^2)^2+k^2(k_1^2+k_2^2)\right)}{24(4R_\nu+15) k^4}(k\tau)^2\psi^{0}_{k_1}\psi^{0}_{k_2}\,,\nonumber\\
E^{(2)}=&-\frac{5\left(9k^4-3(k_1^2-k_2^2)^2+2k^2(k_1^2+k_2^2)\right)}{8(4R_\nu+15) k^4}\tau^2\psi^{0}_{k_1}\psi^{0}_{k_2}\,,\nonumber\\
\delta_{c}^{(2)}=&-\frac{1}{8}\left(3 k^2+5(k_1^2+k_2^2)\right)\tau^2\psi^{0}_{k_1}\psi^{0}_{k_2}\,,\nonumber\\
\delta_{b}^{(2)}=&-\frac{1}{8}\left(3 k^2+5(k_1^2+k_2^2)\right)\tau^2\psi^{0}_{k_1}\psi^{0}_{k_2}\,,\nonumber\\
\delta_{\gamma}^{(2)}=&-\frac{1}{6}\left(3 k^2+5(k_1^2+k_2^2)\right)\tau^2\psi^{0}_{k_1}\psi^{0}_{k_2}\,,\nonumber\\
\delta_{\nu}^{(2)}=&-\frac{1}{6}\left(3 k^2+5(k_1^2+k_2^2)\right)\tau^2\psi^{0}_{k_1}\psi^{0}_{k_2}\,,\\
%v_c^{(2)}=&0\\
v_{\gamma b}^{(2)}=&\frac{1}{72k^2}\left(3k^4+2(k_1^2-k_2^2)^2+7k^2(k_1^2+k_2^2)\right)\tau^3\psi^{0}_{k_1}\psi^{0}_{k_2}\,,\nonumber\\
v_{\nu}^{(2)}=&\frac{23+4R_\nu}{72(4R_\nu+15)k^2}\left(3k^4+2(k_1^2-k_2^2)^2+7k^2(k_1^2+k_2^2)\right)\tau^3\psi^{0}_{k_1}\psi^{0}_{k_2}\,,\nonumber\\
\sigma_\nu^{(2)}=&\frac{\left(9k^4-3(k_1^2-k_2^2)^2+2k^2(k_1^2+k_2^2)\right)}{6(4R_\nu+15) k^4}(k\tau)^2\psi^{0}_{k_1}\psi^{0}_{k_2}\,,\nonumber\\
\Delta_3^{(2)}=&-\frac{37k^4+9(k_1^2-k_2^2)^2-6k^2(k_1^2+k_2^2)}{42(15+4R_\nu)k^4}\tau^3\psi^{0}_{k_1}\psi^{0}_{k_2}\,,\nonumber
\end{align}

These results for the adiabatic case were already known in Poisson gauge \cite{Pettinari:2014vja,Pitrou:2010sn} and one can check that they match ours by using the gauge transformations given in Appendix \ref{gaugetr}. We see here that $\sigma_\nu$ is initially larger (in order of $\tau$) than $v_\nu$. This was not the case at the linear level. However, we also note that $\Delta_3$ is again higher order in $\tau$, giving us confidence that higher-rank tensors can be neglected.

\subsection{Pure cold dark matter isocurvature mode}

For the solution that is sourced by the quadratic dark matter isocurvature first-order modes, the initial evolution is given by:
\begin{align}
\psi^{(2)}=&R_c^2\left(\frac{(\omega\tau)^2}{48}-\frac{(\omega\tau)^3}{72}\right)\delta_{c,k_1}^{0}\delta_{c,k_2}^{0}\,,\nonumber\\
E^{(2)}=&O(\tau^4)\nonumber\\
\delta_{c}^{(2)}=&R_c\left(-\omega\tau+\frac{18+23R_c}{48}(\omega\tau)^2+\frac{16 (k_1^2+k_2^2)-15(6+17R_c)\omega^2}{720}\omega\tau^3\right)\delta_{c,k_1}^{0}\delta_{c,k_2}^{0}\,,\nonumber\\
\delta_{b}^{(2)}=&R_c^2\left(\frac{23}{48}(\omega\tau)^2-\frac{17}{48}(\omega\tau)^3\right)\delta_{c,k_1}^{0}\delta_{c,k_2}^{0}\,,\nonumber\\
\delta_{\gamma}^{(2)}=&R_c^2\left(\frac{3}{4}(\omega\tau)^2-\frac59(\omega\tau)^3\right)\delta_{c,k_1}^{0}\delta_{c,k_2}^{0}\,,\nonumber\\
\delta_{\nu}^{(2)}=&R_c^2\left(\frac{3}{4}(\omega\tau)^2-\frac59(\omega\tau)^3\right)\delta_{c,k_1}^{0}\delta_{c,k_2}^{0}\,,\\
%v_c^{(2)}=&0\\
v_{\gamma b}^{(2)}=&R_c^2\left(-\frac{7\omega^2\tau^3}{144}+\frac{(15 R_b+16 R_\gamma)\omega^3\tau^4}{576R_\gamma}\right)\delta_{c,k_1}^{0}\delta_{c,k_2}^{0}\,,\nonumber\\
v_{\nu}^{(2)}=&R_c^2\left(-\frac{7\omega^2\tau^3}{144}+\frac{\omega^3\tau^4}{36}\right)\delta_{c,k_1}^{0}\delta_{c,k_2}^{0}\,,\nonumber\\
\sigma_\nu^{(2)}=&O(\tau^4)\,,\nonumber\\
\Delta_3^{(2)}=&O(\tau^5)\,.\nonumber
\end{align}

\subsection{Mixture of adiabatic and cold dark matter modes}

When both the adiabatic mode and the dark matter isocurvature are present, a mixed mode is generated, for which the initial evolution is:
\begin{align}
\psi^{(2)}=&R_c\left(\frac{1}{3}\omega\tau-\frac{1}{8}(\omega\tau)^2\right)\delta_{c,k_1}^{0}\psi^{0}_{k_2}\,,\nonumber\\
E^{(2)}=&f_E^{c\psi}(k,k_1,k_2)\omega\tau^3\delta_{c,k_1}^{0}\psi^{0}_{k_2}\nonumber\\
\delta_{c}^{(2)}=&\left(-\frac{1}{4}k_2^2\tau^2 +\frac{1}{180} (-2(k^2-5k_1^2)R_c +k_2^2(9+41R_c))\omega\tau^3\right)\delta_{c,k_1}^{0}\psi^{0}_{k_2}\,,\nonumber\\
\delta_{b}^{(2)}=&-\frac{R_c}{120}\omega\tau^3(3k^2-15k_1^2-29k_2^2)\delta_{c,k_1}^{0}\psi^{0}_{k_2}\,,\nonumber\\
\delta_{\gamma}^{(2)}=&-\frac{R_c}{90}\omega\tau^3(3k^2-15k_1^2-34k_2^2)\delta_{c,k_1}^{0}\psi^{0}_{k_2}\,,\nonumber\\
\delta_{\nu}^{(2)}=&-\frac{R_c}{90}\omega\tau^3(3k^2-15k_1^2-34k_2^2)\delta_{c,k_1}^{0}\psi^{0}_{k_2}\,,\\
%v_c^{(2)}=&0\\
v_{\gamma b}^{(2)}=&\left(\frac{R_c}{12k^2}(k^2+k_1^2-k_2^2)\omega\tau^2-\frac{R_c(R_\gamma+3R_b)}{48R_\gamma k^2}(k^2+k_1^2-k_2^2)\omega^2\tau^3\right)\delta_{c,k_1}^{0}\psi^{0}_{k_2}\,,\nonumber\\
v_{\nu}^{(2)}=&\left(\frac{R_c}{12k^2}(k^2+k_1^2-k_2^2)\omega\tau^2-\frac{R_c}{48k^2}(k^2+k_1^2-k_2^2)\omega^2\tau^3\right)\delta_{c,k_1}^{0}\psi^{0}_{k_2}\,,\nonumber\\
\sigma_\nu^{(2)}=&f^{c\psi}_\sigma(k,k_1,k_2)\omega k^2\tau^3\delta_{c,k_1}^{0}\psi^{0}_{k_2}\,,\nonumber\\
\Delta_3^{(2)}=&O(\tau^4)\,,\nonumber
\end{align}
with the following kernels:
\begin{align}
f^{c\psi}_E=&-\frac{R_c}{576 (15+4R_\nu) (15+2R_\nu) k^4}\left[(225 + 720 R_\nu + 32 R_\nu^2) k^4\nonumber\right.\\
&+ 3 (675 + 240 R_\nu - 32 R_\nu^2) (k_1^2 - k_2^2)^2 \nonumber\\
&\left. +  2 k^2 ((-1125 - 720 R_\nu + 32 R_\nu^2) k_1^2 + (-225 + 240 R_\nu + 32 R_\nu^2) k_2^2)\right]\,,\nonumber\\
f^{c\psi}_\sigma=&-\frac{R_c}{48 (15+4R_\nu) (15+2R_\nu) k^4}\left[(135 + 8 R_\nu) k^4 +3 (5 - 8 R_\nu) (k_1^2 - k_2^2)^2\right.\nonumber\\ 
&\left.+ 2 k^2 ((-75 + 8 R_\nu) k_1^2 + (65 + 8 R_\nu) k_2^2)\right]\,.\nonumber
\end{align}
Note that to get the full results for the mixed mode one would have to add the complementary solution obtained by switching $k_1\leftrightarrow k_2$. We can see that these mixed modes do initially grow (i.e. they are not zero) and are thus not negligible for the evolution of the system. They must be taken into account if one is to have an accurate understanding of the effect of isocurvature modes on non-linear observables. This is even more important in the particular case shown, since this mode includes a contribution from the adiabatic mode, which should make this mixed mode more relevant than the pure isocurvature one, presented before.

\subsection{Pure baryon isocurvature mode}

We now move on to the introduction of the baryon isocurvature mode:
\begin{align}
\psi^{(2)}=&R_b^2\left(\frac{(\omega\tau)^2}{48}-\frac{(\omega\tau)^3}{72}\right)\delta_{b,k_1}^{0}\delta_{b,k_2}^{0}\,,\nonumber\\
E^{(2)}=&O(\tau^4)\nonumber\\
\delta_{c}^{(2)}=&R_b^2\left(\frac{23}{48}(\omega\tau)^2-\frac{17}{48}(\omega\tau)^3\right)\delta_{b,k_1}^{0}\delta_{b,k_2}^{0}\,,\nonumber\\
\delta_{b}^{(2)}=&R_b\omega\tau\left(-1+\frac{18+23R_b}{48}\omega\tau+\frac{16 (k_1^2+k_2^2)+20k^2-15(6+17R_c)\omega^2}{720}\tau^2\right)\delta_{b,k_1}^{0}\delta_{b,k_2}^{0}\,,\nonumber\\
\delta_{\gamma}^{(2)}=&R_b^2\left(\frac{3}{4}(\omega\tau)^2-\frac59(\omega\tau)^3\right)\delta_{b,k_1}^{0}\delta_{b,k_2}^{0}\,,\nonumber\\
\delta_{\nu}^{(2)}=&R_b^2\left(\frac{3}{4}(\omega\tau)^2-\frac59(\omega\tau)^3\right)\delta_{b,k_1}^{0}\delta_{b,k_2}^{0}\,,\\
%v_c^{(2)}=&0\\
v_{\gamma b}^{(2)}=&R_b^2\left(\frac{7R_\nu-16}{144R_\gamma}\omega^2\tau^3+\frac{R_b(69-15R_\nu)+16R_\gamma^2}{576R_\gamma^2}\omega^3\tau^4\right)\delta_{b,k_1}^{0}\delta_{b,k_2}^{0}\,,\nonumber\\
v_{\nu}^{(2)}=&R_b^2\left(-\frac{7}{144}\omega^2\tau^3+\frac{1}{36}\omega^3\tau^4\right)\delta_{b,k_1}^{0}\delta_{b,k_2}^{0}\,,\nonumber\\
\sigma_\nu^{(2)}=&O(\tau^4)\,,\nonumber\\
\Delta_3^{(2)}=&O(\tau^5)\,.\nonumber
\end{align}
This solution is very similar to the ``pure" dark matter isocurvature, as it is already at first order. In this case, however, the application of the compensated isocurvature condition, Eq.~\eqref{CIP}, would not lead to cancellations when this result is summed to the dark matter one, due the quadratic nature of these solutions. Furthermore, some terms are completely different in the two cases, namely the matter densities and the baryon-photon velocity. However, in order to completely analyse the initial evolution of the compensated isocurvature mode, we must still investigate the mixed mode between the baryon and dark matter isocurvatures. 

\subsection{Mixture of baryon and cold dark matter modes}

This mixed mode is given by
\begin{align}
\psi^{(2)}=&\frac{R_bR_c}{48}\left((\omega\tau)^2-\frac23 (\omega\tau)^3\right)\delta_{b,k_1}^{0}\delta_{c,k_2}^{0}\,,\nonumber\\
E^{(2)}=&O(\tau^4)\nonumber\\
\delta_{c}^{(2)}=&R_b\left(-\frac{1}{2}\omega\tau+\frac{9+23 R_c}{48}(\omega\tau)^2 + \frac{16 k_1^2 - 15 (3 + 17 R_c) \omega^2}{720} \omega \tau^3\right)\delta_{b,k_1}^{0}\delta^{0}_{c,k_2}\,,\nonumber\\
\delta_{b}^{(2)}=&R_c\left(-\frac{1}{2}\omega\tau+\frac{9+23 R_b}{48}(\omega\tau)^2 + \frac{10 k^2 - 10 k_1^2 + 26 k_2^2 - 300 \omega^2 + 255 R_c \omega^2}{720} \omega \tau^3 \right)\delta_{b,k_1}^{0}\delta^{0}_{c,k_2}\,,\nonumber\\
\delta_{\gamma}^{(2)}=&R_bR_c\left(\frac{3}{4}(\omega\tau)^2-\frac{5}{9}(\omega\tau)^3\right)\delta_{b,k_1}^{0}\delta_{c,k_2}^{0}\,,\nonumber\\
\delta_{\nu}^{(2)}=&R_bR_c\left(\frac{3}{4}(\omega\tau)^2-\frac{5}{9}(\omega\tau)^3\right)\delta_{b,k_1}^{0}\delta_{c,k_2}^{0}\,,\\
%v_c^{(2)}=&0\\
v_{\gamma b}^{(2)}=&\frac{R_bR_c(9(k_1^2-k_2^2)-(23-14R_\nu)k^2)}{288R_\gamma k^2}\omega^2\tau^3\delta_{b,k_1}^{0}\delta_{c,k_2}^{0}\,,\nonumber\\
v_{\nu}^{(2)}=&-\frac{7R_bR_c}{144}\omega^2\tau^3\delta_{b,k_1}^{0}\delta_{c,k_2}^{0}\,,\nonumber\\
\sigma_\nu^{(2)}=&O(\tau^4)\,,\nonumber\\
\Delta_3^{(2)}=&O(\tau^5)\,.\nonumber
\end{align}
Adding all the matter modes together and applying the compensated isocurvature condition, Eq.~\eqref{CIP}, we can show that again, the compensated isocurvature mode has vanishing initial evolution even at second order. This is not surprising, since, if only these matter isocurvature modes are active and do not evolve at linear order, they would only source the second-order evolution if terms like $\delta_c^2$, $\delta_b^2$ or $\delta_c\delta_b$ existed in the evolution equations. Having concluded that a pure compensated isocurvature mode does not evolve initially, it remains to be seen whether it can mix with the adiabatic mode and generate additional contributions. 

\subsection{Mixture of adiabatic and baryon modes}

To test what happens when one mixes a compensated isocurvature with the adiabatic mode, we first need the mixed mode between the baryon isocurvature and the adiabatic mode:
\begin{align}
\psi^{(2)}=&R_b\left(\frac{1}{3}\omega\tau-\frac{1}{8}\omega^2\tau^2\right)\delta_{b,k_1}^{0}\psi^{0}_{k_2}\,,\nonumber\\
E^{(2)}=&f_E^{b\psi}(k,k_1,k_2)\omega\tau^3\delta_{b,k_1}^{0}\psi^{0}_{k_2}\,,\nonumber\\
\delta_{c}^{(2)}=&-\frac{R_b}{180}\omega\tau^3(-2k^2+10k_1^2+41k_2^2)\delta_{b,k_1}^{0}\psi^{0}_{k_2}\,,\nonumber\\
\delta_{b}^{(2)}=&\left(-\frac{1}{4}k_2^2\tau^2 +\frac{1}{120} ((15k_1^2+29k_2^2-3k^2)R_b +6 k_2^2)\omega\tau^3\right)\delta_{b,k_1}^{0}\psi^{0}_{k_2}\,,\nonumber\\
\delta_{\gamma}^{(2)}=&-\frac{R_b}{90}\omega\tau^3(3k^2-15k_1^2-34k_2^2)\delta_{b,k_1}^{0}\psi^{0}_{k_2}\,,\nonumber\\
\delta_{\nu}^{(2)}=&-\frac{R_b}{90}\omega\tau^3(3k^2-15k_1^2-34k_2^2)\delta_{b,k_1}^{0}\psi^{0}_{k_2}\,,\\
%v_c^{(2)}=&0\\
v_{\gamma b}^{(2)}=&\left(\frac{R_b}{12k^2}(k^2+k_1^2-k_2^2)\omega\tau^2-\frac{R_b(R_\gamma+3R_b)}{48R_\gamma k^2}(k^2+k_1^2-k_2^2)\omega^2\tau^3\right)\delta_{b,k_1}^{0}\psi^{0}_{k_2}\,,\nonumber\\
v_{\nu}^{(2)}=&\left(\frac{R_b}{12k^2}(k^2+k_1^2-k_2^2)\omega\tau^2-\frac{R_b}{48k^2}(k^2+k_1^2-k_2^2)\omega^2\tau^3\right)\delta_{b,k_1}^{0}\psi^{0}_{k_2}\,,\nonumber\\
\sigma_\nu^{(2)}=&f^{b\psi}_\sigma(k,k_1,k_2)\omega k^2\tau^3\delta_{b,k_1}^{0}\psi^{0}_{k_2}\,,\nonumber\\
\Delta_3^{(2)}=&O(\tau^4)\,,\nonumber
\end{align}
with the following kernels:
\begin{gather}
f^{b\psi}_E(k,k_1,k_2)=\frac{R_b}{R_c}f^{c\psi}_E(k,k_1,k_2)\,,\nonumber
\\
f^{b\psi}_\sigma(k,k_1,k_2)=\frac{R_b}{R_c}f^{c\psi}_\sigma(k,k_1,k_2)\,.\nonumber
\end{gather}
It is immediately clear, from the relationship between the kernels for $E$ and $\sigma$, that cancellations will occur when the compensated isocurvature condition, Eq.~\eqref{CIP}, is applied. However, there are some terms that do survive and are given by
\begin{align}
\delta_{c}^{(2)}=&-\frac{1}{20} k_2^2\tau^2(5-\omega\tau)\delta_{\text{CI},k_1}^{0}\psi^{0}_{k_2}\,,\nonumber\\
\delta_{b}^{(2)}=&\frac{R_c}{20R_b} k_2^2\tau^2(5-\omega\tau)\delta_{\text{CI},k_1}^{0}\psi^{0}_{k_2}\,,\\
v_{\gamma b}^{(2)}=&\frac{R_c}{R_\gamma}\frac{k^2+k_1^2-k_2^2}{96k^2}k_2^2\omega\tau^4\delta_{\text{CI},k_1}^{0}\psi^{0}_{k_2}\,,\nonumber
\end{align}
in which $\delta_{\text{CI},k_1}^{0}$ is the initial density contrast of dark matter in the compensated isocurvature mode. We see here that the compensated isocurvature condition is conserved, i.e. $\delta_{b}^{(2)}=-\frac{R_c}{R_b}\delta_c^{(2)}$, but we also see that the common velocity of the baryons and photons is generated in this mixed mode, which was non-existent at linear order. We confirm here that the compensated isocurvature mode does have an effect on the evolution at second order, even at these early times.

\subsection{Pure neutrino density isocurvature mode}

We now introduce the solutions sourced by the neutrino density isocurvature. First we show the results for the ``pure" mode:
\begin{align}
\psi^{(2)}=&f^{\nu\nu}_\psi(k,k_1,k_2) (k\tau)^2\delta_{\nu,k_1}^{0}\delta_{\nu,k_2}^{0}\,,\nonumber\\
E^{(2)}=&f^{\nu\nu}_E(k,k_1,k_2) \tau^2\delta_{\nu,k_1}^{0}\delta_{\nu,k_2}^{0}\,,\nonumber\\
\delta_{c}^{(2)}=&-\frac{R_b R_\nu^2}{320R_\gamma^2}(7k^2-3(k_1^2+k_2^2))\omega\tau^3\delta_{\nu,k_1}^{0}\delta_{\nu,k_2}^{0}\,,\nonumber\\
\delta_{b}^{(2)}=&\frac{R_\nu^2}{32R_\gamma^2}(7k^2-3(k_1^2+k_2^2))\tau^2\delta_{\nu,k_1}^{0}\delta_{\nu,k_2}^{0}\,,\nonumber\\
\delta_{\gamma}^{(2)}=&\frac{R_\nu^2}{12R_\gamma^2}(k^2-k_1^2-k_2^2)\tau^2\delta_{\nu,k_1}^{0}\delta_{\nu,k_2}^{0}\,,\nonumber\\
\delta_{\nu}^{(2)}=&\frac{1}{12}(k^2-k_1^2-k_2^2)\tau^2\delta_{\nu,k_1}^{0}\delta_{\nu,k_2}^{0}\,,\\
%v_c^{(2)}=&0\\
v_{\gamma b}^{(2)}=&\left(\frac{R_\nu^2}{4R_\gamma^2}\tau-\frac{3R_b R_\nu^2}{8R_\gamma^3}\omega\tau^2\right)\delta_{\nu,k_1}^{0}\delta_{\nu,k_2}^{0}\,,\nonumber\\
v_{\nu}^{(2)}=&\frac{1}{4}\tau\delta_{\nu,k_1}^{0}\delta_{\nu,k_2}^{0}\,,\nonumber\\
\sigma_\nu^{(2)}=&f^{\nu\nu}_\sigma(k,k_1,k_2)(k\tau)^2\delta_{\nu,k_1}^{0}\delta_{\nu,k_2}^{0}\,,\nonumber\\
\Delta_3^{(2)}=&f^{\nu\nu}_\Delta(k,k_1,k_2)\tau^3\delta_{\nu,k_1}^{0}\delta^{0}_{\nu,k_2}\,,\nonumber
\end{align}
in which the kernels abbreviated above are given by
\begin{align}
f^{\nu\nu}_\psi(k,k_1,k_2)=&-\frac{R_\nu^2\left[(27 + 68 R_\nu) k^4 -(91 + 4 R_\nu) \left( 3 (k_1^2 - k_2^2)^2 - 2 k^2 (k_1^2 + k_2^2)\right)\right]}{96R_\gamma(4R_\nu+15)^2 k^4}\,,\nonumber\\
f^{\nu\nu}_E(k,k_1,k_2)=&-3 f^{\nu\nu}_\psi(k,k_1,k_2)\,,\nonumber\\
f^{\nu\nu}_\sigma(k,k_1,k_2)=&-\frac{1}{96R_\gamma(4R_\nu+15)^2 k^4}\left[(-225 - 39 R_\nu + 188 R_\nu^2) k^4 \right.\nonumber\\
&\left.+ (225 - 153 R_\nu + 4 R_\nu^2)\left(3  (k_1^2 - k_2^2)^2 - 2  k^2 (k_1^2 + k_2^2)\right)\right]\nonumber\,,\\
f^{\nu\nu}_\Delta(k,k_1,k_2)=&-\frac{R_\nu\left[(-51 + 32 R_\nu) k^4 + (3 + 16 R_\nu)\left(3  (k_1^2 - k_2^2)^2 - \nonumber
 2  k^2 (k_1^2 + k_2^2)\right)\right]}{84R_\gamma(4R_\nu+15)^2 k^4}\,.
\end{align}

\subsection{Mixture of adiabatic and neutrino modes}

The mixed mode between the neutrino density isocurvature and the adiabatic mode is given by
\begin{align}
\psi^{(2)}=&f_\psi^{\nu\psi}(k,k_1,k_2)(k\tau)^2\delta_{\nu,k_1}^{0}\psi^{0}_{k_2}\,,\nonumber\\
E^{(2)}=&f_E^{\nu\psi}(k,k_1,k_2)\tau^2\delta_{\nu,k_1}^{0}\psi^{0}_{k_2}\,,\nonumber\\
\delta_{c}^{(2)}=&\frac{R_bR_\nu}{160R_\gamma}(k^2-5k_1^2-k_2^2)\omega\tau^3\delta_{b,k_1}^{0}\psi^{0}_{k_2}\,,\nonumber\\
\delta_{b}^{(2)}=&-\frac{R_\nu}{16R_\gamma}(k^2-5k_1^2-k_2^2)\tau^2\delta_{\nu,k_1}^{0}\psi^{0}_{k_2}\,,\nonumber\\
\delta_{\gamma}^{(2)}=&-\frac{R_\nu}{12R_\gamma}(k^2-5(k_1^2+k_2^2))\tau^2\delta_{\nu,k_1}^{0}\psi^{0}_{k_2}\,,\nonumber\\
\delta_{\nu}^{(2)}=&\frac{1}{12}(k^2-5(k_1^2+k_2^2))\tau^2\delta_{\nu,k_1}^{0}\psi^{0}_{k_2}\,,\\
%v_c^{(2)}=&0\\
v_{\gamma b}^{(2)}=&\left(\frac{R_\nu(k^2+k_1^2-k_2^2)}{4R_\gamma k^2}\tau-\frac{3R_bR_\nu(k^2+k_1^2-k_2^2)}{16R_\gamma^2k^2}\omega\tau^2\right)\delta_{\nu,k_1}^{0}\psi^{0}_{k_2}\,,\nonumber\\
v_{\nu}^{(2)}=&-\frac{(k^2+k_1^2-k_2^2)}{4k^2}\tau\delta_{\nu,k_1}^{0}\psi^{0}_{k_2}\,,\nonumber\\
\sigma_\nu^{(2)}=&f_\sigma^{\nu\psi}(k,k_1,k_2) (k\tau)^2\delta_{\nu,k_1}^{0}\psi^{0}_{k_2}\,,\nonumber\\
\Delta_3^{(2)}=&f^{\nu\psi}_\Delta(k,k_1,k_2)\tau^3\delta_{\nu,k_1}^{0}\psi^{0}_{k_2}\,.\nonumber
\end{align}
The kernels are given by
\begin{align}
f^{\nu\psi}_\psi=&-\frac{R_\nu\left[(45 + 4 R_\nu) k^4 - 3 (5 + 4 R_\nu) (k_1^2 - k_2^2)^2 + 
 k^2 ((-30 + 8 R_\nu) k_1^2 + 2 (25 + 4 R_\nu) k_2^2)\right]}{24(4R_\nu+15)^2 k^4}\,,\nonumber\\
f^{\nu\psi}_E=&-3f^{\nu\psi}_\psi\,,\nonumber\\
f^{\nu\psi}_\sigma=&-\frac{3}{R_\nu}f^{\nu\psi}_\psi(k,k_1,k_2)\,,\nonumber\\
f^{\nu\psi}_\Delta=&-\frac{1}{336(15+4R\nu)^2 k^6}\left[(1545 + 316 R_\nu) k^6 + 35 (15 + 4 R_\nu) (k_1^2 - k_2^2)^3\right.\nonumber\\
&- 3 k^2 (k_1^2 - k_2^2) (3 (65 + 28 R_\nu) k_1^2 + (225 + 28 R_\nu) k_2^2) \nonumber\\
&\left. + k^4 ((675 + 372 R_\nu) k_1^2 - 5 (147 + 52 R_\nu) k_2^2)\right]\nonumber\,.
\end{align}

\subsection{Mixture of dark matter and neutrino modes}

Now we show the neutrino-dark matter mixed mode:
\begin{align}
\psi^{(2)}=&f_\psi^{\nu c}(k,k_1,k_2)\omega k^2\tau^3\delta_{\nu,k_1}^{0}\delta_{c,k_2}^{0}\,,\nonumber\\
E^{(2)}=&f_E^{\nu c}(k,k_1,k_2)\omega\tau^3\delta_{\nu,k_1}^{0}\delta_{c,k_2}^{0}\,,\nonumber\\
\delta_{c}^{(2)}=&-\frac{R_bR_\nu}{80R_\gamma}k_1^2\omega\tau^3\delta_{\nu,k_1}^{0}\delta_{c,k_2}^{0}\,,\nonumber\\
\delta_{b}^{(2)}=&-\frac{R_\nu R_c}{288R_\gamma}(-5k^2+29k_1^2+5k_2^2)\omega\tau^3\delta_{\nu,k_1}^{0}\delta_{c,k_2}^{0}\,,\nonumber\\
\delta_{\gamma}^{(2)}=&\frac{R_\nu R_c}{R_\gamma}\left(\frac{2}{3}\omega\tau-\frac{1}{4}(\omega\tau)^2\right)\delta_{\nu,k_1}^{0}\delta_{c,k_2}^{0}\,,\nonumber\\
\delta_{\nu}^{(2)}=&\left(-\frac{2R_c}{3}\omega\tau+\frac{R_c}{4}(\omega\tau)^2\right)\delta_{\nu,k_1}^{0}\delta_{c,k_2}^{0}\,,\\
%v_c^{(2)}=&0\\
v_{\gamma b}^{(2)}=&-\frac{R_\nu R_c}{R_\gamma}\left(\frac{k^2+k_1^2-k_2^2}{32k^2}\omega\tau^2+\frac{k^2(9R_b-4R_\gamma)-(k_1^2-k_2^2)(4R_\gamma+15 R_b)}{384R_\gamma k^2}\omega^2\tau^3\right)\delta_{\nu,k_1}^{0}\delta_{c,k_2}^{0}\,,\nonumber\\
v_{\nu}^{(2)}=&\left(\frac{(k^2+k_1^2-k_2^2)R_c}{32k^2}\omega\tau^2-\frac{(k^2+k_1^2-k_2^2)R_c}{96k^2}\omega^2\tau^3\right)\delta_{\nu,k_1}^{0}\delta_{c,k_2}^{0}\,,\nonumber\\
\sigma_\nu^{(2)}=&f_\sigma^{\nu c}(k,k_1,k_2) \omega k^2 \tau^3\delta_{\nu,k_1}^{0}\delta_{c,k_2}^{0}\,,\nonumber\\
\Delta_3^{(2)}=&O(\tau^4)\,,\nonumber
\end{align}
with the following kernels:
\begin{align}
f^{\nu c}_\psi(k,k_1,k_2)=&\frac{R_\nu R_c}{144(2R_\nu+15)^2(4R_\nu+15) k^4}\left[(675 + 90 R_\nu - 6 R_\nu^2) (k^4+ (k_1^2 - k_2^2)^2)\right.\nonumber\\
&\left. - 2 k^2 ((225 + 90 R_\nu + 2 R_\nu^2) k_1^2 - 3 (-75 + 10 R_\nu + 2 R_\nu^2) k_2^2)\right]\nonumber\,,\\
f^{\nu c}_E(k,k_1,k_2)=&-3f^{\nu c}_\psi(k,k_1,k_2)\,,\nonumber\\
f^{\nu c}_\sigma(k,k_1,k_2)=&\frac{R_c}{96(2R_\nu+15)^2(4R_\nu+15) k^4}\left[3 (-1125 - 180 R_\nu + 4 R_\nu^2) (k^4+(k_1^2 - k_2^2)^2) \right.\nonumber\\
&\left. + 2 k^2 ((675 + 300 R_\nu + 4 R_\nu^2) k_1^2 +  3 (525 + 20 R_\nu - 4 R_\nu^2) k_2^2)\right]\nonumber\,.
\end{align}

\subsection{Mixture of baryon and neutrino modes}

Finally, the results for the neutrino-baryon mixed mode are
\begin{align}
\psi^{(2)}=&f_\psi^{\nu b}(k,k_1,k_2)\omega k^2\tau^3\delta_{\nu,k_1}^{0}\delta_{b,k_2}^{0}\,,\nonumber\\
E^{(2)}=&f_E^{\nu b}(k,k_1,k_2)\omega\tau^3\delta_{\nu,k_1}^{0}\delta_{b,k_2}^{0}\,,\nonumber\\
\delta_{c}^{(2)}=&-\frac{R_bR_\nu}{160R_\gamma}(k^2+k_1^2-k_2^2)\omega\tau^3\delta_{\nu,k_1}^{0}\delta_{b,k_2}^{0}\,,\nonumber\\
\delta_{b}^{(2)}=&\frac{R_\nu}{16R_\gamma}(k^2+k_1^2-k_2^2)\tau^2\delta_{\nu,k_1}^{0}\delta_{b,k_2}^{0}\,,\nonumber\\
\delta_{\gamma}^{(2)}=&\frac{R_\nu R_b}{R_\gamma}\left(\frac{2}{3}\omega\tau-\frac{1}{4}(\omega\tau)^2\right)\delta_{\nu,k_1}^{0}\delta_{b,k_2}^{0}\,,\nonumber\\
\delta_{\nu}^{(2)}=&R_b\left(-\frac{2}{3}\omega\tau+\frac{1}{4}(\omega\tau)^2\right)\delta_{\nu,k_1}^{0}\delta_{b,k_2}^{0}\,,\\
%v_c^{(2)}=&0\\
v_{\gamma b}^{(2)}=&\left(\frac{(k^2+k_1^2-k_2^2)R_b R_\nu(R_\nu-4)}{32R_\gamma^2 k^2}\omega\tau^2+f_v^{\nu b} (k,k_1,k_2)\omega^2\tau^3\right)\delta_{\nu,k_1}^{0}\delta_{b,k_2}^{0}\,,\nonumber\\
v_{\nu}^{(2)}=&R_b\left(\frac{(k^2+k_1^2-k_2^2)}{32k^2}\omega\tau^2-\frac{(k^2+k_1^2-k_2^2)}{96k^2}\omega^2\tau^3\right)\delta_{\nu,k_1}^{0}\delta_{b,k_2}^{0}\,,\nonumber\\
\sigma_\nu^{(2)}=&f_\sigma^{\nu b}(k,k_1,k_2) \omega k^2 \tau^3\delta_{\nu,k_1}^{0}\delta_{b,k_2}^{0}\,,\nonumber\\
\Delta_3^{(2)}=&O(\tau^4)\,,\nonumber
\end{align}
with the following kernels:
\begin{align}
f^{\nu b}_\psi=&-\frac{R_bR_\nu}{1440R\gamma(2R_\nu+15)^2(4R_\nu+15) k^4}\left[3 (1125 + 3750 R_\nu + 620 R_\nu^2 - 4 R_\nu^3) k^4 \right.\nonumber\\
&- 30 (225 - 195 R_\nu - 32 R_\nu^2 + 2 R_\nu^3) (k_1^2 - k_2^2)^2 \nonumber\\
&\left.+ k^2 ((14625 + 2700 R_\nu - 860 R_\nu^2 + 8 R_\nu^3) k_1^2 + 3 (-1875 - 3500 R_\nu - 140 R_\nu^2 + 24 R_\nu^3) k_2^2)\right]\,,\nonumber\\
f^{\nu b}_E=&\frac{R_b}{R_c}f^{\nu c}_E\,,\nonumber\\
f^{\nu b}_v=&\frac{R_\nu R_b}{384R_\gamma^3 k^2}\left[k^2(-5+R_\nu+4 R_\nu^2+9R_b(5+R_\nu))\right.\nonumber\\
&\left.+(k_1^2-k_2^2)(-5 + R_b (69 - 15 R_\nu) + R_\nu + 4 R_\nu^2)\right]\,,\nonumber\\
f^{\nu b}_\sigma=&\frac{R_b}{R_c}f^{\nu c}_\sigma\,.\nonumber
\end{align}

We can also analyse here if the compensated isocurvature generates an extra contribution when mixed with the neutrino isocurvature. We find that it does and present below the initial evolution for that mixed mode, showing only the non-zero variables:
\begin{align}
\psi^{(2)}=&\frac{R_\nu R_c}{480 R_\gamma}(k^2+k_1^2-k_2^2)\omega\tau^3\delta_{\nu,k_1}^{0}\delta_{CI,k_2}^{0}\,,\nonumber\\
\delta_{c}^{(2)}=&\frac{R_\nu}{160 R_\gamma}\left(R_c(k^2+k_1^2-k_2^2)-2R_b k_1^2\right)\omega\tau^3\delta_{\nu,k_1}^{0}\delta_{CI,k_2}^{0}\,,\nonumber\\
\delta_{b}^{(2)}=&-\frac{R_\nu R_c}{16 R_\gamma R_b}(k^2+k_1^2-k_2^2)\tau^2\delta_{\nu,k_1}^{0}\delta_{CI,k_2}^{0}\,,\nonumber\\
\delta_{\gamma}^{(2)}=&\frac{R_\nu R_c(6-R_\nu)}{120 R_\gamma^2}(k^2+k_1^2-k_2^2)\omega\tau^3\delta_{\nu,k_1}^{0}\delta_{CI,k_2}^{0}\,,\nonumber\\
\delta_{\nu}^{(2)}=&\frac{R_\nu R_c}{120 R_\gamma}(k^2+k_1^2-k_2^2)\omega\tau^3\delta_{\nu,k_1}^{0}\delta_{CI,k_2}^{0}\,,\\
v_{\gamma b}^{(2)}=&\frac{3 R_\nu R_c}{32 R_\gamma^2}\frac{k^2+k_1^2-k_2^2}{k^2}\omega\tau^2\delta_{\nu,k_1}^{0}\delta_{CI,k_2}^{0}\,,\nonumber\\
v_{\nu}^{(2)}=&-\frac{R_\nu R_c}{1920 R_\gamma}(k^2+k_1^2-k_2^2)\omega\tau^4\delta_{\nu,k_1}^{0}\delta_{CI,k_2}^{0}\,.\nonumber
\end{align}
We see that the mixture of these two modes is far more consequential in this case than it was when the compensated isocurvature mixed with the adiabatic mode. In particular, the compensated isocurvature relation, Eq.~\eqref{CIP}, is not conserved at second order and many other quantities are generated besides the matter density perturbations, in clear contrast to what happens at the linear level.

We also note that in all the solutions above, the hierarchy between $v_\nu$, $\sigma_\nu$ and $\Delta_3$ is maintained, i.e. $v_\nu\gtrsim\sigma_\nu\gtrsim\Delta_3$, in terms of their order in the expansion in $\tau$. This gives us confidence that we can neglect the initial evolution of the higher brightness tensors for all the modes under study.

\section{Conclusion}\label{conclusionCiso}

We have studied the approximate initial solutions for the transfer functions of the most relevant variables used in the initialization of Boltzmann solvers at second order in perturbation theory. In order to do this, we have described the differential system and precisely defined the different modes under study. We have concluded that the number of purely growing modes is smaller at second order, as we have shown that the neutrino velocity mode sources decaying solutions due to its contribution to the total anisotropic stress. Furthermore, we have highlighted the importance of the solutions sourced by multiple modes, which have no first-order counter-part. We show that these solutions exhibit growing behaviour, thus making them essential for the accurate evolution of the cosmological variables.

We also investigated in detail the consequences of a compensated isocurvature mode, the mode which is constrained the least at the linear level. We confirm that a pure compensated isocurvature mode does not generate any evolution both at first and second order in cosmological perturbations. However, we show that, when mixed with other modes, there are additional contributions to many variables, which do not exist at linear order or in the pure compensated mode. In particular, we noted that the mixed adiabatic and compensated isocurvature solution conserves the relation between the baryon and dark matter contrasts given initially, but also causes the compensated density fluctuation to grow, as well as the baryon-photon velocity. Considering the other possible mixture, with the neutrino density isocurvature, we find that the curvature perturbation, density contrasts and velocity perturbations receive a contribution from this mixed mode, but no higher multipoles are affected.

Our results can be applied to initialize second-order Boltzmann codes to evaluate the effects of isocurvatures on a variety of observables. In the future, we aim to apply the same techniques developed here to study the initialization of vector modes, which are known to be sourced when multiple degrees of freedom are present. This would be an interesting application for the mixed modes found in this work.

%\appendix
% % % % % % % % % % % % % % % % % % % % % % % % % % % % 
% chapter.tex - Ian Huston
% Sample chapter layout
% % % % % % % % % % % % % % % % % % % % % % % % % % % % 
% Redefine CVSRevision for this section. 
% If you don't want to use CVS tags comment out this line
\renewcommand{\CVSrevision}{\version$Id: chapter.tex,v 1.3 2009/12/17 18:16:48 ith Exp $}

\definecolor{lightgray}{gray}{0.9}
\newcolumntype{Q}{>{$\displaystyle}l<{$}}
\newcolumntype{q}{>{\columncolor[gray]{0.9}$\displaystyle}l<{$}}
\newcolumntype{R}{>{$\displaystyle}r<{$}}
%\newcolumntype{S}{>{$\displaystyle}c<{$}}
%\newcolumntype{s}{>{\columncolor[gray]{0.9}$\displaystyle}c<{$}}
\newcolumntype{T}{>{\columncolor[gray]{0.9}}c<{}}

\newcommand{\para}[1]{\par\vspace{2mm}\noindent\textbf{{#1}}.---}

% % % % % % % % % % % % % % % % % % % % % % % % % % % % % % % % 
% =========================================================== %
% % % % % % % % % % % % % % % % % % % % % % % % % % % % % % % % 
\chapter{Quantum Quenches in de Sitter}
\label{Ch_quench}
% % % % % % % % % % % % % % % % % % % % % % % % % % % % % % % % 
% =========================================================== %
% % % % % % % % % % % % % % % % % % % % % % % % % % % % % % % % 
\section{Introduction}
\label{sec:introduction}

In this chapter, we study quantum quenches of scalar fields in de Sitter spacetime. Quenches have been used in a cosmological setting by many authors to study phase transitions, both in the flat spacetime approximation~\cite{Tranberg:2006dg,Arrizabalaga:2004iw}, as well as in an inflationary background~\cite{Boyanovsky:1996rw,Boyanovsky:1996sq,Boyanovsky:1996fz,Boyanovsky:1996sv,Boyanovsky:1997cr,Boyanovsky:1997xt,Boyanovsky:2006bf}. In spite of this, this technique had not yet been applied to the study of more general transitions during inflation, such as those arising when the potential has sharp features, which can lead momentarily to violations of the slow-roll approximation. This is the main aim of this study. 

Generally, these fast events occur whenever there are very pronounced slopes in the potential which are traversed during very short times, $\Delta t\ll H^{-1}$. The end result is effectively a transition in the parameters of the potential, such as the masses and couplings of the fields. The interpretation of such features of the potential as quantum quenches is expected to be a good approximation for the description of the system some time after the violent phenomenon has occurred, while not depending on the exact details of the transition, provided that the transition is quicker than the other time scales of the system. The use of quenches to model these features allows for the study of the consequences of different classes of phenomena, based solely on the parameters of the potential before and after the transition has taken place.

We perform this study using the large-$N$ expansion, which we introduce in Section~\ref{sec:largeN}. This method allows one to study a theory with a large number, $N$, of identical fields by expanding the action in powers of $1/N$, instead of the usual expansion in powers of the coupling constant, multiplying the non-linear parts of the potential. Consequently, this is a manifestly non-perturbative method since it allows for studying systems with large couplings. The large-$N$ expansion and other non-perturbative techniques are very useful in describing IR effects in de Sitter, having been used~\cite{Serreau:2011fu}, for example, to show that IR effects and self-interactions force the effective mass of the fields to be strictly positive, something that had already been discussed in the stochastic context~\cite{Starobinsky:1994bd}. This effect, which is proportional to the root of the coupling constant, $\sqrt{g_4}$, would be impossible to obtain using perturbative methods.

We are thus able to study the consequences of the quench for the evolution of the system taking into account IR effects. We compute the two-point function of scalar perturbations generated after the quench in section \ref{sec:quenches}, presenting analytical estimates for the evolution and late-time limit of their effective mass. We also use a numerical approach to verify and correct our analytical calculations. At the end of that section we discuss the effects of the quench on dynamical mass generation, by studying a quench to an initially tachyonic state. We conclude in Section \ref{sec:Discussion}, by enumerating our main results and discussing the advantages of our approach.

\section{Large-$N$ in de Sitter}
\label{sec:largeN}

The action for an $N$-component, $O(N)$ symmetric, $\ph^4$ model in a de Sitter background geometry in $d$ spacetime dimensions is given by
\begin{equation}
S[\ph]=\int{\text{d}^dx\sqrt{-g}\left[-\frac12g^{\mu\nu}\p_\mu\ph^a\p_\nu\ph^a-\frac12\mu^2\ph^a\ph^a-\frac{g_4}{4N}(\ph^a\ph^a)^2 \right]}\,,
\end{equation}
where  $a$ is an $O(N)$ index which labels the field (not to be confused with the scale factor) and repeated indices are summed over as per Einstein's notation. This is a generalization of the action given in Chapter~\ref{Ch_SMC} for multiple fields with a specific potential. As elsewhere in this thesis, the spacetime under study is the FLRW spacetime, whose metric we re-write here, in terms of conformal time $\tau$,
\begin{equation}
g_{\mu\nu}=a(\tau)^2\eta_{\mu\nu}\,,
\end{equation}
in which $\eta_{\mu\nu}$ is the Minkowski metric with mostly plus signature. For exact de Sitter, the solution was given in Eq.~\eqref{Eq_dS1}, and we repeat it here in terms of the Hubble rate, $H$,
\begin{equation}
a(\tau)=-\frac1{H\tau}\,,
\end{equation}
with the conformal time obeying $-\infty<\tau<0$.

We now review the large-$N$ approximation. The general idea is that for a very large number of fields, $N\gg 1$, the action becomes very large, i.e. $S\gg\hbar$. As a consequence, the path integral is dominated by solutions which minimize the action, just as it happens when one takes the classical limit ($\hbar\rightarrow 0$). This simplifies a number of calculations while still keeping contributions of all orders in the couplings of the theory. To see this explicitly, let us start by writing the path integral in the \textit{in-in} formalism~\cite{Weinberg:2005vy} as
\begin{equation}
\mathcal{I}=\int_{\text{CTP}}\mathcal{D}\ph\ e^{iS[\ph]}\,,
\end{equation}
in which CTP is designating the closed-time-path measure one uses to account for the boundary conditions of the \textit{in-in} formalism. We now introduce a new variable defined by
\begin{equation}
\rho\equiv\ph^a\ph^a/N\,,
\end{equation}
whose expectation value is the variance of the fields. We can also change the path-integral by using the identity
\begin{equation}
\mathbf{1}\sim\int{\mathcal{D}\rho \ \delta(\ph^a\ph^a-N \rho)}\sim\int{\mathcal{D}\rho\,\mathcal{D}\xi\ e^{-\frac{i}2\int \text{d}^dx\sqrt{-g}\xi(\ph^a\ph^a-N \rho)}}\,,
\end{equation}
which results in
\begin{equation}
\mathcal{I}=\int_{\text{CTP}}{\mathcal{D}\ph\mathcal{D}\rho\mathcal{D}\xi\ e^{iS[\ph,\rho,\xi]}}\,,
\end{equation}
where the new action $S[\ph,\rho,\xi]$ is given by
\begin{equation}
S[\ph,\rho,\xi]=\int{ \text{d}^dx\sqrt{-g}\frac{1}{2}\left[-g^{\mu\nu}\p_\mu\ph^a\p_\nu\ph^a-(\mu^2+\xi)\ph^a\ph^a-\frac{Ng_4}{2}\rho^2+N\xi\rho \right]}\,.
\end{equation}
It is clear that the action above is simply quadratic in $\ph^a$, which allows one to perform $N$ Gaussian integrals for each field.
Before that, however, it is convenient to change variables to
\begin{equation}
\label{phichirel}
\ph^a \equiv \rchi^a \, \displaystyle{a^{\frac{2-d}{2}}}\,,\ \ \ \ \ \rho \equiv \Pi \, a^{2-d}\,,
\end{equation}
since it is $\rchi^a$ which is the canonically normalized field in a de Sitter spacetime.\footnote{This is equivalent to the Sasaki-Mukhanov variable, $v$, defined in Chapter~\ref{Ch_SMC}, for an exact de Sitter spacetime, generalized to $d$ dimensions.}
Integrating out $N-1$ copies of the $\rchi^a$ fields and substituting for the de Sitter metric, yields the following path integral
\begin{equation}
\mathcal{I}=\int_{\text{CTP}}{\mathcal{D}\Pi\, \mathcal{D}\xi\, \mathcal{D}\sigma\ e^{iS_{\text{eff}}[\Pi,\xi,\sigma]}}\,,
\end{equation}
with
\begin{align}
S_{\text{eff}}[\Pi,\xi,\sigma]=&\ \int{ \text{d}^dx} \left\{\frac12\sigma \left[\p^2+\frac{1}{\tau^2}\left(\frac{d(d-2)}{4}-\frac{\mu^2+\xi}{H^2}\right)\right]\sigma\right.\nonumber\\
&\left.\quad\quad\quad\quad\quad\quad +N\left(\frac{\xi\Pi}{2(H\tau)^2} -\frac{g_4}{4}\Pi^2 (-H\tau)^{d-4}\right)\right\}\nonumber\\
&+(N-1)\frac{i}{2}\text{Tr}\left\{\log{\left[-\p^2-\frac{1}{\tau^2}\left(\frac{d(d-2)}{4}-\frac{\mu^2+\xi}{H^2}\right)\right]}\right\}\,,
\end{align}
in which Tr is the functional trace defined by
\begin{equation}
\text{Tr}[f(x,y)]=\int{ \text{d}^dx \, f(x,x)}\,,\label{tracefxy}
\end{equation}
and $\p^2$ is the Minkowski Laplacian.\footnote{These functional techniques are better understood when a set of basis functions $f_i$ exists, for which a function $g(x)$ is expanded as
\begin{equation}
g(x)=\sum_i{g_i f_i(x)}\ \ \text{with}\ \ \int{\text{d}^dx f_i(x) f_j(x)}=\delta_{ij}\,.
\end{equation}
Then the Laplacian can be written as a matrix with components
\begin{equation}
[\p^2]_{ij}=\int{\text{d}^dx f_i(x) \p^2 f_j(x)}\,.
\end{equation}
Its trace is just the matrix trace, since Eq.~\eqref{tracefxy} becomes
\begin{equation}
\text{Tr}[h(x,y)]=\int{\text{d}^dx \sum_{i,j}h_{ij}f_i(x) f_j(x)}=\sum_{i,j}h_{ij}\delta_{ij}\,.
\end{equation}
} We have not integrated one of the scalar fields, given by $\sigma\equiv\rchi^N=\ph^N/(- H \tau)$, should there be a spontaneous breaking of the $O(N)$ symmetry, in which case $\sigma=O(\sqrt{N})$ instead of $O(1)$, as is assumed for all other field components, $\rchi^a$. Should that be the case, it is clear that all terms in the action are order $N$ and thus, in the large-$N$ limit, one has $S\propto N \gg \hbar$. The path integral can then be evaluated by simply using the stationary phase approximation. Therefore, one must only minimize the action by imposing the following conditions with respect to each of the field species present:
\begin{align}
&\frac{\delta S_{\text{eff}}}{\delta\xi}=0\Rightarrow\\\nonumber
&\frac{\bar\Pi}{(H\tau)^2}-\frac{\bar\sigma^2}{(H\tau)^2}+i\frac{\delta}{\delta\xi}\left.\text{Tr}\left\{\log\left[-\p^2-\frac{1}{\tau^2}\left(\frac{d(d-2)}{4}-\frac{\mu^2+\xi}{H^2}\right)\right]\right\}\right|_{\xi=\bar\xi}=0\,,\\
&\frac{\delta S_{\text{eff}}}{\delta\Pi}=0\Rightarrow \frac{\bar\xi}{(H\tau)^2}-g_4\bar\Pi(-H\tau)^{d-4}=0\,,\\
&\frac{\delta S_{\text{eff}}}{\delta\sigma}=0\Rightarrow\left[\p^2+\frac{1}{\tau^2}\left(\frac{d(d-2)}{4}-\frac{\mu^2+\bar\xi}{H^2}\right)\right]\bar\sigma=0\,.
\end{align}
The barred variables ($\bar\Pi$, $\bar\xi$, $\bar\sigma$) denote the solutions to these equations of motion. For the case of $\bar\sigma$ we also factor out $\sqrt{N}$, for clarity of presentation.\footnote{Please note that should we be dealing with the $O(N)$ symmetric phase, we will simply set $\bar\sigma=0$, as it is assumed to be order $1/\sqrt{N}$ and hence negligible in the large-$N$ limit. For the broken phase, it is order $1$.} One can show that the last term in the first equation above is 
\begin{equation}
i\frac{\delta}{\delta\xi}\left.\text{Tr}\left\{\log\left[-\p^2-\frac{1}{\tau^2}\left(\frac{d(d-2)}{4}-\frac{\mu^2+\xi}{H^2}\right)\right]\right\}\right|_{\xi=\bar\xi}=-i\frac{G(x,x)}{(H\tau)^2}\,,
\end{equation}
where $G(x,x)$ is the Green's function of $\rchi^a$ evaluated at the same spacetime point, $x$, which can be calculated as an integral over the power spectrum:
\begin{equation}
G(x,x)=\int{\frac{\text{d}^{d-1}k}{(2\pi)^{d-1}}\ \tilde G (\tau,\tau,k)}\,.
\end{equation}
Defining the effective mass as $m^2\equiv\mu^2+\bar\xi$, one has the following self-consistent equation for it
\begin{equation}
\label{selfconsdS}
m^2(x)=\mu^2+g_4(-H\tau)^{d-2}\left[\bar\sigma(x)^2+i G(x,x)\right]\,,
\end{equation}
Note that the r.h.s. of Eq.~\eqref{selfconsdS} depends non-trivially on the mass $m^2$ due to contributions from $G$ and $\bar\sigma$, which have a functional dependence on the effective mass. Solving this equation for $m^2$, therefore, allows one to find the effective mass which consistently includes all contributions from the interaction terms. This is due to the fact that the equal time propagator, $\tilde G(\tau,\tau,k)$ encodes the details of the interactions. As a result, the power spectrum will be the main object of focus, not only due to the cosmological implications of our work, but because it encodes all the information necessary to compute the effective mass. Much of the following sections is dedicated to its calculation.

\subsection{No quench}\label{NoQuench}

Before evaluating the consequences of a quench in this system, let us look at the simpler case in which there are no sudden changes in the parameters. This will serve to set some of the notation and also to explain the general procedure.

Our aim is to make use of Eq. \eqref{selfconsdS} to calculate the effective mass in the limit in which the mass is small, i.e. when $m/H\ll1$. This is the interesting case, since the effects of the curved background would disappear should one take the opposite limit.
The first step is the calculation of the Green's function. This can be easily done by expanding the fields in Fourier space, in terms of creation and annihilation operators, $a^\dagger_k$ and $a_k$,\footnote{Note that we are using an unlabeled field, $\rchi$, to represent each of the fields $\rchi^a$. We also omit the $O(N)$ indices everywhere else to avoid clutter.}$^{,}$\footnote{Note that, in general, the expansion of multiple interacting fields in creation and annihilation operators is not diagonal, i.e. each field depends on all of the $N$ pairs of ladder operators and not just on one of them, as seen here. The simplicity of the case presented here is due to the fact that the fields are effectivelly free in the large-$N$ limit, since all the effects of the interactions are contained in the effective mass. Thus it is possible to expand each field with just one pair of creation and annihilation operators, as shown in Eq.~\eqref{expaad}.}
\begin{equation}
\label{expaad}
\chi(\tau,\vec{x})=\int{\frac{\text{d}^{d-1}k}{(2 \pi)^{d-1}}\left(a_k u_k(\tau) e^{i\vec{k}\cdot\vec{x}}+a_k^\dagger (u_k(\tau))^* e^{-i\vec{k}\cdot\vec{x}}\right)}\,,
\end{equation}
in which $a^\dagger_k$ and $a_k$ obey the standard commutation relations:
\begin{equation}
[a_k,a_q^\dagger]=(2\pi)^d\delta^{(d)}(\vec k-\vec q)\,,\ [a_k,a_q]=0\,,\ [a_k^\dagger,a_q^\dagger]=0\,.
\end{equation}
The computation of the two-point function at the same point is straightforward, being given by
\begin{equation}
\label{twopointunquench}
\<0|\chi(\tau,\vec{x})\chi(\tau,\vec{x})|0\>=\int\frac{\text{d}^{d-1}k}{(2 \pi)^{d-1}}\left|u_k(\tau)\right|^2\,.
\end{equation}
This simply depends on the normalized wave-functions $u_k(\tau)$, which can be obtained from the Klein--Gordon equation, assuming the effective mass is constant.\footnote{This assumption is well motivated in a Poincar\'{e}-invariant state, given that in that situation the two-point function for $\ph$ is constant \cite{Serreau:2011fu}, implying that $G(x,x)\propto (H\tau)^{-2}$. The field $\bar\sigma$ has the same behavior in such a state. This is purely a consequence of the de Sitter symmetry~\cite{Weinberg:2010fx}.} 
Choosing the Bunch--Davies vacuum, the wave-functions are given by
\begin{equation}
\label{wavfunc}
u_k(\tau)=-\frac12\sqrt{\frac{\pi}{2}}(1+i)e^{\frac{i\pi\nu}{2}}\sqrt{-\tau}\, H^{(1)}_\nu(-k\tau)\,,
\end{equation}
in which $H^{(1)}_\nu$ is the Hankel function of the first kind and $\nu$ is related to the mass of the field via
\begin{equation}
\nu=\sqrt{\left(\frac{d-1}{2}\right)^2-\frac{m^2}{H^2}}\,.
\end{equation}
Note that this is the same solution as given in Eq.~\eqref{SolSMC3}, but with the effective mass instead of the slow-roll parameters. The self-consistency condition, Eq. \eqref{selfconsdS}, then translates to, in $d=4$,
\begin{equation}
m^2=\mu^2+g_4(-H\tau)^{2}\left[\bar\sigma^2+\int{\frac{\text{d}^{3}k}{(2\pi)^{3}}\frac{\pi}{4}(-\tau)\left|H^{(1)}_\nu(-k\tau)\right|^2 }\right]\,.
\end{equation}
The integral on the r.h.s. is not straightforward to calculate analytically for a general order of the Hankel function. Furthermore, it has UV divergences which need to be regularized. These two issues are discussed, for example, by Serreau \cite{Serreau:2011fu}, and we shall follow the same procedures:
\begin{itemize}
\item{The integral is split into three different parts: $\int_0^\Lambda=\int_0^\kappa+\int_\kappa^{\kappa'}+\int_{\kappa'}^\Lambda$, with $\kappa\ll\kappa'\ll\Lambda$. The IR and UV contributions are calculated by expanding the Hankel function for small and large arguments, respectively. Furthermore, the assumption that the mass is small sets the order $\nu$ to be $\nu=3/2-\varepsilon$ with $\varepsilon\ll1$. This allows for an expansion in $\varepsilon\approx m^2/3H^2$ in all integrals, which for the middle integral, $\int_\kappa^{\kappa'}$, simplifies to setting $\nu=3/2$.}
\item{A change of variables is performed from comoving momentum $k$ to physical momentum $p=k/a$. One then regularizes the integrals with cut-offs in the physical momentum $p$, since this is the choice that respects de Sitter symmetry.}
\end{itemize}
After implementing this procedure, we find for $m^2>0$
\begin{align}
\label{massunrenorm}
\frac{m^2}{g_4}=&\ \frac{\mu^2}{g_4}+(H\tau)^2\bar\sigma^2+\frac{1}{8\pi^2}\left[\Lambda^2+2 H^2 \log\left(\frac{\Lambda}{H}\right)\right]\\
&+\frac{H^2}{8\pi^2}\left(2\gamma_{\text E}-4+2\log2+\frac{3H^2}{m^2}\right)-\frac{m^2}{8\pi^2}\log\left(\frac{\Lambda}{H}\right)\,,\nonumber
\end{align}
in which $\Lambda$ is the UV cut-off in the physical momentum and $\gamma_{\text E}\approx 0.57721$ is the Euler--Mascheroni constant. The divergences are renormalized through\footnote{Note that the term $2\gamma_{\text{E}}-4+2\log 2$ can also be absorbed in the renormalized parameters, without loss of generality.}
\begin{equation}
\frac{1}{g^R_4}=\frac{1}{g_4}+\frac{1}{8\pi^2}\log\left(\frac{\Lambda}{H}\right)\,,\ \ \ \ \frac{\mu^2_R}{g^R_4}=\frac{\mu^2}{g_4}+\frac{1}{8\pi^2}\left[\Lambda^2+2 H^2 \log\left(\frac{\Lambda}{H}\right)\right]\,,
\end{equation}
resulting in
\begin{equation}
\label{unqm}
m^2=\mu^2_R+g_4^R(H\tau)^2\bar\sigma^2+g_4^R\frac{H^2}{8\pi^2}\left(2\gamma_{\text E}-4+2\log2+\frac{3H^2}{m^2}\right).
\end{equation}
This can easily be solved for $m^2$, and one finds solutions which are strictly positive, even when $\mu^2_R\leq0$. This fact is usually referred to as radiative symmetry restoration \cite{Serreau:2011fu}, since the curved spacetime and the interactions forbid the $O(N)$ symmetry of the system from being spontaneously broken. This might not seem surprising given the initial assumption that $m^2>0$, but the existence of positive mass squared solutions is non-trivial when $\mu^2_R\leq0$. Solutions with negative $m^2$ also exist but, in those cases, the two-point function diverges in the IR, giving unphysical results.

In the next sections we will introduce a quench into the dynamics. While this will slightly alter the procedure, the main objective remains the solution of the self-consistent mass equation \eqref{selfconsdS} derived above.

\section{Quantum quenches in de Sitter}
\label{sec:quenches}

As mentioned above, a quench is defined as an instantaneous change in the parameters of a model. In the case under study, that corresponds to a change in the mass parameter, $\mu^2$, and coupling, $g_4$, of the scalar field system. We believe these quenches can arise for a number of different reasons.

In previous studies in de Sitter spacetime~\cite{Boyanovsky:1996rw,Boyanovsky:1997cr,Boyanovsky:1997xt}, the swiftness of the transition is justified by an abrupt change in the temperature of the system, which induces a sudden change in the model parameters. In the context of primordial features, however, one would expect these transitions to be due to the specific form of the scalar potential. Ref.~\cite{Joy:2007na} studies a particular example, in which an interaction between the fields prompts a fast change in the effective mass parameter of the inflaton. The motivation for the present work is the study of similar situations by using the quench approximation. In this work, however, we do not investigate the origin of quenches and they should not depend on specific details of the transitions. Therefore, this work could be applied more generally than to the study of primordial features.

Our starting point assumes exact de Sitter and negligible backreaction of the quantum fluctuations of our system in the background evolution. Furthermore, we assume the system to be in an $O(N)$ invariant state and thus we set $\bar\sigma=0$, except in the discussion of Section \ref{sec:Negative}. This implies that we also do not treat the background evolution of the inflaton. All these contributions would require a fully numerical approach, which we leave for future work. Here we focus on investigating the time evolution of the effective mass as well as its asymptotic behavior. This provides a full description of the system and allows one to study different problems, such as the stationarity of the system at late times and compare it to the flat spacetime case, as studied by Sotiriadis and Cardy \cite{Sotiriadis:2010si}. In that case, the system becomes stationary very soon after the quench, but in the cosmological setting of the de Sitter spacetime, it is possible, in principle, that the contributions to the effective mass vary in time in a different way after the quench.\footnote{In spite of both spacetimes (flat and de Sitter) having a time-like killing vector, the quench breaks the corresponding invariance under time translations of the solution for the scalar field. This is the reason why the results are expected have a different time evolution after the quench.} This is something we investigate in the following sections.

\subsection{Setup}

In order to study the quench, we define an initial state in the pre-quench stage, which is usually taken to be the ground state of the system prior to the quench. Here, we choose exactly that and assume the initial state is the Bunch--Davies vacuum $\left|0\right\rangle_{BD}$. This state is parametrized by the mass before the quench, $\mu_0$. After the quench, the Hamiltonian of the system changes, and hence the initial state is typically now an excited state of the new Hamiltonian. In particular, as will be clear below, the state will be non-Bunch--Davies with respect to the post-quench Hamiltonian.

As the quench happens, the equations the field operator obeys change, due to the change of the parameters themselves. Given that we assume that change to be instantaneous, both the value and first derivative of the field should be continuous across the quench. This implies that at conformal time $\tau_0$, when the quench happens, we have
\begin{align}
\chi^{(\nu_1)}(\tau_0,\vec{x})=\chi^{(\nu_2)}(\tau_0,\vec{x})\,,\\
\frac{d}{d\tau}\chi^{(\nu_1)}(\tau_0,\vec{x})=\frac{d}{d\tau}\chi^{(\nu_2)}(\tau_0,\vec{x})\,,
\end{align}
where the fields have been labeled with $\nu_i$ to emphasize that a set of parameters has changed. Since the initial state $\left|0\right\rangle_{BD}$ is no longer the lowest energy state of the system after the quench, one can therefore define a new vacuum and its corresponding creation and annihilation operators, $b_k^\dagger$ and $b_k$, respectively. Hence, the field is now expanded as
\begin{equation}
\chi^{(\nu_2)}(\tau,\vec{x})=\int{\frac{\text{d}^{d-1}k}{(2 \pi)^{d-1}}\left(b_k u_k^{(\nu_2)}(\tau) e^{i\vec{k}\cdot\vec{x}}+b_k^\dagger (u_k^{(\nu_2)}(\tau))^* e^{-i\vec{k}\cdot\vec{x}}\right)}\,.
\end{equation}
The constraints at $\tau_0$ given above can then be solved by a Bogoliubov transformation\footnote{Equivalently, one could keep the same expansion in $a_k^\dagger$ and $a_k$ and impose the continuity conditions on the wave-function appearing in front. Such wave-functions would be different from $u_k^{(\nu_2)}(\tau)$ and can be derived from the Bogoliubov transformation.}, which is given by
\begin{equation}
b_k=C_k a_{k}+D_k a_{-k}^\dagger\,,
\end{equation}
with
\begin{equation}
C_k=\frac{W\left((u^{(\nu_2)}_k)^*,u^{(\nu_1)}_k\right)}{W\left((u^{(\nu_2)}_k)^*,u^{(\nu_2)}_k\right)}\,,\ \ \ \
D_k=\frac{W\left((u^{(\nu_2)}_k)^*,(u^{(\nu_1)}_k)^*\right)}{W\left((u^{(\nu_2)}_k)^*,u^{(\nu_2)}_k\right)}\,,
\end{equation}
where all the wave-functions are evaluated at $\tau_0$ and $W(f,g)$ is the Wronskian, defined by
\begin{equation}
W(f,g) \equiv \frac{df}{d\tau}g-f\frac{dg}{d\tau}\,.
\end{equation}
It is straightforward to check that should the quench not occur (i.e. if $\nu_1=\nu_2$), one finds $C_k=1$ and $D_k=0$, as expected.

Given the decomposition above, it is now possible to compute the equal-time two-point correlator of the field $\chi$ after the quench. As was discussed in the previous section, this is the quantity which is required for solving the self-consistent mass equation, Eq.~\eqref{selfconsdS}, and it is also that which is observationally constrained. It can be obtained from the general two-point correlator, which is given by
\begin{align}
&_{BD}\<0|\chi(\tau_a,\vec{x})\chi(\tau_b,\vec{y})|0\>_{BD}=\nonumber\\
&\int\frac{\text{d}^{d}k}{(2 \pi)^d}e^{i\vec{k}\cdot(\vec{x}-\vec{y})}\left[C_k D_k u_k^{(\nu_2)}(\tau_a)u_k^{(\nu_2)}(\tau_b)+C_k^* D_k^* u_k^{(\nu_2)*}(\tau_a)u_k^{(\nu_2)*}(\tau_b)+\right.\nonumber\\
&\left.+\left|D_k\right|^2\left(u_k^{(\nu_2)*}(\tau_a)u_k^{(\nu_2)}(\tau_b)+u_k^{(\nu_2)}(\tau_a)u_k^{(\nu_2)*}(\tau_b)\right)+u_k^{(\nu_2)}(\tau_a)u_k^{(\nu_2)*}(\tau_b)\right]\,.\label{bogtrans}
\end{align}
Again, it is clear that in the absence of the quench only the last term survives, which is the result shown in Eq. \eqref{twopointunquench}.

The sections that follow will be dedicated to performing the calculations for different scenarios. For the simplest cases we are able to use analytical methods, which give a general picture of the results. We then complement those estimates with numerical calculations of the time evolution of the mass and interpret the results.

\subsection{Analytical estimates}
\label{sec:Analytical}

Before presenting our results, we make a note of difficulties we encounter and the simplifying assumptions we use in order to make the problem analytically tractable.
As was mentioned above, the state after the quench is no longer the Bunch--Davies vacuum of the system. Therefore, de Sitter invariance is broken and the two-point function of $\ph$ is no longer time-independent, in general. The first approximation we make is related to that: we will assume that time dependence to be negligible, at least in what concerns its effect on the two-point function. By this we mean that we calculate the two-point function assuming the wave-functions, $u_k^{(\nu_i)}$, to be the solutions from the unquenched case (i.e. with constant mass), as given by Eq.~\eqref{wavfunc}. This approximation is necessary given that it is impossible to (analytically) solve the Klein--Gordon equation for a general time-varying mass. Furthermore, as mentioned above, it has been shown that this is a very good approximation in flat spacetime~\cite{Sotiriadis:2010si}, and hence this is a justified approach.

Another difficulty that arises is the calculation of the integral of the power spectrum. It will generally involve integrating four Hankel functions with different arguments, which cannot be done analytically unless the order of the Hankel functions is a half integer. For this reason, we only treat masses close to $0$ or $\sqrt2 H$, due to the simplicity of the corresponding Hankel functions of orders $3/2$ and $1/2$, respectively. This means that, in some cases, we do not explicitly solve the self-consistent mass equation, but instead check if certain transitions are possible and focus on closed form formulae. This does not undermine the generality of the results, although it makes the physical interpretation more transparent. Note, however, that this care is not necessary in flat spacetime, given the analytical simplicity of the wave-functions.

To overcome this, we employ the same procedure as in Section \ref{NoQuench}, by splitting the momentum integral into three parts, which we call the IR, middle and UV integrals. We also change variables to physical momentum, so that UV cut-offs are correctly defined. UV contributions are rather simple to evaluate---they turn out to be the same as in the unquenched case, with the mass $m$ substituted by the mass after the quench.\footnote{This is strictly true only for $\tau>\tau_0$. At the instant in which the quench happens, $\tau=\tau_0$, the continuity of the two-point function implies that the UV contributions are still dependent on the mass before the quench. We disregard that point in time in all calculations.} This is not surprising, as the UV limit should not depend on initial conditions whichever they may be. The UV contribution to the self-consistent mass equation is therefore given by
\begin{equation}
\frac{m^2}{g_4}\supseteq \frac{1}{8\pi^2}\left[\Lambda^2+\left(2 H^2-m^2\right) \log\left(\frac{\Lambda}{H}\right)\right]\,,
\end{equation}
where $m$ denotes again the effective mass after the quench. Renormalization is performed in the same way as in the unquenched case.

\subsubsection{Asymptotic mass}

The first calculation we perform is the limit $x=\tau/\tau_0\rightarrow0$ of the self-consistent mass equation. The mass after the quench is now:
\begin{equation}
m^2_\infty=\mu^2_R+g_4^R(-H\tau)^{2}\int{\frac{\text{d}^{3}k}{(2\pi)^{3}}\frac{\pi}{4}(-\tau)\left|H^{(1)}_{\nu_2^\infty}(-k\tau)\right|^2 }\,,
\end{equation}
where we have also set $\bar\sigma$ to $0$. The integral can actually be calculated without approximations so that the result becomes
\begin{equation}
\label{asympmass}
m^2_\infty=\mu^2_R+\frac{g_4^RH^2}{16 \pi^2} \left(\frac{m^2_\infty}{H^2}-2\right)\left[\log 4 -1-\Psi\left(\nu_2^\infty-1/2\right)-\Psi\left(-\nu_2^\infty-1/2\right)\right]\,,
\end{equation}
where $\Psi(x)$ is the Digamma function, defined as the logarithmic derivative of the Gamma function, $\Psi(x)\equiv\Gamma'(x)/\Gamma(x)$. This result can now be approximated for masses close to $0$ and one would find the same result as in the unquenched case, Eq.~\eqref{unqm}. The point to note in this result is how different it is from the flat spacetime case, in which the system retains some memory of its state before the quench, even in the asymptotic late-time limit. As shown in Ref.~\cite{Hung:2012zr}, the asymptotic mass is a function of the pre-quench mass, $\mu_0$. That does not seem to happen in de Sitter spacetime, given that Eq.~\eqref{asympmass} is independent of the original mass. This is related to the evolution of the cosmological horizon. As was shown in Chapter~\ref{Ch_SMC}, scales $k^{-1}$ larger than the comoving horizon size $(aH)^{-1}$ are enhanced in an accelerating spacetime. These IR scales are the ones that end up dominating the calculation of the two-point function. Given that the horizon shrinks with time, the number of super-horizon scales increases with time. In the presence of a quench, however, the number of scales that exited the horizon before the quench is constant, while the number of modes that are enhanced after the quench increases indefinitely. After sufficient time, the contribution to the integral of the propagator from pre-quench modes becomes negligible in comparison to the scales that became super-horizon after the quench. As a consequence, the dependence of the effective mass on the pre-quench parameters disappears.\footnote{Note, however, that this is only true because $\mu_0^2\leq0$ is not allowed. If it were, IR divergences would appear, and thus the contribution from pre-quench modes would be non-negligible (and infinite).} These effects are not present in flat spacetime and thus the dependence on the initial mass is always present.

This result is not sufficient, on its own, without first making sure that the mass converges in general. While in the flat situation the convergence to a stationary mass is fast enough for one to assume the asymptotic result is valid shortly after the quench, the same is not clear in a curved spacetime, and that is the reason why one must find a more complete time evolution, thus checking both convergence as well as its rate of change.

Note, however, that, should the mass converge to a constant at some time, then the result above must be valid, since for a constant mass, the system is in a de Sitter invariant state, equivalent to the unquenched scenario. Hence, if we can prove that it does converge, we already have the expression for the asymptotic mass, Eq. \eqref{asympmass}.

\subsubsection{Approximate time evolution}\label{ATE}

We now move on to the time evolution. We begin by studying it for specific transitions of masses close to $0$ or $\sqrt2 H$. These cases are interesting for different reasons. Firstly, as mentioned before, they correspond to half-integer orders of the Hankel functions, which simplifies the wave-functions considerably. Furthermore, the $m\approx 0$ case is the relevant situation in inflation, since then the quantum perturbations are enhanced by the accelerated expansion. The other situation, $m=\sqrt2H$, is the conformal case, in which one can completely disregard the cosmic expansion from its evolution---its wave-functions turn out to be equal to those of the massless case in flat spacetime. Furthermore, in a de Sitter-invariant state, its mass does not receive any contributions from the interactions, as can be seen in Eq.~\eqref{asympmass}.

The other main approximation we employ here is the use of the wave-functions obtained for constant masses, i.e., instead of solving the full equation of motion,
\begin{equation}
\label{EOMWF}
u_k''+\left[k^2+\frac{1}{\tau^2}\left(\frac{m^2(\tau)}{H^2}-2\right)\right]u_k=0\,,
\end{equation}
we solve only for $m^2(\tau)=$const. as a first approximation. This will result, in general, in a time-dependent solution of the mass equation, Eq.~\eqref{selfconsdS}, which we label $m_1(\tau)$. Ideally, one could go further in the approximation by substituting the solution $m_1(\tau)$ in the evolution equation, Eq.~\eqref{EOMWF}, and thus finding the second approximation, $m_2(\tau)$, by solving the mass equation once more. Repeating this procedure should result in more and more accurate results with each iteration and convergence to the real effective mass. However, provided the difference between the first iterations is negligible, it is sufficient to use the approximation of constant mass and thus stop at $m(\tau)\approx m_1(\tau)$. We will estimate the size of that difference by comparing the solutions of Eq.~\eqref{EOMWF} for constant mass ($u_0(\tau)$) and for the first approximation $m_1(\tau)$ ($u_1(\tau)$). In particular, we calculate the error, $e_u$, with
\begin{equation}
\label{error}
e_u=\max \left|1-\frac{|u_1(\tau)|^2}{|u_0(\tau)|^2}\right|\,.
\end{equation}
Given that we expect the iterative approach to converge, this error calculation essentially gauges whether the first iteration, $m_1(\tau)$, is sufficiently accurate. An alternative to this procedure would be to check the size of time derivatives of $m_1(\tau)$. A particular test would be the calculation of the following derivative:\footnote{A derivation of this quantity can be made by obtaining the rate of change of the frequency, $\omega^2$ (given in square brackets in Eq.~\eqref{EOMWF}),
\begin{equation}
\frac{d(\omega^2)}{d\tau}=\frac{1}{\tau^2}\left(\frac{d(m^2(\tau)/H^2-2)}{d\tau}-\frac2\tau (m^2(\tau)/H^2-2)\right)\,,\nonumber
\end{equation}
and comparing the contribution from the time-dependent mass (the first term) to the contribution due to the time-dependent background (the second term).}
\begin{equation}
\label{fastslow}
\left|\frac12\frac{d\log\left|m^2(\tau)/H^2-2\right|}{d\log \tau}\right|\ll1\Rightarrow\left|\frac{d\log\left|m^2(\tau)/H^2-2\right|}{dt}\right|\ll 2 H\,,
\end{equation}
where $t$ is cosmic time. Note that the second inequality explicitly shows the connection of this test to the time scale of the problem, the Hubble rate, $H$, thus providing the physical interpretation to how slow the evolution needs to be for the correctness of the constant mass approximation.\footnote{Note that using the opposite inequality in Eq.~\eqref{fastslow} would correspond to the quench itself, in which the transition happens in a much shorter time-scale than $H^{-1}$.} While being more physically intuitive, this method is less accurate in predicting whether the first iteration is sufficiently good, which is why we use the expression given in Eq.~\eqref{error} to estimate the error. 

In the calculations that follow, we begin by assuming the corrections are small, similarly to what occurs under an adiabatic approximation, in which one assumes the evolution of the mass to be slow enough for it not to affect the equations of motion substantially. We will revisit the accuracy of this approximation in Section \ref{sec:Numerical}, thereby justifying our approach.
\\

\noindent\textbf{Transition 1:} $\mu_0\approx 0\ \rightarrow\ m=\sqrt2 H$
\\

The first case we will consider is the transition from $\mu_0\approx 0$ to $m=\sqrt2 H$. By $\mu_0\approx 0$, we mean we use the same approximations as in the unquenched case, i.e. the order of the Hankel function before the quench is $\nu_1=3/2-\varepsilon_1$ with $\varepsilon_1\ll1$ and we expand in powers of $\varepsilon_1\approx \mu_0^2/3H^2$. At lowest order in $\varepsilon_1$, we find
\begin{gather}
\label{Res3212}
2H^2=m^2=\mu_R^2+\frac{g_4^R H^2}{8\pi^2}x^2\left[\left(\frac{1}{\varepsilon_1}-3-2 \log(1-x)\right)(x-2)^2-1\right]\,,
\end{gather}
in which $x=\tau/\tau_0$. We can see that this result does converge to a constant at late times ($x\rightarrow 0$), and becomes $m^2=\mu_R^2$, in agreement with our estimate from Eq.~\eqref{asympmass}.

The conclusion seems to be that should we have $\mu_R^2=2H^2$, a transition does exist from $\mu_0\ll H$ to $m\approx\sqrt2 H$, given that the time evolving part is very small, when compared to $2H^2$. Should that not be the case, not only is it not guaranteed that the evolution is slow enough, but the result is not even consistent with the original assumption. Recall that we are checking whether the transition exists by assuming the final mass is $m=\sqrt2 H$ and attempting to find parameters $\mu_R^2$, $g_4^R$ and $\varepsilon_1$ for which the solution is consistent. If we find the time dependent part to be very large, consistency is violated and our result for the two-point function could no longer be valid. We check this in Section \ref{sec:Numerical} using numerical calculations and find no such problems.
\\

\noindent\textbf{Transition 2:} $\mu_0=\sqrt2 H\ \rightarrow\ m\approx 0$
\\

We now look into the inverse transition, $\mu_0=\sqrt2 H\ \rightarrow\ m\approx 0$. We use the same approximations as in the previous case, but expand now in $\varepsilon_2\approx m^2/3H^2$. Again, at first order in this parameter, we find
\begin{equation}
\label{q32p}
m^2=\mu_R^2+\frac{g_4^RH^2}{16\pi^2}\left[4C_1 +4x+x^4+4 \log(1-x)-4\log x\right]\,,
\end{equation}
in which we introduced the constant, $C_1$, defined by $C_1\equiv\gamma_{\text E}-\frac54+\log 2$, to simplify the notation. This result does not match our original predictions for the final masses, due to an apparent divergence when $x\rightarrow 0$. This is re-analyzed in Section \ref{sec:Numerical}, and the numerical results show no divergences, indicating that this is a problem owing to the expansion in $\varepsilon$.
\\

\noindent\textbf{Transition 3:} $\mu_0\approx 0\ \rightarrow\ m\approx 0$
\\

The final case we deal with here is the transition $\mu_0\approx 0\ \rightarrow\ m\approx 0$, now expanded both in $\varepsilon_1\approx \mu_0^2/3H^2$ and $\varepsilon_2\approx m^2/3H^2$. The self consistency condition for this case is
\begin{equation}
\label{q3232}
m^2=\mu_R^2+\frac{g_4^RH^2}{8\pi^2}\left[C_2+\frac{1}{\varepsilon_1}\left(1+(\varepsilon_2-\varepsilon_1)\left(\frac23-\frac23x^3+2\log x\right)\right)\right]\,,
\end{equation}
with the constant $C_2$ given by , $C_2=2\gamma_{\text E}-4+2\log 2$. We can see that the late-time limit ($x\rightarrow0$) again results in a divergence, unless there is no quench, i.e. $\varepsilon_2=\varepsilon_1$. The logarithmic divergences are now slightly more complicated, with one term being identical to that of \textbf{transition 2}, while the other is dependent on $\varepsilon_2$. Again, for this case, it will be made clear in the next Section that the problem comes from the expansion in $\varepsilon_1$ and $\varepsilon_2$, rather than being symptomatic of a ``dynamical impossibility".

\subsection{Numerical and re-summed results}
\label{sec:Numerical}

In this section we perform the calculations from the previous section again but using numerical techniques. Instead, this allows one to see that the full results from the previous calculations do now match the final mass estimates from Eq.~\eqref{asympmass} once we implement a re-summation technique and that most of the other issues are solved. However, we do still use the same approximation, in which we take the mass to be constant for the purposes of calculating the integrals. We remind the reader that we have defined the parameters $\varepsilon_1$ and $\varepsilon_2$ as
\begin{equation}
\label{epsilons}
\varepsilon_1 \equiv \frac32-\sqrt{\frac94-\frac{\mu_0^2}{H^2}}\approx \frac{\mu_0^2}{3H^2}\ \ \textrm{and} \
\ \varepsilon_2 \equiv \frac32-\sqrt{\frac94-\frac{m^2}{H^2}}\approx \frac{m^2}{3H^2}\, ,
\end{equation}
respectively. Recall as well that conformal time is defined in the range $-\infty<\tau<0$, so that $x=\tau/\tau_0$ is positive and approaches $x\rightarrow 0$ in the far future.
\\

\noindent\textbf{Transition 1:} $\mu_0\approx 0\ \rightarrow\ m=\sqrt2 H$
\\

\begin{figure}[h]
    \centering
    \includegraphics[scale=0.8]{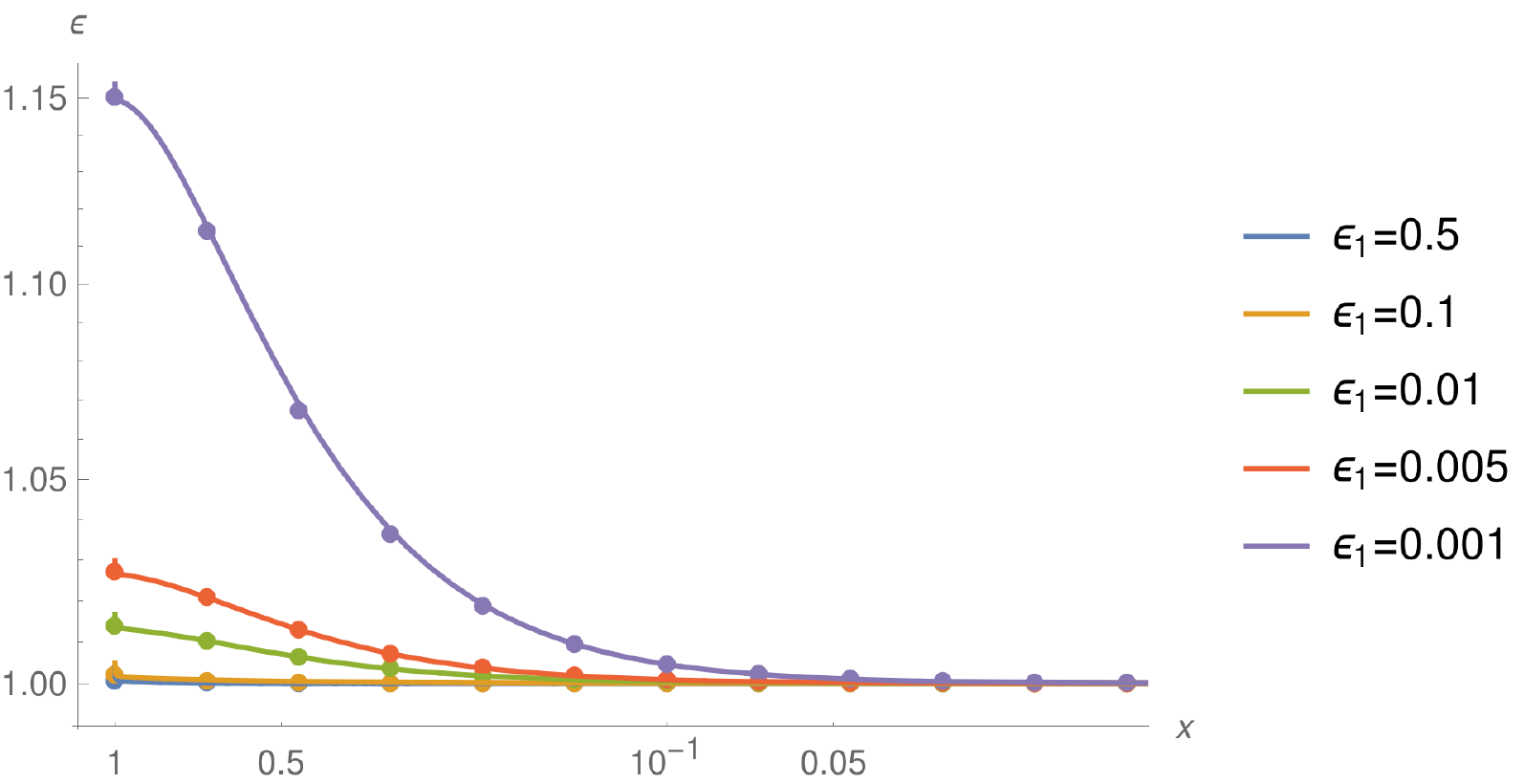}
    \caption{Evolution of $\varepsilon_2(x)$ for transition 1 (dotted) as compared to the analytical result (solid) for $g_4^R=0.01$, varying $\varepsilon_1$. }
    \label{fig3212e1}
\end{figure}

\begin{figure}[h]
    \centering
    \includegraphics[scale=0.8]{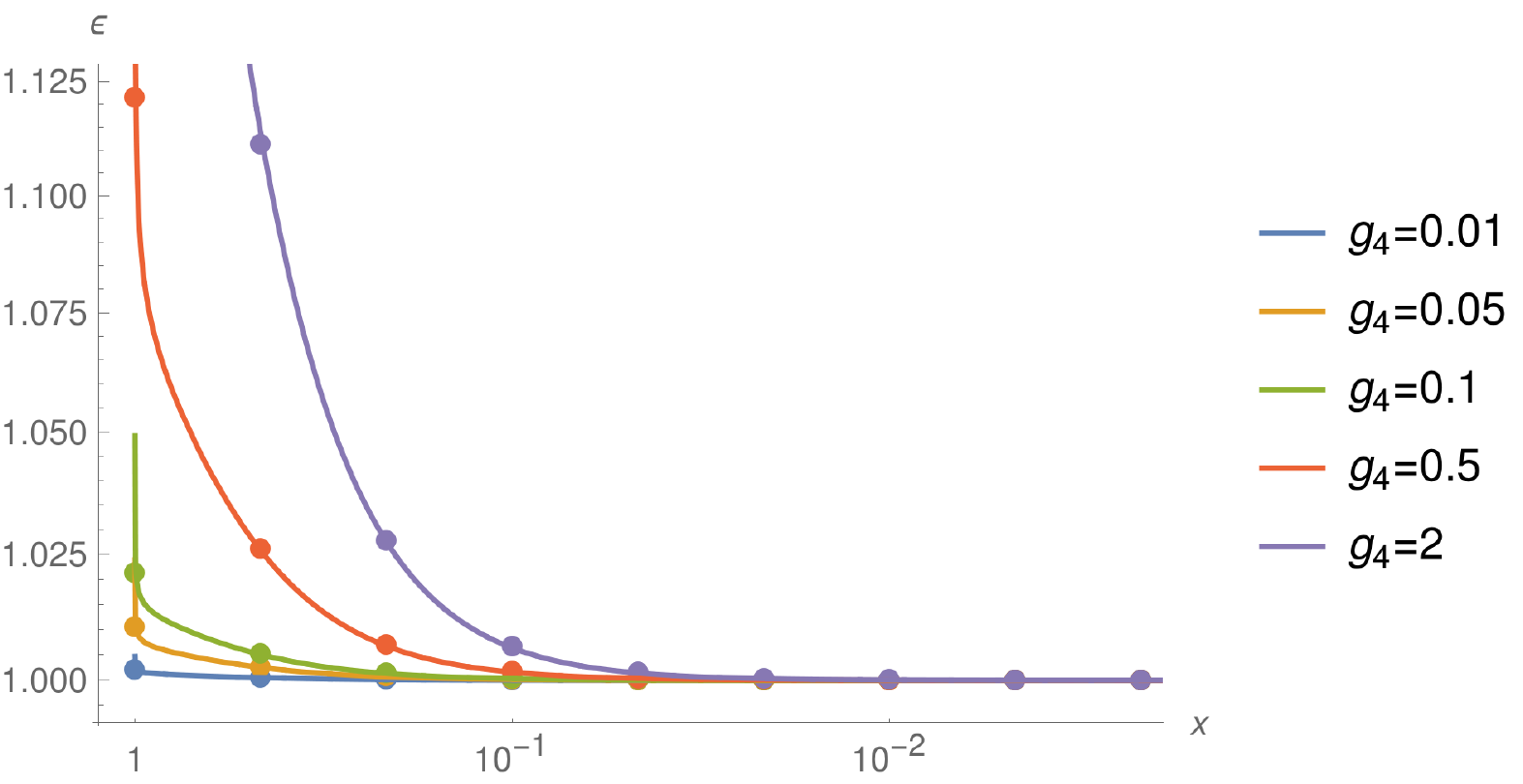}
    \caption{Evolution of $\varepsilon_2(x)$ for transition 1 (dotted) as compared to the analytical result (solid) for $\varepsilon_1=0.1$, varying $g_4^R$. }
    \label{fig3212g4}
\end{figure}

Let us follow the same order as before and start with the case $\mu_0\approx 0\rightarrow m=\sqrt2 H$. We have seen that, in order for this transition to occur, one must have $\mu_R^2=2H^2$, so we choose that value for the mass parameter. We demonstrate the dependence on the remaining parameters by plotting $\varepsilon_2$ as a function of $x=\tau/\tau_0$ for different values of the original mass, $\mu_0$ (labeled by $\varepsilon_1$), and the coupling strength, $g_4^R$ in two different plots, in Figs. \ref{fig3212e1} and \ref{fig3212g4}.

Firstly, we notice that the analytical expression obtained above in Eq.~\eqref{Res3212} is a very good approximation to the numerical solution in all situations and for all values of $x$. This is somewhat surprising, given that that expression was derived for a specific final mass. Furthermore, from Fig. \ref{fig3212e1}, we see that even when $\varepsilon_1$ is not so small, as exemplified by the case $\varepsilon_1=0.5$, our original approximation almost reproduces the numerical results, with only a small deviation of less than $0.01\%$ around $x=1/3$. It would fail for larger values of $\varepsilon_1$, but those cases are somewhat less interesting, since the initial and final masses are too similar.

As expected, evolution is faster and more pronounced in the cases in which the coupling strength, $g_4^R$, is larger. The dependence on the initial mass, $\varepsilon_1$, seems to indicate that there is less evolution for larger initial masses, which is to be expected given the terms with $H/\mu_0$ present in Eq.~\eqref{Res3212}.
\\\\\\

\noindent\textbf{Transition 2:} $\mu_0=\sqrt2 H\ \rightarrow\ m\approx 0$
\\

\begin{figure}[h]
    \centering
    \includegraphics[scale=0.8]{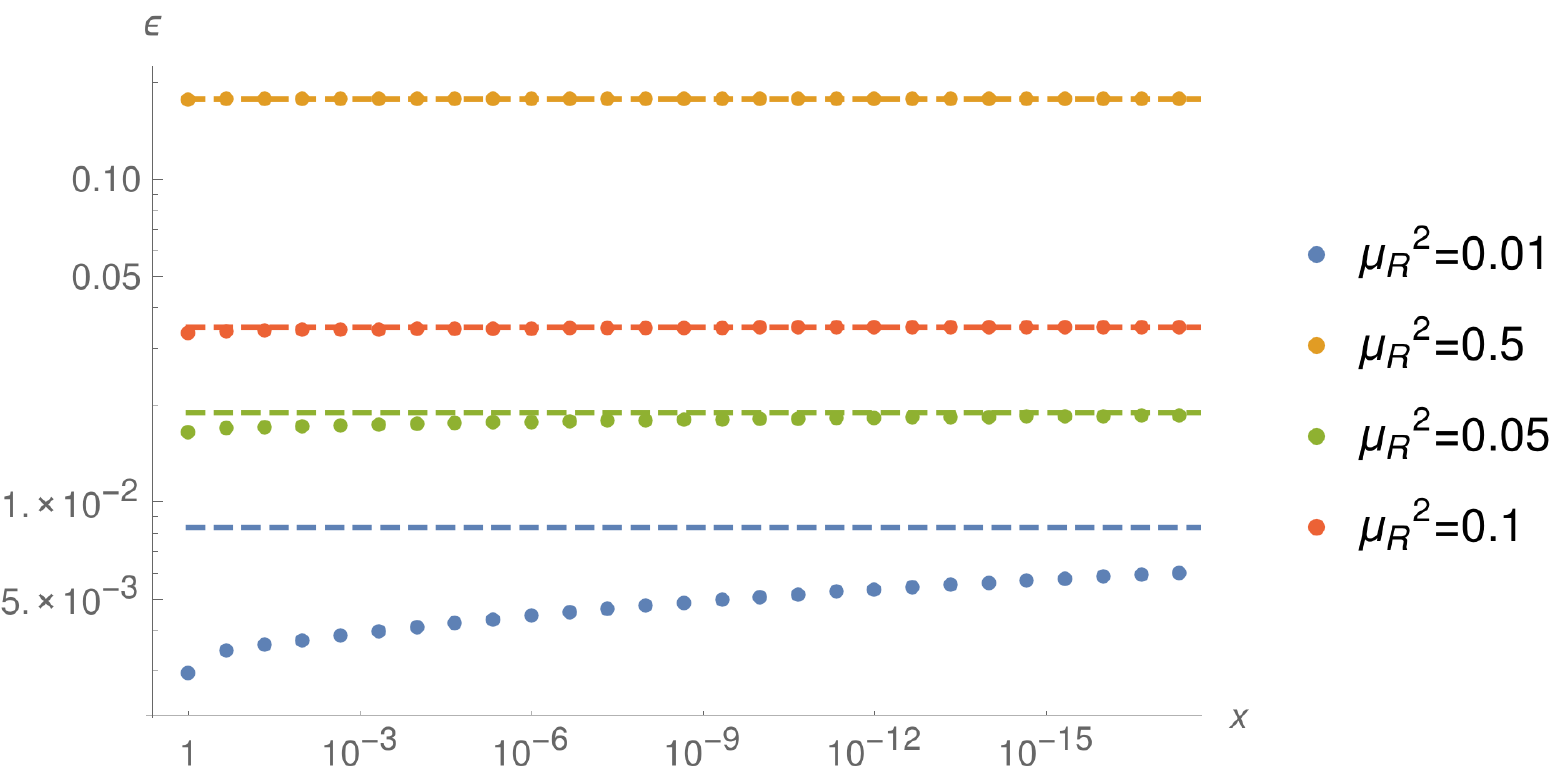}
    \caption{Numerical evolution of $\varepsilon_2(x)$ for transition 2 (dotted), showing the asymptotic mass (dashed) for $g_4^R=0.01$, varying $\mu_R^2$ (shown in units of $H^2$). }
    \label{fig1232g4}
\end{figure}

Moving now to the results for the inverse transition, $\mu_0=\sqrt2 H\rightarrow m\approx 0$, we are interested again in showing that this transition is possible under our approximations. Our analytical result from the previous section hinted at convergence problems in the late-time limit, and here we check whether those issues are present when \emph{no expansion} in $\varepsilon_2$ is made. Given that we are checking \textbf{transition 2}, we set the initial mass to $\mu_0=\sqrt2 H$, or equivalently $\varepsilon_1=1$. We begin by showing the results for $\varepsilon_2$ by varying the mass parameter, $\mu_R^2$, in Fig.~\ref{fig1232g4}. We also plot the asymptotic value (dashed curve) as predicted by Eq.~\eqref{asympmass}.

We note that convergence is indeed achieved and that it agrees with the expectation for the asymptotic mass from Eq.~\eqref{asympmass}. Furthermore, we note that in the analytical result for the evolution, Eq.~\eqref{q32p}, the r.h.s. did not depend on the final mass, $m$ (or $\varepsilon_2$), which would imply that the time-evolving part of the solution for $\varepsilon_2$ would not change among different choices of $\mu_R^2$. It is clear from Fig.~\ref{fig1232g4}, however, that the evolution is different from case to case, which emphasizes the need for an extension to that analytical result.

It turns out that one can improve the analytical estimate substantially, by changing the divergent $\log x$ term into a dynamical renormalization group (DRG) inspired expression~\cite{Burgess:2009bs, Dias:2012qy}. The resulting mass equation becomes
\begin{equation}
\label{mass1232corr}
m^2=\mu_R^2+\frac{g_4^RH^2}{16\pi^2}\left[4C_1 +4x+x^4+4 \log(1-x)+\frac{2}{\varepsilon_2} \left(1-x^{2\varepsilon_2}\right) e^{-\frac{3\varepsilon_2}{2}}\right],
\end{equation}
where the last term has been added. It is easy to show that this term is equal to $-4 \log x$ in the limit $\varepsilon_2\rightarrow 0$, as required. Given the similarity with the DRG method, we also call this new expression the re-summation of the previous one, given that one understands this correction as the sum of infinite terms with different powers of $\log x$.\footnote{A similar problem was detected in scattering calculations in kinematic regions where there is a large hierarchy of scales, the so-called Sudakov region~\cite{Sudakov:1954sw}, for which the Kinoshita--Lee--Nauenberg theorem \cite{Kinoshita:1962ur,Lee:1964is} is not valid. Re-summation of the large logarithms that appear is then required to make sense of the result. The techniques used for that case offered inspiration to the solution to very similar problems in inflationary correlation function calculations~\cite{Dias:2012qy} dealing with secular divergences \cite{Seery:2010kh}. The logarithms that appear in the present work are also, in fact, due to an IR divergence arising because of the evolution of the system towards a massless state. After re-summation, it is clear that the presence of a finite mass resolves the divergence.}

\begin{figure}[h]
    \centering
    \includegraphics[scale=0.8]{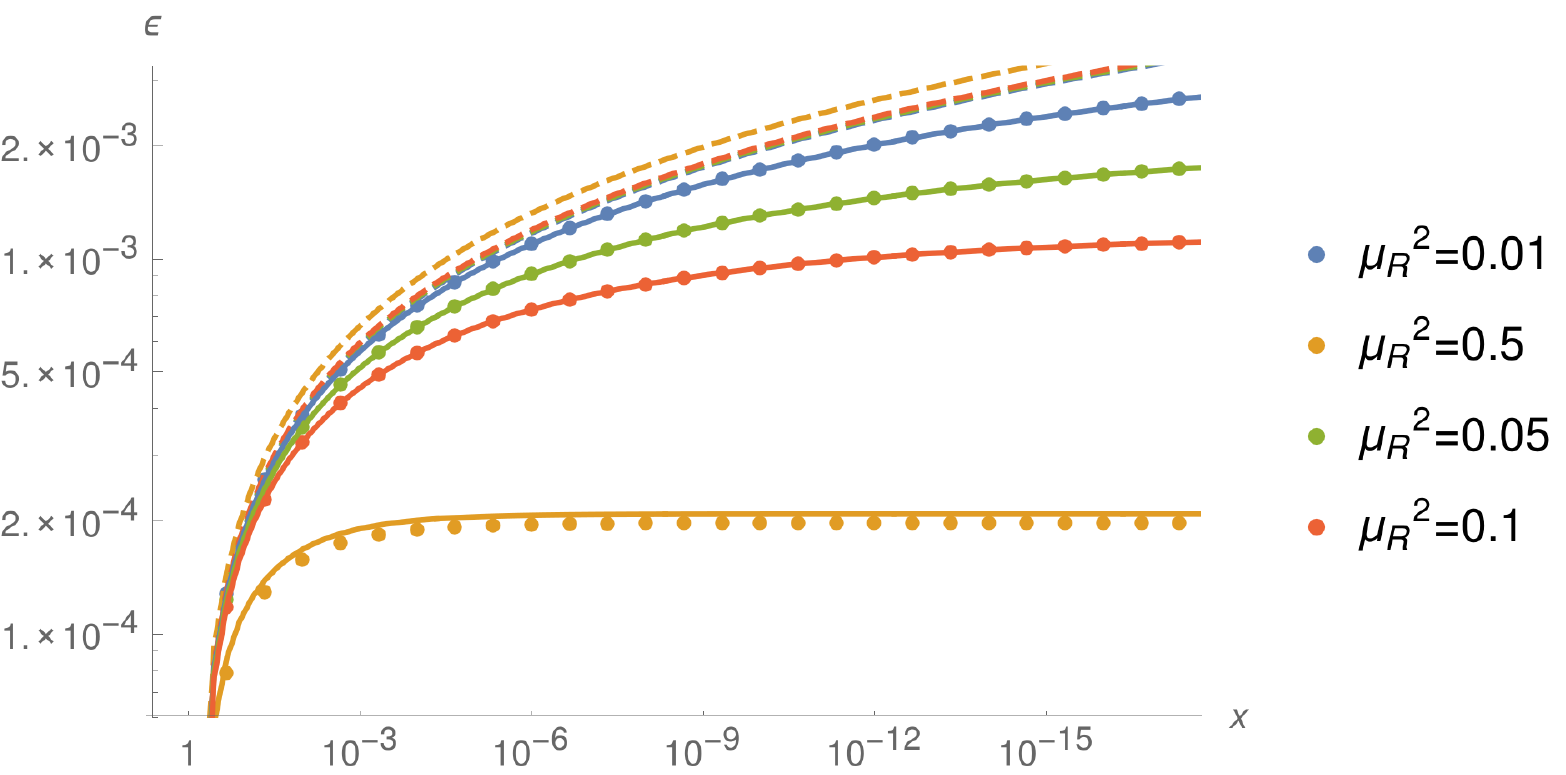}
    \caption{Numerical evolution of $\varepsilon_2(x)$ for transition 2 (dotted) as compared to both the corrected (solid) and uncorrected (dashed) analytical results, for $g_4^R=0.01$, varying $\mu_R^2$ (shown in units of $H^2$) and rescaled by $\mu_R^2$. }
    \label{fig1232g4Re}
\end{figure}

The improvement the re-summation brings to the result can be seen in the plot of Fig.~\ref{fig1232g4Re}, in which the results have been rescaled according to $\mu_R^2$ and we plot both the numerical results and the solution to the new mass equation, Eq.~\eqref{mass1232corr}.

The uncorrected result of Eq.~\eqref{q32p} is also shown in dashed lines. In spite of there being a substantial improvement, there is still a visible discrepancy for the case with the higher mass. This is expected, as the analytical result was derived for small masses, $m^2\ll H^2$ and the heavier example is already at $m^2\approx H^2/2$.

\begin{figure}[h]
    \centering
    \includegraphics[scale=0.8]{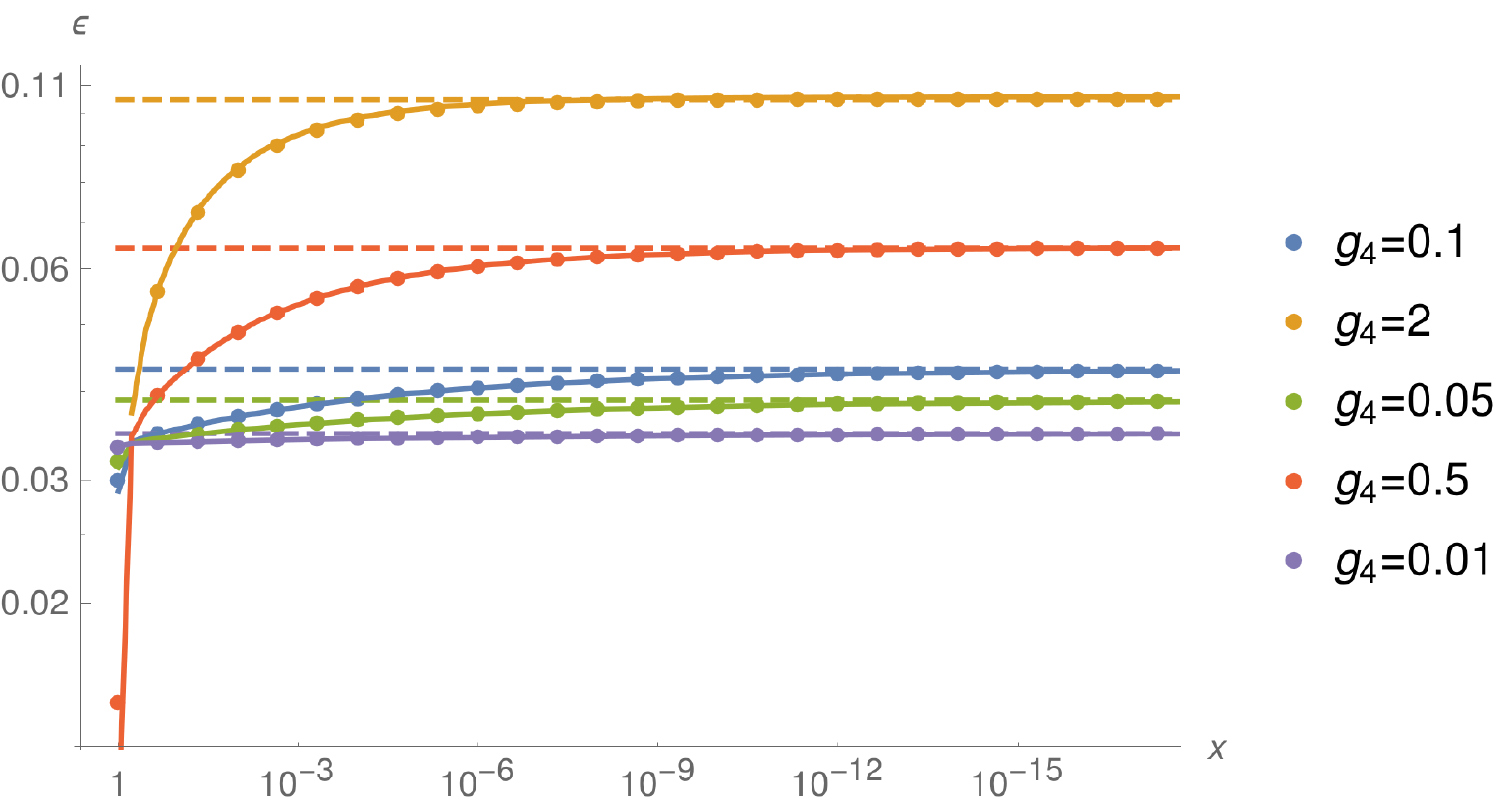}
    \caption{Numerical evolution of $\varepsilon_2(x)$ for transition 2 (dotted) as compared to the corrected analytical results (solid) and showing the asymptotic mass (dashed), with $\mu_R^2/H^2=0.1$, varying $g^R_4$. }
    \label{fig1232mu}
\end{figure}

All the cases presented in Figs.~\ref{fig1232g4} and \ref{fig1232g4Re} have $g_4^R=0.01$ and the contribution from the time evolution parts to the final result was not very large. The results presented in Fig.~\ref{fig1232mu} show the dependence on $g_4^R$ for higher values of the coupling. We see that, once again, the corrected result does very well in all cases and that it converges to the asymptotic result of Eq.~\eqref{asympmass}.

We note that when $g_4^R$ becomes large, the initial evolution can become quite fast, as expected, given the effect of the interaction in Eq.~\eqref{mass1232corr}. A quick analysis of that equation shows that the evolution is slower for larger $\mu_R^2$, since in those cases the interaction terms become almost negligible in comparison to $\mu_R^2$.
\\

\noindent\textbf{Transition 3:} $\mu_0\approx 0\ \rightarrow\ m\approx 0$
\\

\begin{figure}[h]
    \centering
    \includegraphics[scale=0.8]{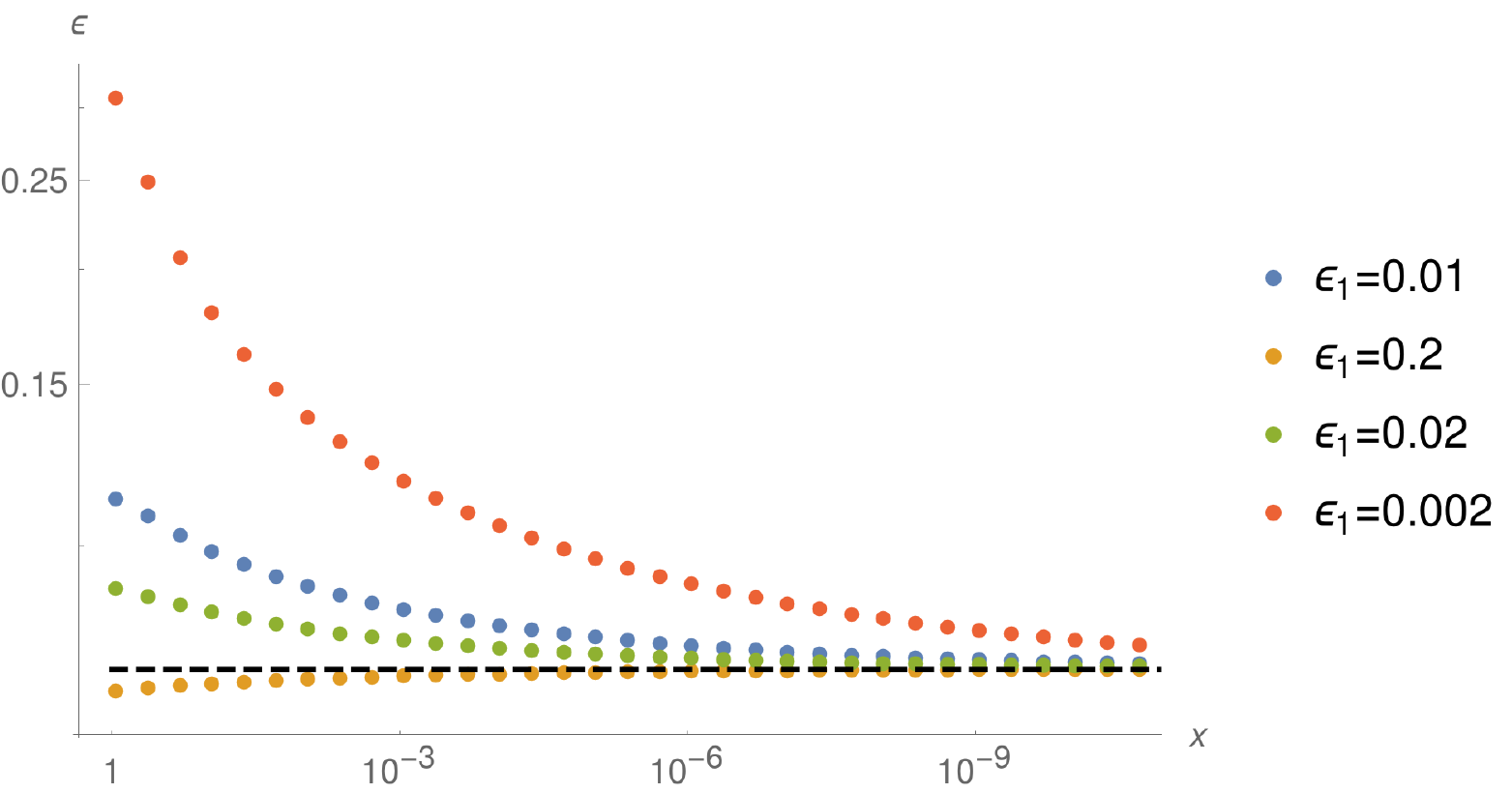}
    \caption{Numerical evolution of $\varepsilon_2(x)$ for transition 3 (dotted) and showing the asymptotic mass (dashed) with $\mu_R^2/H^2=0.2$, $g^R_4=0.1$, varying $\varepsilon_1$. }
    \label{fig3232e1}
\end{figure}

Let us now look at the more general case in which no mass is fixed. We focus on the cases in which the masses are small in order to compare with our results for the transition $\mu_0\approx 0\ \rightarrow\ m\approx 0$. One of the conclusions following from the expression for the asymptotic mass, Eq.~\eqref{asympmass}, was that, when $x\rightarrow 0$, the mass after the quench, $m$, should not depend on the mass before the quench, $\mu_0$. Fig.~\ref{fig3232e1} shows the time evolution of $\varepsilon_2$ for the quench with parameters given by $\mu_R^2=0.2$, $g_4^R=0.1$ and varying $\varepsilon_1$.

\begin{figure}[h]
    \centering
    \includegraphics[scale=0.8]{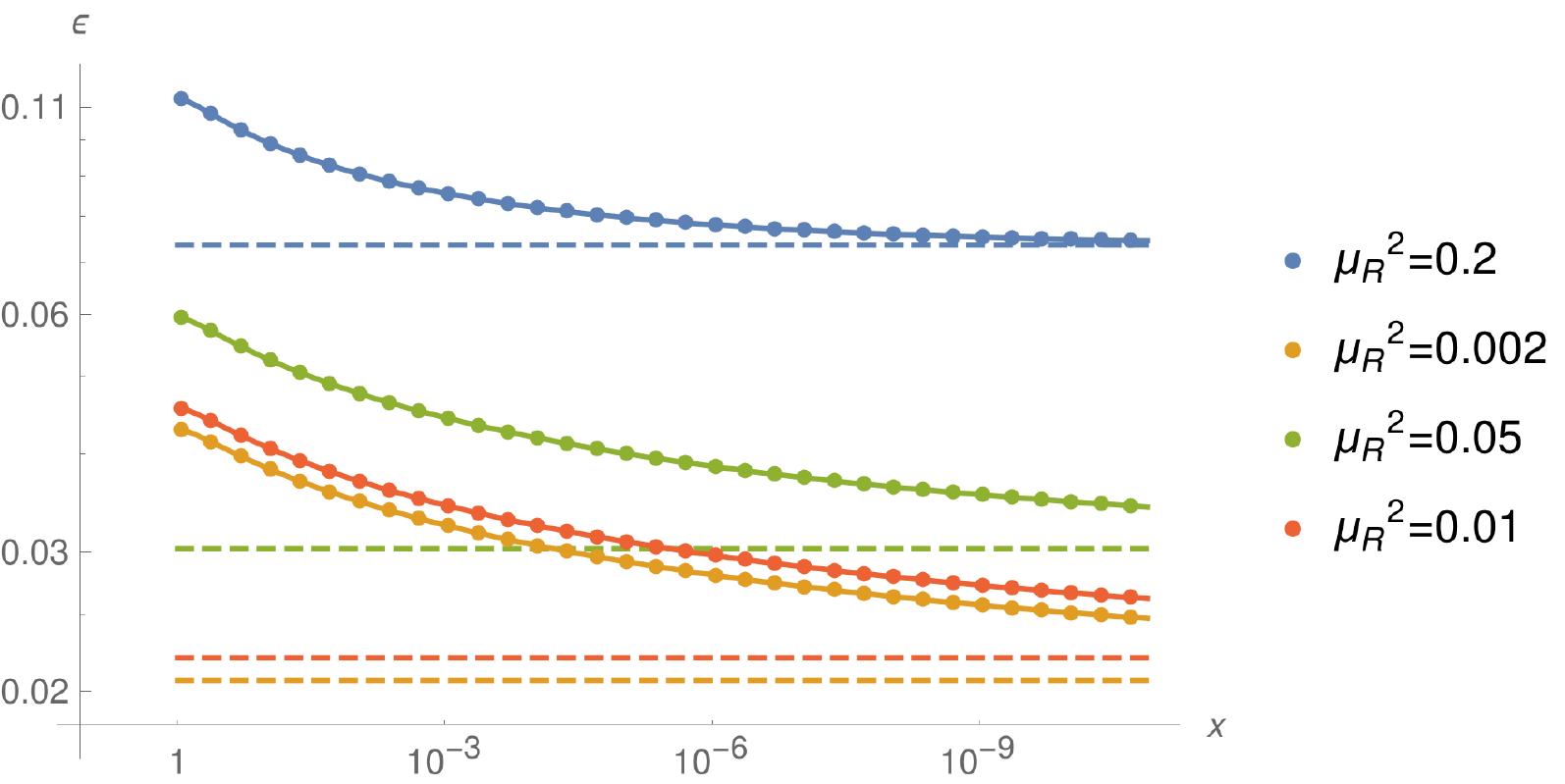}
    \caption{Numerical evolution of $\varepsilon_2(x)$ for transition 3 (dotted) as compared to the corrected analytical results (solid) and showing the asymptotic mass (dashed), with $\varepsilon_1=0.01$, $g^R_4=0.1$, varying $\mu_R^2$ (shown in units of $H^2$). }
    \label{fig3232mu}
\end{figure}

\begin{figure}[h]
    \centering
    \includegraphics[scale=0.8]{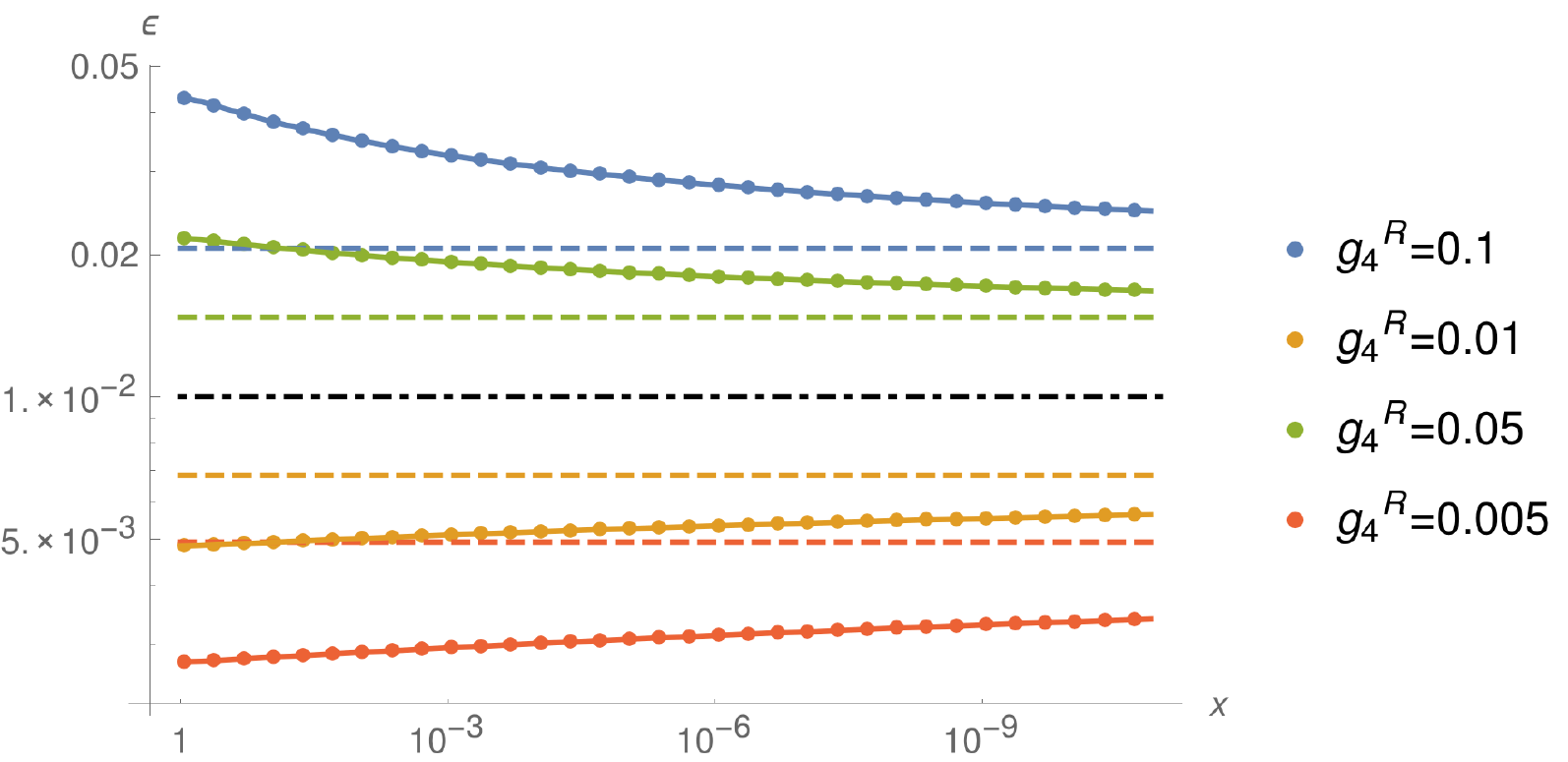}
    \caption{Numerical evolution of $\varepsilon_2(x)$ for transition 3 (dotted) as compared to the corrected analytical results (solid) and showing the asymptotic mass (dashed), with $\varepsilon_1=0.01$ (also shown as dot-dashed in the middle), $\mu_R^2/H^2=0.002$, varying $g^R_4$. }
    \label{fig3232g4}
\end{figure}

It is clear that, in spite of the previous analysis of Eq.~\eqref{q3232} indicating a divergent behavior at late times, the masses converge to the same value---that given by Eq.~\eqref{asympmass}. Again, in this case, it is possible to find a better approximation to the results, by drawing inspiration from dynamical renormalization group techniques \cite{Dias:2012qy,Burgess:2009bs} and applying them to Eq~\eqref{q3232}. This amounts to exponentiating the divergent terms, which results in the following expression
\begin{equation}
\label{q3232ren}
m^2=\mu_R^2+\frac{g_4^RH^2}{8\pi^2}\left[C_2+\frac{1}{\varepsilon_2}+\frac{\varepsilon_2-\varepsilon_1}{\varepsilon_1\varepsilon_2}x^{2\varepsilon_2}e^{\frac{2\varepsilon_2}{3}(1-x^3)}\right]\,.
\end{equation}
It is now clear that this solution has the correct asymptotic limit up to $O(\varepsilon_2)$ corrections, given by
\begin{equation}
m^2=\mu_R^2+\frac{g_4^RH^2}{8\pi^2}\left[C_2+\frac{1}{\varepsilon_2}\right]\,.
\end{equation}
This is equivalent to the result for the unquenched situation, Eq.~\eqref{unqm}, as expected from our previous analysis. We can see that this matches the numerical results very well in the plots that follow. We show both the effect of varying $\mu_R^2/H^2$ in Fig.~\ref{fig3232mu} and the dependence on $g_4^R$ in Fig.~\ref{fig3232g4}. Again, we show that, asymptotically, there is convergence towards the values given by Eq.~\eqref{asympmass}.

We see that the difference in mass $\varepsilon_2-\varepsilon_1\approx (m^2-\mu_0^2)/3H^2$ is very relevant for the evolution as it controls the slope of $\varepsilon_2(x)$. We can see this clearly in Fig.~\ref{fig3232g4}, in which the final mass appears to be attracted to the initial mass, approaching it until the asymptotic value of Eq.~\eqref{asympmass} is reached. The results plotted in Fig.~\ref{fig3232g4} also reveal that this behavior towards the initial mass is not symmetric about that value, i.e. the rate of change of the mass is larger for larger $g_4^R$. Hence, for larger asymptotic masses, the convergence to the final value is much faster than for the results below the initial mass. Furthermore, we notice some similarities between this transition and the others, as one sees a faster evolution for smaller $\varepsilon_1$ and for smaller $\mu_R^2$. However, the effect is slightly different, since a smaller $\varepsilon_1$ essentially contributes to a fast evolution through the terms $\propto 1/\varepsilon_1$, but a smaller $\mu_R^2$ removes part of the constant contribution to the mass. This affects the rate of change of the mass somewhat differently as well as the convergence towards the asymptotic mass.
\\

From these numerical results, we were able to find new expressions for the effective mass, which are far more reliable than those obtained in the previous section, given the absence of divergences at late times. In all cases, the results converge to the asymptotic mass, given by Eq.~\eqref{asympmass} and evolve differently depending on the parameters of the system before and after the quench. We also estimate the error in our constant-mass approximation below for the cases under study and conclude that, in spite of the large deviations existing for many situations, there are many relevant parameter values for which one can trust the approximation, which concludes the proof of concept we proposed to do.
\\

\noindent\textbf{Critical analysis of the constant mass approximation}
\\

Concerning our constant mass approximation, we analyze its error in terms of the quantity defined in Eq.~\eqref{error}, $e_u$, by calculating it for all the transitions studied here. We do not expect our results to be trustworthy for all of the cases presented, given the fast evolution of the mass in many. However, we also find several situations in which the error estimate is small, thus making our results reliable.

Regarding \textbf{transition 1}, we find the error to be approximately described by $e_u=3g_4^R/2\varepsilon_1$ (in \%), such that a few of the results plotted in Figs.~\ref{fig3212e1} and \ref{fig3212g4} have an error of less than $1\%$, while all except the largest have an error smaller than $10\%$. These case studies justify the approach we have adopted from the beginning.

For \textbf{transition 2}, however, we find that most of the results in Fig.~\ref{fig1232mu} have errors larger than $10\%$. For a value of $\mu_R^2/H^2=0.1$, the error is only smaller than $1\%$ when $g_4^R<2.7\times 10^{-3}$. This changes to $g_4^R<3.7\times 10^{-5}$ for $\mu_R^2/H^2=0.01$. This difference is not surprising, given that we had found a more substantial evolution of the mass for smaller values of $\mu_R^2$. This is also why the results with the smallest error in Fig.~\ref{fig1232g4} are those which have a higher value of $\mu_R^2$. The case $\mu_R^2=H^2/2$, for example, has an error of only $e_u=0.16\%$. The general trend is similar to that of \textbf{transition 1}, with smaller errors for larger $\mu_R^2$ and smaller $g_4^R$.

In the case of \textbf{transition 3}, we report similar error estimates as for the other transitions, again consistent with the error being smaller whenever the evolution is slower. It is possible to find errors smaller than $1\%$ for situations with very small coupling, $g_4^R$, or for large $\varepsilon_1$ and $\mu_R^2$. For example, the cases with the rather large $\varepsilon_1=0.1$, $\mu_R^2/H^2=0.1$, have errors $e_u<1\%$ if $g_4^R<4.3\times10^{-3}$. An effect that was not present in \textbf{transitions 1} and \textbf{2} takes place here when the difference of masses, or equivalently $\varepsilon_2-\varepsilon_1$, turns out to be small. In those cases there is a sharp decrease of the error, since the time-dependent terms are suppressed. For example, for $\varepsilon_1=0.01$, $\mu_R^2/H^2=0.01$, one finds the error to be $e_u\approx 1\%$ for $g_4^R=1.6\times10^{-2}$, while it is larger than $10\%$ for $g_4^R=10^{-3}$. Other similar examples exist, including situations in which $g_4^R$ is non-perturbative, i.e. of order 1. This is not entirely surprising, given that when $\varepsilon_2-\varepsilon_1$ is very small, the quench is nearly non-existent.

Furthermore, there is an important point that must be made with respect to the reliability of our approximation. Given that the parameter values for which the error is small are those for which the evolution is suppressed, one could wonder whether our results for the time dependence are accurate at all, i.e. whether they are an improvement to simply saying that, after the quench, one has a constant mass equal to the asymptotic mass. To answer this question, we compute the error with two versions of $u_0(\tau)$. We use the same expression in both cases, but in one we keep the value of the mass constant, while for the other version we substitute for the first approximation of the time dependent mass, $m_1(\tau)$. In all cases studied here, the error is smaller for the second version, indicating that our approximation is converging towards the real evolution of the mass, which is essential for the reliability of the method. Thus, we confirm that we are indeed finding a first approximation to the evolution of the mass and not just its asymptotic value.

\subsection{Negative $m^2$ and symmetry breaking}
\label{sec:Negative}

In this section we study whether non-positive values for $m^2$ are possible and what is the consequence for the spontaneous breaking of the $O(N)$ symmetry of the system.

We begin by re-stating the fact that, in a de Sitter invariant state, IR effects force the effective mass squared, $m^2$, to be strictly positive. This occurs regardless of the sign of $\mu_R^2$, since there always exist solutions to the mass equation for which $m^2$ is positive. This implies that the $O(N)$ symmetry of the system cannot be spontaneously broken, i.e. the only minimum of the effective potential is at $\ph^a=0$.

\begin{figure}[h]
    \centering
    \includegraphics[scale=0.8]{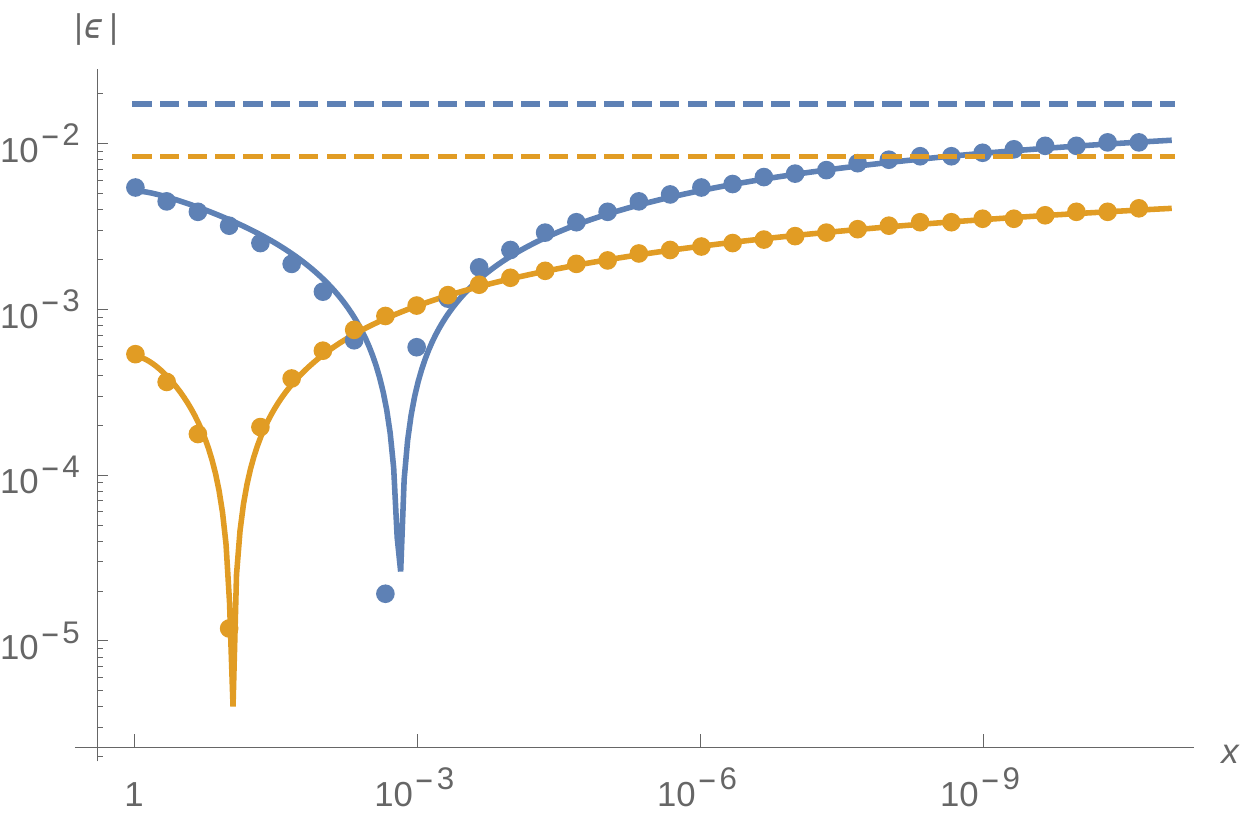}
    \caption{Numerical evolution of $|\varepsilon_2(x)|$ for negative $\mu_R^2$ (dotted) as compared to the analytical results (solid) and showing the asymptotic mass (dashed).  Initially $\varepsilon_2(x)$ is negative, but evolves towards positive values after some time.}
    \label{fig_negm2}
\end{figure}

For the case of a quench, the scalars are no longer in a de Sitter invariant state, and thus their mass squared may not be strictly positive. While it is true that, asymptotically, the mass squared always converges to the positive value given by the solution of Eq.~\eqref{asympmass}, there is a possibility that it is not always positive throughout the evolution. An analysis of Eq.~\eqref{q3232ren}, for example, reveals that, for values of $\mu_R^2$ that are sufficiently negative, one cannot find solutions for the effective mass squared which are positive. These solutions have been verified with numerical integration and are found to match the analytical results for a negative $m^2$, as shown in Fig.~\ref{fig_negm2}. The absence of IR divergences is due to the quench, as the IR part of the integrals of the power spectrum is dominated by the mass before the quench, $\mu_0^2$, which is positive. The influence of the state before the quench is gradually washed out and thus the mass squared is forced once again to become positive. Therefore, these results indicate that $m^2$ can be negative over the course of the evolution, but only temporarily.

Should the mass squared be negative, however, one expects the $O(N)$ symmetry to be broken and thus that the minimum of the effective potential to change to a non-zero value, i.e. one would have, $\bar\ph^N=H^2\tau^2\bar\sigma\neq0$. The discussion above neglected this factor, which has to be taken into account in the mass equation, Eq.~\eqref{selfconsdS}. We re-write it below in terms of $\ph$ instead of $\bar\sigma$ and $\chi$,
\begin{equation}
m^2=\mu^2+g_4\left[\bar\ph^N(x)^2+i G_\ph(x,x)\right]\,.
\end{equation}
This equation, will, in general, have a different solution due the extra contribution of the term $g_4 \bar\ph^N(x)^2$. Such a contribution is, however, not expected to be present immediately as the quench happens, at $\tau=\tau_0$, as the continuity of the fields imposes $\bar\ph^N=0$ at that time. Thus, the solution to the gap equation at $\tau_0$ remains the same as the one we obtained above, with the extra effect of the background field increasing in time as it evolves towards the minimum of the effective potential. This evolution is difficult to predict within our framework, but it seems clear that the effective mass will approach $m^2(\tau)=0$, as the term $g_4 \bar\ph^2$ cancels the negative $\mu^2$. However, the mass is not expected to remain at this value. If it did, then both the background field, $\bar\ph^N$, and the two-point function $G_\ph(x,x)$ would have to be constant, a situation which only happens in a de Sitter invariant state. But, one already knows from previous arguments that, in such a state, the mass squared must be strictly positive, which it would not be. Therefore, the mass should keep evolving, becoming positive again and eventually reaching the asymptotic value given by Eq.~\eqref{asympmass}, since, in that late-time limit, the background field $\bar\ph^N$ will once again have stabilized at $\bar\ph^N=0$. These arguments are somewhat in disagreement with the results of Ref.~\cite{Boyanovsky:1996rw}, which states that the system should be massless in the late-time limit. Nevertheless, should the mass be zero, it is not clear how one would avoid the IR divergences.

Given the arguments above, we conclude there is the possibility of a transient period in which the mass squared is negative, the duration of which should be calculable from a full numerical evolution of the entire system. We leave that for future work. During this period, the $O(N)$ symmetry of the system is broken to $O(N-1)$, but it is subsequently restored.

\section{Discussion and conclusions}
\label{sec:Discussion}

In this work, we have studied a quantum quench of an $O(N)$ scalar field theory in the background of a de Sitter spacetime. We have obtained the approximate evolution of the effective mass, in the regime in which it is slowly varying. In particular we have derived an expression for the mass in the late-time limit, Eq.~\eqref{asympmass}, which is an accurate limit for the effective mass, even in the general situation not covered by the present approximation. We reproduce that here:
\begin{equation}
m^2_\infty=\mu^2_R+\frac{g_4^RH^2}{16 \pi^2} \left(\frac{m^2_\infty}{H^2}-2\right)\left(\log 4 -1-\Psi\left(\nu_2^\infty-1/2\right)-\Psi\left(-\nu_2^\infty-1/2\right)\right)\nonumber\,,
\end{equation}
with $\nu^\infty_2=\sqrt{9/4-m_\infty^2/H^2}$. Analyzing that limit, we notice that it is independent of the initial mass prior to the quench, in contrast to a similar result in flat spacetime \cite{Hung:2012zr}.

Furthermore, we have obtained analytical expressions for the evolution of the mass, which we summarize in table \ref{tab1} for \textbf{transitions 1}, \textbf{2} and \textbf{3}.

\begin{table}[htpb]

    \heavyrulewidth=.08em
    \lightrulewidth=.05em
    \cmidrulewidth=.03em
    \belowrulesep=.65ex
    \belowbottomsep=0pt
    \aboverulesep=.4ex
    \abovetopsep=0pt
    \cmidrulesep=\doublerulesep
    \cmidrulekern=.5em
    \defaultaddspace=.5em
    \renewcommand{\arraystretch}{1.6}
    \begin{center}
        \small
        \begin{tabular}{Q|q}

            \toprule
            \textrm{Effective mass equation}
            &
            \multicolumn{1}{c}{ Transition}
            \\
            \cmidrule{1-2}
%            \rowcolor[gray]{0.9}
     %\displaystyle
						m^2=\mu_R^2+\frac{g_4^R H^2}{8\pi^2}x^2\left[\left(\frac{1}{\varepsilon_1}-3-2 \log(1-x)\right)(x-2)^2-1\right] &
            \multicolumn{1}{c}{$\ \displaystyle \frac{\mu_0}{H}\ll 1\rightarrow \frac{m}{H}\approx\sqrt2$}
            \\[2mm]

            \cmidrule{1-2}
%            \rowcolor[gray]{1.0}

            \displaystyle m^2=\mu_R^2+\frac{g_4^RH^2}{4\pi^2}\left(C_1 +x+\frac{x^4}{4}+\log(1-x)+\frac{1-x^{2\varepsilon_2}}{2\varepsilon_2} e^{-\frac{3\varepsilon_2}{2}}\right)  &
            \multicolumn{1}{c}{$\ \displaystyle \frac{\mu_0}{H}=\sqrt2\rightarrow \frac{m}{H}\ll1$}
            \\[2mm]

            \cmidrule{1-2}
 %           \rowcolor[gray]{0.9}
            \displaystyle m^2=\mu_R^2+\frac{g_4^RH^2}{8\pi^2}\left(C_2+\frac{1}{\varepsilon_2}+\frac{\varepsilon_2-\varepsilon_1}{\varepsilon_1\varepsilon_2}x^{2\varepsilon_2}e^{\frac{2\varepsilon_2}{3}(1-x^3)}\right) & 
            \multicolumn{1}{c}{$\displaystyle \frac{\mu_0}{H}\ll 1\rightarrow \frac{m}{H}\ll1$}
            \\
             \bottomrule
   
        \end{tabular}
    \end{center}
    \caption{Summary of the solutions to the self-consistent mass in different transitions.}\label{tab1}
    \end{table}

In the table, $x=\tau/\tau_0$ is the ratio between the current value of conformal time, $\tau$, and the initial value, $\tau_0$, at which the quench happened. The parameters $\varepsilon_1$ and $\varepsilon_2$ are proportional to the initial and final masses and are given by $\varepsilon_1\approx \mu_0^2/3H^2$ and $\varepsilon_2\approx m^2/3H^2$, respectively. 
In all cases, we report an evolution of the effective mass in the direction of the value of the mass before the quench, until it approaches a strictly positive asymptotic value. We confirm this result within our constant mass approximation for many values of the parameters of the system, by showing that the error in the approximation is small. In all other situations, in which the evolution is too fast, we can only be certain about the direction of the initial evolution of the mass and its final value, as per the assumptions of our calculations.

We have also evaluated the possibility of a transition to a negative mass squared and consequent symmetry breaking. We have argued that, should the parameters of the system be such that spontaneous symmetry breaking happens, this stage will be
transient, with the symmetry being restored after a certain time. Within our approximations, that time interval cannot be calculated and hence its evaluation is left for future work.

\para{Implications for cosmology}One of our original motivations was the direct application of the quench to fast transitions during inflation. If one interprets the scalars under study here as the perturbations 
of the inflaton field, the effect of the quench can be seen by calculating the power spectrum from Eq.~\eqref{bogtrans}.

\begin{figure}[h]
    \centering
    \includegraphics[scale=0.9]{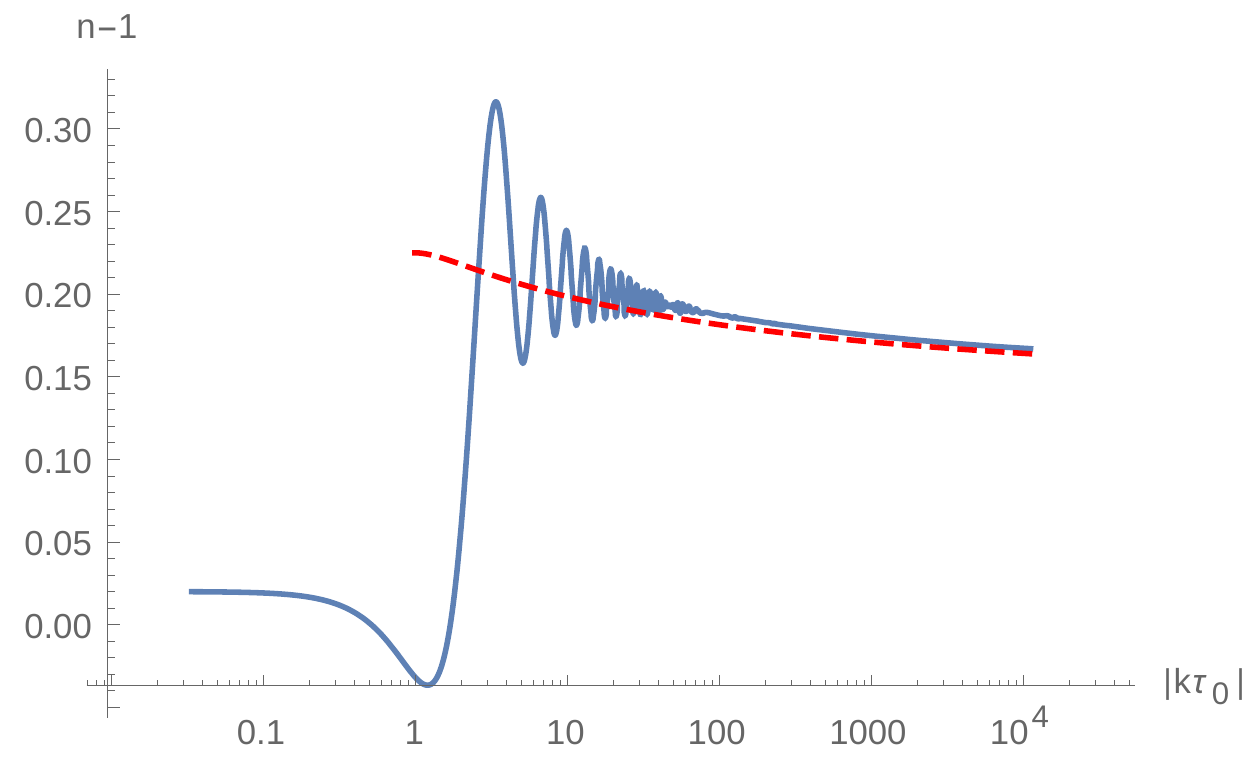}
    \caption{The spectral index (solid) 
    as a function of scale $k$ for the transition with parameters $\varepsilon_1=0.01$, $g_4^R=0.1$, $\mu^2_R=0.2$, shown at a time in which all scales have become super-horizon. Also shown is the value of $2\varepsilon_2$ (dashed), representing the effective mass via $\varepsilon_2\approx m^2/3H^2$ and evaluated at the time each scale exited the horizon, $\tau=k^{-1}$.}
    \label{figns}
\end{figure}

Another key quantity is the spectral index, which can be derived from the power spectrum, $\mathcal{P}=k^3\langle\ph^2\rangle$, via\footnote{Note that this is the same definition as in Eq.~\eqref{nsC3}, but we use a different notation. This is to distinguish the spectral index of the curvature perturbation from this one, defined in terms of the scalar field fluctuations.}
\begin{equation}
    n-1=\frac{{\text d}\log \mathcal{P}}{{\text d}\log k} \ .
\end{equation}
Evaluating the spectral index at the end of inflation, one would see an abrupt change in its value, occurring approximately at the scale $k_0\sim\tau_0^{-1}$, accompanied by small oscillations for $k>k_0$, as can be seen in Fig.~\ref{figns}. This is because the spectral index depends on the mass of the field at the time a certain scale left the horizon, and therefore will be sensitive to when the quantum quench occurs.

This situation is quite similar to what is described in Ref.~\cite{Joy:2007na}. However, given that our results do not take slow-roll into account, nor do we attribute the accelerated expansion to the effects of our scalars, the tendencies described here may not be realized in practice.

In any case, we have shown that it is possible to solve for the dynamics of a scalar field theory after a quantum quench in de Sitter spacetime, which is a very important first step towards the application to inflation. Beyond what we have done here, a full numerical evolution of the mode equation, Eq.~\eqref{EOMWF}, would be required, as well as the solution of the background equations, as those are also affected by the quench.

Another interesting application would be to the study the effect of spectator fields in inflation. It would be particularly interesting to study the quench to a negative mass, described in Section \ref{sec:Negative}, to check if that period can last for long enough to destabilize the slow-roll expansion and potentially end inflation.\\

\para{Summary}We have introduced a new method to study fast transitions in de Sitter spacetime using the large-$N$ technique. We 
have obtained an approximate solution to the dynamics of the system, which we believe to include most of the relevant features of the full solution, including the time dependence of the mass and its asymptotic value. We have also pointed to future directions, including a more direct application to inflation using numerical methods.

% % % % % % % % % % % % % % % % % % % % % % % % % % % % 
% chapter.tex - Ian Huston
% Sample chapter layout
% % % % % % % % % % % % % % % % % % % % % % % % % % % % 
% Redefine CVSRevision for this section. 
% If you don't want to use CVS tags comment out this line
\renewcommand{\CVSrevision}{\version$Id: chapter.tex,v 1.3 2009/12/17 18:16:48 ith Exp $}

% % % % % % % % % % % % % % % % % % % % % % % % % % % % % % % % 
% =========================================================== %
% % % % % % % % % % % % % % % % % % % % % % % % % % % % % % % % 
\chapter{Testing multi-field cosmological attractors in Palatini and metric gravity}
\label{Ch_palatini}
% % % % % % % % % % % % % % % % % % % % % % % % % % % % % % % % 
% =========================================================== %
% % % % % % % % % % % % % % % % % % % % % % % % % % % % % % % % 

\section{Introduction}
\label{introduction}

In this chapter, we concentrate on models of inflation with multiple fields that couple non-minimally to the gravity sector of the theory. We study couplings of the type $\xi_I (\ph^I)^n g^{\mu\nu}R_{\mu\nu}$, where $\xi_I$ are coupling constants and we take $n>0$. With a suitable potential, single-field models of this type universally approach a single set of predictions, which are approximately equal to those of Starobinsky inflation, shown at the end of Section~\ref{Sec_Inf}. Here, we verify that that is the case also for multi-field models of inflation and we test whether this similarity with the single-field case also remains true in two different formulations of gravity, the metric and the Palatini formulations. We aim also to investigate what multi-field effects appear in these scenarios and whether they are different for the two gravitational theories.

The chapter is organized as follows: in Section \ref{inflation}, we present the multi-field models we are considering and perform the conformal transformation to the Einstein frame where the non-minimal couplings vanish. In Section \ref{results}, we present the numerical set-up and the results, discussing observational ramifications and demonstrating the influence of multi-field effects on the inflationary dynamics. Finally, in Section \ref{conclusions}, we summarize our findings.

This chapter is based on work in collaboration with John Ronayne, Tommi Tenkanen and David Mulryne. My main contribution was in the theoretical aspects of this work, as well as in performing some of the analytical estimates required to better understand the numerical results.

\section{Multi-field inflation with non-minimal couplings to gravity}
\label{inflation}

We consider a theory with multiple scalar fields, all of which are non-minimally coupled to gravity. We generalize the action in Eq.~\eqref{phch_act} to account for that, and write it here in the Jordan frame, in which the non-minimal coupling is explicit:
\be
\label{nonminimal_action}
S_J = \int d^4x \sqrt{-g}\left(\frac{1}{2} \delta_{IJ}g^{\mu\nu}\partial_{\mu}\ph^I\partial_{\nu}\ph^J -\frac{M_{\text P}^2}{2}\left(1 + f(\ph^I)\right) g^{\mu\nu}R_{\mu\nu}(\Gamma) + V(\ph^I)\right) ,
\ee
where we have explicitly written the Ricci tensor, $R_{\mu\nu}$, as a function of the connection $\Gamma$, to make that dependence clear. We are, once more, using the Einstein summation convention also in the field-space indices labelled by capital letters ($I,J$), for which the sum runs over the total number of fields. The potential $V(\ph^I)$ is at this point completely general and could, in principle, contain all possible mass and interaction terms of the scalar fields allowed by the underlying symmetries of the theory. The non-minimal coupling function $f(\ph^I)$ is also unspecified in the action, but will, in the following, generally take the form
\be
\label{nmc_function}
f(\ph^I)=\sum_I{\xi_I^{(n)}\left(\frac{\ph^I}{M_{\text P}}\right)^n}\,,
\ee
with $\xi_I^{(n)}$ the dimensionless non-minimal coupling parameters.\footnote{Note that the superscript $(n)$ used throughout this chapter is only a label, meant to distinguish the parameters of different models, and is unrelated to the order of perturbation theory, for which the same notation was used in Chapters~\ref{Ch_CPT}--\ref{Ch_iso2}.} The most well studied of these couplings is the one generated by quantum corrections of a quartic scalar theory in a curved spacetime, for which $n=2$. For example, this is the case for the usual (single-field) Higgs inflation \cite{Bezrukov:2007ep}.

In the metric formulation of gravity, the connection $\Gamma$ is determined uniquely as a function of the metric tensor, i.e. it is $\bar{\Gamma}(g_{\mu\nu})$, the Levi-Civita connection, as given in Eq.~\eqref{connectionC2}. In the Palatini formalism both $g_{\mu\nu}$ and $\Gamma$ are treated as independent variables, and the only assumption is that the connection is torsion-free, $\Gamma^\lambda_{\alpha\beta}=\Gamma^\lambda_{\beta\alpha}$. The application of the variational principle then gives rise to an extra equation for the connection, in addition to the one for the metric. For the Einstein-Hilbert action, the extra equation forces the connection to have the usual Levi-Civita form, but in more general theories of gravity, such as $f(R)$ theories, or in the presence of non-minimal couplings, this is no longer true in the Jordan frame.
 	
However, the non-minimal couplings in the Jordan frame action \eqref{nonminimal_action} can be removed by a conformal transformation to the Einstein frame, 
\begin{equation}
\label{Omega}
g_{\mu\nu} \to \Omega^{-1}(\ph^I)g_{\mu\nu}, \hspace{.5cm} \Omega(\ph^I)\equiv 1+f(\ph^I) \, .
\end{equation}
Note that in the Palatini case, the connection is unchanged by this transformation, since it is independent of the metric. After this transformation, the action \eqref{nonminimal_action} becomes
\be
S_{\text E} = \int d^4x \sqrt{-g}\bigg(\frac{1}{2}G_{IJ}(\ph^I){\partial}_{\mu}\ph^I{\partial}^{\mu}\ph^J -\frac{1}{2}M_{\text P}^2R + V(\ph^I)\Omega^{-2}(\ph^I)  \bigg),
\label{EframeS1}
\ee
where $R = g^{\mu\nu}R_{\mu\nu}(\bar{\Gamma})$, i.e. in the Einstein frame we retain the standard Levi-Civita connection regardless of the chosen theory of gravity, and the scalars have acquired a non-trivial field-space metric, given by
\be
\label{fieldmetric}
G_{IJ}=\Omega^{-1}\delta_{IJ}+\frac32 \upsilon M_{\text P}^2\Omega^{-2}\frac{\partial\Omega}{\partial \ph^I}\frac{\partial\Omega}{\partial \ph^J}\,,
\ee
where $\upsilon=1$ in the metric case and $\upsilon=0$ in the Palatini case. With this conformal transformation, we have therefore transferred the dependence on the choice of gravitational degrees of freedom from the connection to the field-space metric. 

The existence of a non-trivial field-space metric has the consequence that a Levi-Civita connection can now be defined via the field-space equivalent of Eq.~\eqref{connectionC2}, and the field-space can have a non-zero Riemann curvature, $R^{\ A}_{\text{fs}\ BCD}$. This can have several consequences for the evolution of the fields in this space, such as causing an equivalent to geodesic deviation~\cite{Turzynski:2014tza,Renaux-Petel:2015mga,Garcia-Saenz:2018ifx}. We shall check below, what is the effect of curvature for the models under study here.

In the following, we will analyse inflation in both cases, metric and Palatini. For simplicity, we study two-field models with the potential
\be
\label{potential}
V(\ph,\sigma) = \lambda_\ph^{(2n)}  M_{\text P}^{4-2n} \ph^{2n} + \lambda_\sigma^{(2n)}  M_{\text P}^{4-2n} \sigma^{2n},
\ee
where $n>0$, $\lambda_\ph^{(2n)}$ and $\lambda_\sigma^{(2n)}$ are dimensionless coupling constants, and $M_{\text P}^{4-2n}$ has been introduced to have a scalar potential with a mass dimension equal to four. Later on, in Sec. \ref{3fieldcase}, we will also discuss the case where more than two fields take part in inflationary dynamics.

In metric gravity, the above models are cosmological attractors, i.e. their predictions for observables asymptote to those of $R^2$ or Starobinsky inflation in the limit of strong non-minimal coupling $\xi$, see Eq. \eqref{StaroC3}. This is, however, known not to be true for the single-field case in the Palatini scenario~\cite{Jarv:2017azx}, and we will test it also in a multi-field case.

For the potential \eqref{potential}, the Einstein frame potential is
\be
U(\ph,\sigma) = \Omega(\ph,\sigma)^{-2} V(\ph,\sigma)= \frac{\lambda_\ph^{(2n)}  M_{\text P}^{4-2n} \ph^{2n} + \lambda_\sigma^{(2n)}  M_{\text P}^{4-2n} \sigma^{2n}}{\left(1+\xi_\ph^{(n)}\left(\frac{\ph}{M_{\text P}}\right)^n+\xi_\sigma^{(n)} \left(\frac{\sigma}{M_{\text P}}\right)^n\right)^2}\,.
\ee

For this and all other models in this formulation, the potential $U$ is the same for both metric and Palatini gravity. The major difference between the two is the Einstein frame field-space metric, $G_{IJ}$. We will therefore focus mostly on the parameters appearing in $G_{IJ}$ in our analysis, namely the non-minimal couplings, $\xi_I^{(n)}$. The overall amplitude of the parameters $\lambda_I^{(2n)}$ in Eq. \eqref{potential} can be fixed by requiring that the dimensionless curvature power spectrum, defined as
\be
\label{powerspec}
\mathcal{P}_{\zeta}(k)=\frac{k^3}{2\pi^2}P_\zeta(k)\,,
\ee
has the measured amplitude,  $\mathcal{P}_{\zeta}=(2.141\pm 0.052)\times 10^{-9}$ (at the $68\%$ confidence level) \cite{Ade:2015xua}. Their ratio, however, is unconstrained and does play a role in the dynamics, as we will show.

In the following, we calculate the predictions for observables in this type of model. We compute the usual spectral index of curvature perturbations, $n_s$, defined in Eq.~\eqref{nsC3}, the tensor-to-scalar ratio, $r$, given in Eq.~\eqref{rC3} and the amount of non-Gaussianity, measured via the amplitude of the reduced bispectrum in the equilateral configuration
\begin{equation}
\label{fnldef}
f_{\text NL}=\frac{5}{18}\frac{B_\zeta(k,k,k)}{P_\zeta(k)^2}\,,
\end{equation}
in which $B_\zeta(k_1,k_2,k_3)$ is the bispectrum, defined via Eq.~\eqref{bispC3}.

All of the above variables are evaluated at horizon crossing of the Planck pivot scale, $k=0.05~\text{Mpc}^{-1}$, which we take to correspond to modes which crossed the horizon $60$ e-folds before the end of inflation. We explore the parameter space of the models under consideration by varying all parameters of the scalar potential and the field-space metric, as well as the initial conditions for the evolution during inflation. In order to compute the predictions, we employ the transport method \cite{Dias:2016rjq} (see Refs.~\cite{Mulryne:2013uka,Anderson:2012em,Seery:2012vj,Mulryne:2010rp,Mulryne:2009kh,Dias:2011xy,Dias:2014msa,Dias:2015rca} for earlier related work) and the open source PyTransport code\footnote{The package is available at \href{https://github.com/jronayne/PyTransport}{github.com/jronayne/PyTransport}.}~\cite{Mulryne:2016mzv}. The results and the set-up for finding initial conditions are presented in the next section. The transport approach 
evolves the two and three-point function of field fluctuations from initial conditions set in the quantum regime on sub-horizon scales (as well as the two point function of tensor perturbations), and includes all tree-level contributions. It then uses these correlations to calculate the power spectrum and bispectrum of $\zeta$. It was recently extended to include a non-trivial field-space metric in Refs.~\cite{DavidJohn2,Butchers:2018hds} (and is also the basis of another open source package CppTransport \cite{Seery:2016lko}).  

%%%%%%%%%%%%%%%%%%%%%%%%%%%%%%%%%%%%%%%%%%%%%%%%%%%%%%%%%%%%%%%%%%%%%%%%%%%%%%%%%%%%%%%%%%

\section{Results}
\label{results}

\subsection{Numerical Set-up}

For a given set of model parameters, we explore the initial condition space by first calculating an approximate position in field-space corresponding to 73 e-folds before the end of inflation\footnote{The number $N=73$ is chosen to start the evolution so that the modes which cross the horizon 60 e-folds before the end of inflation are accurately evolved in the sub-horizon stage.}. Before sampling, we transform our fields to polar form. Then we sample an angle from a uniform distribution. Following that we incrementally increase the radial distance from the minimum of the potential until a coordinate in field space is found for which inflation lasts 73 e-folds under the assumption of slow-roll initial conditions. Sampling over the full distribution of angles would reveal an approximate 73 e-fold surface in the field space. Next we transform our fields back to their Cartesian form and numerically evolve the background equations forward in time until the end of inflation. This provides a set of evolutions of roughly $73$ e-folds. For each set of model parameters the process is repeated with a new random angle. Finally, we evaluate the observables of interest -- $n_s$, $r$ and $f_{\text NL}$ as defined above -- at the scale which left the horizon 60 e-folds before the end of inflation. We repeat this procedure for a representative set of values of the model parameters focusing mostly on the effect of the non-minimal couplings, $\xi_I$.

\begin{figure}
	\centering
	        \includegraphics[width=0.45\textwidth]{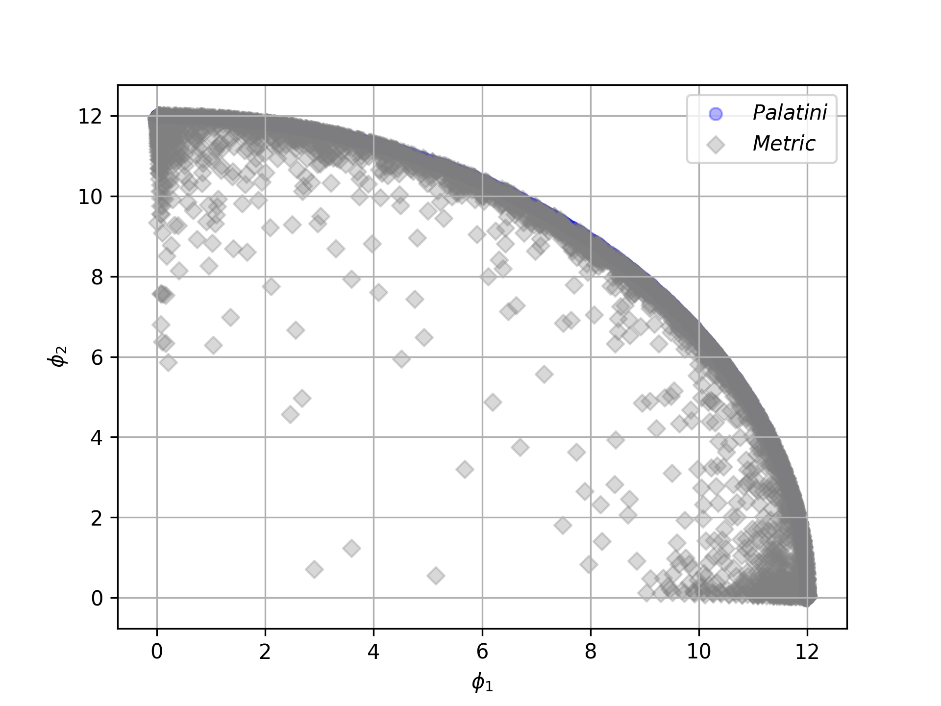}
	        \includegraphics[width=0.45\textwidth]{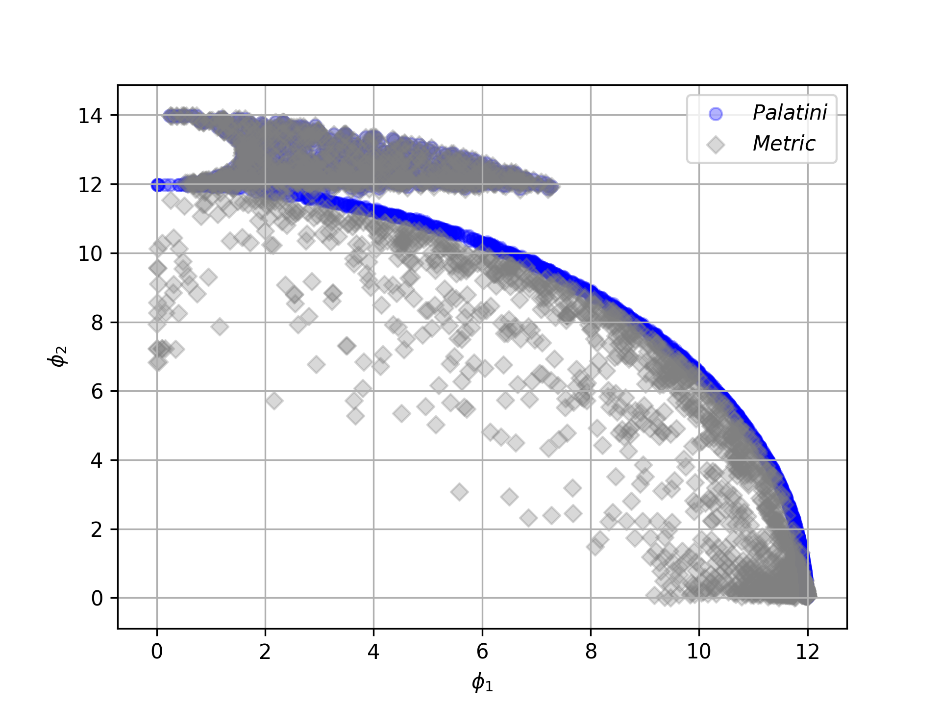}
        	\includegraphics[width=0.45\textwidth]{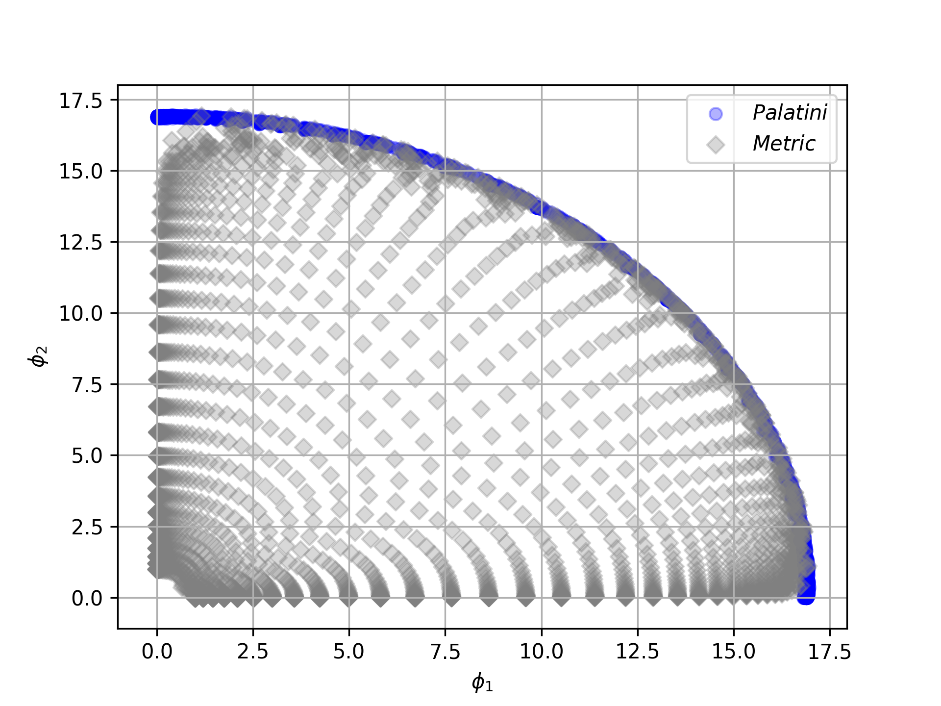}
        	\includegraphics[width=0.45\textwidth]{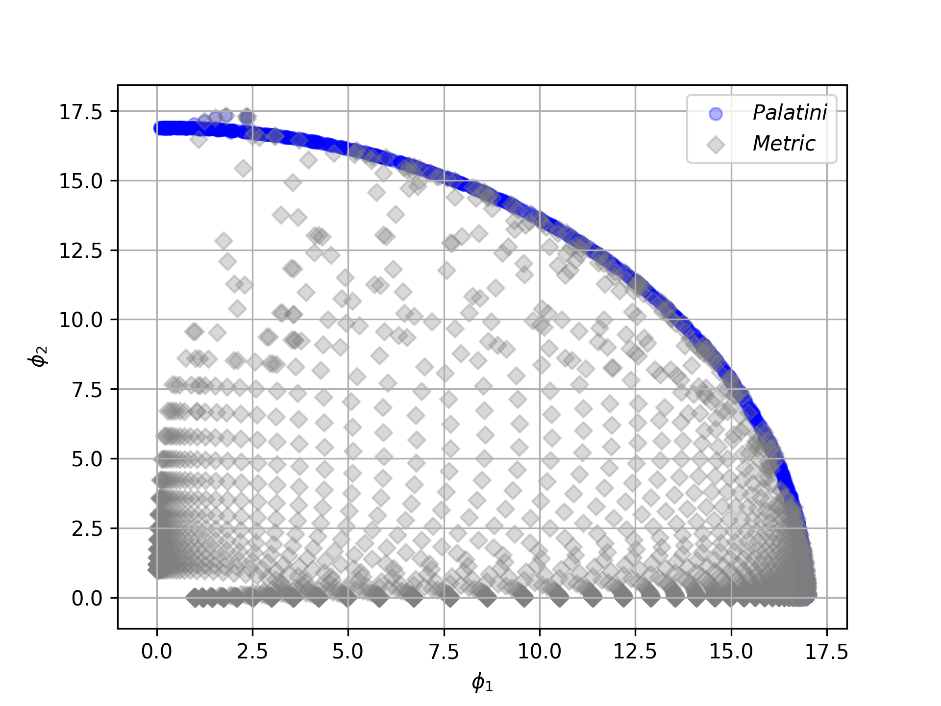}
        	\includegraphics[width=0.45\textwidth]{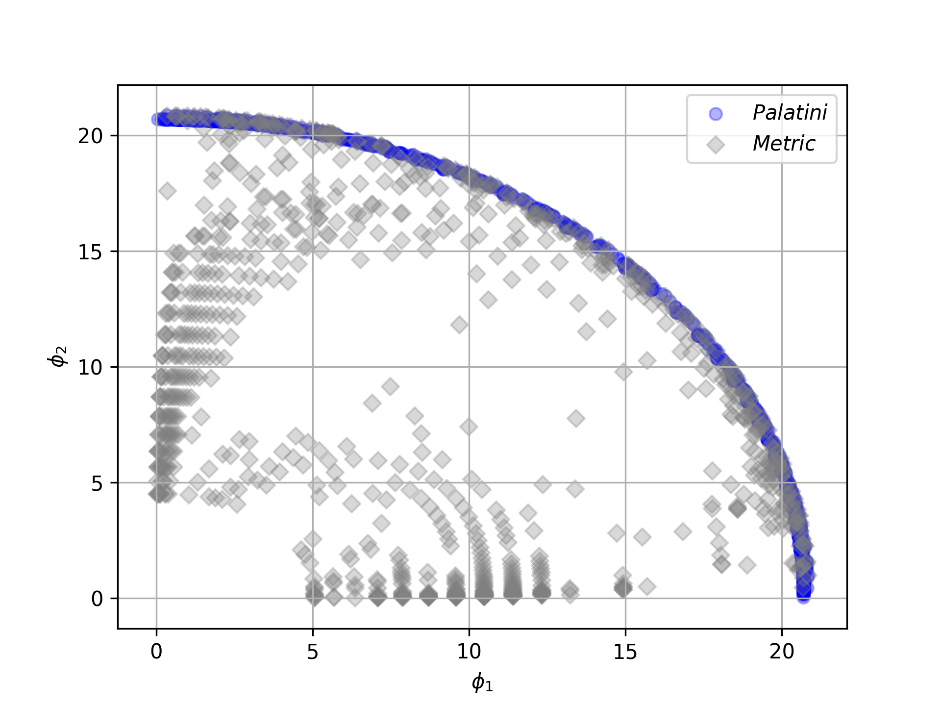}
 	        \includegraphics[width=0.45\textwidth]{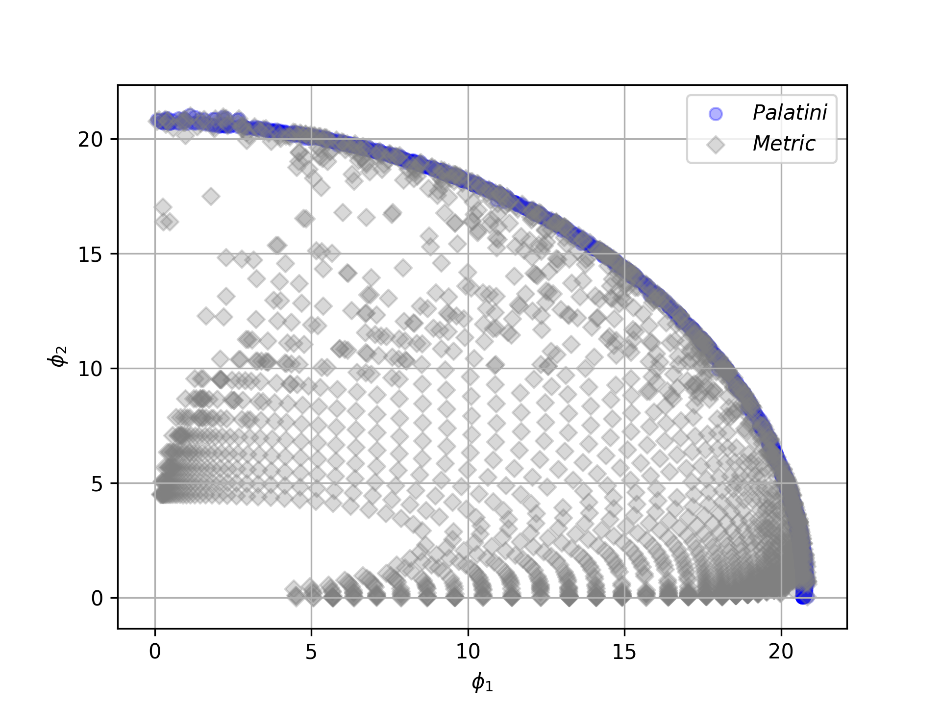}
        	\includegraphics[width=0.45\textwidth]{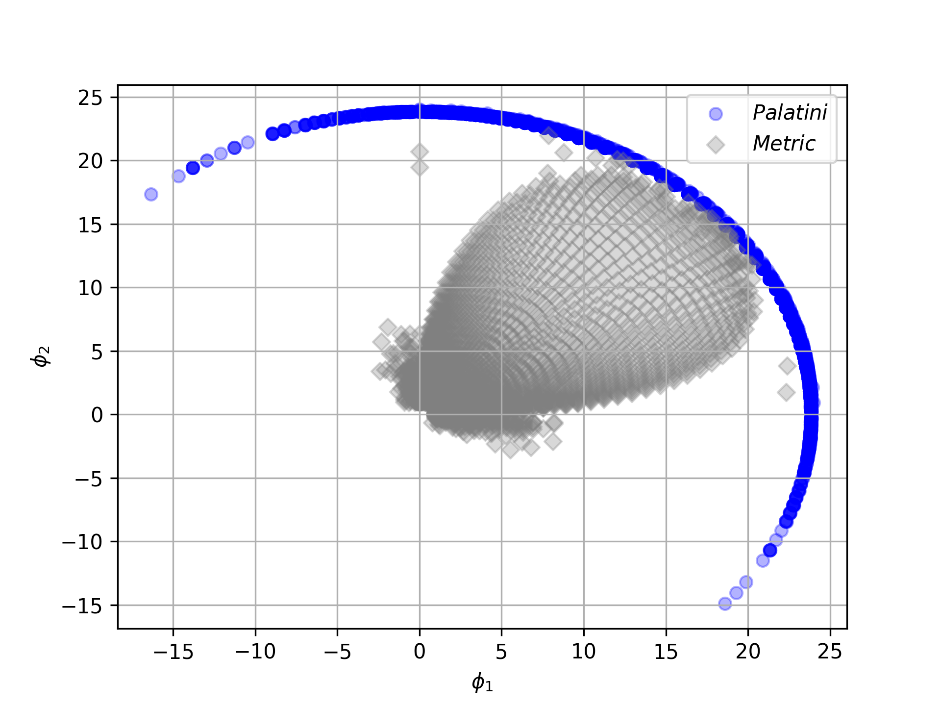}
        	\includegraphics[width=0.45\textwidth]{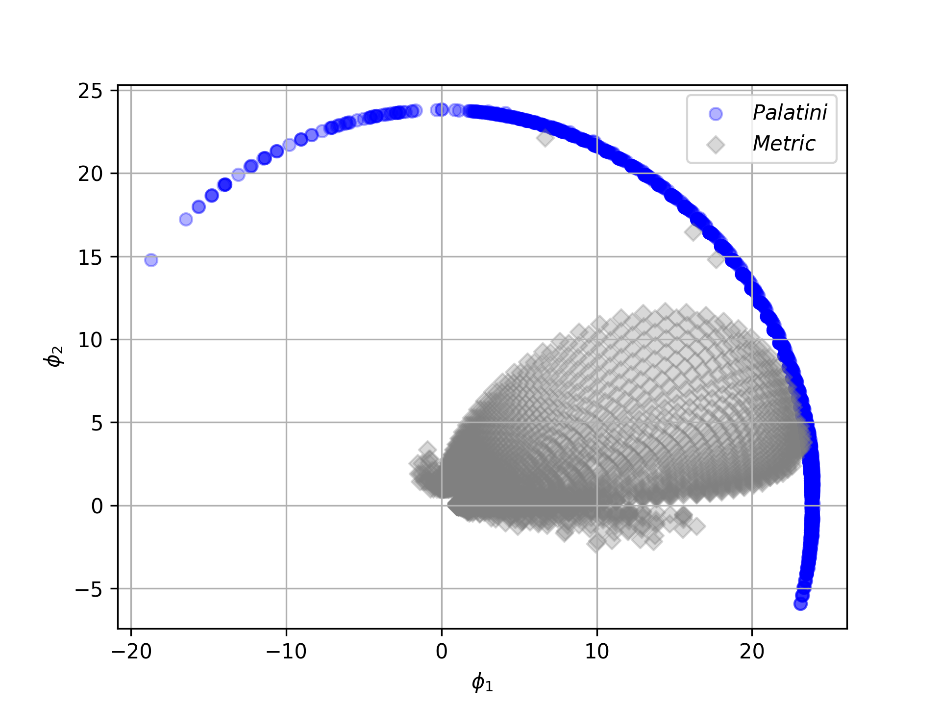}
	\caption{Sampling of initial conditions for $\phi_1=\ph$ and $\phi_2=\sigma$ for metric (grey) and Palatini gravity (blue), $n=(1/2,1,3/2,2)$ from top to bottom. The left and right panels show the scenarios for different parameter ratios: $\lambda_\sigma/\lambda_\ph=19/14$ (left) and $\lambda_\sigma/\lambda_\ph=95/14$ (right). In all cases $\xi$ is varied between $(10^{-3}, 10)$.}\label{fig:ICs}
\end{figure}

Already at the background level, the evolution is different between metric and Palatini gravity. We can clearly see this in Fig.~\ref{fig:ICs}, which shows the initial conditions corresponding to 73 e-folds of inflation for both metric and Palatini gravity, with varying strengths of the non-minimal couplings. For Palatini gravity, the initial condition surface is independent of the value of the non-minimal coupling for nearly all cases, while for metric gravity the distance from the origin decreases with $\xi_I$ regardless of the value of $n$.

One can understand this by using the slow-roll approximation introduced in Chapter~\ref{Ch_SMC}. We first note that the Klein--Gordon equations for the fields $\ph^I$ are given by
\begin{equation}
\Box\ph^I+\Gamma^I_{JK}\p_\nu\ph^J\p^\nu\ph^K=G^{IL}\p_L U\,,
\end{equation}
in which $\Gamma^I_{JK}$ are the components of the Levi-Civita connection associated with the field-space metric $G_{IJ}$ and $\p_L U$ is the derivative in the direction of the field $\ph^L$. This implies that the background equations are
\begin{equation}
\ddot{\ph}^I+3H\dot{\ph}^I+\Gamma^I_{JK}\dot{\ph}^J\dot{\ph}^K=-G^{IL}\p_L U\,,
\end{equation}
which, under slow-roll, reduce to
\begin{equation}
3H\dot{\ph}^I\approx-G^{IL}\p_L U\,.
\end{equation}
Projecting this equation in the inflationary direction, by contracting with $\dot{\ph}_I$, results in
\begin{equation}
3H\dot{\ph}^2\approx-\p_\ph U\,,
\end{equation}
in which $\dot{\ph}^2=\dot{\ph}_I\dot{\ph}^I=G_{IJ}\dot{\ph}^I\dot{\ph}^J$ is the norm of the field velocity and $\p_\ph U$ is the derivative of $U$ in the inflationary direction. It is clear that this is equation is the same as the one used in the single-field case, which implies that the number of e-folds can be written as in Eq.~\eqref{Nev}, which in this notation is given by
\begin{equation}
N= \int_{\ph_{\text{e}}}^{\ph_{\text{i}}} \frac{U}{\p_\ph U}\text{d}\ph\,.
\end{equation}
We now assume that the background trajectories are approximately radial. Writing the fields in polar coordinates, $(\rho,\psi)$, as\footnote{Note that the varible $\rho$ is unrelated to the energy density defined elsewhere in this thesis. The angle $\psi$ is also not to be confused with the curvature perturbation defined in Chapter~\ref{Ch_CPT}. In the current chapter, the gauge-invariant perturbation, $\zeta$, is always used to represent the curvature perturbation.}
\begin{equation}
\ph=\rho\cos \psi \,,\ \ \sigma=\rho\sin\psi\,,
\end{equation}
the number of e-folds can be approximated by assuming that $\p_\ph U\approx\p_\rho U \p\rho/\p\ph$ and $\p\ph/\p\rho= \dot{\ph}/\dot\rho\approx\sqrt{G_{\rho\rho}}$. This results in
\begin{equation}
N\approx \int_{\rho_{\text{e}}}^{\rho_{\text{i}}} \frac{U}{\p_\rho U}G_{\rho\rho}\text{d}\rho\,.
\end{equation}
All of the quantities in the integrand above can be calculated straightforwardly, given the field-space metric, $G_{IJ}$, and the Einstein frame potential, $U$. To further simplify the notation, we also write the non-minimal couplings in polar coordinates as
\begin{align}
\xi_\ph=\xi \cos \theta\,,\ \ \xi_\sigma=\xi \sin \theta\,.\label{polarxi}
\end{align}
For Palatini gravity, the number of e-folds is independent of $\xi$ and given by
\begin{equation}
N\approx \frac{\rho_\text{i}^2-\rho_\text{e}^2}{4nM_{\text P}^2}\,,
\end{equation}
in which $\rho_\text{i}$ is the value of $\rho$ when the mode of interest exists the horizon and $\rho_\text{e}$ is the value at the end of inflation. $N$ is thus the number of e-folds of expansion between the times in which $\rho$ took those values. Interestingly, this is exactly the same result as for $\xi=0$, which is why the initial conditions for the Palatini case coincide with those for the metric case at low $\xi$. For metric gravity, the result is rather long in the general case and we choose to show it only for large values of $\xi$, and by performing an expansion in $\xi^{-1}$. The leading order result for the number of e-folds is
\begin{equation}
N\approx \xi F_n(\psi,\theta)\frac{\rho_\text{i}^n-\rho_\text{e}^n}{M_{\text P}^n}\,,
\end{equation}
which shows that to keep the number of e-folds constant, one requires smaller $\rho_\text{i}$ for larger $\xi$, as indeed is the case in Fig.~\ref{fig:ICs}. The function $F_n(\psi,\theta)$ is a well defined functions of the angular variables, which we do not show here, for brevity. It simplifies to the single-field result when $\psi=\theta=0$ or $\psi=\theta=\pi/2$, which, for $n=2$, is $F_2(0,0)=3/4$, matching the result in Ref.~\cite{Bauer:2008zj}.

We see that this approximation works generically very well, except when the parameter ratio is large in certain directions in the field-space. This is because the approximation of radial trajectories fails in those cases, rendering the above approximate result inapplicable. This emphasizes the importance of accurate numerical analysis of multi-field models, to which we now turn.

\subsection{Attractor models}

\begin{figure}
	\centering
		\includegraphics[width=0.45\textwidth]{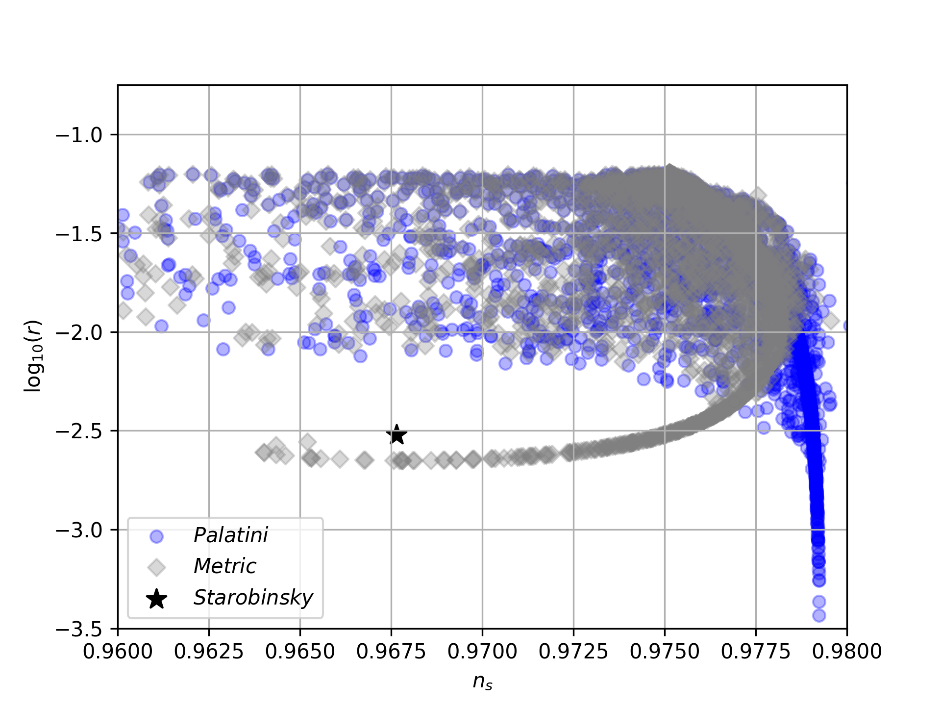}
        	\includegraphics[width=0.45\textwidth]{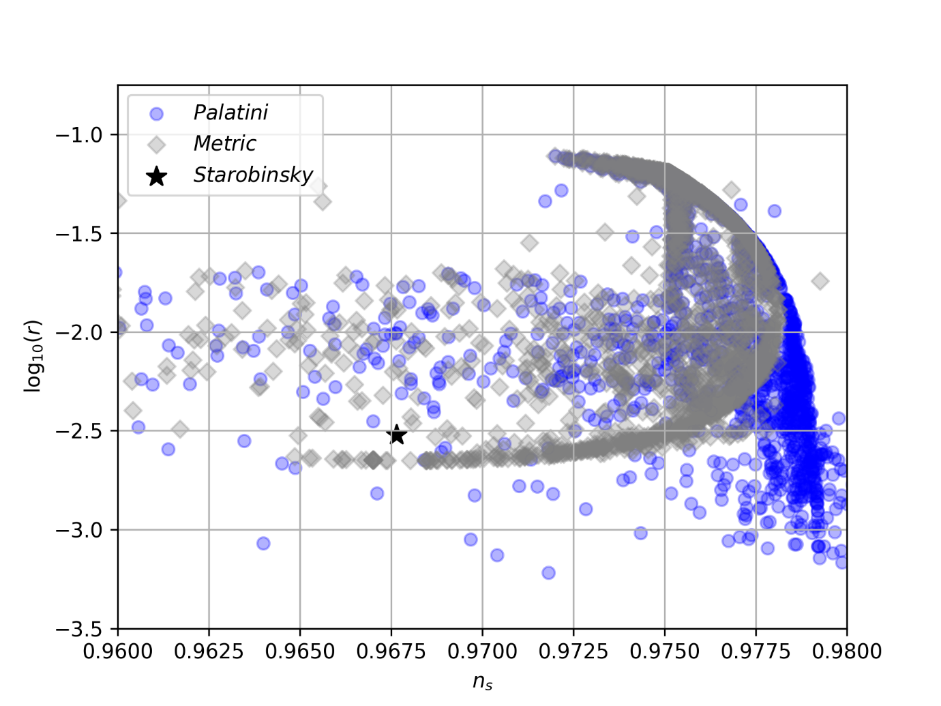}
        	\includegraphics[width=0.45\textwidth]{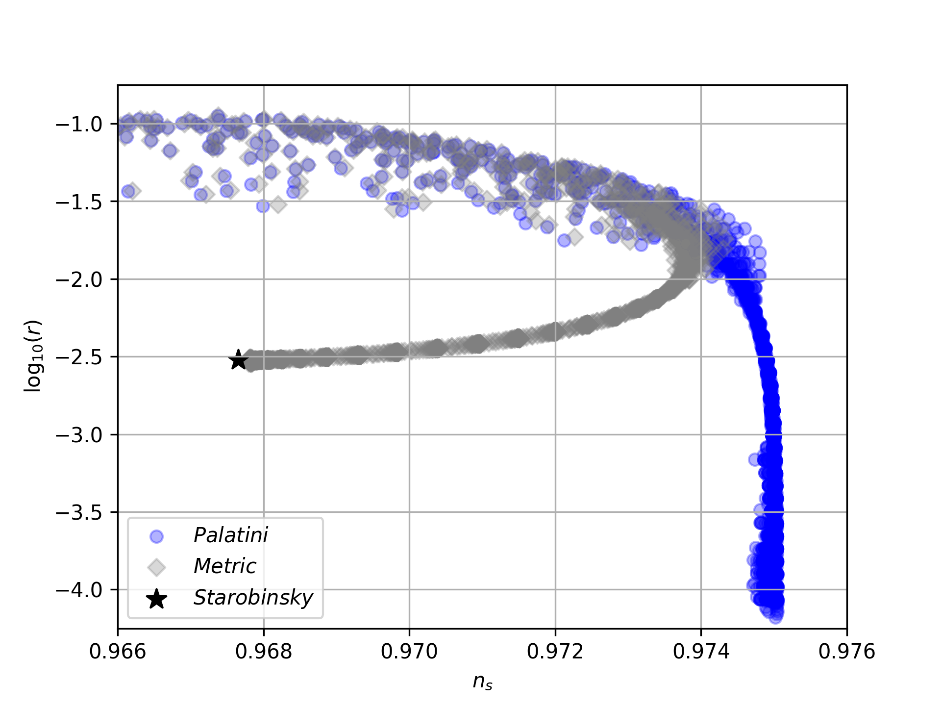}
        	\includegraphics[width=0.45\textwidth]{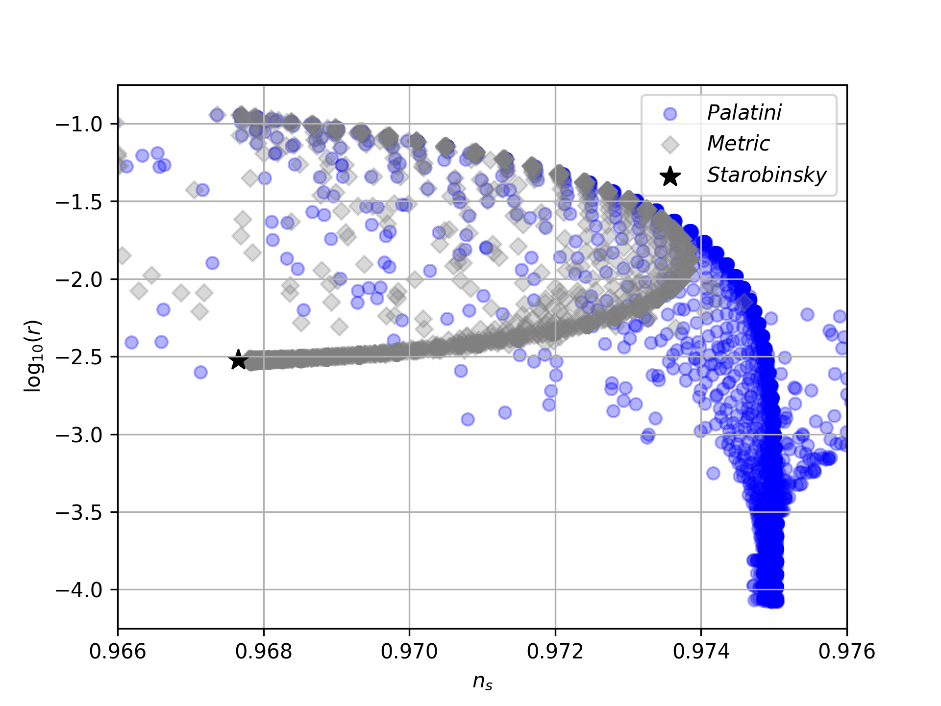}
        	\includegraphics[width=0.45\textwidth]{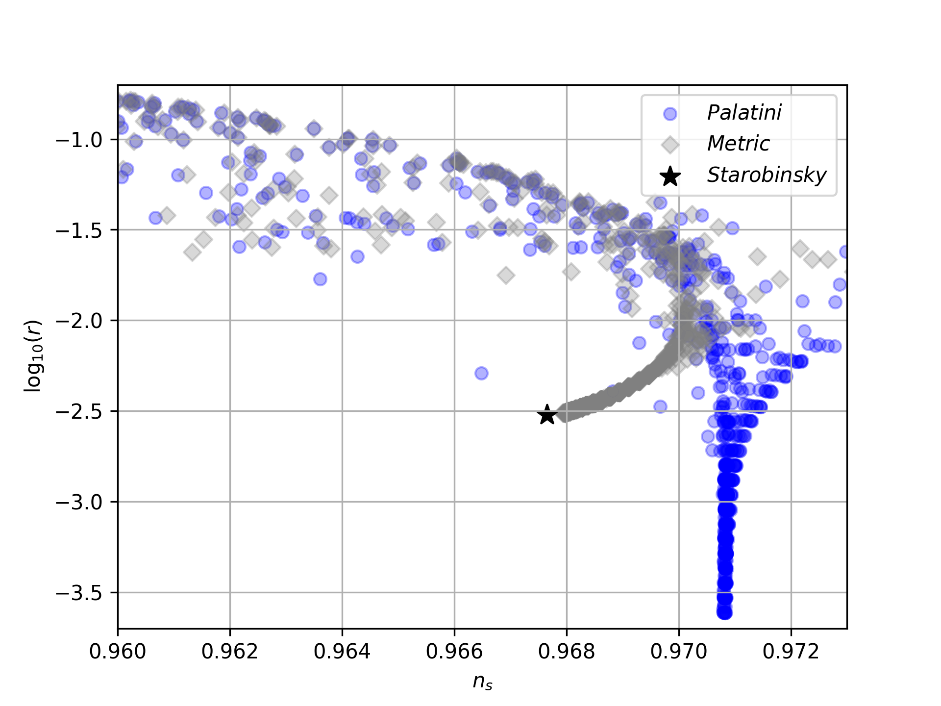}
        	\includegraphics[width=0.45\textwidth]{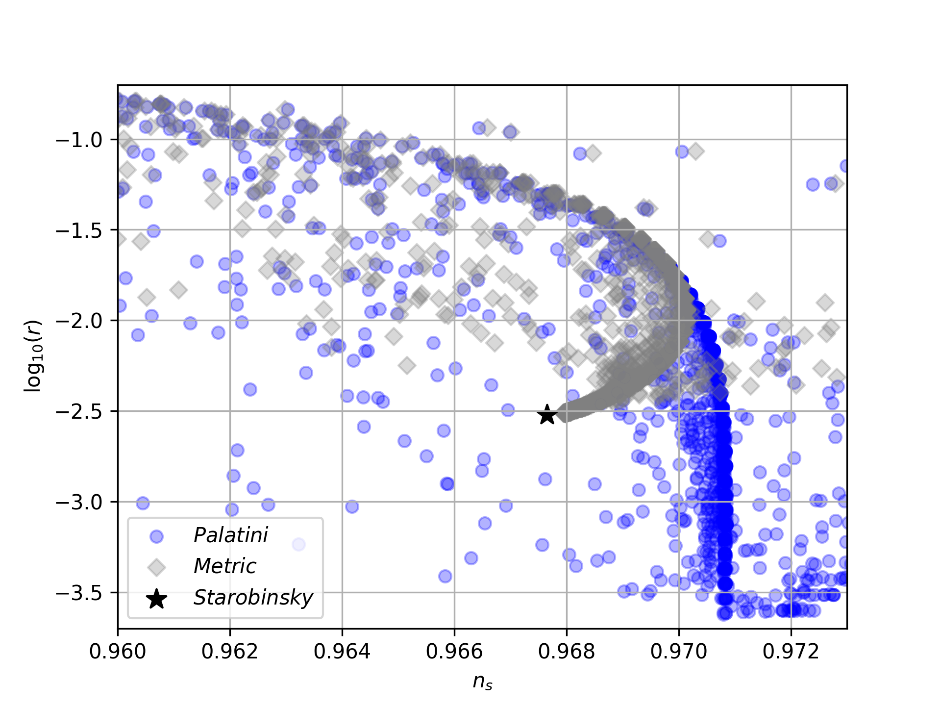}
        	\includegraphics[width=0.45\textwidth]{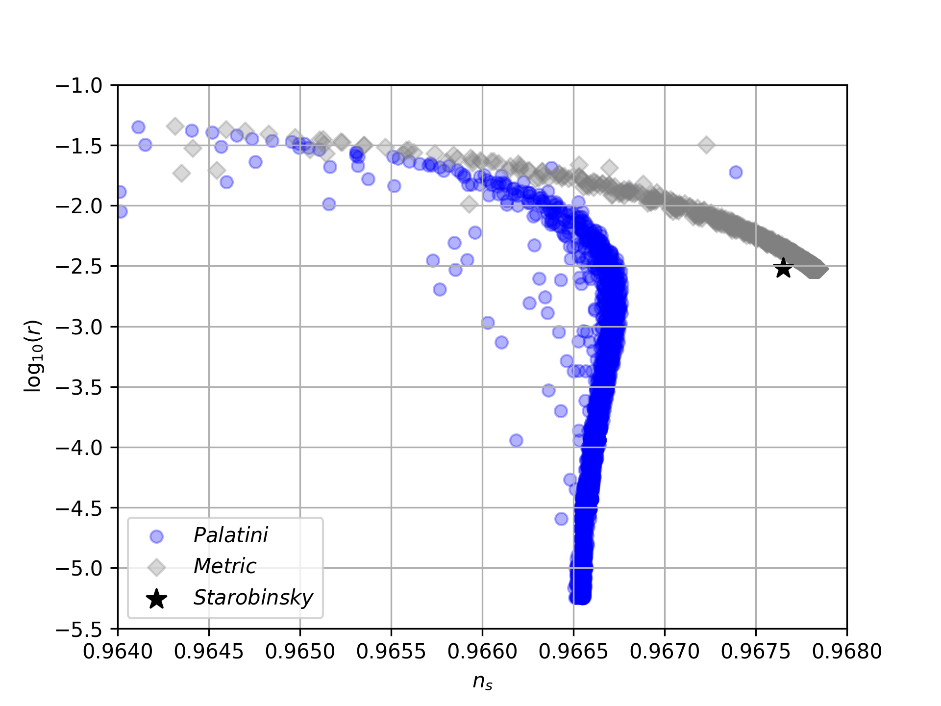}
        	\includegraphics[width=0.45\textwidth]{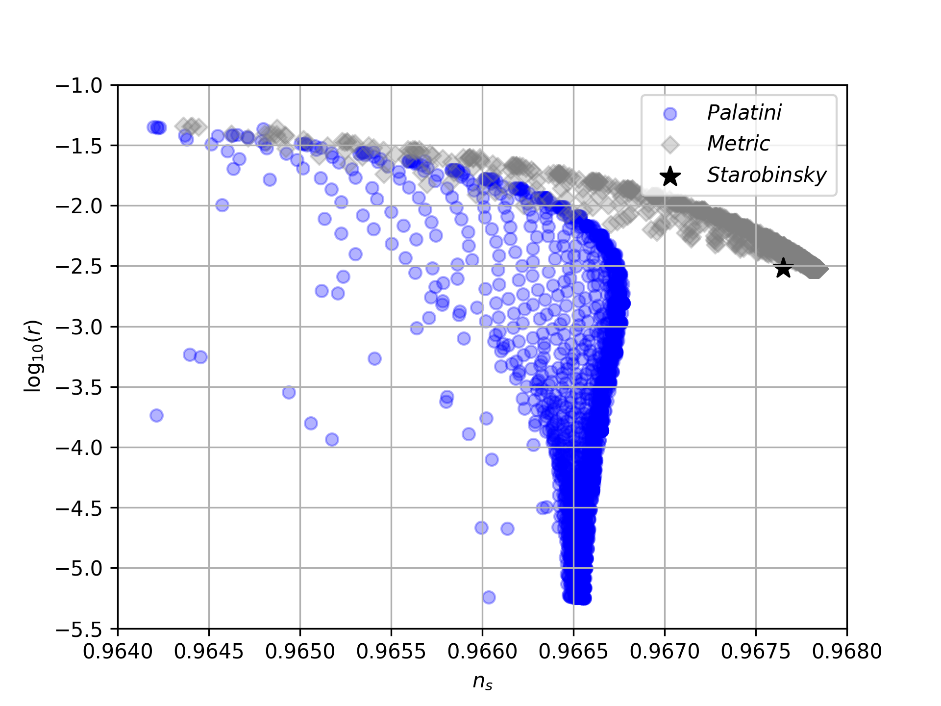}
	\caption{Predictions for $n_s$ and $r$ in metric (grey) and Palatini gravity (blue). The panels are the same as in Fig. \ref{fig:ICs}.}\label{nsr}
\end{figure}

Moving now to the observables, we study the cases for which $n=(1/2,1,3/2,2)$ in Eqs. \eqref{nmc_function} and \eqref{potential}. We show the results for $n_s$ and $r$ in Fig. \ref{nsr}. We see here a clear difference between the formulations of gravity at large values of $\xi_I$, with the results for the metric case asymptoting to those of Starobinsky inflation given in Eq.~\eqref{StaroC3} \cite{Starobinsky:1980te},
\begin{align}
\label{starobinsky_nsr}
n_s^{\text{M}} &\simeq 1 - \frac{2}{N} 
, \\
r^{\text{M}} &\simeq \frac{12}{N^2} ,
\end{align}
while those for Palatini do not. The Palatini case approaches vanishing $r$ at strong coupling, asymptoting to the single-field case \cite{Jarv:2017azx}
\begin{align}
n_s^{\text{P}} &\simeq 1 - \left( 1+\frac{n}{2} \right) \frac{1}{N} \,, \\
r^{\text{P}} &\simeq 0 \, ,
\end{align}

\begin{figure}
	\centering
	        \includegraphics[width=0.45\textwidth]{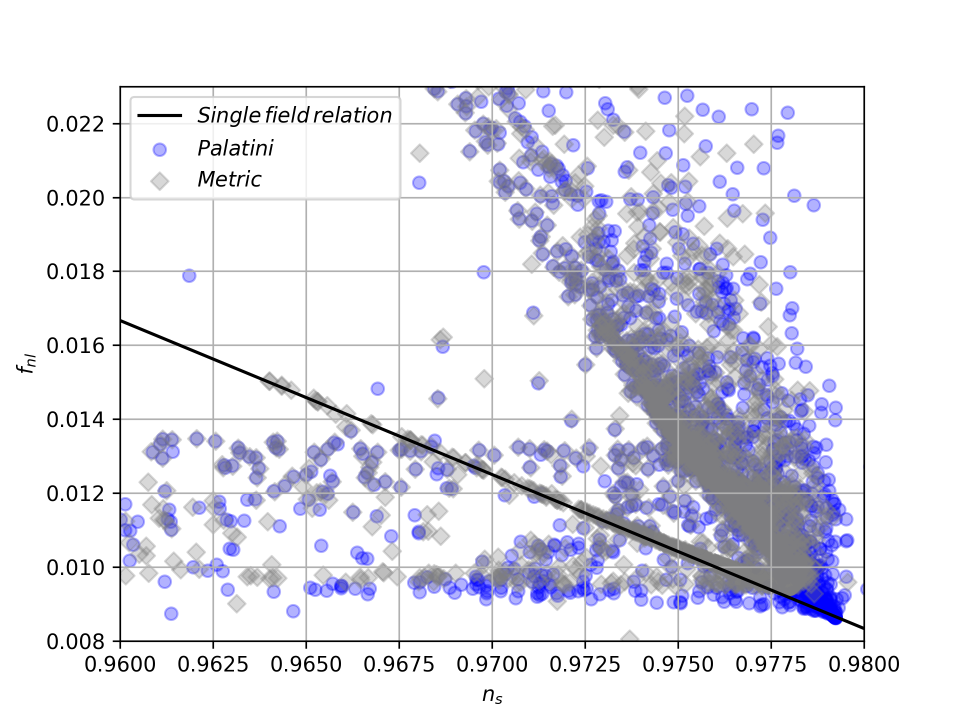}
        	\includegraphics[width=0.45\textwidth]{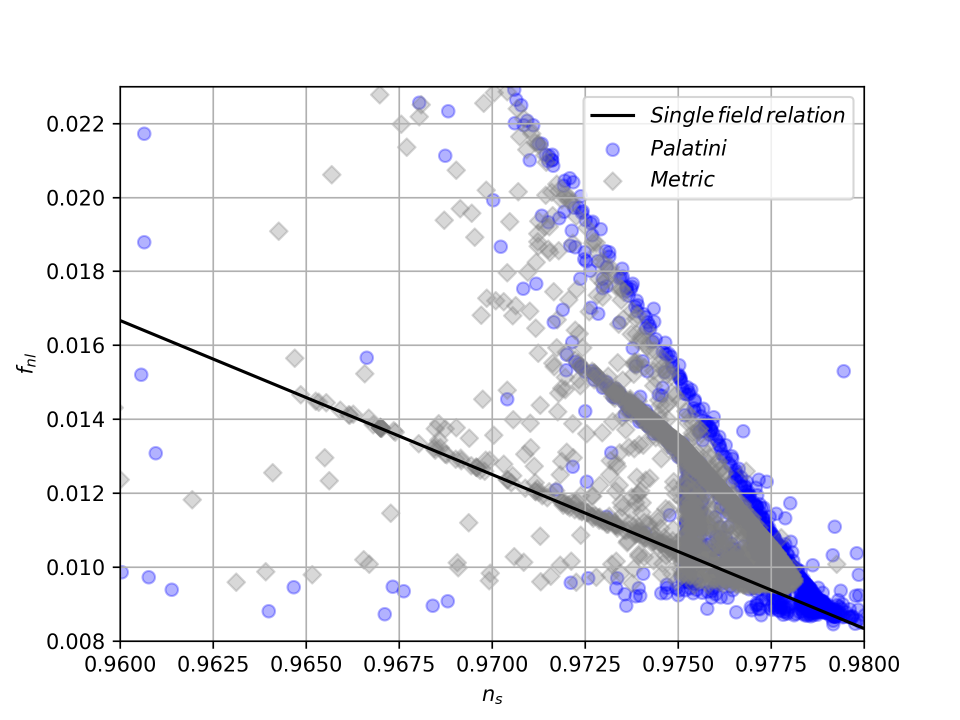}
        	\includegraphics[width=0.45\textwidth]{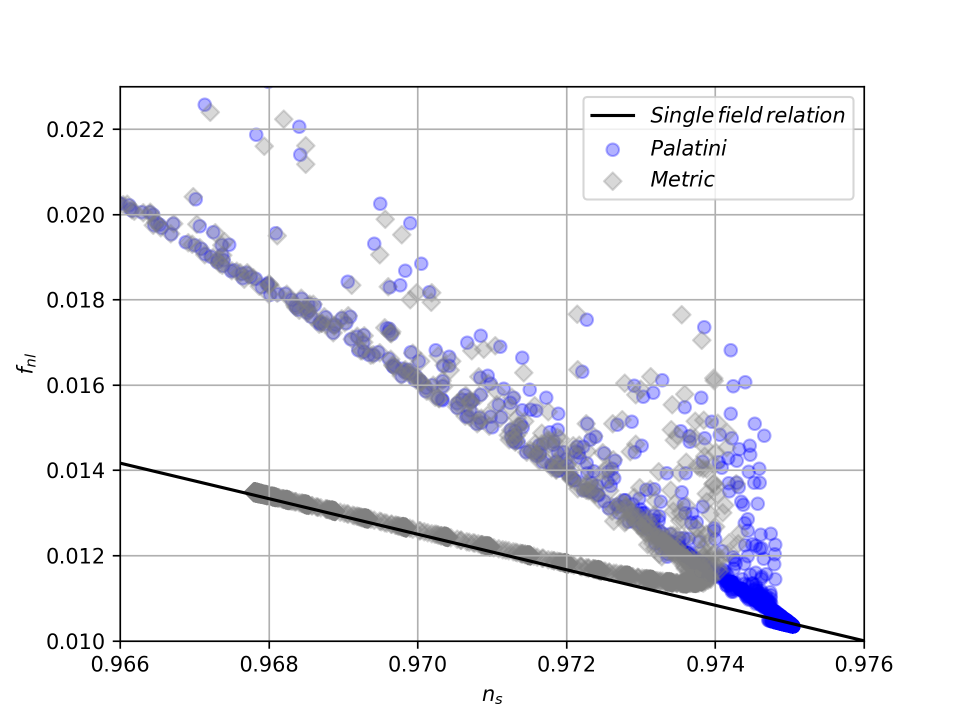}
        	\includegraphics[width=0.45\textwidth]{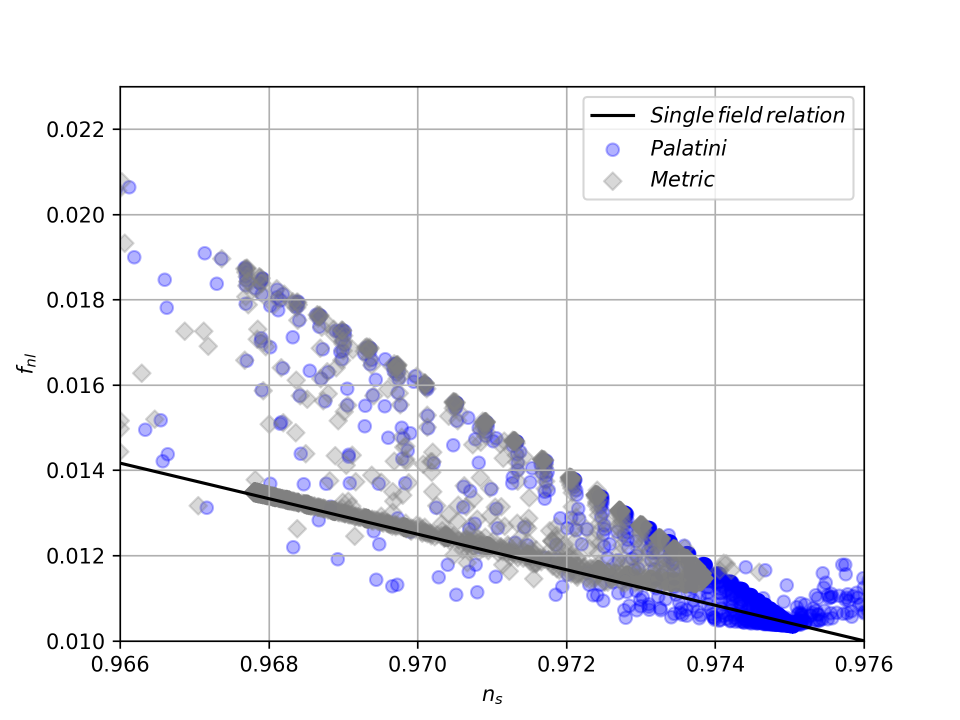}
        	\includegraphics[width=0.45\textwidth]{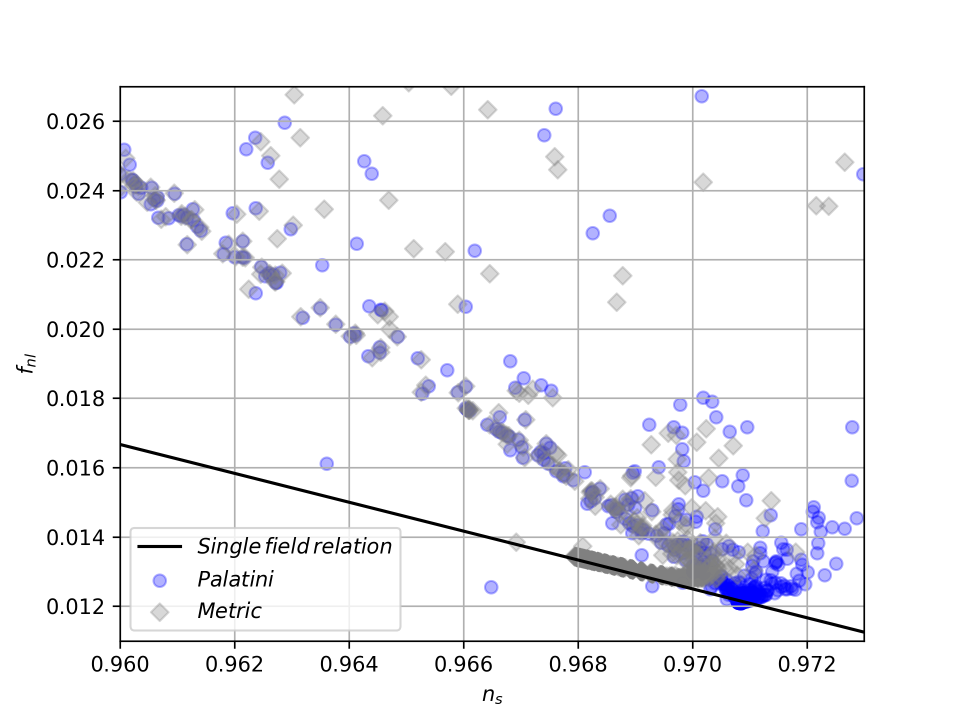}
        	\includegraphics[width=0.45\textwidth]{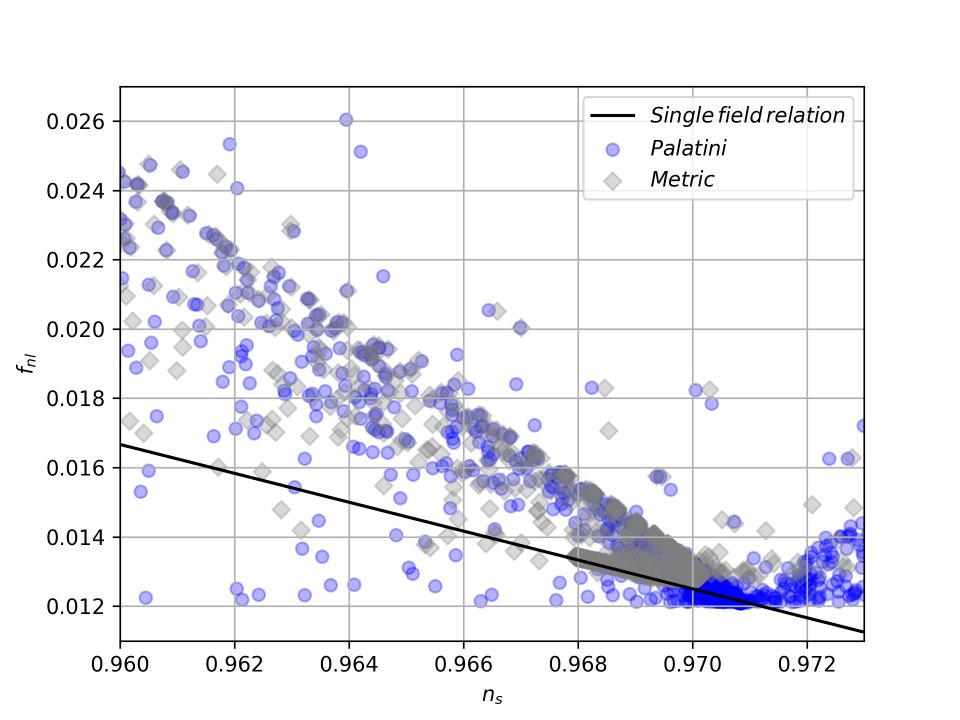}
        	\includegraphics[width=0.45\textwidth]{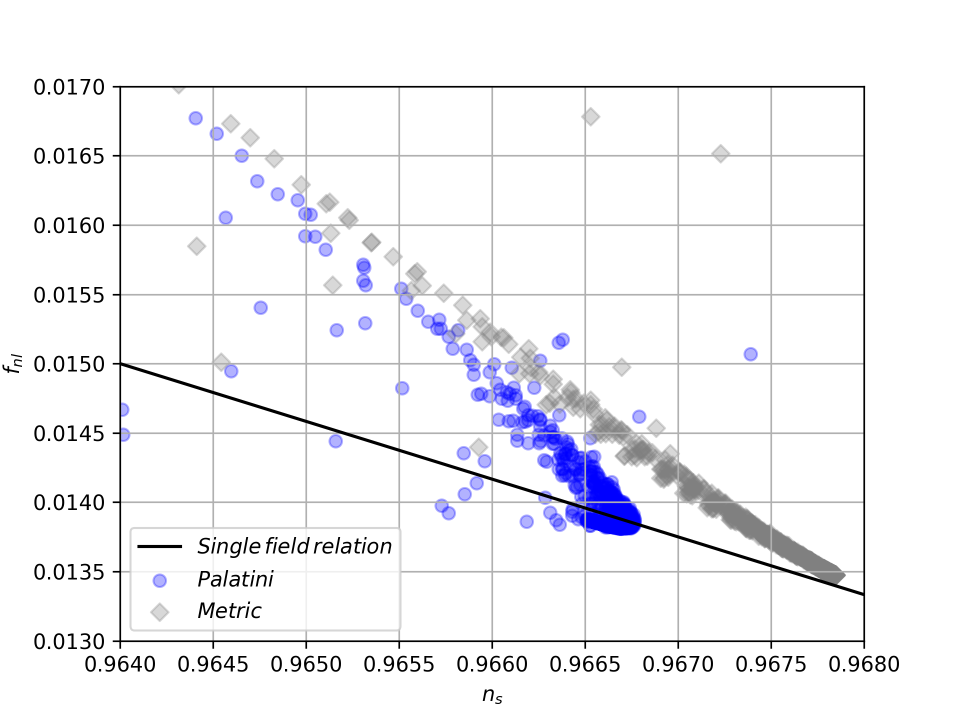}
        	\includegraphics[width=0.45\textwidth]{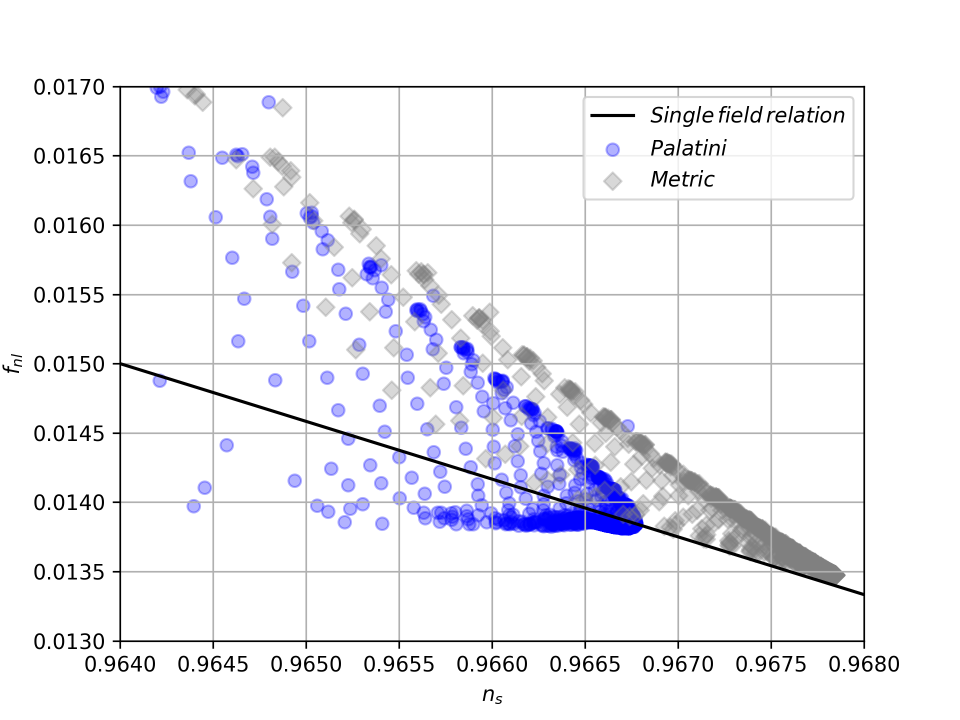}
	\caption{Predictions for $n_s$ and $f_{\text NL}$ in metric (grey) and Palatini gravity (blue). The panels are the same as in Fig. \ref{fig:ICs}.}\label{nsfnl}
\end{figure}

\begin{figure}
	\centering
	        \includegraphics[width=0.45\textwidth]{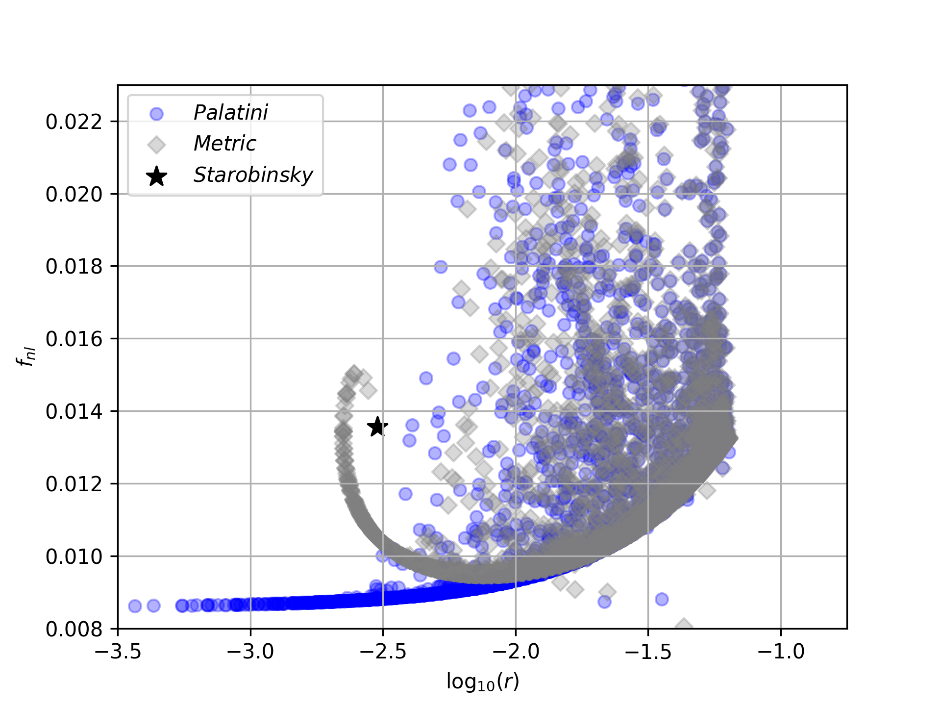}
        	\includegraphics[width=0.45\textwidth]{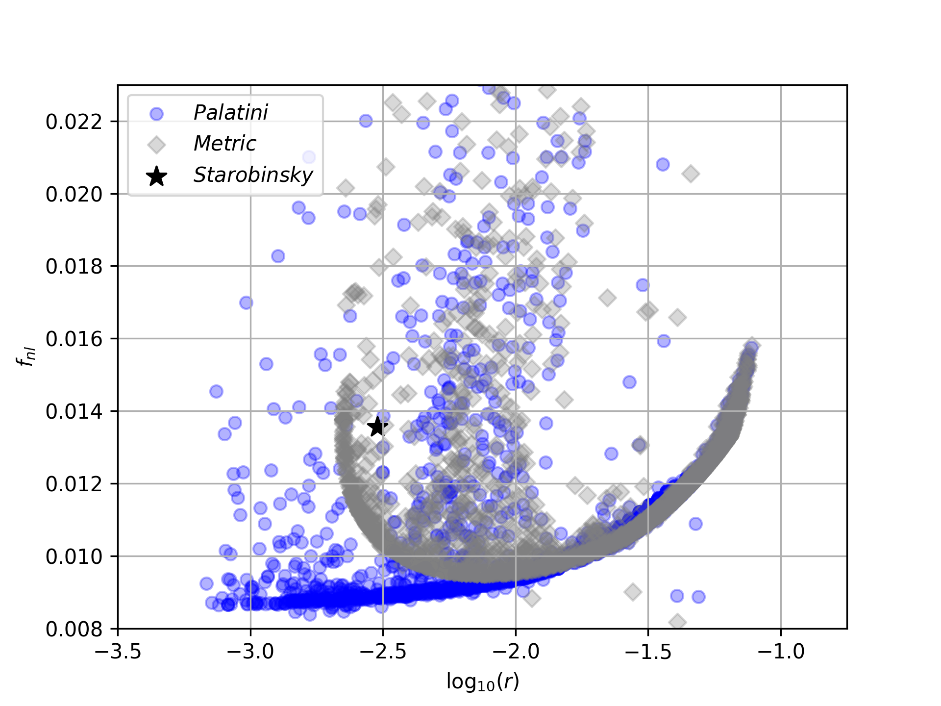}
        	\includegraphics[width=0.45\textwidth]{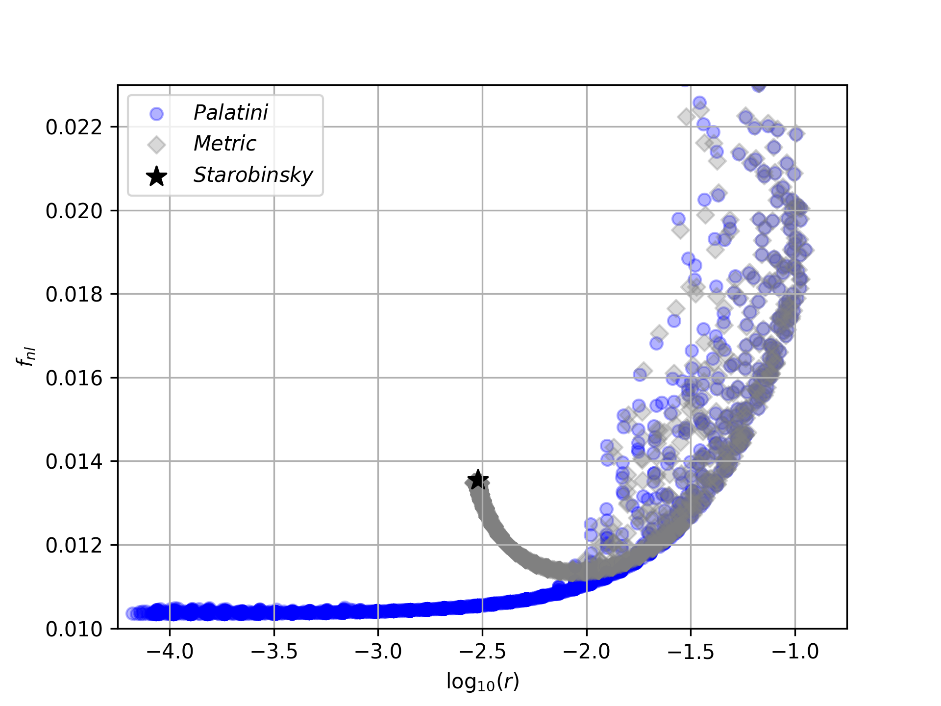}
        	\includegraphics[width=0.45\textwidth]{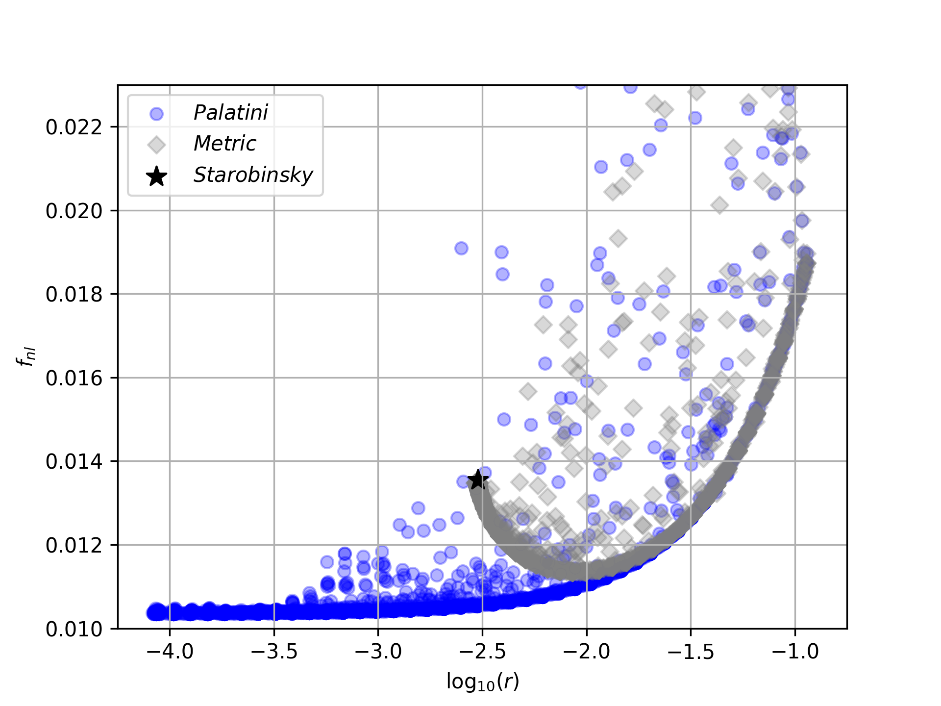}
        	\includegraphics[width=0.45\textwidth]{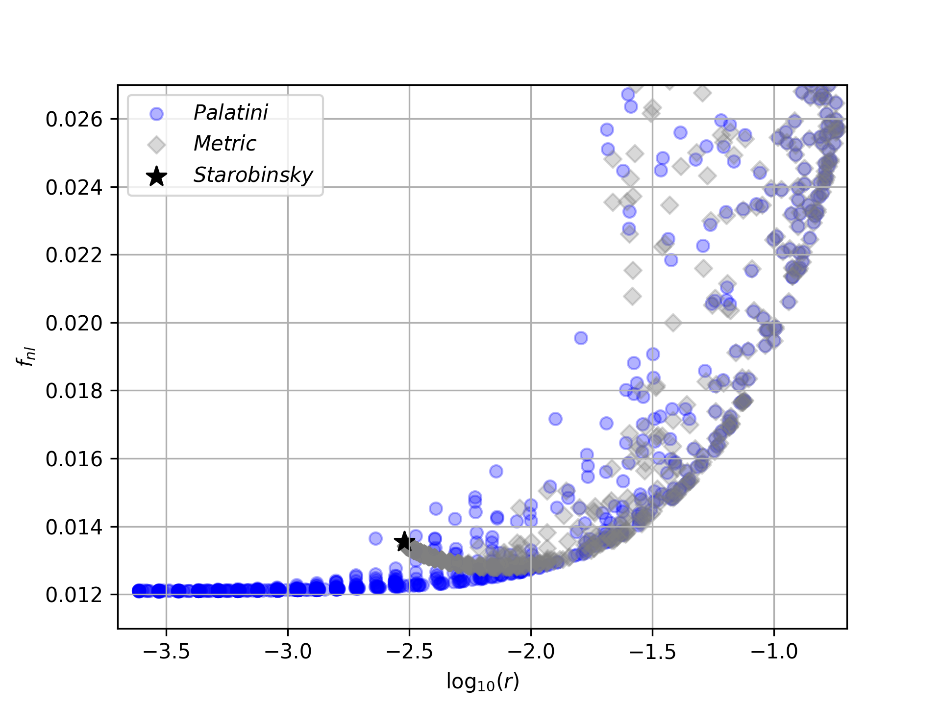}
        	\includegraphics[width=0.45\textwidth]{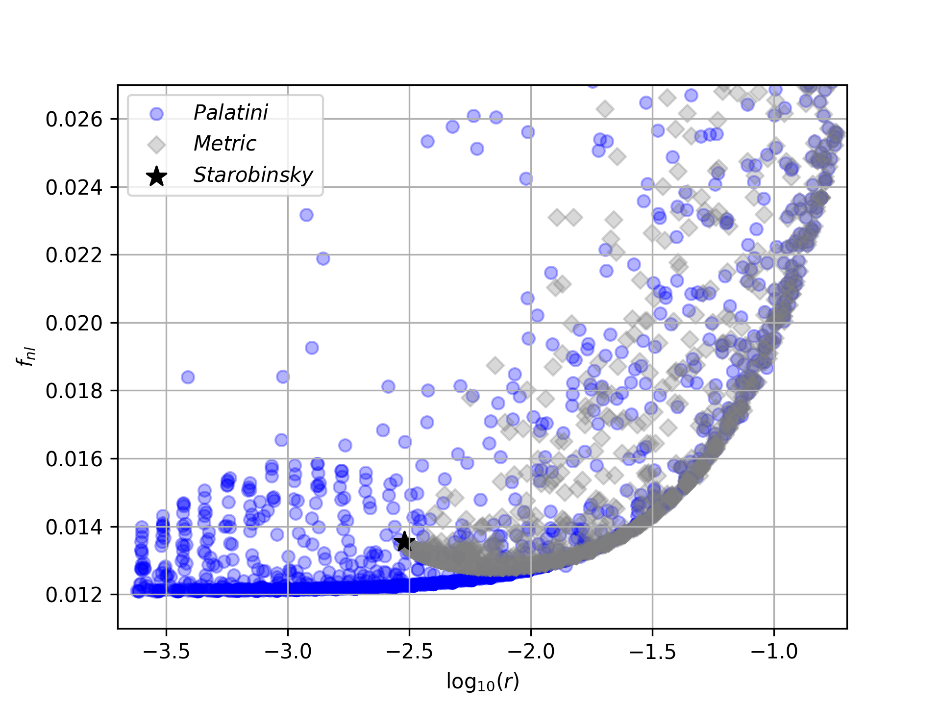}
        	\includegraphics[width=0.45\textwidth]{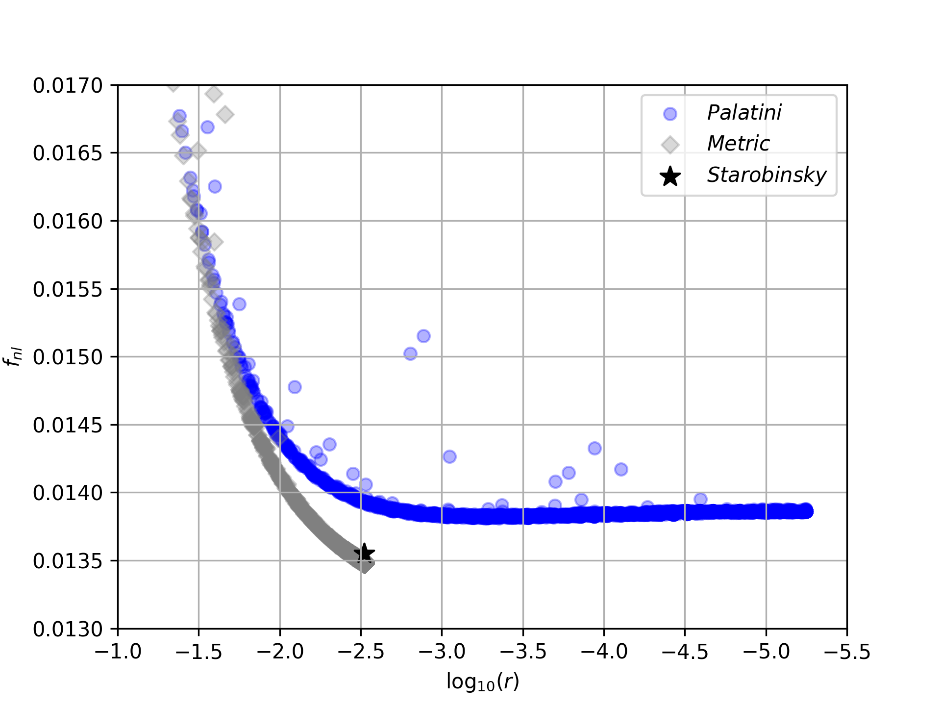}
        	\includegraphics[width=0.45\textwidth]{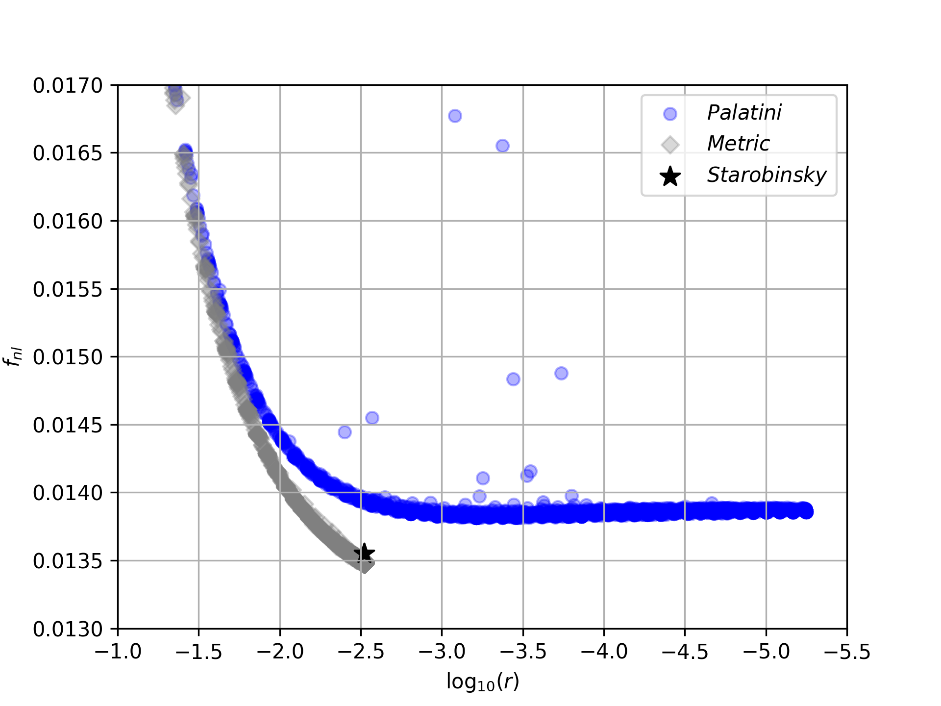}
	\caption{Predictions for $r$ and $f_{\text NL}$ in metric (grey) and Palatini gravity (blue). The panels are the same as in Fig. \ref{fig:ICs}.}\label{rfnl}
\end{figure}

However, we find that in the Palatini case the results converge to a non-zero value of $f_{\text NL}$, which is different from that of the metric case. The results are shown in Figs. \ref{nsfnl} and \ref{rfnl} along with lines corresponding to the Maldacena's consistency relation $f_{\text NL}=5/12(1-n_s)$ \cite{Maldacena:2002vr} for the single-field case\footnote{As discussed in section~\ref{Secgenfluc}, one expects Maldacena's relation to hold for squeezed configurations of the reduced bispectrum (Eq.~\eqref{maldacenaC3}), while here we are plotting the reduced bispectrum in the equilateral limit. However, in canonical single-field models in which $\epsilon \ll \eta$, which is the case for the single-field limit here, the bispectrum is very close to local and the reduced bispectrum is almost the same in all configurations. This is why our plot for $f_{\text NL}$ against $n_s$ follows so closely the Maldacena relation.}. We see that the values of $f_{\text NL}$ converge to the single-field result at strong coupling for both Palatini and metric gravity, confirming the general trend that the multi-field results mimic those of the single-field case in the strong coupling limit.

We see that all multi-field models considered in this chapter reduce to an effective single-field model at the limit of strong coupling. In the metric case this generalizes the earlier findings in the literature \cite{Kaiser:2013sna}\footnote{Outside the context of inflation, similar single-field behaviour has been found in other scenarios with non-minimally coupled multi-field models~\cite{Hohmann:2016yfd}.}, whereas in the Palatini case the results are entirely new. We elaborate on the reason for this behaviour in the next subsection.  

\subsection{Multi-field effects}
\label{fs_curvature}

Having discussed the general trends in the previous sections, we now discuss some of the effects of having multiple fields. The first effect we study is the dependence on the hierarchy between the values for the non-minimal couplings. In order to do that, we use the polar coordinates in parameter space introduced in Eq.~\eqref{polarxi} and test the evolution of the observables depending on $\theta$.

\begin{figure}
	\centering
        	\includegraphics[width=0.45\textwidth]{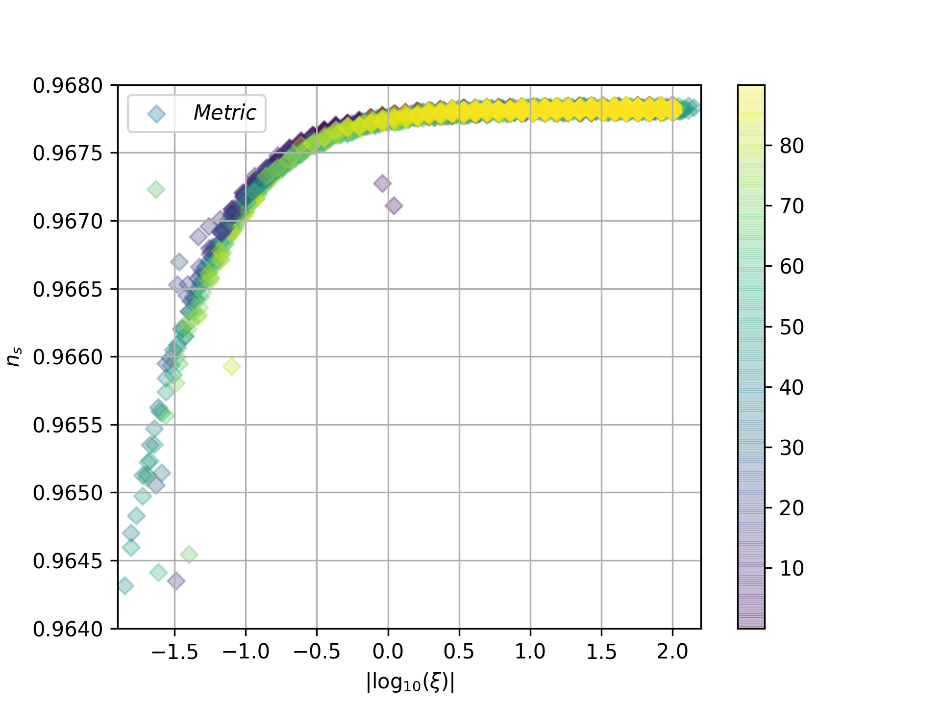}
        	\includegraphics[width=0.45\textwidth]{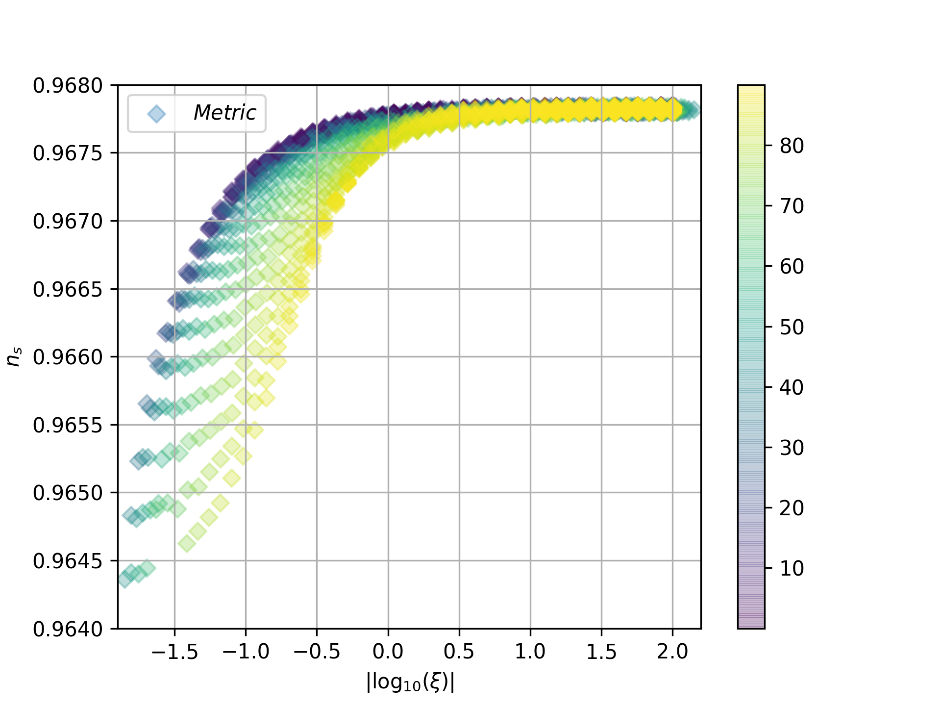}
        	\includegraphics[width=0.45\textwidth]{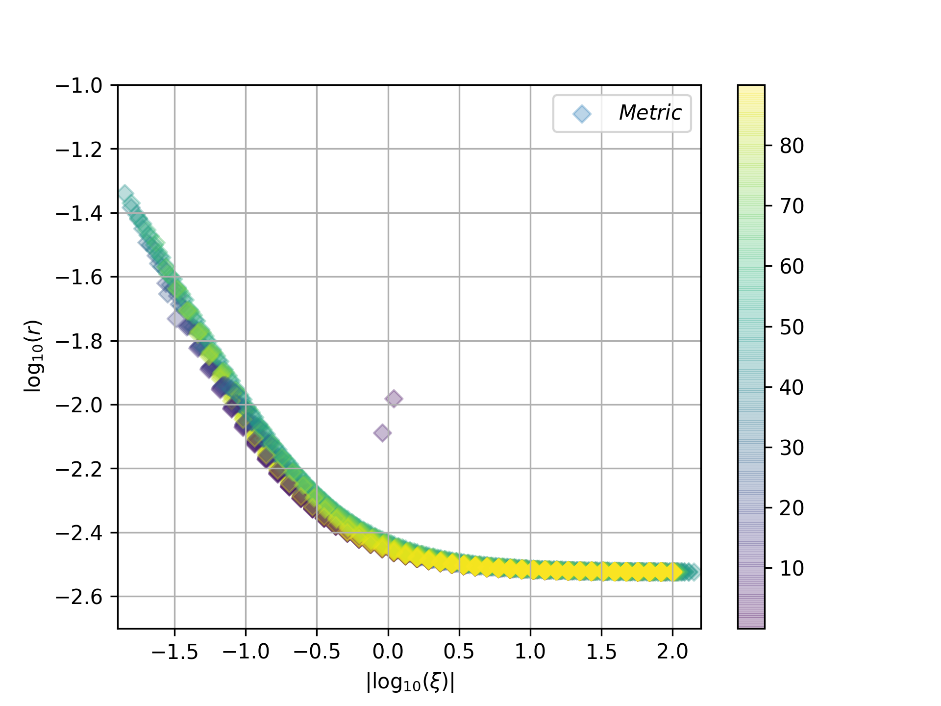}
        	\includegraphics[width=0.45\textwidth]{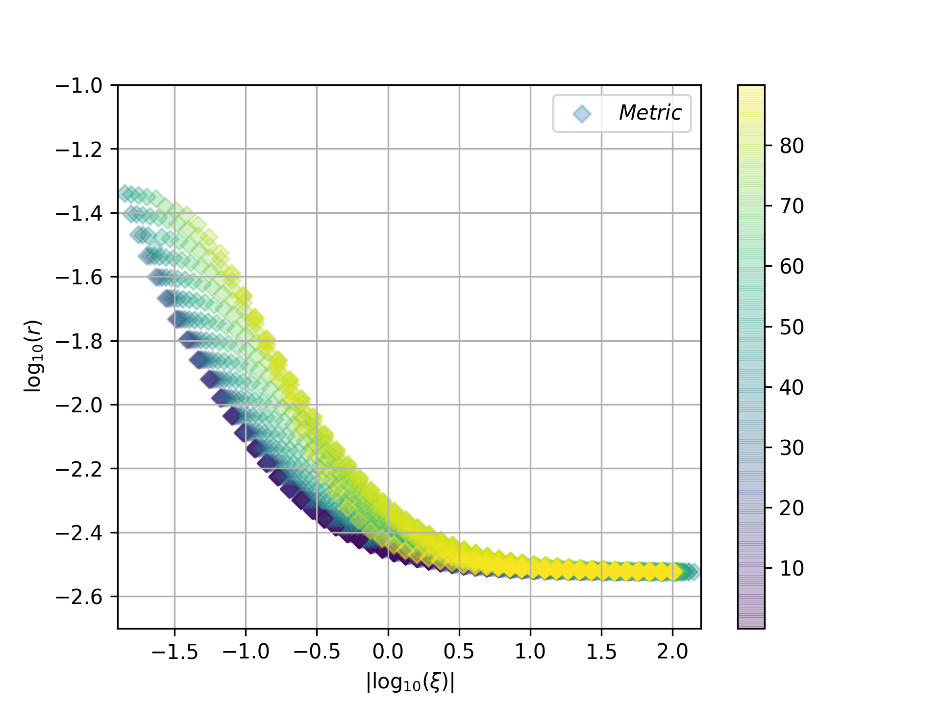}
		\includegraphics[width=0.45\textwidth]{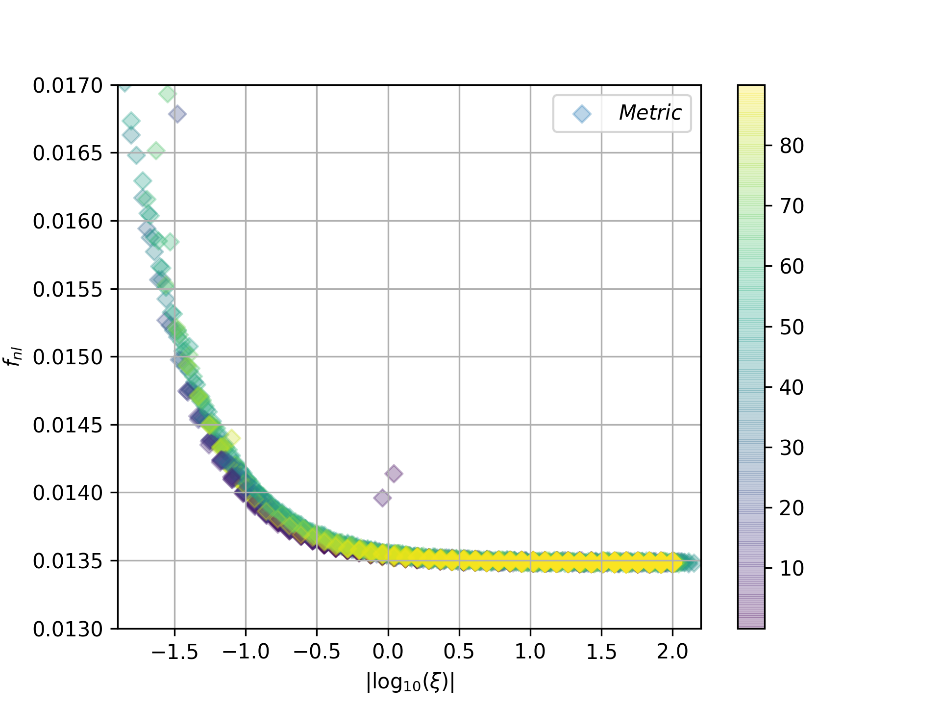}
		\includegraphics[width=0.45\textwidth]{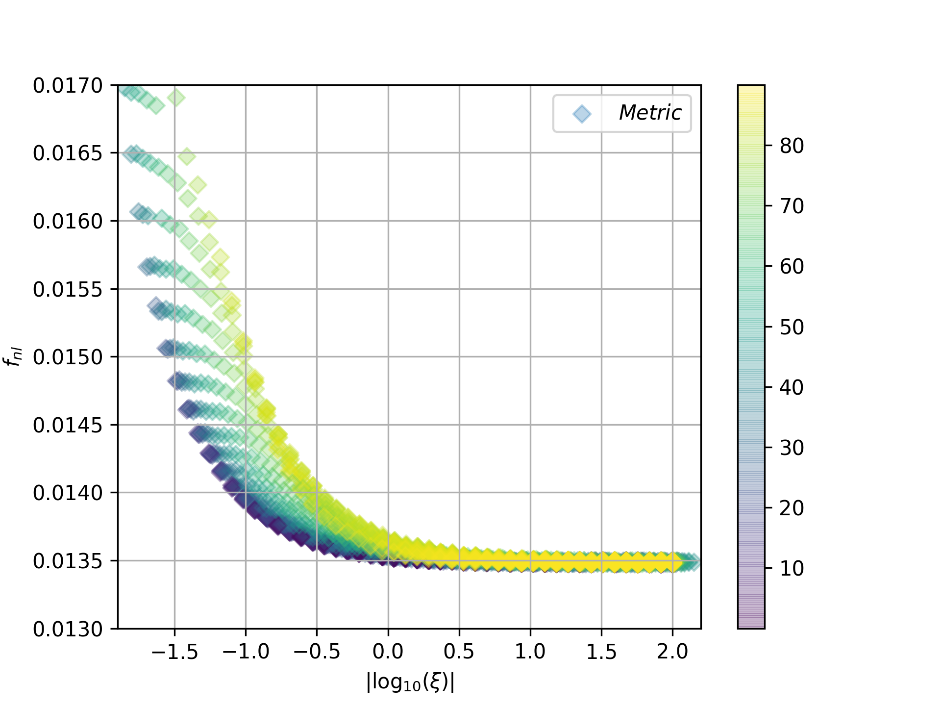}
	\caption{Predictions for $n_s$ (top) $r$ (middle) and $f_{\text NL}$ (bottom) as a function of $\xi$ along the $x$-axis and $\theta=\tan^{-1}(\xi_\sigma/\xi_\ph)$ (illustrated by the color gradient in degrees) in metric gravity for $n=2$ and for the same $\lambda_\sigma/\lambda_\ph$ ratios as in Fig. \ref{fig:ICs}.}\label{vartheta}
\end{figure}

We see in Fig.~\ref{vartheta} that the results depend crucially on the ratio of the parameters in the potential, $\lambda_I$. When the parameters for both fields are similar, the observables quickly approach a single limiting value corresponding to the single-field case, while for the larger $\lambda_I$ ratio 
the predictions are substantially broadened throughout the entire $\xi$ range, with a clear dependence on the angular parameter $\theta$. The trajectories in $(n_s,r)$ space as a function of $\xi$ are also broadened, as is also clear in Fig.~\ref{nsr}. The predictions are thus somewhat different from the single-field case for low and intermediate values of $\xi$, but converge to the same limit for sufficiently large $\xi$.

Having now analysed the dependence on both $\xi$ and $\theta$, we confirm that the results resemble the single-field case for both metric and Palatini gravity. The differences between single-field and multi-field that do arise are apparent in the spread in the results for low values of $\xi$. This spread is due to a larger dependence on the initial conditions of the fields and on the direction in $\xi_I$ parameter space. At strong coupling, all the results found asymptote to the single-field ones. This similarity may be somewhat surprising, given that in the multi-field case the field-space can be curved. We now show the reasons why this additional multi-field effect is not affecting the results at strong coupling.

\begin{figure}
	\centering
	       \includegraphics[width=0.45\textwidth]{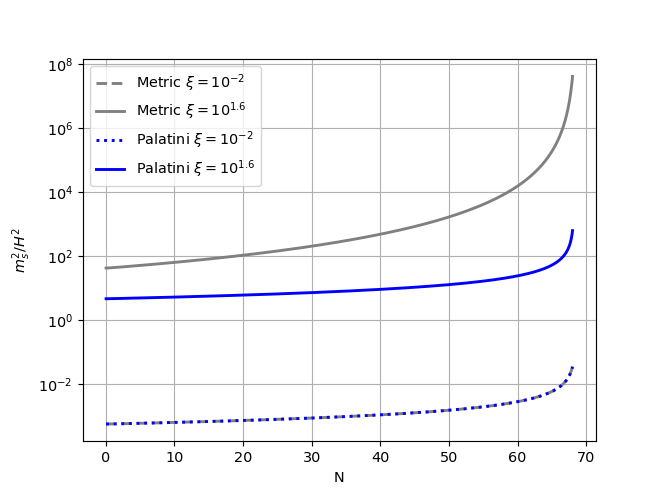}
	       \includegraphics[width=0.45\textwidth]{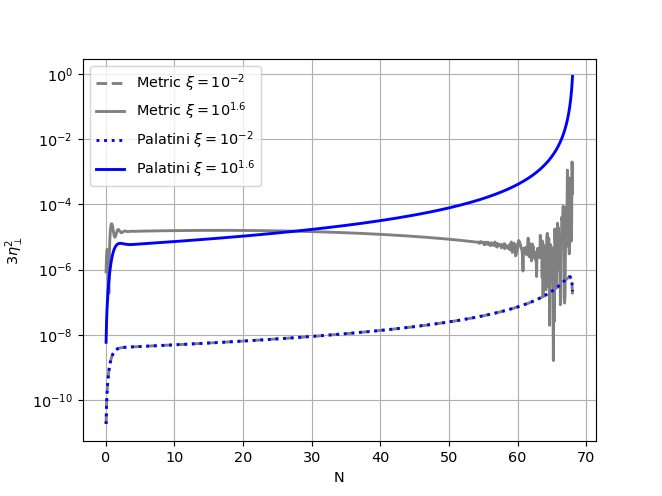}
        	\includegraphics[width=0.45\textwidth]{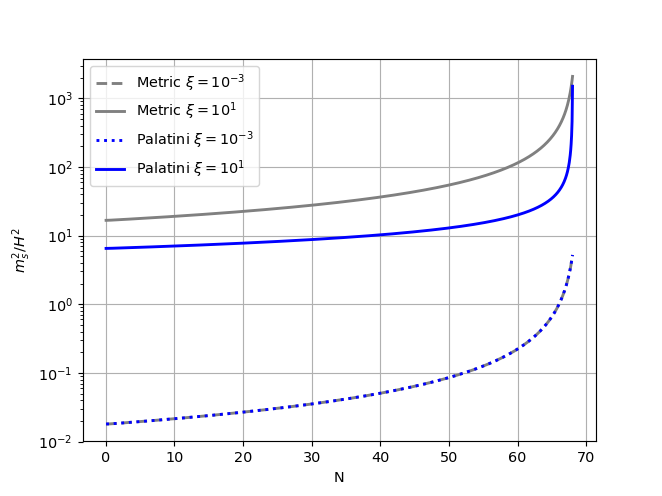}
        	\includegraphics[width=0.45\textwidth]{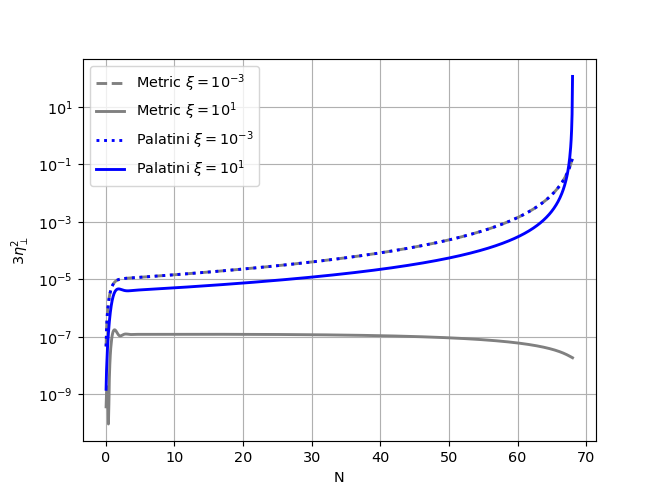}
		\includegraphics[width=0.45\textwidth]{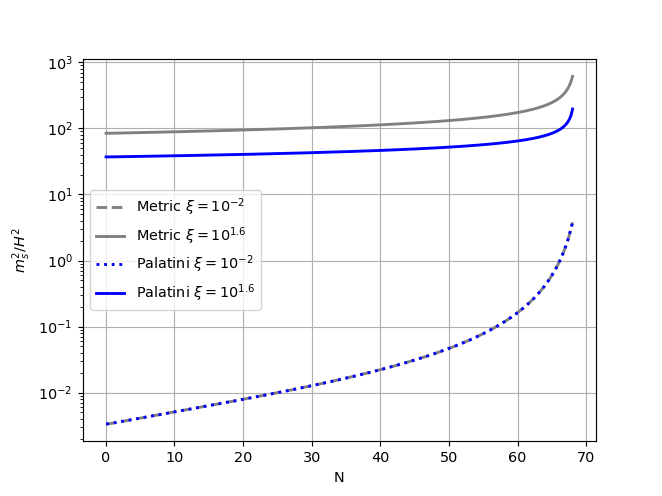}
	       \includegraphics[width=0.45\textwidth]{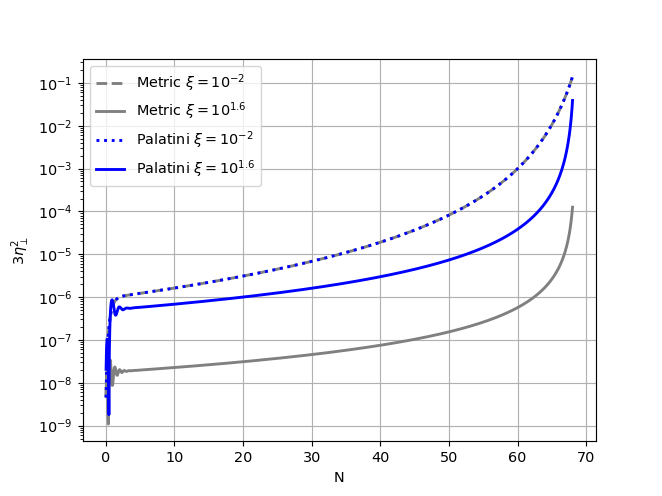}
        	\includegraphics[width=0.45\textwidth]{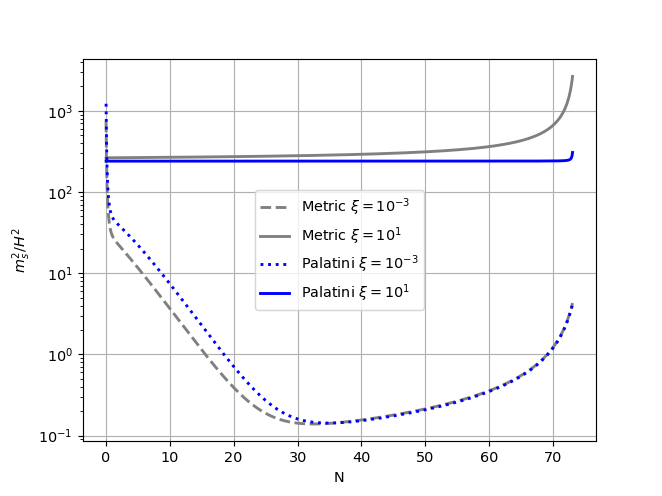}
        	\includegraphics[width=0.45\textwidth]{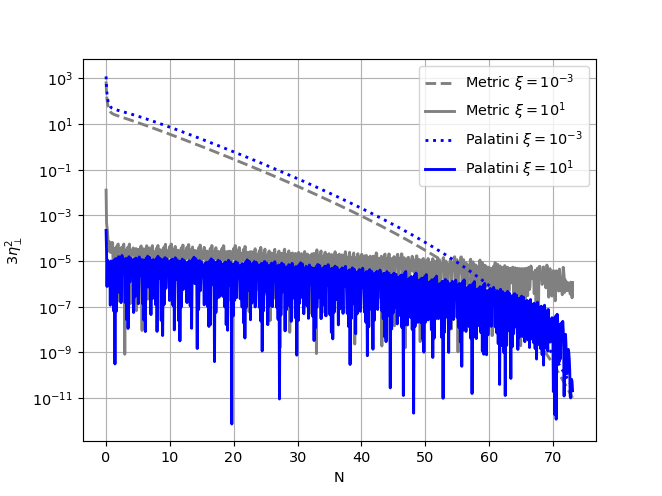}
	\caption{Evolution of the effective mass, $m_s^2$, normalized to $H^2$ and bending parameter $\eta_\perp^2$ for metric (grey) and Palatini gravity (blue), $n=(1/2,1,3/2,2)$ from top to bottom. The dashed lines represent a sample with a small magnitude of the coupling parameters $\xi$ whereas the solid lines represents one with a large coupling. }\label{fig:EfMs}
\end{figure}

We first note that field-space curvature, $R^{\ A}_{\text{fs}\,BCD}$, does not directly affect the evolution of the field fluctuations in the inflationary direction. This is because the field-space Riemann tensor appears in the effective mass matrix of the fluctuations, $m^I_L$, in the following term~\cite{DavidJohn2}
\begin{equation}
m^I_L\supset R^{\ I}_{\text{fs}\,JKL} \dot{\ph}^J\dot{\ph}^K\,.
\end{equation}
To obtain the term relevant for the fluctuations in the inflationary direction, one must multiply $m^I_L$ with $\dot{\ph}^L$, which always results in zero for the term shown above, given the symmetries of the Riemann tensor. There is, however, an effect on the entropy perturbations, as they are sensitive to the perpendicular projection of the effective mass matrix. For the two-field case, the total effective mass for those fluctuations is given by the generalization of Eq.~\eqref{EffmassC3} to the curved case,
\begin{equation}
\label{eff-mass_eq}
\frac{m_s^2}{H^2}= \frac{\n_s\p_s U}{H^2}+3\eta_\perp^2+\epsilon R_{\text{fs}}\,,
\end{equation}
in which $\eta_\perp=\p_s U/H\dot\ph$ is a measure of the bending of the trajectory, proportional to $\theta'$ defined in Eq.~\eqref{thetapC3}, $\dot\ph=\sqrt{G_{IJ}\dot{\ph}^I\dot{\ph}^J}$, $s$ is the field coordinate in the entropic direction --- the direction perpendicular to $\dot{\ph}^I$ --- and $R_{\text{fs}}$ is the Ricci scalar of the field-space manifold. The effect of the curvature is somewhat less relevant if $R_{\text{fs}}$ is positive, as it simply contributes to a smaller amplitude of the  entropy perturbations. If it is negative, however, it reduces the effective mass and may even render it tachyonic should it be large enough~\cite{Renaux-Petel:2015mga}, thus dangerously enhancing the entropy fluctuations. Our numerical results seem to indicate that this never occurs, given their similarity with the single-field results, for which the curvature is not present. We can verify this by checking whether the condition $m_s^2>0$ is always verified in our numerical results. We can see this in Fig.~\ref{fig:EfMs}, in which we show that the effective mass is always positive for all values of $n$ studied above. When $\xi$ is large, the effective mass is also large, with the dominant contribution coming from the first term on the right-hand side of Eq.~\eqref{eff-mass_eq}, the Hessian of the potential. Specifically, the effective mass values calculated in the metric and Palatini cases are equivalent for small $\xi$ and consequently the resulting observables ($n_s$, $r$ and $f_{\text NL}$) are affected in similar ways in both cases. Where the observables deviate between the two cases, i.e. for large $\xi$, the effective masses also deviate with an overall larger effective mass in the metric case.

The evolution of the entropy modes is independent of the adiabatic modes on large scales, and thus only depends on the effective mass. They can, however, source curvature perturbations \cite{Turzynski:2014tza,Wands:2000dp,Carrilho:2015cma}. To see this, we rewrite Eq.~\eqref{z1evo} as in Chapter~\ref{Ch_SMC}, using now the variables introduced in this chapter:
\be
\label{dotzeta}
\dot\zeta\approx \sqrt{2}H\eta_\perp\frac{H}{M_{\text P}\sqrt{\epsilon}}\frac{\delta S}{H}\,,
\ee
Thus, following the same arguments as in Section~\ref{multfieldC3}, we conclude that we can recover the single-field results if $\eta_\perp$ is sufficiently small. In more detail, those arguments go as follows. We can estimate the entropy fluctuations via their variance $\delta S \sim H^2/m_s$.\footnote{This formula arises from the calculation of the two-point function of $\delta S$ in de Sitter space by assuming it is a spectator field. This has been done, for example, in Chapter~\ref{Ch_quench}, and the result, while not quoted explicitly, can be seen in Eq.~\eqref{massunrenorm}, in which the renormalized part of the variance of $\rchi$ is 
\begin{equation}
i G_{\text{Ren}}(x,x)=a^2\frac{3 H^4}{8\pi^2 m^2}\,.
\end{equation}
Since the relation between $\delta S$ and $\rchi$ can be deduced from Eq.~\eqref{phichirel} as $\delta S=a^{-1}\rchi$, we find $\delta S\sim a^{-1}\sqrt{iG_{\text{Ren}}}\sim H^2/m$.} Furthermore, we note that ${H}/\left({M_{\text P}\sqrt{\epsilon}}\right)$ is approximately the value of $\zeta$ at horizon crossing, $\zeta_*$, and that the typical time scale associated to its variation is $H$, making $H\zeta_*$ the natural size of $\dot\zeta$, should it vary considerably. Given these arguments, we can rewrite Eq.~\eqref{dotzeta} as
\be
\label{dotzeta2}
\frac{\dot\zeta}{H\zeta_*} \sim \frac{\eta_\perp H}{m_s}\,,
\ee
and conclude that if the right-hand side of Eq.~\eqref{dotzeta2} is much smaller than 1, the evolution of $\zeta$ is negligible. Therefore, to determine the importance of entropy fluctuations in the evolution of adiabatic ones, we must only calculate $\eta_\perp H/m_s$. In the right panel of Fig.~\ref{fig:EfMs}, we show the size of $\eta_\perp^2$ during inflation. Comparison with the effective mass shown in the left panel demonstrates that the bending parameter is sub-dominant relative to the effective mass. For example, for the $n=1$ metric case the ratio $\eta_\perp^2 H^2/m^2_s\sim 10^{-3}$ when $\xi$ is small and for large $\xi$, $\eta_\perp^2 H^2/m^2_s\sim 10^{-8}$, demonstrating that the entropy fluctuations are negligible at strong coupling. Comparing the metric and Palatini case for small $\xi$ we see that the results for the evolution of $\eta_\perp$ are the same. For large $\xi$, the evolutions diverge and $\eta_\perp$ in the metric case decays, while it grows in the Palatini case.

\subsection{Extension to scenarios with higher number of fields}
\label{3fieldcase}

We have also extended our calculations to the three-field case for $n_s$ and $r$. We found that the results resemble those for the two-field case, converging to the same limit in the strong coupling approximation for both metric and Palatini gravity. The main difference is the spread in observable space, which is substantially larger than in the two-field case. This is a consequence of the increased number of possible background field trajectories that result in successful inflation in higher field-space dimensions as well as the larger number of free parameters. This can affect the ability of distinguishing between different models, with some results for the Palatini model giving the same observables as those for the metric case, even at strong coupling for the latter. The strongly coupled Palatini case is still distinctive, given its very low tensor-to-scalar ratio prediction.

With an even larger number of fields, these predictions are expected to broaden further, but may ultimately converge again, in a statistical sense, 
as such a behaviour has been demonstrated in other scenarios with random potentials and very large numbers of fields~\cite{Aazami:2005jf,Easther:2005zr,Easther:2013rva,Dias:2016slx,Dias:2017gva,Bjorkmo:2017nzd}.
%%%%%%%%%%%%%%%%%%%%%%%%%%%%%%%%%%%%%%%%%%%%%%%%%%%%%%%%%%%%%%%%%%%%%%%%%%%%%%%%%%%%%%%%%%

\section{Conclusions}
\label{conclusions}

We studied multi-field inflation in scenarios where the fields are coupled non-minimally to gravity via $\xi_I(\ph^I)^n g^{\mu\nu}R_{\mu\nu}$. We concentrated on the so-called $\alpha$-attractor models with the potential $V=\lambda_I^{(2n)}  M_{\text P}^{4-2n} (\ph^I)^{2n} $ in two formulations of gravity: in the usual metric case where $R_{\mu\nu}=R_{\mu\nu}(g_{\mu\nu})$, and in the Palatini formulation where also the connection $\Gamma$ and hence also $R_{\mu\nu}=R_{\mu\nu}(\Gamma)$ are independent variables. 

As the main result, we showed that the curvature of the field-space in the Einstein frame has no influence on the inflationary dynamics at the limit of large $\xi_I$, and one effectively retains the single-field case  regardless of the underlying theory of gravity. In the metric case this means that multi-field models approach the single-field $\alpha$-attractor limit, whereas in the Palatini case the attractor behaviour is lost also in the case of multi-field inflation.

A point must be made here about the differences in the phenomenology being due to the distinct formulations of gravity: metric vs. Palatini. We note that if one considered a scenario in which the Jordan frame action already included non-canonical kinetic terms of a specific kind, one could construct models with the same phenomenology as in the present cases, while still working only in the metric case. For example, the models that we consider in the Palatini formalism are equivalent to non-canonical scalar-tensor theories in the metric formalism~\cite{Sotiriou:2006hs}. Had we started initially in the Einstein frame, we could also have used just the metric formalism. Given these arguments, one could suggest that our emphasis on distinguishing theories of gravity is unrealistic and that we are just testing different models of inflation. While this is correct, we argue that the models we explore are simpler when written in terms of the Palatini formalism --- with simple kinetic terms, potentials and non-minimal couplings in the Jordan frame --- than the equivalent model would be, in metric gravity, but with non-standard kinetic terms. Furthermore, it should be noted that non-minimal couplings to gravity should be seen not as an \emph{ad-hoc} addition to inflationary models but as a generic requirement for the consistency of a theory, since they are always generated by quantum corrections in a curved space-time. It is because of these reasons that one can say that the differences observed between the cases which we call `metric' and `Palatini' are indeed in the underlying theory of gravity, i.e. whether the space-time connection was determined by the metric only, or both the metric and the inflaton field(s). This study therefore reveals an interesting subtlety in a broad class of models where the scalar potential is multidimensional and the fields are non-minimally coupled to gravity.

Alternatively, one can view this work as a more detailed way to answer the question `What are the predictions of a given model of inflation?'. As shown in this chapter, predictions clearly depend on the choice of the gravitational degrees of freedom, even though usually such a choice is not considered to be part of models of inflation. It is therefore important to investigate all possibilities concerning the physics at high energies, as one cannot distinguish between the metric and Palatini formalisms at late times. Detailed studies of non-minimally coupled models are therefore interesting not only from the inflationary point of view, but also because they may provide for a way to distinguish between different formulations of gravity.

% % % % % % % % % % % % % % % % % % % % % % % % % % % % 
% chapter.tex - Ian Huston
% Sample chapter layout
% % % % % % % % % % % % % % % % % % % % % % % % % % % % 
% Redefine CVSRevision for this section. 
% If you don't want to use CVS tags comment out this line
\renewcommand{\CVSrevision}{\version$Id: chapter.tex,v 1.3 2009/12/17 18:16:48 ith Exp $}

% % % % % % % % % % % % % % % % % % % % % % % % % % % % % % % % 
% =========================================================== %
% % % % % % % % % % % % % % % % % % % % % % % % % % % % % % % % 
\chapter{Discussion and Outlook}
\label{Ch_conclusions}
% % % % % % % % % % % % % % % % % % % % % % % % % % % % % % % % 
% =========================================================== %
% % % % % % % % % % % % % % % % % % % % % % % % % % % % % % % % 

In this thesis, we have investigated several instances in which non-linear dynamics affects the evolution of the Universe. We focused particularly on the early Universe and discovered a new version of the curvature perturbation that is conserved non-linearly; showed new mixed modes generated at second order; calculated the effects of quenches in the early Universe and demonstrated the equivalence between single-field and multi-field attractor models of inflation. 
In spite of the emphasis of the thesis being on the early Universe, many of the techniques used here can be applied in far more general settings in cosmology.

We began our review of cosmology in Chapter~\ref{Ch_CPT}, by describing the main aspects of general relativity. We gave equations for the evolution of the geometry as well as that of matter, including different ways to describe the matter content. We then reviewed relativistic perturbation theory in detail, accounting for the gauge issue and establishing general formulas for the later parts of the thesis. Finally, we applied those results to the case of interest --- perturbations around an FLRW spacetime. We gave the equations for general gauge transformations up to second order in fluctuations for all quantities of interest. We also wrote down the perturbed field equations up to the same order, which were later applied in all other chapters of this thesis.

In Chapter~\ref{Ch_SMC}, we described the evolution of the Universe according to standard theory, from its very early stages when inflation is believed to have occurred, until the time in which the cosmic microwave background was produced. We reviewed the basic theory of single-field slow-roll inflation, giving the main results explicitly. We also briefly surveyed alternative cases, such as multi-field inflation and non-slow-roll scenarios. This built the necessary framework for the presentation of our results in the rest of the thesis. In the latter part of Chapter~\ref{Ch_SMC}, we discussed the generation of the different matter species currently present in the Universe by briefly describing the evolution from reheating to neutrino decoupling and nucleosynthesis. After that, we described the epoch leading up to recombination and photon decoupling, and gave estimates for the temperature and redshift of those events. We also mentioned the evolution of perturbations during that stage and briefly reviewed the methods for calculating the anisotropies of the CMB using Boltzmann solvers.
\\

Chapter~\ref{Ch_zeta2} included the research published in Ref.~\cite{Carrilho:2015cma}. In it, we computed the second-order evolution of different versions of the gauge-invariant curvature perturbation, $\zeta$. We took into account all possible contributions to that evolution, including vector and tensor modes, as well as anisotropic stress. This calculation was performed for six distinct versions of that curvature perturbation, based on various decompositions of the metric and different choices of gauge fixing conditions. In particular, two of these version are completely new. They were found by using a new way to split the metric, based on the decomposition of the spatial part of the inverse metric. These new versions were also shown to be related to the perturbation of the extrinsic curvature in Appendix~\ref{app3}, giving them a more geometric interpretation.

A key objective of this study was to find the conditions for conservation of the curvature perturbation on super-horizon scales. Our results showed that both vector and tensor modes can have an effect on the evolution of the curvature perturbation at non-linear orders, but only for some versions of $\zeta$. In other cases, the definition of $\zeta$ already includes such perturbations and, for that reason, the evolution equation is simpler and the conditions for conservation are easier to achieve. This is the case for one of the new versions of the curvature perturbation, $\zeta_{I(v)}$, for which the conditions for conservation on large scales are simply that non-adiabatic pressure vanishes along with anisotropic stress.

These conclusion are also argued to be valid for a general theory of gravity, as long as the stress-energy tensor is covariantly conserved. For the particular case of GR, the results go even further, showing that all versions of $\zeta$ are effectively conserved, since both vector and tensor modes are either constant or decay with time too fast to alter the evolution appreciably.

Future work on this topic could take different directions. One of those would be to test which of the different versions of $\zeta$ studied here has a more direct link with quantities which are later observed. On large scales, this is not expected to be important, given that different versions only differ by the presence of vectors and tensors, which would leave too small an imprint to be noticeable. However, when observations probe quantities that correlate large and small scales, such as the squeezed bispectrum, the full form of $\zeta$ may be important, as the different versions have different scalar contributions. Furthermore, this may be related to the vanishing of the bispectrum due to observer effects in single-field inflation~\cite{Tanaka:2011aj,Pajer:2013ana,Bravo:2017gct} and it would be interesting to investigate if any of the versions of $\zeta$ given here can mimic what is seen by an observer in that case.
\\

In Chapter~\ref{Ch_iso2}, we investigated isocurvature initial conditions to be applied to the initialization of second-order Boltzmann solvers, exposing the work published in Ref.~\cite{Carrilho:2018mqy}. We reviewed the general differential system of equations ruling the evolution and investigated the number and properties of growing modes at second order. We found this number to be reduced by one, when compared to the linear case, due to the neutrino velocity mode being non-regular in the second-order case.

We then calculated the approximate initial solutions for all the combinations of linear growing modes, showing our results in synchronous gauge in the main text, as well as those for Poisson gauge in Appendix~\ref{gaugetr}. We detailed the results for a compensated isocurvature mode, showing that this mode can have non-vanishing evolution, when mixed with other modes, such as the adiabatic and neutrino mode. This is in contrast to the linear order result, for which no evolution arises.

Further work is ongoing to extend these results to compute the initial conditions for vector and tensor modes sourced by scalars. An application of that work that is already underway is the study of magnetogenesis from vortical currents in the pre-recombination epoch. This is motivated by the fact that magnetic fields and other vectors are expected to receive considerable contributions from the non-linear mixing of adiabatic and isocurvature modes. Some results from that project have already been produced using the second order Boltzmann code SONG~\cite{Pettinari:2014vja}, but have not yet been published. 

The initial conditions found in Chapter~\ref{Ch_iso2} can be applied to many more situations. In the future, we aim to use them to calculate the intrinsic bispectrum of the CMB sourced by isocurvature modes. This could provide the community with an alternative way to constrain isocurvatures. In particular, this could be an additional way to probe the compensated isocurvature mode, given our result regarding the evolution of the mixed solution between adiabatic and compensated modes.
\\

In Chapter~\ref{Ch_quench}, we studied quantum quenches of a scalar field system in de Sitter spacetime as detailed in Ref.~\cite{Carrilho:2016con}. We used the non-perturbative large-$N$ expansion to account for non-linear effects of the largest scales (IR effects). We reviewed the large-$N$ formalism applied to the case of de Sitter spacetime and discussed some fundamental issues regarding spontaneous symmetry breaking in a curved spacetime. We then calculated the evolution of the system via the evaluation of the time-dependence of the effective mass of the scalar fields. We found approximate algebraic equations for the effective mass as a function of time elapsed after the quench for three distinct fast transitions of the parameters. We then used numerical methods and techniques inspired by the renormalisation group to resolve divergent behaviour due to our approximations, thus obtaining accurate analytic formulas for the evolution of the system. 

We also discussed the importance of our results for cosmology and calculated the power spectrum of the scalar field perturbations as well as their spectral index. We showed that the quench would generate a sudden jump in the spectral index, followed by oscillations similar to those found in the literature for features of the same type. Our approach is somewhat advantageous, since we can analytically estimate the jump in the spectral index as well as its oscillation-averaged scale dependence.

In spite of the advantages of our approach, it is quite rudimentary, due to being valid only in exact de Sitter and not accounting for the effect of the scalar fields in the evolution of inflation. Improving on those shortcomings could be pursued in the future. It would require the application of less transparent numerical techniques, which would spoil the advantages of our simple results, but would be more reliable and represent more realistic situations. 

Another exploration detailed in Chapter~\ref{Ch_quench} regards a quench to a tachyonic state, i.e. a state with $m^2<0$. As we argued, such a state may exist temporarily right after the quench, but the system always evolves back to a situation with $m^2>0$ due to the effects of long wavelength modes. A future direction could be to explore this further with numerical methods and analyse whether the tachyonic instability is stronger than we are led to believe by our simple arguments. If our results are confirmed, another avenue of research would be to investigate more complicated $\text{models}$, such as those with a negative field-space curvature, which give rise to geometrical instabilities~\cite{Turzynski:2014tza,Renaux-Petel:2015mga,Garcia-Saenz:2018ifx}, and check if IR effects are strong enough to avoid the instability.
\\

Finally, in Chapter~\ref{Ch_palatini}, we investigated multi-field models of inflation with a non-minimal coupling to gravity~\cite{Carrilho:2018ffi}. These models were studied in two different formulations of gravity, the metric and the Palatini formulations. We analysed particular models with power-law potentials and non-minimal coupling functions, which were known to present attractor behaviour in their single-field version. We calculated predictions of these multi-field models for the spectral index, tensor-to-scalar ratio and non-Gaussianity parameter for several different potentials and explored the parameter space in each case. We concluded that the multi-field scenario is very similar to the single-field case for both metric and Palatini gravity, particularly in what concerns the attractor behaviour of observables. We describe the reasons for this similarity by ruling out multi-field effects such as those generated by a curved field-space. 

Future work could include the study of models with a more generic non-minimal coupling to gravity and further test whether the choice of gravity formulation has an effect on observables. Furthermore, other gravity formulations, such as teleparallel gravity, could also be included in that test. Another interesting avenue of research would be the study of other models with an attractor behaviour in the single-field case and test whether that behaviour is significantly affected by multi-field dynamics. This could also allow for a more complete understanding of the relation between the attractor behaviour and effective single-field behaviour of the corresponding multi-field model.

% % % % % % % % % % % % % % % % % % % % % % % % % % % % 
% chapter.tex - Ian Huston
% Sample chapter layout
% % % % % % % % % % % % % % % % % % % % % % % % % % % % 
% Redefine CVSRevision for this section. 
% If you don't want to use CVS tags comment out this line
\renewcommand{\CVSrevision}{\version$Id: chapter.tex,v 1.3 2009/12/17 18:16:48 ith Exp $}

% % % % % % % % % % % % % % % % % % % % % % % % % % % % % % % % 
% =========================================================== %
% % % % % % % % % % % % % % % % % % % % % % % % % % % % % % % % 
\begin{appendices}

\chapter{On intrinsic and extrinsic curvature}\label{app3}

In this appendix, we aim to clarify the relation between the different definitions of $\psi$ defined in Chapter \ref{Ch_zeta2} and the perturbation to both the intrinsic and extrinsic curvature of hypersurfaces of constant time. 

We begin by looking at the intrinsic curvature scalar. It is given by
\be
^{(3)}R=R+R_{\mu\nu}n^\mu n^\nu-K^2+K^{\mu\nu}K_{\mu\nu}\,,
\ee
in which $R_{\mu\nu}$ and $R$ are the 4D Ricci tensor and scalar, respectively, $n^\mu$ is the unit normal to the hypersurface, $K_{\mu\nu}$ is the extrinsic curvature and $K$ is its trace. The latter are given by
\begin{align}
K_{\mu\nu}=-\frac12[\L_n\gamma]_{\mu\nu}\,,\ \ \ \ K=-\n_\mu n^\mu\,,
\end{align}
with $\gamma$ the induced metric, given by $\gamma_{\mu\nu}=g_{\mu\nu}+n_\mu n_\nu$. The normal, $n^\mu$, is perpendicular to all vectors in the tangent space of the hypersurface. It is therefore often convenient to choose coordinates such that $n_i=0$. However, this specific coordinate choice means that the usual gauge transformation rules are not obeyed, and for this reason, we shall also compute these quantities using a generic time-like unit 4-vector, $u^\mu$, instead of $n^\mu$. In the latter situation, we will denote quantities with the subscript $(u)$ and often call this vector a 4-velocity, since it can be used to represent the 4-velocity vector of a set of observers. Note, however, that those quantities do not always have the same geometric meaning, since an hypersurface orthogonal to $u^\mu$ can only be defined in the absence of vorticity. 

We now present the calculations of these quantities up to second order in cosmological perturbation theory. The intrinsic curvature scalar is found to be
\begin{align}
\delta^{(3)}R^{(1)}&=\frac{4}{a^2}\n^2\psi^{(1)}\,,\\
\delta{}^{(3)}R^{(2)}&=\frac{1}{a^2}\Big[
4\nabla^2\psi^{(2)}-8C_{km}^{(1)~,m}C_{~~~~,n}^{(1)\,kn}
+6C_{mn}^{(1)~,k}C^{(1)\,mn}_{~~~~,k} -2C^{(1)}_{~,n}C^{(1),n}
\nonumber\\
& +8C^{(1)\,mn}\left(
C_{mn,k}^{(1)~~,k}+C^{(1)}_{~,mn}-C_{mk,n}^{(1)~,k}-C_{kn,m}^{(1)~,k}
\right)\nonumber\\
&  +4\left(C^{(1)}_{~,j} C^{(1)\,jn}_{~~~~~,n}
+C_{jk}^{(1),j}C^{(1),k} -C^{(1)k}_{~~~n,m}C^{(1)\,mn}_{~~~~~,k}
\right)\Big]\,,
\end{align}
where $C_{ij}$ is the perturbation to the spatial part of the metric and $C$ (without indices) is its trace. The relation between $\psi$ and the intrinsic curvature is clear at first order, as they are related linearly. This is the reason why the perturbation $\psi$ is called the curvature perturbation. However, this is only true for the original version of $\psi$, as given by the definition \eqref{gij}, since all other definitions include a contribution from the metric potential $E$. In any case, at second order, this simple connection between the intrinsic curvature and $\psi$ is lost, as there is no simple relation between any of our definitions of the curvature perturbation and $\delta{}^{(3)}R^{(2)}$.

Performing the same calculation using the 4-velocity to define the quantities above, one finds instead a connection to the curvature perturbation on comoving gauge, $\mathcal{R}$, since the first order result for $^{(3)}R_{(u)}$ is\footnote{Note that this quantity has the expected gauge transformation properties, since, being $0$ at the background level (because of the assumption of flatness), the Stewart-Walker lemma \cite{Stewart:1974uz,Stewart:1990fm} dictates it to be gauge invariant at first order. Notice that this does not happen in the calculation with $n$.}
\be
\delta^{(3)}R^{(1)}_{(u)}=\frac{4}{a^2}\n^2\left[\psi^{(1)}-\H(v^{(1)}+B^{(1)})\right]=\frac{4}{a^2}\n^2\mathcal{R}^{(1)}\,.
\ee
At second order, however, the result is no longer related to the second order comoving curvature perturbation $\mathcal{R}^{(2)}$ in a simple way, i.e. $\delta^{(3)}R^{(2)}_{(u)}\neq\frac{4}{a^2}\n^2\mathcal{R}^{(2)}$. This can be seen by evaluating $\delta^{(3)}R^{(2)}_{(u)}$ in comoving gauge ($v=B=v_V^i=0$) and comparing it with $\frac{4}{a^2}\n^2\psi^{(2)}$. In this gauge, the intrinsic curvature is given by
\begin{align}
\delta^{(3)}R^{(2)}_{(u)}=\delta^{(3)}R^{(2)}+S^{(1)\,i}w_i+S^{(1)\,i,j}w_{ij}\,,
\end{align}
where $w_i$ and $w_{ij}$ are linear functions of the metric potentials. It is clear that this is not equal to $\frac{4}{a^2}\n^2\psi^{(2)}$, as there are no further cancellations that would recover that result. Therefore, one must conclude that none of our definitions of $\psi$ has a straightforward interpretation as the perturbation to the intrinsic curvature at an order higher than first.

Moving now to the scalar extrinsic curvature, we start by noting that it is proportional to the local expansion $\n_\mu n^\mu$ (or $\n_\mu u^\mu$, when choosing the velocity 4-vector to define it). It is well known that the integral of the expansion along world lines, with respect to proper time $s$, can be used to define a local scale factor \cite{LV,Lyth:2004gb}.
This integral is defined as
\be
\alpha=\frac13 \int \n_\mu n^\mu ds=-\frac13 \int K ds\,,
\ee
and the local scale factor is given by $e^\alpha$. This interpretation is further supported by the fact that, at the background level, one has $\alpha'=\H$. At first order, one finds
\be
\delta\alpha^{(1)\prime}=-\psi^{(1)\prime}-\frac13\n^2(B^{(1)}-E^{(1)\prime})\,.
\ee
This variable has some similarity with our definition of $\psi_T$, but still has a contribution from $B$, which is not present in any of our versions of the curvature perturbation at first order. Turning now to the situation with $u^\mu$ as the defining vector, the first order result is
\be
\delta\alpha^{(1)\prime}_{(u)}=-\psi^{(1)\prime}+\frac13\n^2(v^{(1)}+E^{(1)\prime})\,.
\ee
While this is still not equal to any version of $\psi$ directly, $\delta\alpha^{(1)}_{(u)}$ is, in fact, equal to $\zeta_{T}^{(1)}$, when the latter is evaluated using a uniform density slicing. Going to second order, we find
\begin{align}
\delta\alpha^{(2)\prime}_{(u)}=&-\psi^{(2)\prime}+\frac13\n^2(v^{(2)}+E^{(2)\prime})\nonumber\\
&+\frac13\left(-4C^{(1)}_{ij} C^{(1)\,ij\prime}+2\phi^{(1)}\n^2v^{(1)}+2 (v_{V}^{(1)\,i}+v^{(1),i})\left(\phi^{(1)}+C^{(1)}\right)_{,i}\right.\nonumber\\
&\left. +\left[(v_{V}^{(1)\,i}+v^{(1),i}+B^{(1),i}-S^{(1)\,i})(v^{(1)}_{Vi}+v^{(1)}_{,i}+B^{(1)}_{,i}-S^{(1)}_{i})\right]'\right)\\
& - 2 (v_{V}^{(1)\,i}+v^{(1),i})\left(-\psi^{(1)}+\frac13\n^2\int(E^{(1)\prime}+v^{(1)})d\tau\right)_{,i}\nonumber\,.
\end{align}
Again, this variable is not equal to any version of $\psi$, but it becomes exactly $\psi_I$, when evaluated using a comoving threading ($v=v_V^i=0$). This is equivalent to saying that, by applying the same procedure to this quantity, one would obtain a gauge invariant quantity that is equal to $\zeta^{(2)}_{I(v)}$. This is not surprising, given the results of Refs. \cite{LV,LV2,LV3,Enqvist:2006fs}, which found similar evolution equations for gauge invariants defined from the expansion scalar, $\Theta=\n_\mu u^\mu$.

We conclude our exposition of this appendix by noting that, even though the connection between the intrinsic curvature and $\psi$ is lost at second order, it is still possible to find a definition of $\psi$ which closely matches the expansion scalar in a gauge with a comoving threading, which can still be interpreted as a perturbation to the scale factor. The reason why the version of $\psi$ that resembles $\Theta$ is the one arising from the determinant of $g^{ij}$ can be explained by a relation between the determinant of the metric and the covariant divergence of a 4-vector. This is given by
\be
\n_\mu u^\mu=\partial_\mu u^\mu+\Gamma^\mu_{\nu\mu}u^\nu=\partial_\mu u^\mu+u^\nu\partial_\nu \log\left(\sqrt{-g}\right)\,,
\ee
in which $g=\det[g_{\mu\nu}]$. Furthermore, it can be shown that $\det[g^{ij}]$ is related to $g$ by
\be
g=g_{00}\left(\det[g^{ij}]\right)^{-1}\,,
\ee
and thus the previous relation becomes
\be
\label{nablaU}
\n_\mu u^\mu=\partial_\mu u^\mu+u^\nu\partial_\nu \log\left(\sqrt{-g_{00}}\right)-u^\nu\partial_\nu \log\left(\sqrt{\det[g^{ij}]}\right)\,.
\ee
Choosing a comoving threading is equivalent to setting $u^i=0$ and in that case it is straightforward to show that $u^0=\sqrt{-g_{00}}$. This implies that the first two terms on the r.h.s. of Eq.~\eqref{nablaU} cancel, and one finds 
\be
\left(\n_\mu u^\mu\right)_{\text{com}}=-u^\nu\partial_\nu \log\left(\sqrt{\det[g^{ij}]}\right)=3\frac{d}{ds}\left(\log a+\psi_I\right)\,,
\ee
in which we substituted $\det[g^{ij}]$ by the definition of $\psi_I$. Equivalently, one has
\be
-\left(\frac13 \int \Theta ds\right)_{\text{com}}=\log a+\psi_I\,.
\ee
This shows $\psi_I$ to be the perturbation to the integrated expansion when written using a comoving threading. This result is valid at all orders and provides a clear interpretation to this perturbation derived from the determinant of the spatial part of the inverse metric.

%%%%%%%%%%%%%%%%%%%%%%%%%%%%%%%%%%%%%%%%%%%%%%%%%%%%%%%%%%%%%%%%%%%%%%%%%%%%%%%%%%%%%%%%%%%%%%%%%%%%%%%%%%%%%%%%%%%%%%%%%%%%%%%%%%%%%%%%%%%%%%%%%%%%%%%%%%5555555

\chapter{Gauge transformations to Poisson gauge}\label{gaugetr}

In this appendix, we describe the gauge transformation of the results of Chapter~\ref{Ch_iso2} into Poisson gauge, so as to allow the application of those results in that popular gauge.

The Poisson gauge is specified by the following choices
\begin{equation}
\wt{E}=\wt{B}=0\,,\ \ \ \ \ \wt{F^i}=0\,,
\end{equation}
which implies that the gauge generator components are, at first order,
\begin{align}
&\alpha^{(1)}=B^{(1)}-E^{(1)\prime}\,,\\
&\beta^{(1)}=-E^{(1)}\,,\\
&\gamma^{(1)\,i}=-F^{(1)\,i}\,.
\end{align}
In this appendix, we are interested in a transformation from synchronous to Poisson gauge, thus we may simply re-write the first equation above as $\alpha^{(1)}_{\text{S2P}}=-E^{(1)\prime}_{\text{S}}$. Therefore the gauge transformations for the scalars depend only on the metric potential $E^{(1)}$. For that reason, the difference between variables on both gauges depends on the size of $E$ in each mode, in orders of $\tau$. For example, in the CDM isocurvature mode shown in Chapter~\ref{Ch_iso2} in Eq.~\eqref{matteriso1}, the metric potential $E$ is $O(\tau^3)$. However, it enters $\alpha$ with a time derivative and is usually multiplied by $\Hh$, thus the gauge transformation will make a difference of order $O(\tau)$ in most variables. At leading order in $\tau$, the CDM isocurvature mode is now given in Poisson gauge by
\begin{align}
\psi=&-\frac{R_c(4R_\nu+15)}{8(15+2R_\nu)}\omega\tau\delta_c^0\,,\nonumber\\
\phi=&\frac{R_c(4R_\nu-15)}{8(15+2R_\nu)}\omega\tau\delta_c^0\,,\nonumber\\
\delta_{c}=&\left(1-\frac{3R_c(4R_\nu+15)}{8(15+2R_\nu)}\omega\tau\right)\delta_c^0\,,\nonumber\\
\delta_{b}=&-\frac{3R_c(4R_\nu+15)}{8(15+2R_\nu)}\omega\tau\delta_c^0\,,\nonumber\\
\delta_{\gamma}=&-\frac{R_c(4R_\nu+15)}{2(15+2R_\nu)}\omega\tau\delta_c^0\,,\nonumber\\
\delta_{\nu}=&-\frac{R_c(4R_\nu+15)}{2(15+2R_\nu)}\omega\tau\delta_c^0\,,\\
v_c=&\frac{R_c(15-4R_\nu)}{24(15+2R_\nu)}\omega\tau^2\delta_c^0\,,\nonumber\\
v_{\gamma b}=&\frac{15 R_c}{8(15+2R_\nu)}\omega\tau^2\delta_c^0\,,\nonumber\\
v_{\nu}=&\frac{15R_c}{8(15+2R_\nu)}\omega\tau^2\delta_c^0\,,\nonumber\\
\sigma_\nu=&-\frac{R_c}{6(15+2 R_\nu)}k^2\omega\tau^3\delta_c^0\,.\nonumber
\end{align}

In other modes, the transformation is similar, but can introduce additional issues. For example, in the case of the neutrino velocity isocurvature, some variables will have decaying solutions already at linear order, as $E$ is $O(\tau)$ in that case. This is described, for example, in Ref.~\cite{Shaw:2009nf}, in which the potentials $\phi$ and $\psi$ are given in Poisson gauge for all five linear growing modes. We do not comment further on this issue, as we do not study the neutrino velocity mode at second order, for the reasons explained in the main text.

At second order, the transformation rules are given in Chapter~\ref{Ch_CPT} and, as before, one can calculate the form of the gauge generators required to transform from synchronous gauge to Poisson gauge. Applying those transformations to the results in the main text, we find the results for Poisson gauge, which we show in the same order as before, starting with the adiabatic sourced solution. Note, however, that the defining variables (e.g. $\psi^{0}_{k_1}\psi^{0}_{k_2}$) still refer to those variables in synchronous gauge.

\section{Pure adiabatic mode}

\begin{align}
\psi^{(2)}=&f^{\psi\psi}_{\psi,P}\psi^{0}_{k_1}\psi^{0}_{k_2}\,,\nonumber\\
\phi^{(2)}=&\left(\frac{20(35+8 R_\nu)}{(15+4R_\nu)^2}-2 f^{\psi\psi}_{\psi,P}\right)\psi^{0}_{k_1}\psi^{0}_{k_2}\,,\nonumber\\
\delta_{c}^{(2)}=&\delta_{b}^{(2)}=\left(-\frac{15(35+16 R_\nu)}{(15+4R_\nu)^2}+3 f^{\psi\psi}_{\psi,P}\right)\psi^{0}_{k_1}\psi^{0}_{k_2}\,,\nonumber\\
\delta_{\gamma}^{(2)}=&\delta_{\nu}^{(2)}=\left(-\frac{40(15+8 R_\nu)}{(15+4R_\nu)^2}+4 f^{\psi\psi}_{\psi,P}\right)\psi^{0}_{k_1}\psi^{0}_{k_2}\,,\\
v_c^{(2)}=&v_{\gamma b}^{(2)}=v_{\nu}^{(2)}=\left(-\frac{40(10+3 R_\nu)}{(15+4R_\nu)^2}+f^{\psi\psi}_{\psi,P}\right)\tau\psi^{0}_{k_1}\psi^{0}_{k_2}\,,\nonumber\\
\sigma_\nu^{(2)}=&-\frac{9 k^4 - 3 (k_1^2 - k_2^2)^2 + 2 k^2 (k_1^2+k_2^2)}{3(15+4R_\nu)k^4}(k\tau)^2\psi^{0}_{k_1}\psi^{0}_{k_2}\,,\nonumber
\end{align}
with,
\begin{align}
f^{\psi\psi}_{\psi,P}=&\frac{5}{(15 + 4 R_\nu)^2 k^4} \left[(25 + 9 R_\nu) k^4 -(5+R_\nu)\left(3 (k_1^2 - k_2^2)^2 - 2 k^2 (k_1^2+k_2^2)\right)\right]\,.\nonumber
\end{align}

We can very easily verify that the adiabatic condition at second order, given in Eq.~\eqref{Adconds2}, is indeed verified in this gauge, as it must. Furthermore, we can now directly compare these results to those given in Refs.~\cite{Pettinari:2014vja,Pitrou:2010sn}. They do not exactly match, due to a different choice of defining variable --- we choose $\psi=-\zeta$, while they choose $\zeta_D=\zeta+\zeta^2$, as defined in Chapter~\ref{Ch_zeta2}. Applying this transformation to the general solution in terms of transfer functions, we find
\begin{align}
X(\tau,k)&=\mathcal{T}^{(1)} \psi^0(k)+\frac12\int_{k_1,k_2}\mathcal{T}^{(2)} \psi^0(k_1)\psi^0(k_2)\\
&=-\mathcal{T}^{(1)} \zeta_D^0(k)+\frac12\int_{k_1,k_2}\lb2\mathcal{T}^{(1)}+\mathcal{T}^{(2)}\rb \zeta_D^0(k_1)\zeta_D^0(k_2)\,,
\end{align}
which shows that, in terms of $\zeta_D$, the second-order transfer functions receive an extra contribution of twice the linear transfer function. This is exactly the difference we find between our results and those of Refs.~\cite{Pettinari:2014vja,Pitrou:2010sn}, confirming the match between all results. Care must be taken, however, when these results are applied to situations in which one assumes the initial conditions to be Gaussian. In that case, one must make clear which of the variables has that property, since should $\zeta_D$ be Gaussian, $\zeta$ will not be and vice versa.

\section{Pure cold dark matter isocurvature mode}

\begin{align}
\psi^{(2)}=&\left(-\frac{5(15 + 4 R_\nu)^2}{ 64 (15 + 2 R_\nu)^2} R_c^2+\frac13f^{cc}_{b,P}\right)(\omega\tau)^2\delta^{0}_{c,k_1}\delta^{0}_{c,k_2}\,,\nonumber\\
\phi^{(2)}=&\left(\frac{8325 + 2280 R_\nu + 272 R_\nu^2}{ 64 (15 + 2 R_\nu)^2} R_c^2-\frac43f^{cc}_{b,P}\right)(\omega\tau)^2\delta^{0}_{c,k_1}\delta^{0}_{c,k_2}\,,\nonumber\\
\delta_{c}^{(2)}=&\left(-\frac{3(15+4R_\nu)}{4(15+2R_\nu)}R_c\omega\tau+\left(\frac{3(675 + 230 R_\nu + 8 R_\nu^2)}{ 16 (15 + 2 R_\nu)(25+2 R_\nu)} R_c+f^{cc}_{b,P}\right)(\omega\tau)^2\right)\delta^{0}_{c,k_1}\delta^{0}_{c,k_2}\,,\nonumber\\
\delta_{b}^{(2)}=&f^{cc}_{b,P}(\omega\tau)^2\delta^{0}_{c,k_1}\delta^{0}_{c,k_2}\,,\nonumber\\
\delta_{\gamma}^{(2)}=&\delta_{\nu}^{(2)}=\left(\frac{(15+4R_\nu)^2}{ 16 (15 + 2 R_\nu)^2} R_c^2+\frac43f^{cc}_{b,P}\right)(\omega\tau)^2\delta^{0}_{c,k_1}\delta^{0}_{c,k_2}\,,\\
v_c^{(2)}=&\left(-\frac{5(1305+360R_\nu+32R_\nu^2)}{ 192 (15 + 2 R_\nu)^2} R_c^2+\frac13f^{cc}_{b,P}\right)\omega^2\tau^3\delta^{0}_{c,k_1}\delta^{0}_{c,k_2}\,,\nonumber\\
v_{\gamma b}^{(2)}=&v_{\nu}^{(2)}=\left(-\frac{2925+780R_\nu+64R_\nu^2}{ 64 (15 + 2 R_\nu)^2} R_c^2+\frac13f^{cc}_{b,P}\right)\omega^2\tau^3\delta^{0}_{c,k_1}\delta^{0}_{c,k_2}\,,\nonumber\\
\sigma_\nu^{(2)}=&f^{cc}_{\sigma,P}R_c^2\omega^2k^2\tau^4\delta^{0}_{c,k_1}\delta^{0}_{c,k_2}\,,\nonumber
\end{align}
with,
\begin{align}
f^{cc}_{b,P}=&\frac{3R_c^2}{128 (25 + 2 R_\nu) (15 + 2 R_\nu)^2 k^4} \left[(88875 + 42150 R_\nu + 6160 R_\nu^2 + 256 R_\nu^3) k^4 \right.\nonumber\\
&\left.+ 15(-225 + 110 R_\nu + 16 R_\nu^2)\left(3 (k_1^2 - k_2^2)^2 - 2 k^2 (k_1^2 + k_2^2)\right)\right]\,,\nonumber\\
f^{cc}_{\sigma,P}=&-\frac{5(855+138R_\nu+4R_\nu^2)k^4+(825+70R_\nu-4R_\nu^2)\left(3(k_1^2-k_2^2)^2-2k^2(k_1^2+k_2^2)\right)}{48(15+2R_\nu)^2(25+2R_\nu)}\,.\nonumber
\end{align}

\section{Mixture of adiabatic and cold dark matter modes}

\begin{align}
\psi^{(2)}=&f^{c\psi}_{\psi,P}\omega\tau\delta^{0}_{c,k_1}\psi^{0}_{k_2}\,,\nonumber\\
\phi^{(2)}=&\left(\frac{75+8 R_\nu(20+3 R_\nu)}{2(15+2R_\nu)(15+4R_\nu)}R_c-3f^{c\psi}_{\psi,P}\right)\omega\tau\delta^{0}_{c,k_1}\psi^{0}_{k_2}\,,\nonumber\\
\delta_{c}^{(2)}=&\left(-\frac{15}{15+4R_\nu}+\left(-\frac{3\left(75(1+R_c)+4 R_\nu(20+(35+8 R_\nu)R_c)\right)}{8(15+2R_\nu)(15+4R_\nu)}+3f^{c\psi}_{\psi,P}\right)\omega\tau\right)\delta^{0}_{c,k_1}\psi^{0}_{k_2}\,,\nonumber\\
\delta_{b}^{(2)}=&\left(-\frac{3(5+8 R_\nu)}{8(15+2R_\nu)}R_c+3f^{c\psi}_{\psi,P}\right)\omega\tau\delta^{0}_{c,k_1}\psi^{0}_{k_2}\,,\nonumber\\
\delta_{\gamma}^{(2)}=&\delta_{\nu}^{(2)}=\left(-\frac{4 R_\nu}{15+2R_\nu}R_c+4f^{c\psi}_{\psi,P}\right)\omega\tau\delta^{0}_{c,k_1}\psi^{0}_{k_2}\,,\\
v_c^{(2)}=&\left(-\frac{(35+8R_\nu)(k^2+k_1^2-k_2^2)}{24(15+4R_\nu)k^2}R_c+f^{c\psi}_{v,P}\right)\omega\tau^2\delta^{0}_{c,k_1}\psi^{0}_{k_2}\,,\nonumber\\
v_{\gamma b}^{(2)}=&v_{\nu}^{(2)}=f^{c\psi}_{v,P}\omega\tau^2\delta^{0}_{c,k_1}\psi^{0}_{k_2}\,,\nonumber\\
\sigma_\nu^{(2)}=&f^{c\psi}_{\sigma,P}\omega k^2\tau^3\delta^{0}_{c,k_1}\psi^{0}_{k_2}\,,\nonumber
\end{align}
with,
\begin{align}
f^{c\psi}_{\psi,P}=&\frac{R_c}{16 (15 + 2 R_\nu) (15 + 4 R_\nu) k^4} \left[(375 + 315 R_\nu + 64 R_\nu^2) k^4\right.\nonumber\\
&\left. - 45 (-5 + R_\nu) (k_1^2 - k_2^2)^2 + 30 k^2 ((-5 + 3 R_\nu) k_1^2 - (5 + R_\nu) k_2^2)\right]\,,\nonumber
\end{align}
\begin{align}
f^{c\psi}_{v,P}=&\frac{5R_c}{16 (15 + 2 R_\nu) (15 + 4 R_\nu) k^4} \left[(135 + 19 R_\nu) k^4 \right.\nonumber\\
&\left.- 9 (-5 + R_\nu) (k_1^2 - k_2^2)^2 + k^2 ((30 + 38 R_\nu) k_1^2 - 2 (45 + 13 R_\nu) k_2^2)\right]\,,\nonumber\\
f^{c\psi}_{\sigma,P}=&\frac{R_c}{12 (15 + 2 R_\nu) (15 + 4 R_\nu) k^4} \left[(75 + 4 R_\nu) k^4 \right.\nonumber\\
&\left.- 3 (-5 + 4 R_\nu) (k_1^2 - k_2^2)^2 + k^2 ((-70 + 8 R_\nu) k_1^2 + 2 (25 + 4 R_\nu) k_2^2)\right]\,.\nonumber
\end{align}

\section{Pure baryon isocurvature mode}

\begin{align}
\psi^{(2)}=&\left(-\frac{5(15+ 4 R_\nu)^2}{ 64 (15 + 2 R_\nu)^2} R_b^2+\frac13f^{bb}_{c,P}\right)(\omega\tau)^2\delta^{0}_{b,k_1}\delta^{0}_{b,k_2}\,,\nonumber\\
\phi^{(2)}=&\left(\frac{8325 + 2280 R_\nu + 272 R_\nu^2}{ 64 (15 + 2 R_\nu)^2} R_b^2-\frac43f^{bb}_{c,P}\right)(\omega\tau)^2\delta^{0}_{b,k_1}\delta^{0}_{b,k_2}\,,\nonumber\\
\delta_{c}^{(2)}=&f^{bb}_{c,P}(\omega\tau)^2\delta^{0}_{b,k_1}\delta^{0}_{b,k_2}\,,\nonumber\\
\delta_{b}^{(2)}=&\left(-\frac{3(15+4R_\nu)}{4(15+2R_\nu)}R_b\omega\tau+\left(\frac{3(675 + 230 R_\nu + 8 R_\nu^2)}{ 16 (15 + 2 R_\nu)(25+2 R_\nu)} R_b+f^{bb}_{c,P}\right)(\omega\tau)^2\right)\delta^{0}_{b,k_1}\delta^{0}_{b,k_2}\,,\nonumber\\
\delta_{\gamma}^{(2)}=&\delta_{\nu}^{(2)}=\left(\frac{(15+4R_\nu)^2}{ 16 (15 + 2 R_\nu)^2} R_b^2+\frac43f^{bb}_{c,P}\right)(\omega\tau)^2\delta^{0}_{b,k_1}\delta^{0}_{b,k_2}\,,\\
v_c^{(2)}=&\left(-\frac{5(1305+360R_\nu+32R_\nu^2)}{ 192 (15 + 2 R_\nu)^2} R_b^2+\frac13f^{bb}_{c,P}\right)\omega^2\tau^3\delta^{0}_{b,k_1}\delta^{0}_{b,k_2}\,,\nonumber\\
v_{\gamma b}^{(2)}=&\left(-\frac{3825-1905R_\nu-700R_\nu^2-64R_\nu^3}{ 64 R_\gamma(15 + 2 R_\nu)^2} R_b^2+\frac13f^{bb}_{c,P}\right)\omega^2\tau^3\delta^{0}_{b,k_1}\delta^{0}_{b,k_2}\,,\nonumber\\
v_{\nu}^{(2)}=&\left(-\frac{2925+780R_\nu+64R_\nu^2}{ 64 (15 + 2 R_\nu)^2} R_b^2+\frac13f^{bb}_{c,P}\right)\omega^2\tau^3\delta^{0}_{b,k_1}\delta^{0}_{b,k_2}\,,\nonumber\\
\sigma_\nu^{(2)}=&f^{cc}_{\sigma,P}R_b^2\omega^2k^2\tau^4\delta^{0}_{b,k_1}\delta^{0}_{b,k_2}\,,\nonumber
\end{align}
with,
\begin{align}
f^{bb}_{c,P}=&\frac{R_b^2}{R_c^2}f^{cc}_{b,P}\,.\nonumber
\end{align}

\section{Mixture of baryon and cold dark matter modes}

\begin{align}
\psi^{(2)}=&f^{bc}_{\psi,P}(\omega\tau)^2\delta^{0}_{b,k_1}\delta^{0}_{c,k_2}\,,\nonumber\\
\phi^{(2)}=&\left(\frac{3(1275 - 40 R_\nu - 16 R_\nu^2)}{ 64 (15 + 2 R_\nu)^2} R_bR_c-4f^{bc}_{\psi,P}\right)(\omega\tau)^2\delta^{0}_{b,k_1}\delta^{0}_{c,k_2}\,,\nonumber\\
\delta_{c}^{(2)}=&\left[-3R_b\frac{15+4R_\nu}{15+2R_\nu}\omega\tau+\left(f^{bc}_\delta(R_b)+3f^{bc}_{\psi,P}\right)(\omega\tau)^2\right]\delta^{0}_{b,k_1}\delta^{0}_{c,k_2}\,,\nonumber\\
\delta_{b}^{(2)}=&\left[-3R_c\frac{15+4R_\nu}{15+2R_\nu}\omega\tau+\left(f^{bc}_\delta(R_c)+3f^{bc}_{\psi,P}\right)(\omega\tau)^2\right]\delta^{0}_{b,k_1}\delta^{0}_{c,k_2}\,,\nonumber\\
\delta_{\gamma}^{(2)}=&\delta_{\nu}^{(2)}=\left(-\frac{3(15+4R_\nu)^2}{ 8 (15 + 2 R_\nu)^2} R_bR_c+4f^{bc}_{\psi,P}\right)(\omega\tau)^2\delta^{0}_{b,k_1}\delta^{0}_{c,k_2}\,,\\
v_c^{(2)}=&\left(-\frac{5(315-8R_\nu^2)}{ 96(15 + 2 R_\nu)^2} R_bR_c+f^{bc}_{\psi,P}\right)\omega^2\tau^3\delta^{0}_{b,k_1}\delta^{0}_{c,k_2}\,,\nonumber\\
v_{\gamma b}^{(2)}=&\left(-\frac{(1125-750R_\nu-94R_\nu^2+8R_\nu^3)k^2-(15+2R_\nu)^2(k_1^2-k_2^2)}{ 32 R_\gamma(15 + 2 R_\nu)^2k^2} R_bR_c+f^{bc}_{\psi,P}\right)\omega^2\tau^3\delta^{0}_{b,k_1}\delta^{0}_{c,k_2}\,,\nonumber\\
v_{\nu}^{(2)}=&\left(-\frac{450+45R_\nu-4R_\nu^2}{ 16 (15 + 2 R_\nu)^2} R_bR_c+f^{bc}_{\psi,P}\right)\omega^2\tau^3\delta^{0}_{b,k_1}\delta^{0}_{c,k_2}\,,\nonumber\\
\sigma_\nu^{(2)}=&f^{bc}_{\sigma,P}R_b^2\omega^2k^2\tau^4\delta^{0}_{b,k_1}\delta^{0}_{c,k_2}\,,\nonumber
\end{align}
with,
\begin{align}
f^{bc}_{\psi,P}=&-\frac{R_bR_c}{128(25+2R_\nu)(15+2R_\nu)^2k^4}\left[(-32625 - 7650 R_\nu + 240 R_\nu^2 + 64 R_\nu^3) k^4 \right.\nonumber\\\nonumber
&\left.- 15 (-225 + 110 R_\nu + 16 R_\nu^2)\left(3 (k_1^2 - k_2^2)^2 -2 k^2 (k_1^2 + k_2^2)\right)\right]\,,\\
f^{bc}_\delta(R_x)=&\frac{3R_x\left(48375-5R_x(25+2R_\nu)(15+4R_\nu)^2+2R_\nu(13425+4R_\nu(545+24R_\nu))\right)}{64(25+2R_\nu)(15+2R_\nu)^2}\,,\nonumber\\
f^{bc}_{\sigma,P}=&\frac{R_bR_c}{48(25+2R_\nu)(15+2R_\nu)^2k^4}\left[5 (855 + 138 R_\nu + 4 R_\nu^2) k^4 \right.\nonumber\\\nonumber
&\left.- (-825 - 70 R_\nu + 4 R_\nu^2)\left(3 (k_1^2 - k_2^2)^2 - 2 k^2 (k_1^2 + k_2^2)\right)\right]\,.
\end{align}

\section{Mixture of adiabatic and baryon modes}

\begin{align}
\psi^{(2)}=&f^{b\psi}_{\psi,P}\omega\tau\delta^{0}_{b,k_1}\psi^{0}_{k_2}\,,\nonumber\\
\phi^{(2)}=&\left(\frac{75+8 R_\nu(20+3 R_\nu)}{2(15+2R_\nu)(15+4R_\nu)}R_b-3f^{b\psi}_{\psi,P}\right)\omega\tau\delta^{0}_{b,k_1}\psi^{0}_{k_2}\,,\nonumber\\
\delta_{c}^{(2)}=&\left(-\frac{3(5+8 R_\nu)}{8(15+2R_\nu)}R_b+3f^{b\psi}_{\psi,P}\right)\omega\tau\delta^{0}_{b,k_1}\psi^{0}_{k_2}\,,\nonumber\\
\delta_{b}^{(2)}=&\left(-\frac{15}{15+4R_\nu}+\left(-\frac{3\left(75(1+R_c)+4 R_\nu(20+(35+8 R_\nu)R_b)\right)}{8(15+2R_\nu)(15+4R_\nu)}+3f^{b\psi}_{\psi,P}\right)\omega\tau\right)\delta^{0}_{b,k_1}\psi^{0}_{k_2}\,,\nonumber\\
\delta_{\gamma}^{(2)}=&\delta_{\nu}^{(2)}=\left(-\frac{4 R_\nu}{15+2R_\nu}R_c+4f^{b\psi}_{\psi,P}\right)\omega\tau\delta^{0}_{b,k_1}\psi^{0}_{k_2}\,,\\
v_c^{(2)}=&\left(-\frac{(35+8R_\nu)(k^2+k_1^2-k_2^2)}{24(15+4R_\nu)k^2}R_b+f^{b\psi}_{v,P}\right)\omega\tau^2\delta^{0}_{b,k_1}\psi^{0}_{k_2}\,,\nonumber\\
v_{\gamma b}^{(2)}=&v_{\nu}^{(2)}=f^{b\psi}_{v,P}\omega\tau^2\delta^{0}_{b,k_1}\psi^{0}_{k_2}\,,\nonumber\\
\sigma_\nu^{(2)}=&f^{b\psi}_{\sigma,P}\omega k^2\tau^3\delta^{0}_{b,k_1}\psi^{0}_{k_2}\,,\nonumber
\end{align}
with,
\begin{align}
f^{b\psi}_{\psi,P}=&\frac{R_b}{R_c}f^{c\psi}_{\psi,P}\,,\nonumber\\
f^{b\psi}_{v,P}=&\frac{R_b}{R_c}f^{c\psi}_{v,P}\,,\nonumber\\
f^{b\psi}_{\sigma,P}=&\frac{R_b}{R_c}f^{c\psi}_{\sigma,P}\,.\nonumber
\end{align}

\section{Mixture of adiabatic and compensated modes}

\begin{align}
\delta_{c}^{(2)}=&-\frac{R_b}{R_c}\delta_{b}^{(2)}=\left(-\frac{15}{15+4R_\nu}-\frac{15\left(15+16 R_\nu\right)}{8(15+2R_\nu)(15+4R_\nu)}\omega\tau\right)\delta^{0}_{CI,k_1}\psi^{0}_{k_2}\,,\nonumber\\
v_{\gamma b}^{(2)}=&\frac{\left(k^2-k_1^2+k_2^2\right)R_c}{96R_\gamma k^2}k_2^2\omega\tau^4 \delta^{0}_{CI,k_1}\psi^{0}_{k_2}\,.
\end{align}

\section{Pure neutrino density isocurvature mode}

\begin{align}
\psi^{(2)}=&f^{\nu\nu}_{\psi,P}\delta^{0}_{\nu,k_1}\delta^{0}_{\nu,k_2}\,,\nonumber\\
\phi^{(2)}=&\left(\frac{4R_\nu^2}{(15+4R_\nu)^2}-2 f^{\nu\nu}_{\psi,P}\right)\delta^{0}_{\nu,k_1}\delta^{0}_{\nu,k_2}\,,\nonumber\\
\delta_{c}^{(2)}=&\delta_{b}^{(2)}=\left(\frac{15R_\nu^2}{(15+4R_\nu)^2}+3 f^{\nu\nu}_{\psi,P}\right)\delta^{0}_{\nu,k_1}\delta^{0}_{\nu,k_2}\,,\nonumber\\
\delta_{\gamma}^{(2)}=&\left(-\frac{8R_\nu^2(12+7R_\nu)}{R_\gamma(15+4R_\nu)^2}+4 f^{\nu\nu}_{\psi,P}\right)\delta^{0}_{\nu,k_1}\delta^{0}_{\nu,k_2}\,,\nonumber\\
\delta_{\nu}^{(2)}=&\left(\frac{8R_\nu(15+7R_\nu)}{(15+4R_\nu)^2}+4 f^{\nu\nu}_{\psi,P}\right)\delta^{0}_{\nu,k_1}\delta^{0}_{\nu,k_2}\,,\\
v_c^{(2)}=&\left(\frac{2R_\nu^2}{(15+4R_\nu)^2}+f^{\nu\nu}_{\psi,P}\right)\tau\delta^{0}_{\nu,k_1}\delta^{0}_{\nu,k_2}\,,\nonumber\\
v_{\gamma b}^{(2)}=&\left(\frac{R_\nu^2(233+8R_\nu(13+3R_\nu))}{4R_\gamma^2(15+4R_\nu)^2}+f^{\nu\nu}_{\psi,P}\right)\tau\delta^{0}_{\nu,k_1}\delta^{0}_{\nu,k_2}\,,\nonumber\\
v_{\nu}^{(2)}=&\left(\frac{3(75+8R_\nu(5+R_\nu))}{4(15+4R_\nu)^2}+f^{\nu\nu}_{\psi,P}\right)\tau\delta^{0}_{\nu,k_1}\delta^{0}_{\nu,k_2}\,,\nonumber\\
\sigma_\nu^{(2)}=&f^{\nu\nu}_{\sigma,P}(k\tau)^2\delta^{0}_{\nu,k_1}\delta^{0}_{\nu,k_2}\,,\nonumber
\end{align}
with,
\begin{align}
f^{\nu\nu}_{\psi,P}=&-\frac{R_\nu^2\left((1-96R_\nu)k^4+285(k_1^2-k_2^2)^2-190k^2(k_1^2+k_2^2)\right)}{16  R_\gamma (15 + 4 R_\nu)^2 k^4} \,,\nonumber\\
f^{\nu\nu}_{\sigma,P}=&\frac{1}{48 R_\gamma (15 + 4 R_\nu)^2 k^4} \left[(-225 - 39 R_\nu + 188 R_\nu^2) k^4 \right.\nonumber\\
&\left.+  (225 - 153 R_\nu + 4 R_\nu^2)\left(3 (k_1^2 - k_2^2)^2 - 2 k^2 (k_1^2 + k_2^2)\right)\right]\,.\nonumber
\end{align}

\section{Mixture of adiabatic and neutrino modes}

\begin{align}
\psi^{(2)}=&f^{\nu\psi}_{\psi,P}\delta^{0}_{\nu,k_1}\psi^{0}_{k_2}\,,\nonumber\\
\phi^{(2)}=&\left(-\frac{16R_\nu(5+R_\nu)}{(15+4R_\nu)^2}+2 f^{\nu\psi}_{\psi,P}\right)\delta^{0}_{\nu,k_1}\psi^{0}_{k_2}\,,\nonumber\\
\delta_{c}^{(2)}=&\delta_{b}^{(2)}=\left(\frac{3R_\nu(5+8R_\nu)}{(15+4R_\nu)^2}+3 f^{\nu\psi}_{\psi,P}\right)\delta^{0}_{\nu,k_1}\psi^{0}_{k_2}\,,\nonumber\\
\delta_{\gamma}^{(2)}=&\left(\frac{4R_\nu(75+4R_\nu(7-2R_\nu))}{R_\gamma(15+4R_\nu)^2}+4 f^{\nu\psi}_{\psi,P}\right)\delta^{0}_{\nu,k_1}\psi^{0}_{k_2}\,,\nonumber\\
\delta_{\nu}^{(2)}=&\left(-\frac{4(75+4R_\nu(5-2R_\nu))}{(15+4R_\nu)^2}+4 f^{\nu\psi}_{\psi,P}\right)\delta^{0}_{\nu,k_1}\psi^{0}_{k_2}\,,\\
v_c^{(2)}=&f^{\nu\psi}_{v,P}\tau\delta^{0}_{\nu,k_1}\psi^{0}_{k_2}\,,\nonumber\\
v_{\gamma b}^{(2)}=&\left(\frac{R_\nu(k^2+k_1^2-k_2^2)}{4R_\gamma k^2}+f^{\nu\psi}_{v,P}\right)\tau\delta^{0}_{\nu,k_1}\psi^{0}_{k_2}\,,\nonumber\\
v_{\nu}^{(2)}=&\left(-\frac{k^2+k_1^2-k_2^2}{4 k^2}+f^{\nu\psi}_{v,P}\right)\tau\delta^{0}_{\nu,k_1}\psi^{0}_{k_2}\,,\nonumber\\
\sigma_\nu^{(2)}=&f^{\nu\psi}_{\sigma,P}(k\tau)^2\delta^{0}_{\nu,k_1}\psi^{0}_{k_2}\,,\nonumber
\end{align}
with,
\begin{align}
f^{\nu\psi}_{\psi,P}=&\frac{R_\nu\left((-55 - 32 R_\nu) k^4 + 45 (k_1^2 - k_2^2)^2 + 10 k^2 (-7 k_1^2 + k_2^2)\right)}{4 (15 + 4 R_\nu)^2 k^4} \,,\nonumber\\
f^{\nu\psi}_{v,P}=&\frac{R_\nu\left((85 + 16 R_\nu) k^4 + 45 (k_1^2 - k_2^2)^2 + 2 k^2 ((-5 + 8 R_\nu) k_1^2 - (25 + 8 R_\nu) k_2^2)\right)}{4 (15 + 4 R_\nu)^2 k^4} \,,\nonumber\\
f^{\nu\psi}_{\sigma,P}=&-\frac{1}{4 (15 + 4 R_\nu)^2 k^4} \left[(45 + 4 R_\nu) k^4 - 3 (5 + 4 R_\nu) (k_1^2 - k_2^2)^2 \right.\nonumber\\
&\left. + k^2 ((-30 + 8 R_\nu) k_1^2 + 2 (25 + 4 R_\nu) k_2^2)\right]\,.\nonumber
\end{align}

\section{Mixture of dark matter and neutrino modes}

\begin{align}
\psi^{(2)}=&f^{\nu c}_{\psi,P}\omega\tau\delta^{0}_{\nu,k_1}\delta^{0}_{c,k_2}\,,\nonumber\\
\phi^{(2)}=&\left(\frac{R_\nu R_c(105+8R_\nu)}{4(15+2R_\nu)(15+4R_\nu)}-3 f^{\nu c}_{\psi,P}\right)\omega\tau\delta^{0}_{\nu,k_1}\delta^{0}_{c,k_2}\,,\nonumber\\
\delta_{c}^{(2)}=&\left(\frac{3R_\nu}{15+4R_\nu}+\left(\frac{3R_\nu(5R_b(15+4R_\nu)-105-16R_\nu)}{8(15+2R_\nu)(15+4R_\nu)}+3 f^{\nu c}_{\psi,P}\right)\omega\tau\right)\delta^{0}_{\nu,k_1}\delta^{0}_{c,k_2}\,,\nonumber\\
\delta_{b}^{(2)}=&\left(-\frac{15R_\nu R_c}{8(15+2R_\nu)}+3 f^{\nu c}_{\psi,P}\right)\omega\tau\delta^{0}_{\nu,k_1}\delta^{0}_{c,k_2}\,,\nonumber\\
\delta_{\gamma}^{(2)}=&\left(\frac{R_\nu R_c(9+10R_\nu)}{2 R_\gamma (15+2R_\nu)}+4 f^{\nu c}_{\psi,P}\right)\omega\tau\delta^{0}_{\nu,k_1}\delta^{0}_{c,k_2}\,,\nonumber\\
\delta_{\nu}^{(2)}=&\left(-\frac{5R_c(3+2R_\nu)}{2(15+2R_\nu)}+4 f^{\nu c}_{\psi,P}\right)\omega\tau\delta^{0}_{\nu,k_1}\delta^{0}_{c,k_2}\,,\\
v_c^{(2)}=&\left(-\frac{R_c\left((45+8R_\nu)k^2+(45+16R_\nu)(k_1^2-k_2^2)\right)}{96(15+4R_\nu)k^2}+f^{\nu c}_{v,P}\right)\omega\tau^2\delta^{0}_{\nu,k_1}\delta^{0}_{c,k_2}\,,\nonumber\\
v_{\gamma b}^{(2)}=&\left(-\frac{R_c(k^2+k_1^2-k_2^2)}{32R_\gamma k^2}+f^{\nu c}_{v,P}\right)\omega\tau^2\delta^{0}_{\nu,k_1}\delta^{0}_{c,k_2}\,,\nonumber\\
v_{\nu}^{(2)}=&f^{\nu c}_{v,P}\omega\tau^2\delta^{0}_{\nu,k_1}\delta^{0}_{c,k_2}\,,\nonumber\\
\sigma_\nu^{(2)}=&f^{\nu c}_{\sigma,P}(k\tau)^2\delta^{0}_{\nu,k_1}\delta^{0}_{c,k_2}\,,\nonumber
\end{align}
with,
\begin{align}
f^{\nu c}_{\psi,P}=&\frac{3R_\nu R_c}{64 (15 + 2 R_\nu)^2 (15 + 4 R_\nu) k^4} \left[(975 + 550 R_\nu + 64 R_\nu^2) k^4 \right.\nonumber\\
&\left.- 15 (135 + 22 R_\nu) (k_1^2 - k_2^2)^2 + 10 k^2 ((105 + 26 R_\nu) k_1^2 + 3 (55 + 6 R_\nu) k_2^2)\right]\,,\nonumber\\
f^{\nu c}_{v,P}=&-\frac{15 R_c}{64 (15 + 2 R_\nu)^2 (15 + 4 R_\nu) k^4} \left[3 (-150 + 55 R_\nu + 14 R_\nu^2) k^4 \right.\nonumber\\
&+ 3 R_\nu (135 + 22 R_\nu) (k_1^2 - k_2^2)^2 \nonumber\\
&\left.- 2 k^2 (15 (15 + 11 R_\nu + 2 R_\nu^2) k_1^2 + (-225 + 105 R_\nu + 14 R_\nu^2) k_2^2)\right]\,,\nonumber\\
f^{\nu c}_{\sigma,P}=&-\frac{R_c}{48 (15 + 2 R_\nu)^2 (15 + 4 R_\nu) k^4} \left[3 (-1125 - 150 R_\nu + 8 R_\nu^2) (k^4 + (k_1^2 - k_2^2)^2) \right.\nonumber\\\nonumber
&\left.- 2 k^2 ((-675 - 210 R_\nu + 8 R_\nu^2) k_1^2 + (-1575 - 90 R_\nu + 8 R_\nu^2) k_2^2)\right]\,.
\end{align}

\section{Mixture of baryon and neutrino modes}

\begin{align}
\psi^{(2)}=&f^{\nu c}_{\psi,P}\omega\tau\delta^{0}_{\nu,k_1}\delta^{0}_{b,k_2}\,,\nonumber\\
\phi^{(2)}=&\left(\frac{R_\nu R_b(105+8R_\nu)}{4(15+2R_\nu)(15+4R_\nu)}-3 f^{\nu b}_{\psi,P}\right)\omega\tau\delta^{0}_{\nu,k_1}\delta^{0}_{b,k_2}\,,\nonumber\\
\delta_{c}^{(2)}=&\left(-\frac{15R_\nu R_b}{8(15+2R_\nu)}+3 f^{\nu b}_{\psi,P}\right)\omega\tau\delta^{0}_{\nu,k_1}\delta^{0}_{b,k_2}\,,\nonumber\\
\delta_{b}^{(2)}=&\left(\frac{3R_\nu}{15+4R_\nu}+\left(-\frac{3R_\nu(5R_b(15+4R_\nu)+30-4R_\nu)}{8(15+2R_\nu)(15+4R_\nu)}+3 f^{\nu b}_{\psi,P}\right)\omega\tau\right)\delta^{0}_{\nu,k_1}\delta^{0}_{b,k_2}\,,\nonumber\\
\delta_{\gamma}^{(2)}=&\left(\frac{R_\nu R_b(9+10R_\nu)}{2 R_\gamma (15+2R_\nu)}+4 f^{\nu b}_{\psi,P}\right)\omega\tau\delta^{0}_{\nu,k_1}\delta^{0}_{b,k_2}\,,\nonumber\\
\delta_{\nu}^{(2)}=&\left(-\frac{5R_b(3+2R_\nu)}{2(15+2R_\nu)}+4 f^{\nu b}_{\psi,P}\right)\omega\tau\delta^{0}_{\nu,k_1}\delta^{0}_{b,k_2}\,,\\
v_c^{(2)}=&\left(-\frac{R_bR_\nu\left(5(21+4R_\nu)k^2+2(15+2R_\nu)(k_1^2-k_2^2)\right)}{12(15+2R_\nu)(15+4R_\nu)k^2}+f^{\nu b}_{\psi,P}\right)\omega\tau^2\delta^{0}_{\nu,k_1}\delta^{0}_{b,k_2}\,,\nonumber\\
v_{\gamma b}^{(2)}=&\left(f^{\nu b}_{v,P}+f^{\nu b}_{\psi,P}\right)\omega\tau^2\delta^{0}_{\nu,k_1}\delta^{0}_{b,k_2}\,,\nonumber\\
v_{\nu}^{(2)}=&\left(-\frac{3R_b\left((2R_\nu(35+8R_\nu)-75)k^2-5(15+2R_\nu)(k_1^2-k_2^2)\right)}{32(15+2R_\nu)(15+4R_\nu)k^2}+f^{\nu b}_{\psi,P}\right)\omega\tau^2\delta^{0}_{\nu,k_1}\delta^{0}_{b,k_2}\,,\nonumber\\
\sigma_\nu^{(2)}=&f^{\nu b}_{\sigma,P}(k\tau)^2\delta^{0}_{\nu,k_1}\delta^{0}_{b,k_2}\,,\nonumber
\end{align}
with,
\begin{align}
f^{\nu b}_{\psi,P}=&\frac{3R_b R_\nu}{64 (15 + 2 R_\nu)^2 (15 + 4 R_\nu) k^4} \left[(975 + 550 R_\nu + 64 R_\nu^2) k^4\right.\nonumber\\\nonumber
&\left. - 15 (135 + 22 R_\nu) (k_1^2 - k_2^2)^2 +  10 k^2 ((105 + 26 R_\nu) k_1^2 + 3 (55 + 6 R_\nu) k_2^2)\right]\,,\\
f^{\nu b}_{v,P}=&-\frac{R_bR_\nu((1200+R_\nu(2R_\nu(65+24R_\nu)-409))k^2-(15+2R_\nu)(7R_\nu-64)(k_1^2-k_2^2))}{32R_\gamma^2 (15+2R_\nu)(15+4R_\nu)k^2}\nonumber\\
f^{\nu b}_{\sigma,P}=&\frac{R_b}{R_c}f^{\nu c}_{\sigma,P}\,.\nonumber
\end{align}

\section{Mixture of compensated and neutrino modes}

\begin{align}
\psi^{(2)}=&\frac{R_\nu R_c(25+2R_\nu)}{192 R_\gamma (75+4R_\nu)}(k^2+k_1^2-k_2^2)\omega\tau^3\delta_{\nu,k_1}^{0}\delta_{CI,k_2}^{0}\,,\nonumber\\
\phi^{(2)}=&\frac{R_\nu R_c(25-2R_\nu)}{192 R_\gamma (75+4R_\nu)}(k^2+k_1^2-k_2^2)\omega\tau^3\delta_{\nu,k_1}^{0}\delta_{CI,k_2}^{0}\,,\nonumber\\
\delta_{c}^{(2)}=&-\frac{R_b}{R_c}\delta_{b}^{(2)}=\left(\frac{3R_\nu}{15+4R_\nu}+\frac{3R_\nu(-15+2R_\nu)}{4(15+2R_\nu)(15+4R_\nu)}\omega\tau\right)\delta^{0}_{\nu,k_1}\delta^{0}_{CI,k_2}\,,\nonumber\\
\delta_{\gamma}^{(2)}=&\frac{R_\nu R_c(175-15R_\nu-2R_\nu)}{48 R_\gamma^2(75+4R_\nu)}(k^2+k_1^2-k_2^2)\omega\tau^3\delta_{\nu,k_1}^{0}\delta_{CI,k_2}^{0}\,,\nonumber\\
\delta_{\nu}^{(2)}=&\frac{R_\nu R_c(25+2R_\nu)}{48 R_\gamma (75+4R_\nu)}(k^2+k_1^2-k_2^2)\omega\tau^3\delta_{\nu,k_1}^{0}\delta_{CI,k_2}^{0}\,,\\
v_{\gamma b}^{(2)}=&\frac{3R_\nu R_c(k^2+k_1^2-k_2^2)}{32R_\gamma^2k^2}\omega\tau^2\delta^{0}_{\nu,k_1}\delta^{0}_{CI,k_2}\,,\nonumber\\
v_{\nu}^{(2)}=&-\frac{25 R_\nu R_c}{384 R_\gamma^2}(k^2+k_1^2-k_2^2)\omega\tau^4\delta_{\nu,k_1}^{0}\delta_{CI,k_2}^{0}\,,\nonumber\\
v_{c}^{(2)}=&\frac{R_\nu R_c(25-2R_\nu)}{960 R_\gamma (75+4R_\nu)}(k^2+k_1^2-k_2^2)\omega\tau^4\delta_{\nu,k_1}^{0}\delta_{CI,k_2}^{0}\,.\nonumber
\end{align}

\end{appendices}

% % % % % % % % % % % % % % % % % % % % % % % % % 

% If you want to list all the todos that you have put in the document (using
% \addtodo{}) then uncomment the next line:
% \listoftodos

% % % % % % % % % % % % % % % % % % % % % % % % % % 

% Start single space again for bibliography
\begin{singlespace}
% % % % % % % % % % % % % % % % % % % % % % % % % 

% Bibliography
% Put your bibliography file here
%\renewcommand\bibname{References} %cth added. Changes "bibliography" back to "References"
\bibliography{Thesis_Biblio}
% 
% This bibtex style file puts entries in alphabetical order and treats arxiv
% references correctly.
%\bibliographystyle{abbrvnat}
%\bibliographystyle{unsrt}
%\bibliographystyle{./utphys-ih}
%\bibliographystyle{ieeetr} %cth one I found in order of appearence
%\bibliographystyle{plainnat}

% % % % % % % % % % % % % % % % % % % % % % % % % 

\end{singlespace}

\listoffigures %cth commented
\listoftables %cth commented

% % % % % % % % % % % % % % % % % % % % % % % % % 
\end{document}